\documentclass[11pt, a4paper, twoside, openright]{report}

\renewcommand*{\title}{Entanglement and Excitations in Gauge/Gravity Duality}

\usepackage[backend=bibtex, style=numeric-comp,sorting=none,maxbibnames=10]{biblatex}

\usepackage{amsmath}
\usepackage{amsfonts}
\usepackage{amssymb}
\usepackage{xcolor}
\usepackage{titling}
\usepackage{graphicx}
\usepackage{subcaption}
\usepackage{cases}
\usepackage{tabularx}
\usepackage{xifthen} 
\usepackage{siunitx}
\usepackage{bbm} 

\usepackage{placeins}

\usepackage{hyperref}
\hypersetup{
    colorlinks,
    linkcolor=black,
    citecolor=black,
    urlcolor=black
} 

\usepackage{geometry}
\usepackage{setspace}
\onehalfspace

\newcommand\frontmatter{%
    \cleardoublepage
  \pagenumbering{roman}}

\newcommand\mainmatter{%
    \cleardoublepage
  \pagenumbering{arabic}}

\renewcommand{\le}{\left}
\newcommand{\ri}{\right}
\newcommand{\diff}{\mathrm{d}} 
\newcommand{\p}{\partial} 
\DeclareMathOperator{\tr}{tr} 
\newcommand{\vol}{\mathrm{Vol}} 
\newcommand{\dMatrix}{\hat{\rho}} 
\renewcommand{\Re}{\operatorname{Re}} 
\renewcommand{\Im}{\operatorname{Im}} 
\newcommand{\id}{\mathbbm{1}} 

\newcommand{\ads}[1][]{\ifmmode \mathrm{AdS}_{#1} \else AdS\(_{#1}\)\fi}
\newcommand{\aads}[1][] {\ifmmode \mathrm{aAdS}_{#1} \else aAdS\(_{#1}\)\fi}
\newcommand{\sph}[1][]{\ifmmode \mathrm{S}^{#1} \else S\(^{#1}\)\fi}

\newcommand{\SO}{\mathrm{SO}}
\newcommand{\U}{\mathrm{U}}
\newcommand{\SU}{\mathrm{SU}}
\newcommand{\E}{\mathrm{E}}

\newcommand{\PSU}{\mathrm{PSU}}

\newcommand{\gym}{g_\mathrm{YM}} 
\newcommand{\gn}{G_\mathrm{N}} 
\newcommand{\lp}{\ell_\mathrm{P}} 

\newcommand{\sgrav}{S_\mathrm{grav}}

\newcommand*{\C}[2][]{
\ifthenelse{\isempty{#1}}{
{C_{#2}}
}{
	{C_{#2,\,#1}}
}
} 
\newcommand*{\CF}[2][]{
\ifthenelse{\isempty{#1}}{
{F_{#2}}
}{
	{F_{#2,\,#1}}
}
} 

\renewcommand{\a}{\alpha}
\renewcommand{\b}{\beta}
\newcommand{\g}{\gamma}
\renewcommand{\d}{\delta}
\newcommand{\e}{\epsilon}
\newcommand{\ve}{\varepsilon}
\newcommand{\f}{\phi}
\newcommand{\m}{\mu}
\newcommand{\n}{\nu}
\newcommand{\y}{\psi}
\renewcommand{\l}{\lambda}
\renewcommand{\t}{\tau}
\newcommand{\h}{\eta}
\renewcommand{\c}{\chi}
\newcommand{\q}{\theta}

\renewcommand{\r}{\rho}
\newcommand{\w}{\omega}

\newcommand{\s}{\sigma}
\newcommand{\z}{\zeta}
\newcommand{\G}{\Gamma}
\newcommand{\D}{\Delta}
\newcommand{\F}{\Phi}
\newcommand{\Q}{\Theta}
\renewcommand{\O}{\Omega}

\newcommand{\cO}{\mathcal{O}}
\newcommand{\cN}{\mathcal{N}}

\newcommand{\cA}{\mathcal{A}}
\newcommand{\cD}{\mathcal{D}}
\newcommand{\cB}{\mathcal{B}}

\newcommand{\cH}{\mathcal{H}}
\newcommand{\cS}{\mathcal{S}}

\newcommand{\cQ}{\mathcal{Q}}

\newcommand{\tent}{T_\mathrm{E}}

\newcommand{\dcrit}{d_{\textrm{crit}}}

\newcommand{\talpha}{\tilde{\alpha}}

\newcommand{\see}{S_\mathrm{E}}
\newcommand*{\tfive}{T_\mathrm{M5}}
\newcommand*{\ttwo}{T_\mathrm{M2}}

\newcommand*{\mt}{\tilde{m}}

\newcommand*{\ket}[1]{|#1\rangle}
\newcommand*{\bra}[1]{\langle#1|}

\newcommand*{\vev}[1]{\langle#1\rangle}

\begin{document}

\frontmatter
    

\pagestyle{empty}
\addtocounter{page}{-2}

\begin{center}
    {\large

    \href{https://www.southampton.ac.uk/}{\textbf{ University of Southampton}}

    \vspace{1em}

    Faculty of Engineering and Physical Sciences

    \vspace{1em}

    Physics and Astronomy
    }

    \vfill

    {\large
        \textbf{Entanglement and Excitations in Gauge/Gravity Duality}
        \\[1em]
        by
        \\[1em]
        \textbf{Ronald James Rodgers}
    }

    \vfill

    {\large
        Thesis for the degree of Doctor of Philosophy
        \\[2em]
        September 2019

    }
\end{center}

\cleardoublepage

\pagestyle{plain}

\begin{center}
    University of Southampton
    \\[1em]
    \underline{Abstract}
    \\[1em]
    Faculty of Engineering and Physical Sciences
    \\[0em]
    Physics and Astronomy
    \\[1em]
    \underline{Thesis for the degree of Doctor of Philosophy}
    \\[1em]
    Entanglement and Excitations in Gauge/Gravity Duality
    \\[0em]
    by Ronald James Rodgers
\end{center}

\vspace{2em}

\noindent Gauge/gravity duality, also known as holography, relates quantum field theories to theories of gravity. When one theory is strongly coupled, and therefore difficult to study directly, the other is weakly coupled. In this thesis, we study a variety of phenomena in strongly coupled quantum field theories by performing calculations in their gravitational~duals.

\noindent~~We compute entanglement entropy in a variety of holographic systems, paying particular attention to its long-distance behaviour, characterised by a term proportional to surface area. This term is known to decrease along Lorentz-invariant renormalisation group flows, suggesting that it may count massless degrees of freedom. We find that more general deformations may \textit{increase} this area term, possibly indicating an enhanced number of long-distance degrees of freedom. We observe a correlation between this enhancement and the emergence of new scaling symmetry at long distances.

\noindent~~Next, we study the spectrum of collective excitations in a holographic model of a non-Fermi liquid. At high temperatures, the spectrum of collective excitations includes hydrodynamic sound waves. As in similar models, we observe that sound-like modes also exist at low temperatures. Such modes are known as holographic zero sound.  We study the changing properties of holographic zero sound and the emergence of hydrodynamic behaviour at high temperatures as we vary the parameters of the model. We find that for certain values of the parameters, the temperature-dependence of holographic zero sound qualitatively resembles that of a normal Fermi liquid.

\noindent~~Finally, we study the entanglement entropy contribution of surface defects in a six-dimensional quantum field theory of relevance to M-theory, which is a candidate theory of quantum gravity. We find that the entanglement entropy does not montonically decrease along renormalisation group flows on these defects, ruling it out as a potential measure of degrees of freedom. On the other hand, we find that two of the contributions of the defect to the Weyl anomaly of the quantum field theory decrease along all of the flows that we study.

\cleardoublepage

\tableofcontents

\cleardoublepage

\listoffigures
\addcontentsline{toc}{chapter}{\listfigurename}

\listoftables
\addcontentsline{toc}{chapter}{\listtablename}

\cleardoublepage

\chapter*{Declaration of Authorship}
\addcontentsline{toc}{chapter}{Declaration of Authorship}

I, Ronald James Rodgers, declare that the thesis, entitled \textit{\title}, and the work presented in it are both my own and have been generated by me as the result of my own original research.
\\[0.5em]
I confirm that:
\begin{enumerate}
    \item This work was done wholly or mainly while in candidature for a research degree at this University;
    \item Where any part of this thesis has previously been submitted for a degree or any other qualification at this University or any other institution, this has been clearly stated;
    \item Where I have consulted the published work of others, this is always clearly attributed;
    \item Where I have quoted from the work of others, the source is always given. With the exception of such quotations, this thesis is entirely my own work;
    \item I have acknowledged all main sources of help;
    \item Where the thesis is based on work done by myself jointly with others, I have made clear exactly what was done by others and what I have contributed myself;
    \item Parts of this work have been published as references~\cite{Gushterov:2017vnr,Gushterov:2018spg,Rodgers:2018mvq}. I have also collaborated on references~\cite{OBannon:2016exv,Estes:2018tnu,Jensen:2018rxu}, the contents of which are not covered in this thesis.
\end{enumerate}

\vspace{2em}

\noindent Signed:
\\[2em]
Date:

\cleardoublepage

\chapter*{Acknowledgements}
\addcontentsline{toc}{chapter}{Acknowledgements}

I would first like to thank my supervisors, Andy O'Bannon and Nick Evans for their guidance and insight over the past four years, which has shaped the way that I think about theoretical physics. I am also very grateful to my collaborators, John Estes, Nikola Gushterov, Kristan Jensen, Darya Krym, Jonas Probst, Brandon Robinson, and Christoph Uhlemann, for all that they taught me during our time spent working together, and to Matthew Russell for helpful comments on a draft of this thesis.

Over the last four years I have shared offices with many of my fellow PhD students, who kept my working days enjoyable with many amusing conversations. (Sometimes we even talked about physics!) There are too many to name here, but special mention must be made for Azaria Coupe, Simon King, Andrew Lawson, James Richings, and Sam Rowley. Without their friendship, my time in Southampton would not have been nearly so fun.

Finally, I would like to thank Sophie, who has been my companion for almost all of my physics career, and my parents. This thesis would not have been possible without their constant encouragement and support.

\cleardoublepage

\mainmatter

\part{Introduction}

\chapter{Motivation}

Quantum field theory (QFT) is a framework which successfully models many systems in particle and condensed matter physics. However, exact calculations in all but the simplest QFTs are impossible. To make progress, the traditional tool is perturbation theory, an expansion in the coupling constants of the theory, which works extremely well for many theories. Some QFTs, though, have large coupling constants, in which case one cannot hope to obtain a good approximation by truncating the perturbative series at a finite number of terms. Such QFTs are known as \textit{strongly coupled}.

Examples of strongly coupled systems in nature are the quantum chromodynamics (QCD) sector of the Standard Model, and condensed matter systems such as ultracold Fermi gases and cuprate high-temperature superconductors. To study such systems, tools other than perturbation theory are needed.

One approach is numerical calculation, in particular lattice Monte Carlo simulation. This is an active and productive field of research. For example, lattice QCD calculations accurately reproduce the spectrum of light hadronic particles~\cite{Durr:2008zz}, and are essential in the computation of hadron decay constants~\cite{Aoki:2019cca}. However, precise lattice simulations are computationally expensive. In addition, the sign problem\footnote{See ref.~\cite{Aarts:2015tyj} for a review.} makes it difficult to simulate systems at non-zero chemical potential, limiting applications to strongly interacting condensed matter systems, heavy ion collisions, and neutron stars.

Gauge/gravity duality \cite{Maldacena:1997re} provides a complementary tool for studying strongly interacting QFTs. Also known as the anti-de Sitter/conformal field theory (AdS/CFT) correspondence or holography, it relates QFTs in \(d\)-dimensional spacetime to quantum gravity in \((d+1)\)-dimensional asymptotically \ads\ spacetime. When the QFT is strongly coupled, the corresponding theory of gravity is weakly coupled, and therefore amenable to perturbative methods. In chapters~\ref{chap:entanglement_density} and~\ref{chap:zero_sound} we will use the duality to study models of strongly coupled condensed matter systems.
 
One obstacle to this programme is that quantum gravity is poorly understood. Straightforward attempts to quantize general relativity fail, as the theory is non-renormalisable. Even string theory, an ultraviolet-finite theory of quantum gravity, is not well understood on spacetime with \ads\ asymptotics. In order to carry out calculations, one therefore usually takes the classical limit of the gravitational theory.

The resulting dual QFT has an unrealistically large number of degrees of freedom, in a sense that will be made more precise in chapter~\ref{chap:ads_cft}. Thus, at present, gauge/gravity duality may not be used to make quantitative predictions for observables in real physical systems. However, one can attempt to make qualitative predictions about strongly coupled systems by looking for  phenomena which are universal in holographic models, which may then be generic to strongly coupled systems.

Perhaps the most striking success of this approach comes from the hydrodynamics of holographic fluids. When the gravitational side of the duality is classical Einstein gravity, possibly coupled to matter, the ratio of the shear viscosity \(\h\) to entropy density \(s\) in holographic models takes the universal value \(\h/s = 1/4\pi \approx 0.08\)~\cite{Policastro:2001yc,Kovtun:2003wp,Buchel:2003tz,Kovtun:2004de,Starinets:2008fb}.\footnote{Throughout this thesis I use natural units in which the reduced Planck constant \(\hbar\), the speed of light \(c\), and Boltzmann's constant \(k_B\) are all set equal to one. In this case \(\h/s\) is dimensionless. Otherwise, the dimensionless ratio is \(\hbar \h/k_B s\).} This ratio is very small compared to that for familiar fluids. For example, at a temperature of 300K and atmospheric pressure, \(\h/ s \approx 12\) or 300 for water or nitrogen, respectively.\footnote{These values were calculated using tables found at~\cite{nist_thermophysical_table}.}
On the other hand, \(\h/s \approx 0.5\) has been observed in ultracold Fermi gases~\cite{Schafer:2007pr,2008JLTP..150..567T,Adams:2012th}, and heavy ion collision experiments place an upper bound of \(\h/s \lesssim 0.15\) for the quark-gluon plasma~\cite{Bernhard:2016tnd}. The universal holographic result may therefore be viewed as a successful prediction that \(\h/s\) may be small in strongly coupled fluids, of which the ultracold gases and quark-gluon plasma are examples.

In chapter~\ref{chap:probe_m5} we will use gauge/gravity duality for a different purpose: to learn about quantum gravity. The strong-coupling limit of type IIA string theory is an eleven-dimensional theory called M-theory, which is yet to be fully formulated~\cite{Witten:1995ex}. M-theory contains extended six-dimensional objects, M5-branes, which are described by a QFT that appears to have no classical limit. This makes M5-branes very challenging to study. In chapter~\ref{chap:probe_m5}, we will use gauge/gravity duality to calculate observables in the M5-brane QFT.

\chapter{Gauge/gravity duality}
\label{chap:ads_cft}

In this section we provide a brief review of gauge/gravity duality. For more details we refer to the books~\cite{Becker:2007zj,Freedman:2012zz,Ammon:2015wua} and review~\cite{Aharony:1999ti}. Before defining the duality in section~\ref{sec:duality_statement}, we will review aspects of conformal field theory, string theory, and \ads\ space, which will be needed for the statement of the duality, and for understanding later chapters of this thesis.

\section{Conformal field theory}
\label{sec:cft}

Many examples of holographic dualities involve conformal field theories (CFTs), which are field theories invariant under conformal transformations. A conformal transformation is a coordinate transformation \(x \to x'\) such that the metric changes by an overall scale factor \(\Omega^2(x)\)~\cite{Ginsparg:1988ui, DiFrancesco:1997nk}
\begin{equation} \label{eq:conformal_transformation}
    g_{\m\n}(x) \to g'_{\m\n}(x') = \Omega^2(x) g_{\m\n}(x),
\end{equation}
followed by a Weyl transformation which removes the scale factor. Conformal transformations change the lengths of vectors while preserving the angles between them.

In \(d\)-dimensional Minkowski space, where \(g_{\m\n} = \h_{\m\n}\), for \(d\geq3\) conformal transformations form the group \(\SO(d,2)\). They consist of:
\begin{enumerate}
    \item Poincar\'e transformations: translations \(x^\m \to x^\m + a^\m\) and Lorentz transformations \(x^\m \to {\Lambda^\m}_{\n} x^\n\), with \(a \in \mathbb{R}^d\) and \(\Lambda \in \SO(d,1)\). The Minkowski metric is invariant under such coordinate transformations, i.e. \(\Omega = 1\) in equation~\eqref{eq:conformal_transformation}.
    \item Dilatations \(x^\m \to \lambda x^\m\), with \(\l \in \mathbb{R}_+\). For these transformations \(\Omega = \l^{-1}\).
    \item Special conformal transformations, \(x^\m \to \frac{x^\m + b^\m x^2}{1 + 2 b \cdot x + b^2 x^2}\), where \(b \in \mathbb{R}^{d-1,1}\). These satisfy \(\Omega =1 + 2 b \cdot x + b^2 x^2\).
\end{enumerate}
For \(d=2\), the conformal group is infinite-dimensional and has \(\SO(2,2)\) as a subgroup.

\subsection{The Weyl anomaly}
\label{sec:weyl_anomaly}

The trace of the stress tensor \(T_{\m\n}\) in a classical CFT vanishes when the equations of motion are satisfied. To see this, consider an infinitesimal conformal transformation with \(\O(x) = 1 - \e(x)\), with \(\e(x) \ll 1\). By definition, the action of a CFT is invariant under such a transformation. A coordinate transformation is not a physical transformation, so any change in the action must come from the Weyl transformation, which acts as \(g_{\m\n} \to (1 + 2\e) g_{\m\n}\). The Weyl transformation also acts on the set of fields \(\{\F_a\}\) as \(\Phi_a \to (1 - \e \D_a) \F_a\), where \(\D_a\) is a constant, called the conformal dimension of the field \(\F_a\). The change in the action \(S\) is then
\begin{equation} \label{eq:traceless_stress_tensor_intermediate}
   \d S = \int \diff^d x   \le[
    2\frac{\d S}{\d g_{\m\n}} g_{\m\n}  - \sum_a \frac{\d S}{\d \F_a} \D_a \F_a
   \ri] \e(x)= \int \diff^d x \sqrt{-g} \, T^{\m\n} g_{\m\n} \e(x),
\end{equation}
where to obtain the second equality we used the equations of motion to set \(\d S/\d \F_a = 0\), and the definition of the stress tensor \(T^{\m\n} = \frac{2}{\sqrt{-g}} \frac{\d S}{\d g_{\m\n}}\). By definition, the variation~\eqref{eq:traceless_stress_tensor_intermediate} vanishes in a CFT, implying that on shell
\begin{equation}
    T^{\m\n} g_{\m\n} \equiv {T^\m}_\m = 0.
\end{equation}

In QFT, the vanishing trace of the stress tensor is the Ward identity of conformal invariance. Conformal invariance often becomes anomalous upon quantization of a theory; if any couplings have non-vanishing beta functions then scale invariance is broken by the renormalisation scale. For example, four-dimensional \(\U(1)\) Yang-Mills theory coupled to massless charged particles is classically a CFT. In the quantum theory, one finds~\cite{Peskin:1995ev}
\begin{equation}
    \vev{{T^\m}_{\m}} = \frac{\b(e)}{2 e^3} F_{\m\n}F^{\m\n},
\end{equation}
where \(\b(e)\) is the beta function for the gauge coupling \(e\), and \(F_{\m\n}\) is the background value of the field strength.

Even when conformal invariance is preserved upon quantization, \(\vev{{T^\m}_\m}\) generically does not vanish when the theory is placed on a curved manifold, with contributions coming from contractions of the Riemann curvature. In this case, non-vanishing \(\vev{{T^\m}_\m}\) is known as the Weyl anomaly. Since the scaling dimensions of the stress tensor and curvature  are respectively \([T] = d\) and \([R] = 2\), by dimensional analysis the Weyl anomaly vanishes when \(d\) is odd. The general form of the anomaly for even \(d\) is determined by solving the Wess-Zumino consistency condition~\cite{Wess:1971yu}, which states that the commutator of two Weyl transformations acting on the generating functional vanishes. The result is that a theory with \(\vev{{T^\m}_\m} = 0\) in flat space has~\cite{Deser:1993yx}
\begin{equation} \label{eq:weyl_anomaly}
    \vev{{T^\m}_\m} = a E_d + \sum_n c_n I_n,
\end{equation}
in curved space, where \(E_d\) is the Euler class in \(d\) dimensions, the integral of which gives the Euler characteristic.\footnote{Note that~\eqref{eq:weyl_anomaly} implies that correlation functions of \({T^\m}_{\m}\) with one or more copies of the stress tensor may be non-zero even in flat space, since these are constructed by taking functional derivatives of \(\vev{{T^\m}_\m}\) with respect to the metric.
} The \(I_n\) are contractions of products of the Riemann tensor and the Laplacian that transform homogeneously under conformal transformations as \(I_n \to \O^{-d} I_n\) when \(g_{\m\n} \to \O^2 g_{\m\n}\). The anomaly is parameterised by theory-dependent coefficients \(a\) and \(c_n\). Ref.~\cite{Deser:1993yx} classified the contributions to the Weyl anomaly into types A and B. Type A contributions to \(\sqrt{-g} \vev{{T^\m}_\m}\) transform as total derivatives under Weyl transformations, while type B contributions are Weyl invariant. In~\eqref{eq:weyl_anomaly}, the Euler density term is type A, while the \(I_n\) terms are type B.\footnote{In general, there may also be total derivative terms in~\eqref{eq:weyl_anomaly}. However, these may be removed by the addition of local counterterms to the action.}

For example, in two dimensions the Euler class is proportional to the Ricci scalar, \(E_2 = R/4\pi\), and there are no \(I_n\). The Weyl anomaly in two dimensions is then~\cite{francesco1997conformal}
\begin{equation}
    \vev{{T^\m}_\m} = - \frac{c}{24 \pi} R,
\end{equation}
where \(c = -6a\) is the central charge of the CFT.

Roughly speaking, \(c\) counts the degrees of freedom in the CFT. For example, the thermodynamic entropy of a two-dimensional CFT on an interval of length \(\ell\) at temperature \(T \gg 1/\ell\) is \(S = \pi c T \ell / 3\)~\cite{Cardy:1986ie, Affleck:1986bv}, which is proportional to \(c\). Consider a two-dimensional RG flow between ultraviolet (UV) and infrared (IR) fixed points, with central charges \(c_\mathrm{UV}\) and \(c_\mathrm{IR}\), respectively. The central charges satisfy a monotonicity theorem, the \(c\)-theorem~\cite{Zamolodchikov:1986gt}, which implies that there is a function which interpolates between \(c_\mathrm{UV}\) and \(c_\mathrm{IR}\), decreasing monotonically along the RG flow. Intuitively, this function decreases as degrees of freedom are integrated out along the RG flow.

Similarly, in four dimensions there is the \(a\)-theorem~\cite{Osborn:1989td,Jack:1990eb,Komargodski:2011vj,Komargodski:2011xv}, a monotonicity theorem that guarantees that \(a\), the type A anomaly coefficient in~\eqref{eq:weyl_anomaly}, decreases along RG flows. In odd dimensions there is no Weyl anomaly. However, in three dimensions the free energy on \sph[3] is smaller in the IR than in the UV~\cite{Jafferis:2011zi,Klebanov:2011gs,Casini:2012ei} (although it is not always monotonic along the whole RG flow~\cite{Taylor:2016kic}). This is known as the \(F\)-theorem.  Monotonicity theorems are useful, for example, for ruling out possible IR fixed points as the end points of RG flows~\cite{Grover:2012sp}.

\subsection{Conformal defects}
\label{sec:defects_background}

A defect in a QFT is a distinguished submanifold. For example, given a QFT with no defect, one could insert a defect by imposing boundary conditions for the fields of the QFT on a submanifold, or introducing new fields localised to it. A defect conformal field theory (DCFT) is a CFT with a defect that preserves dilatation invariance (a conformal defect).

In chapter~\ref{chap:probe_m5}, we will study properties of planar defects in a particular CFT. The presence of an \(n\)-dimensional planar conformal defect breaks the conformal group \(\SO(d,2)\) into the \(\SO(n,2) \times \SO(d-n)\) subgroup that leaves the position of the defect invariant. The \(\SO(n,2)\) factor consists of conformal transformations along the defect, while the \(\SO(d-n)\) factor consists of rotations about the defect. For the defects studied in chapter~\ref{chap:probe_m5}, \(n=2\).

The Weyl anomaly of a defect CFT may include a contribution localised to the defect. The general form of this contribution is not known for arbitrary \(n\), however for \(n=2\) one finds~\cite{Graham:1999pm,Henningson:1999xi,Schwimmer:2008yh}
\begin{equation} \label{eq:defect_weyl_anomaly}
    \vev{{T^\m}_\m} = \mathcal{A}_\mathrm{bulk} - \frac{\d^{(d-2)}(x_\perp)}{24\pi} \le[
        b R(\g) + d_1 \mathring{\Pi}^\m_{ab} \mathring{\Pi}_\m^{ab} - d_2 \g^{ac} \g^{bd} W_{abcd}
    \ri],
\end{equation}
where \(\mathcal{A}_\mathrm{bulk}\) is the contribution~\eqref{eq:weyl_anomaly} from the bulk of the CFT, \(\d^{(d-2)}(x_\perp)\) is a Dirac delta function which localises to the defect, \(\g\) is the induced metric on the defect, \(R(\g)\) is the Ricci scalar computed from \(\g\), \(\mathring{\Pi}^\m_{ab}\) is the traceless part of the extrinsic curvature of the defect, and \(W_{abcd}\) is the bulk Weyl tensor pulled back to the defect. The defect's contribution to the Weyl anomaly is parameterised by the three theory-dependent coefficients \(b\), \(d_1\), and \(d_2\).\footnote{The bulk Weyl tensor vanishes identically in three dimensions, so there is no \(d_2\) for a two-dimensional defect in a three-dimensional CFT.} In the classification of ref.~\cite{Deser:1993yx}, \(R(\g)\) is type A, while \(\mathring{\Pi}^\m_{ab} \mathring{\Pi}_\m^{ab}\) and \(\g^{ac} \g^{bd} W_{abcd}\) are type B. The type A coefficient \(b\) in~\eqref{eq:defect_weyl_anomaly} satisfies a monotonicity theorem, the \(b\)-theorem~\cite{Jensen:2015swa}, which implies that \(b\) decreases along RG flows triggered by the source for a relevant operator localised to the defect.\footnote{See also ref.~\cite{Casini:2018nym} for an alternative proof of the \(b\)-theorem for the special case of two-dimensional boundaries.}

\section{String theory, M-theory and supergravity}

\subsection{Superstring theory and supergravity}

String theory was first developed as a possible theory of the strong interaction. The latter was eventually realised to instead be described by quantum chromodynamics, but interest in string theory has continued since it provides a rare example of a consistent quantum theory of gravity. Later it was realised that string theory is a limit of another theory, called M-theory~\cite{Witten:1995ex,Horava:1995qa}. The original conjecture of gauge/gravity duality came from string theory, and string and M-theory provide several concrete examples of the duality~\cite{Maldacena:1997re}. We now review aspects of these theories relevant to this thesis. For further details, we refer to refs.~\cite{Polchinski:1998rq,Polchinski:1998rr,Becker:2007zj}.

The fundamental objects of string theory are strings, two-dimensional (one space and one time) objects. The fluctuations of strings are quantized, with different particles corresponding to different fluctuation modes. The spectrum of fluctuations includes a massless spin-2 field, the graviton, so string theory is a quantum theory of gravity. 

There are five different supersymmetric string (superstring) theories, corresponding to different ways of implementing supersymmetry: types I, IIA, IIB, and \(\SO(32)\) and \(\E_8 \times \E_8\) heterotic. Each of these theories must be formulated in ten-dimensional spacetime to avoid an anomaly that would render the theory inconsistent~\cite{Polchinski:1998rr}.

String theory contains a single dimensionful parameter, the Regge slope \(\a'\), of mass dimension \([\a']=-2\). The massive string fluctuations have masses proportional to \(1/\sqrt{\a'}\). If we are interested in processes at energy scales small compared to \(1/\sqrt{\a'}\), we can therefore consider only the massless string modes, the massive modes having decoupled. The resulting low-energy effective theory is ten-dimensional supergravity (SUGRA). There are multiple ten-dimensional supergravity theories, corresponding to low-energy limits of the different superstring theories. Of particular interest in this thesis are type IIA and IIB SUGRA, each the low-energy limit of the string theory with the same name, which contain the massless bosonic fields:
\begin{itemize}
    \item the metric \(g_{\m\n}\),
    \item a two-form gauge field \(B_{\m\n}\),
    \item a scalar field \(\f\), called the dilaton, and
    \item \(n\)-form gauge fields \(\C{n}\),  with odd (even) \(n\) in type IIA (IIB) SUGRA,
\end{itemize}
in addition to the associated fermionic ten-dimensional \(\cN = 2\) superpartners.

String theory contains dynamical extended objects, called D-branes. A D\(p\)-brane is a \((p+1)\)-dimensional extended object, electrically charged under the gauge field \(\C{p+1}\),\footnote{This means that the D\(p\)-brane action contains a term proportional to the integral of \(\C{p+1}\) over the world volume of the brane.} on which open strings can end. The world volume of a D-brane supports a \(\U(1)\) gauge field. The endpoint of an open string is charged under this gauge field. For \(N\) coincident D-branes, the gauge symmetry is enhanced to \(\U(N)\), and the endpoints of strings transform in the fundamental representation.

In type IIA (IIB) string theory, D\(p\)-branes with even (odd) \(p\) are stable. In supergravity, these D-branes are solitonic solutions of the classical equations of motion. For example the solution of type IIB supergravity corresponding to a flat stack of \(N\) coincident D3-branes is~\cite{Duff:1991pea}
\begin{align}
    \diff s^2 &= \le(1 + \frac{L^4}{r^4}\ri)^{-1/2} \h_{\m\n} \diff x^\m \diff x^\n + \le(1 + \frac{L^4}{r^4}\ri)^{1/2} \le( \diff r^2 + r^2 \diff s_{\sph[5]}^2 \ri),
    \nonumber
    \\
    \CF5 &= (1 + *) \diff \tilde{\C4}, \quad
    \tilde{\C4} \equiv \le( 1 + \frac{L^4}{r^4} \ri)^{-1} \diff x^0 \wedge \diff x^1 \wedge \diff x^2 \wedge \diff x^3,
    \label{eq:d3_brane_solution}
\end{align}
with the dilaton \(\f\) and axion \(\C0\) constant and all other fields vanishing. In the solution~\eqref{eq:d3_brane_solution}, \(\h_{\m\n}\) is the four-dimensional Minkowski metric, \(\diff s_{\sph[5]}^2\) is the metric on a unit, round \sph[5], \(*\) denotes the Hodge star, and \(\CF5 \equiv \diff \C4\) and \(L^4 = 4 \pi \a'^2 g_s N\), where \(g_s\) is the string coupling.

\subsection{M-theory and eleven-dimensional supergravity}

It has been observed by Witten~\cite{Witten:1995ex}, based partly on the work of Hull and Townsend~\cite{Hull:1994ys}, that at strong coupling, type IIA string theory looks like the compactification of an eleven-dimensional theory on a circle, with a radius that grows with the string coupling. The eleven-dimensional theory has become known as M-theory~\cite{Horava:1995qa}. In addition to type~IIA string theory, the four other superstring theories arise from M-theory in different limits.

The ``M'' in M-theory is often taken to stand for membrane, because the theory describes \((2+1)\)-dimensional membranes, called M2-branes.\footnote{The ``M'' is often also taken to stand for mysterious, magical, or matrix~\cite{Becker:2007zj}.} In the compactification of M-theory on a circle, type IIA fundamental strings arise from M2-branes wrapping the compact extra dimension. M-theory also contains \((5+1)\)-dimensional extended objects, called M5-branes.

The low energy limit of M-theory is eleven-dimensional supergravity (11D SUGRA). The fields of 11D SUGRA are the metric \(G\), a three-form gauge field \(\C3\), and a gravitino. The requirements of diffeomorphism invariance, local Lorentz invariance, local supersymmetry, and invariance under gauge transformations of \(\C3\) completely fix the action, the bosonic part of which is~\cite{Becker:2007zj}
\begin{equation} \label{eq:11d_sugra_action}
    S = \frac{1}{16 \pi \gn} \int \diff^{11}x \sqrt{-\det G} \le( R - \frac{1}{2} |\CF4|^2 \ri) - \frac{1}{96 \pi \gn} \int \C3 \wedge \CF4 \wedge \CF4,
\end{equation}
where \(R\) is the Ricci scalar, \(\CF4 \equiv \diff\C3\), and
\[|\CF4|^2 \equiv \frac{1}{4!} G^{M_1 N_1} \dots G^{M_4 N_4} \CF[M_1 \dots M_4]{4} \CF[N_1 \dots N_4]{4}.
\]
One can define a six-from gauge field \(\C6\) by \(\diff \C6 - \frac{1}{2} \diff \C3 \wedge \C3 = \CF7 \equiv * \CF4\), where \(*\) denotes the Hodge dual. M2-branes couple electrically to \(\C3\), while M5-branes couple electrically to \(\C6\).

The equations of motion of 11D SUGRA admit solutions solutions similar to the D3-brane solution~\eqref{eq:d3_brane_solution}, describing a flat stack of \(N_p\) M\(p\)-branes~\cite{Duff:1990xz,Gueven:1992hh},
\begin{align}
    \diff s^2 &= \le(1 + \frac{L^{8-p}}{r^{8-p}} \ri)^{-(8-p)/9} \h_{\m\n} \diff x^\m \diff x^\n + \le(1 + \frac{L^{8-p}}{r^{8-p}} \ri)^{(p+1)/9} \le( \diff r^2 + r^2 \diff s_{\sph[9-p]}^2 \ri),
    \nonumber
    \\
    \C{p+1} &= \le( 1 + \frac{L^{8-p}}{r^{8-p}} \ri)^{-1} \diff x^0 \wedge  \dots \wedge \diff x^p,
    \label{eq:m_brane_solution}
\end{align}
where \(p = 2\) or \(5\). The parameter \(L\) is determined by the number of branes and the Planck length \(\lp\); \(L^6 = 32 \pi^2 N_2 \lp^6\) for M2-branes and \(L^3 = \pi N_5 \lp^3\) for M5-branes.

\section{Anti-de Sitter space}

Anti-de Sitter (\ads) space is the maximally symmetric spacetime of constant negative curvature. It arises in general relativity as a solution to the vacuum Einstein equations with a negative cosmological constant. This spacetime is crucially important in gauge/gravity duality, so we will briefly review some of its properties. The discussion follows that of~\cite{Ammon:2015wua}.

One way to obtain \((d+1)\)-dimensional \ads\ (\ads[d+1]) is by an embedding in \(\mathbb{R}^{2,d}\). Denoting the coordinates and metric of \(\mathbb{R}^{2,d}\) as \(\tilde X^M\) and \(\tilde \h_{MN}\) respectively, \ads[d+1] of radius \(L\) is the hypersurface \(\tilde \h_{MN} \tilde X^M  \tilde X^N = -L^2\). The isometry group of \ads[d+1], made manifest by this embedding, is \(\SO(d,2)\). This is the \(d\)-dimensional conformal group.

In this thesis we will typically work in a coordinate patch of \ads[d+1] called the Poincar\'e patch, the metric on which may written as
\begin{equation} \label{eq:ads_metric}
    \diff s^2 = \frac{L^2}{z^2} \le( \diff z^2 + \h_{\m\n} \diff x^\m \diff x^\n \ri),
\end{equation}
where \(\h_{\m\n}\) is the \(d\)-dimensional Minkowski metric and \(z \in [0,\infty)\). \ads\ has a conformal boundary, consisting of a copy of \(d\)-dimensional Minkowski space at \(z \to 0\), with the addition of the point at \(z \to \infty\)~\cite{Witten:1998qj}. This latter point is known as the Poincar\'e horizon. At the boundary, the \(\SO(d,2)\) isometries of \ads[d+1] act on the \(x^\m\) coordinates in the same way as the conformal transformations listed in section~\ref{sec:cft}.

\ads\ arises in supergravity as the near horizon region of the stacks of coincident branes that were discussed in the previous section. For example, in the near-horizon (\(r \ll L\)) region of the D3-brane solution~\eqref{eq:d3_brane_solution}, the metric and gauge field take the form
\begin{align}
    \diff s^2 &= \frac{r^2}{L^2} \h_{\m\n} \diff x^\m \diff x^\n + \frac{L^2}{r^2} \diff r^2 + L^2 \diff s_{\sph[5]}^2,
    \nonumber
    \\
    \tilde{\C4} &= \frac{r^4}{L^4} \diff x^0 \wedge \diff x^1 \wedge \diff x^2 \wedge \diff x^3,
    \label{eq:d3_brane_near_horizon}
\end{align}
Defining a new coordinate \(z \equiv L^2/r\), we recognise the metric in~\eqref{eq:d3_brane_near_horizon} as that of \(\ads[5]\times\sph[5]\), where both \ads[5] and \sph[5] have radius \(L\) and the metric of \ads[5] is in the form~\eqref{eq:ads_metric}. Similarly, the near horizon limit of the eleven-dimensional M\(p\)-brane solution~\eqref{eq:m_brane_solution} is \(\ads[p+2] \times \sph[9-p]\).

Also relevant for gauge/gravity duality is asymptotically locally \ads\ (\aads) spacetime, in which one may choose a coordinate system in which the metric takes the Fefferman-Graham form~\cite{AST_1985__S131__95_0}
\begin{equation} \label{eq:general_asymptotic_ads_metric}
    \diff s^2 = G_{MN} \diff X^M \diff X^N = \frac{L^2}{z^2} \le[ \diff z^2 + g_{\m\n}(x,z) \diff x^\m \diff x^\n \ri],
\end{equation}
with the boundary at \(z = 0\). Near the boundary, the matrix \(g_{\m\n}\) has an expansion
\begin{equation} \label{eq:metric_near_boundary_expansion}
    g(x,z) = g^{(0)}(x) + z^2 g^{(2)} (x) + \dots,
\end{equation}
where the dots indicate terms of higher order in \(z\). Poincar\'e \ads~\eqref{eq:ads_metric} is recovered by setting \(g^{(0)} = \h_{\m\n}\) and \(g^{(n\geq2)} = 0\).

\section{Statement of gauge/gravity duality}
\label{sec:duality_statement}

Gauge/gravity duality is the equivalence between a \(d\)-dimensional QFT and a theory of gravity on \aads[d+1] spacetime.\footnote{The fact that the QFT lives in one dimension fewer than the gravity theory is why gauge/gravity duality is also known as holography.} This means there is a map between physical (gauge-invariant) quantities in the two theories, and that the generating functionals of the two theories are the same. Since quantum gravity is poorly understood, one normally takes the classical limit on the gravity side, replacing the generating functional by its saddle-point approximation. The duality is then~\cite{Gubser:1998bc, Witten:1998qj}
\begin{equation} \label{eq:duality_statement}
    Z_\mathrm{QFT}[\f_{(0)}] = Z_\mathrm{grav}[\f_{(0)}] \approx e^{- I_\mathrm{grav}^\star[\f_{(0)}]} \quad \text{(classical limit)}.
\end{equation}

The generating functional for the QFT, \(Z_\mathrm{QFT}\), is  a functional of the sources for the operators in the QFT, collectively denoted by \(\f_{(0)}\). Each gauge invariant operator in the QFT is dual to field in the gravitational theory. On the gravity side, the \(\f_{(0)}\) are boundary conditions for these fields. The right hand side of~\eqref{eq:duality_statement} is the classical saddle point approximation to the gravity generating functional \(Z_\mathrm{grav}[\f_{(0)}]\), where \(I^\star_\mathrm{grav}[\f_{(0)}]\) is the Euclidean-signature on-shell action, i.e. the action evaluated on the solutions to the classical equations of motion. Throughout this thesis we use \(I\) and \(S\) to denote actions in Euclidean and Lorentzian signatures, respectively. In Lorentzian signature, the right hand side of~\eqref{eq:duality_statement} reads \(e^{i S_\mathrm{grav}^\star}\).

The gravitational action \(I_\mathrm{grav}\) must in general include counterterms evaluated on the \aads\ boundary. These terms remove divergences in the on-shell action, which would otherwise render the classical variational problem and the duality~\eqref{eq:duality_statement} ill-defined~\cite{deHaro:2000vlm,Skenderis:2002wp,Papadimitriou:2004ap,Papadimitriou:2004rz}. The addition of these counterterms is known as holographic renormalisation, and is the holographic analogue of renormalisation in the dual QFT.

One of the first examples of gauge/gravity duality was the conjectured equivalence of \(\mathcal{N}=4\) supersymmetric Yang-Mills theory (SYM) with gauge group \(\SU (N)\) and type IIB string theory on \(\ads[5]\times\sph[5]\) \cite{Maldacena:1997re}. We will use this to illustrate some important aspects of the duality.

The correspondence between these two theories arises from consideration of the low energy excitations of a stack of \(N\) coincident D3-branes in asymptotically flat space. When the closed string coupling constant \(g_s\) is small, there are two decoupled sets of low energy modes, low energy closed strings propagating far from the branes, described by type IIB SUGRA, and massless open strings ending on the branes. The latter are described by four-dimensional \(\cN = 4\) SYM theory with gauge group \(\SU(N)\), and coupling \(g_\mathrm{YM} = \sqrt{4\pi g_s}\).\footnote{The gauge group describing \(N\) D3-branes is actually \(\U(N)\), however a \(\U(1)\) subgroup corresponding to the centre-of-mass position of the branes decouples, and is ignored.}

\begin{figure}
    \begin{center}
        \includegraphics{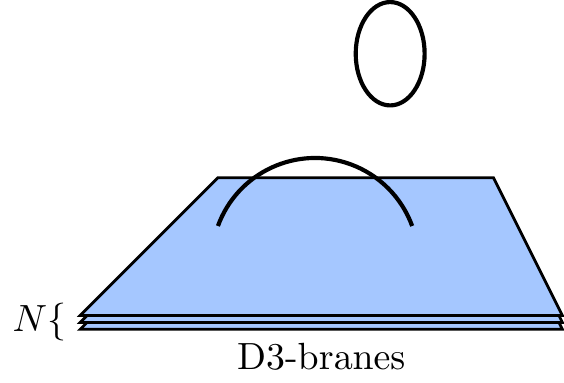}
        \caption[Cartoon of a stack of D3-branes.]{Cartoon of the D3-brane stack that gives rise to the duality between \(\cN = 4\) SYM and type IIB string theory. The arc is an open string, with both ends on the stack. Such strings give rise to fields transforming in the adjoint representation of \(\SU(N)\). The ellipse represents a closed string.}
    \end{center}
    \label{fig:d3_stack}
\end{figure}

Alternatively, when \(g_s\) is large, the appropriate description of the D3-branes is entirely in terms of closed strings. For large \(g_s\), the D3-branes curve spacetime, producing a geometry with a horizon. Near the horizon is an \( \ads[5] \times \sph[5] \) throat. Once more there are two decoupled sets of low energy excitations. One set is again massless closed string modes far from the branes, described by type IIB SUGRA. The other set of low energy excitations consists of closed string modes moving in the near-horizon \( \ads[5] \times \sph[5] \) region. These modes appear highly redshifted to an observer at infinity, so one must include the full spectrum of near-horizon string states in the count of low energy excitations.

Now imagine we begin from the weakly coupled description in terms of weakly coupled SYM (we neglect the decoupled closed strings from now on), and dial up \(g_s\) (and therefore \(g_\mathrm{YM}\)). Yang-Mills theory is well defined for all values of \(g_\mathrm{YM}\). Hence, there is no clear reason for this description to break down at large \(g_s\), where we expect the D-branes to be described by strings in \(\ads[5] \times \sph[5]\). The conjecture of ref.~\cite{Maldacena:1997re} is that that both descriptions are equally valid, so that \(\cN = 4\) SYM with gauge group \(\SU(N)\) is equivalent to type IIB string theory on \(\ads[5] \times \sph[5]\). The radius of both \ads[5] and \sph[5] are determined by the rank of the gauge group,
\begin{equation} \label{eq:ads_5_radius}
    L = (4 \pi \a'^2 g_s N)^{1/4},
\end{equation}
where \(\a'\) is the Regge slope.

As described above, if we wish to be able to perform calculations on the string theory side of the correspondence then we must take a classical limit. In addition, string theory on curved backgrounds is poorly understood, so it is usual to take the limit that string effects are small, in which case string theory is well approximated by type IIB SUGRA.

The classical limit is that of weak coupling \(g_s \ll 1\). Gauge/gravity duality relates the Yang-Mills coupling \(\gym\) to the string coupling as
\(
    \gym^2 = 4 \pi g_s,
\)
so the classical limit implies \(\gym \ll 1\).\footnote{The string coupling constant is \(g_s = e^\f\), where \(\f\) is the constant value of the dilaton in the D3-brane solution~\eqref{eq:d3_brane_solution}. A non-zero axion \(\C0\) in the D3-brane solution corresponds to a non-zero theta term \(\frac{C_0}{8 \pi} \int F \wedge F\) in the \(\cN=4\) SYM action, where \(F\) is the field strength~\cite{Becker:2007zj}.} Since \(\a'\) determines the typical length scale of stringy effects, the supergravity limit is \(L \gg \sqrt{\a'}\) (i.e. the curvature is small in string units). From~\eqref{eq:ads_5_radius} this implies that \(g_s N \gg 1\).  To satisfy both conditions we require \(N \gg 1\). In this limit the strength of the coupling between fields in Yang-Mills theory is better described by the 't Hooft coupling, \(\l = \gym^2 N\), in terms of which the appropriate limits are:
\begin{align}
    N &\to \infty,~\l~\text{fixed}
    && \text{(classical limit)}
    \nonumber \\
    \l &\to \infty
    && \text{(supergravity limit)}
    \label{eq:holographic_limits}
\end{align}
The SUGRA limit therefore implies that the dual Yang-Mills theory is strongly coupled.

When \(N\) is very large, \(\cN = 4\) SYM has an extremely large number of degrees of freedom. The field content of the theory is \(N^2 - 1\) copies of the vector multiplet of \(\cN = 4\) SUSY, one for each generator of \(\mathfrak{su}(N)\), so at large \(N\) the number of degrees of freedom is proportional to \(N^2\). For example, the thermodynamic entropy density \(s\) of \(\cN=4\) SYM at temperature \(T\) and large \(N\) is  proportional to \(N^2\), explicitly \(s = \frac{1}{2} \pi^2 N^2 T^3\) at large \(\l\)~\cite{Gubser:1996de}. The Weyl anomaly coefficients \(a\) and \(c\) (in \(d=4\) there is only one type B anomaly) are also proportional to \(N^2\) at large \(N\)~\cite{Henningson:1998gx,Henningson:1998ey}.

Since \(\cN = 4\) SYM and type IIB SUGRA on \( \ads[5] \times \sph[5]\) are expected to be equivalent descriptions of the same physics, the symmetries of the two theories should be the same, and indeed they are. The bosonic symmetries of \(\cN = 4\) SYM form the group \(\SU(2,2) \times \SU(4)\). The \(\SU(4)\) factor is the R-symmetry of \(\cN=4\) SUSY, while \(\SU(2,2)\) arises because the theory is conformally invariant, \(\SU(2,2)\) being the double cover of the four-dimensional conformal group \(\SO(4,2)\).

On the gravity side, the bosonic symmetries are the isometries of \(\ads[5] \times \sph[5]\), which form the group \(\SO(4,2) \times \SO(6)\). Since type IIB SUGRA contains fermions, we should again replace these groups with their double covers, yielding \(\SU(2,2) \times \SU(4)\), precisely the bosonic symmetry of \(\cN=4\) SYM. One can show that on accounting for the fermionic symmetries, the full superalgebra is \(\PSU(2,2|4)\) on both sides of the duality~\cite{Becker:2007zj}.

This illustrates a couple of general principles of gauge/gravity duality. One is that gravity on \ads\ (as opposed to asymptotically \ads) is holographically dual to a \textit{conformal} field theory, since the isometries of \ads[d+1] form the \(d\)-dimensional conformal group. The other is that gauge symmetries on the gravity side are dual to global symmetries on the QFT side; large gauge transformations,\footnote{Large gauge transformations are those that act non-trivially on the boundary.} in this case isometries, act as global transformations on the boundary.

The duality between \(\cN=4\) SYM and type IIB string theory has not been proven, but a large body of evidence has accumulated in favour of its existence, with quantities which can be computed on both sides of the duality matching exactly. Examples include three point functions of half-BPS operators~\cite{Lee:1998bxa,Freedman:1998tz} and the Weyl anomaly~\cite{Henningson:1998gx,Henningson:1998ey}.

As discussed above, the near horizon regions of M2- and M5-brane stacks are also of the form of \(\ads \times \sph\). Gauge/gravity duality is therefore expected to relate 11D SUGRA on \(\ads[4] \times \sph[7]\) and \(\ads[7]\times \sph[4]\) to QFTs that describe the low-energy excitations of M2- and M5-branes, respectively~\cite{Maldacena:1997re}. In the case of M2-branes, this QFT is maximally supersymmetric Chern-Simons-matter theory, known as ABJM theory after Aharony, Bergman, Jafferis and Maldacena~\cite{Bagger:2006sk,Gustavsson:2007vu,Bagger:2007jr,Bagger:2007vi,Aharony:2008ug}. The QFT describing a stack of M5-branes has not been formulated, but gauge/gravity duality provides a powerful tool for its study. In chapter~\ref{chap:probe_m5} we will use the duality to study defects in the M5-brane theory.

\section{The holographic dictionary}
\label{sec:holographic_dictionary}

In this subsection we review the relevant entries of the holographic dictionary, the map between quantities in a QFT and its dual gravitational theory.

The radial coordinate \(z\) in~\eqref{eq:general_asymptotic_ads_metric} is identified with the renormalisation scale \(\m\) in the QFT as \(\m \sim 1/z\). Approaching the boundary at \(z = 0\) therefore corresponds to sending \(\m \to \infty\), i.e. approaching the UV of the dual QFT, while \(z\to\infty\) corresponds to the IR of the dual QFT. Since a QFT is usually defined by its UV behaviour, it is common to think of the \(d\)-dimensional QFT as ``living'' at the boundary of the \aads[d+1] spacetime.

Equation~\eqref{eq:duality_statement} provides the recipe for computing correlation functions in a QFT from its gravity dual. Given a set of operators \(\cO_i\) in the QFT, dual to fields \(\f_i\), the \(n\)-point correlation function of the \(\cO_i\) is determined by functional differentiation of the on-shell gravity action with respect to the boundary values \(\f_{i,(0)}\) of the \(\f_i\)~\cite{Gubser:1998bc, Witten:1998qj},
\begin{equation} \label{eq:correlator_formula}
    \vev{\cO_1(x_1) \dots \cO_n(x_n)} = \le(- \frac{\d}{\d \f_{1,(0)} (x_1)} \ri) \dots \le(- \frac{\d}{\d \f_{n,(0)}  (x_n)} \ri) e^{-I_\mathrm{grav}^\star[\f_0]},
\end{equation}
where \(x_i\) is the location of the operator insertion \(\cO_i\). This equation is  for Euclidean signature. In Lorentzian signature one should replace \(- I_\mathrm{grav}^\star\) with \(i S_\mathrm{grav}^\star\), and multiply each of the functional derivatives by \(i\). One must also place the operator insertions on the appropriate part of a Schwinger-Keldysh contour to obtain the desired time ordering~\cite{Son:2002sd,Herzog:2002pc,Skenderis:2008dg,Skenderis:2008dh}.

Consider a bulk field \(\f\) --- for simplicity we consider a scalar field --- dual to some operator \(\cO\) of the boundary theory. The equation of motion for \(\f\), which follows from the action \(I_\mathrm{grav}\), will be second order in \(z\) derivatives, so two boundary conditions are required. A series solution around \(z = 0\) therefore takes the generic form
\begin{equation} \label{eq:generic_scalar_expansion}
    \f(x,z) = z^{d - \D}   \f_{(0)}(x) \le(1 + \dots \ri)  + z^{\D}  \f_{(2\D - d)}(x) \le(1 + \dots \ri) 
\end{equation}
where the ellipses denote terms with higher powers of \(z\), and the coefficients \(\f_{(0)}\) and \(\f_{(2\D - d)}\) are determined by the boundary conditions. The quantity \(\D\) is determined by the mass \(m\) of the scalar field; it is the largest root of
\begin{equation} \label{eq:mass_delta_relation}
    m^2 L^2 = \D (\D - d).
\end{equation}

To derive~\eqref{eq:mass_delta_relation}, suppose that the part of \(I_\mathrm{grav}\) depending on \(\f\) is
\begin{equation}
    I_\mathrm{grav} \supset \int \diff^{d+1}x \sqrt{G} \le[ \frac{1}{2}G^{MN} \p_M \f \p_N \f + V(\f) \ri],
\end{equation}
with some potential \(V(\f)\). The Euler-Lagrange equation for \(\f\) is then
\begin{equation} \label{eq:scalar_eom}
    \frac{z^{d+1}}{\sqrt{g}} \p_z \le(\sqrt{g} \frac{\p_z \f }{z^{d-1}} \ri) 
    +  \frac{z^{d+1}}{\sqrt{g}}\p_\m  \le(\sqrt{g} \frac{g^{\m\n} \p_\n \f }{z^{d-1}} \ri) = L^2 V'(\f) =m^2 L^2 \f + L^2\sum_{n=2}^\infty c_n \f^n ,
\end{equation}
where on the left we have used the Fefferman-Graham gauge~\eqref{eq:general_asymptotic_ads_metric}, and on the right we have expanded the potential in powers of \(\f\), with constants \(c_n\), and separated out the mass term.

Let as assume that the scalar field has the near-boundary series solution \(\f(x,z) = z^{\D_-} \sum_{n=0}^\infty \f_{(n)}(x) z^n\), obtained by solving~\eqref{eq:scalar_eom} order by order in small \(z\). The leading powers of \(z\) in each of the terms in~\eqref{eq:scalar_eom} are
\begin{equation}
    \frac{z^{d+1}}{\sqrt{g}} \p_z \le(\sqrt{g} \frac{\p_z \f }{z^{d-1}} \ri) \sim z^{\D_-},
    \quad
    \frac{z^{d+1}}{\sqrt{g}}  \p_\m  \le(\sqrt{g} \frac{g^{\m\n} \p_\n \f }{z^{d-1}} \ri) \sim z^{{\D_-} + 2},
    \quad
    \f^n \sim z^{n\D_-}.
\end{equation}
Assuming \(\D_- > 0\), the leading order term in the series expansion of~\eqref{eq:scalar_eom} is then the \(\cO(z^{\D_-})\) term, which reads,
\begin{equation}
    z^{d+1} \p_z \le[ \frac{\p_z (\f_{(0)}(x) z^{\D_-}) }{z^{d-1}} \ri] 
    = m^2 L^2 \f_{(0)}(x) z^{\D_-} .
\end{equation}
Evaluating the \(z\) derivatives, one finds that this is satisfied provided \(\D_-\) is one of the roots of~\eqref{eq:mass_delta_relation}.

We previously identified \(\f_{(0)}\) with the source for the operator \(\cO\). Dimensional analysis shows that its mass dimension is \([\f_{(0)}] = d - \D\), which is the appropriate dimension for the source of \(\cO\) if the operator has scaling dimension \(\D\). Dimensional analysis also implies that \([\f_{(2\D-d)}] = \D\). A natural guess is that this coefficient is related to the one-point function of \(\cO\), and indeed applying the prescription~\eqref{eq:correlator_formula}, one finds~\cite{deHaro:2000vlm,Skenderis:2002wp}
\begin{equation}
    \vev{\cO(x)} = - (2 \D - d) \f_{(2\D - d)}(x) + f[\f_{(0)}(x)],
\end{equation}
where \(f\) is a function of \(\f_{(0)}(x)\) and its derivatives, the form of which depends on the action \(I_\mathrm{grav}\).

The requirement that \(\D\) is the \textit{largest} root of~\eqref{eq:mass_delta_relation} is due to the CFT unitarity bound.\footnote{In a unitary CFT, no scalar operator except the identity operator has \(\D < \frac{d}{2} - 1\).} For masses in the range \(-\frac{d^2}{4} < m^2 L^2 \leq 1 - \frac{d^2}{4}\) one can instead choose \(\D\) to be the smallest root of~\eqref{eq:mass_delta_relation} while still satisfying the unitary bound~\cite{Klebanov:1999tb}.\footnote{In \(\ads\) a scalar field may be stable with negative mass-squared, provided \(m^2 L^2 \geq - \frac{d^2}{4}\). This is known as the Breitenlohner-Freedman bound~\cite{Breitenlohner:1982bm,Breitenlohner:1982jf}.}

An important operator is the stress tensor \(T^{\m\n}\), the conserved current associated to translational invariance. In gauge/gravity duality, translational invariance of the boundary is dual to diffeomorphism invariance of the bulk, and so the stress tensor is dual to the metric, \(G_{MN}\). In Fefferman-Graham gauge \eqref{eq:general_asymptotic_ads_metric}, where the metric takes the near boundary expansion~\eqref{eq:metric_near_boundary_expansion}, we identify the leading coefficient \(g_{(0)\m\n}\) as the source for the stress tensor, i.e. the metric in the boundary theory. Correlation functions of the stress tensor are then computed by functional differentiation with respect to this boundary value. For example~\cite{deHaro:2000vlm}
\begin{equation}
    \vev{T_{\m\n}} = - \frac{2}{\sqrt{\det g_{(0)}}} \frac{\d \ln Z}{\d g_{(0)}^{\m\n}} = \frac{d}{16 \pi \gn} g_{(d)\m\n} + \dots\;,
\end{equation}
where the ellipsis denotes terms which depend on \(g_{(0)}\) and other sources.

We note that there is an inherent ambiguity in the value of the boundary metric \(g_{(0)\m\n}\). For example, defining a new radial coordinate \(z' = \O z\) with constant \(\O\), transforms the metric~\eqref{eq:general_asymptotic_ads_metric} into
\begin{equation}
    \diff s^2 = \frac{L^2}{z'^2} \le( \diff z'^2 + g'_{\m\n} \diff x^\m \diff x^\n \ri),
\end{equation}
where \(g'_{\m\n} = \O^2 g_{\m\n} = \O^2 g^{(0)}_{\m\n} + z'^2 g^{(2)}_{\m\n} + \dots\)\;. In this coordinate system we would naturally read off the  boundary metric as \(\O^2 g^{(0)}_{\m\n}\), in other words we have implemented a constant Weyl transformation. General Weyl transformations may be performed by allowing \(\O\) to depend on \(x\) and \(z\).

Because of this ambiguity, the ``boundary metric'' \(g^{(0)}\) is really a representative of a conformal class of metrics. The choice of representative is encoded in a defining function \(f(x,z)\), which may be any positive function with a simple zero at \(z =0\). The boundary metric is taken to be
\begin{equation}
    \lim_{z \to 0} f(x,z)^2 G.
\end{equation}
Changing the defining function holographically implements a Weyl transformation on the boundary. Normally one chooses a defining function which is natural in a given coordinate system. For example in the coordinate system~\eqref{eq:general_asymptotic_ads_metric} we chose \(f = z/L\), leading to the identification of \(g^{(0)}\) as the boundary metric.

\subsection{Gauge/gravity duality and thermodynamics}
\label{sec:thermodynamics}

In this thesis we will study several models of systems at non-zero temperature and chemical potential. The holographic dual to a QFT at finite temperature \(T\) is a gravitational solution with the same temperature~\cite{Witten:1998zw}. The free energy \(F\) is determined from the generating functional, \(Z = \exp(-F/T)\). Often, the appropriate gravitational solution is a black hole or black brane, with Hawking temperature \(T\). In this case, the thermodynamic entropy \(S\) in the CFT is given in the semiclassical limit by the Bekenstein-Hawking entropy of the black hole,
\begin{equation} \label{eq:bh_entropy}
    S = \frac{A}{4 \gn},
\end{equation}
where \(A\) is the area of the horizon, and \(\gn\) is Newton's constant. To study thermodynamics in the QFT, we Wick rotate to imaginary time, with periodic time coordinate \(\t \sim \t + 1/T\).

Suppose the boundary theory has a \(\U(1)\) global symmetry, with corresponding conserved current \(J^\m\). The current is dual to a \(\U(1)\) gauge field \(A_M\) in the bulk. In particular, the time component of the gauge field \(A_t\) is dual to the charge density \(J^t\). In \aads[d+1], \(A_t\) has the near boundary expansion
\begin{equation}
    A_t = \m + j z^{d-2} + \dots\;.
\end{equation}
Applying the holographic dictionary, the leading coefficient \(\m\) is identified as the source for the charge density, which is by definition the chemical potential. The subleading coefficient \(j\) is proportional to the charge density.

\begin{table}
\begin{center}
    \begin{tabularx}{0.85\textwidth}{X | X}
        QFT in \(d\) dimensions & Gravity on \aads[d+1]
        \\
        \hline
        \\[-0.75em]
        Operator & Field
        \\
        \quad -- Source \(\f_{(0)}\) & \quad -- Boundary value \(\f_{(0)}\)
        \\
        \quad -- Scaling dimension \(\D\) & \quad -- Mass \(m^2 L^2 = \D (\D - d) \)
        \\[0.5em]
        Stress tensor \(T^{\m\n}\) & Metric \(G_{MN}\)
        \\
        \quad -- Source (metric) \(g_{(0)\m\n}\) & \quad -- Boundary value in~\eqref{eq:metric_near_boundary_expansion}
        \\[0.5em]
        \(\U(1)\) current \(J^\m\) & Gauge field \(A_M\)
        \\
        \quad -- Chemical potential \(\m\) & \quad -- \(A^{(0)}_t = \m\)
        \\[0.5em]
        Thermodynamics & Black hole thermodynamics
        \\
        \quad -- Temperature \(T\) & \quad -- Hawking temperature
        \\
        \quad -- Free energy \(F\) & \quad -- On-shell action \(I_\mathrm{grav}^\star = F/T\)
        \\
        \quad -- Entropy \(S\) & \quad -- Bekenstein-Hawking entropy
        \\[0.5em]
    \end{tabularx}
\end{center}
    \caption[The holographic dictionary.]{Important entries in the holographic dictionary. The relationship between \(\D\) and \(m^2\) is given for a scalar field; a complete list for fields of different spin may be found in ref.~\cite{Ammon:2015wua}. The expression for \(\m\) in terms of \(A_t^{(0)}\) assumes a gauge where \(A_t\) vanishes at the horizon.}
    \label{tab:holographic_dictionary}
\end{table}
The entries in the holographic dictionary that we have discussed so far are summarised in table~\ref{tab:holographic_dictionary}.

\subsection{Two-point correlation functions and quasinormal modes}
\label{sec:quasinormal_modes}

In chapter~\ref{chap:zero_sound} we will use holography to study two-point correlation functions of a QFT at non-zero temperature, in particular the retarded Green's function~\cite{Bellac:2011kqa}
\begin{equation}
    G_{ab} (x_1 - x_2) = \q(t_1 - t_2) \vev {\cO_a(x_1) \cO_b(x_2)} + \q(t_2 - t_1) \vev{\cO_b (x_2) \cO_a(x_1)},
\end{equation}
where \(\q\) is the Heaviside step function, \(t_{1,2}\) is the time component of \(x_{1,2}\), and \((a,b)\) index the operators of the theory. In writing \(G_{ab}\) as a function of the separation \(x_1 - x_2\), we have assumed that theory is invariant under translations in space and time, as will be the case for the model of chapter~\ref{chap:zero_sound}.

The retarded Green's function \(G_{ab}\) is physically significant in that it determines the linear response of the operator \(\cO_a\) to a small change in the source of \(\cO_b\)~\cite{Kapusta:2006pm},
\begin{equation}
    \d \vev{\cO_a(x)} \approx \int \diff^d y \, G_{ab}(x - y) \d \f_{(0),a}(y).
\end{equation}
Let us take the Fourier transform of the retarded Green's function,
\begin{equation}
G_{ab}(\w,\mathbf{k}) = \int \diff^d x \,  e^{i \w t - i \mathbf{k} \cdot \mathbf{x}} G_{ab}(t,\mathbf{x}),
\end{equation}
where \(x = (t, \mathbf{x})\). For a given momentum \(\mathbf{k}\),  \(G_{ab}(\w,\mathbf{k})\) will have poles in the complex \(\w\) plane, at frequencies we denote \(\w = \w_*(\mathbf{k})\). These poles correspond to unstable propagating modes in the QFT, with frequency \(\Re \w_*\) and decay rate \(-\Im \w_*\).\footnote{We assume that \(\Im \w_* < 0\). If instead \(\Im \w_\star > 0\), the mode grows exponentially in time, indicating an instability of the system.}

In gauge/gravity duality, poles in the retarded Green's functions of the boundary theory are dual to the frequencies of \textit{quasinormal modes} in the gravitational theory~\cite{Son:2002sd,Kovtun:2005ev}, which are defined as follows.

Consider small fluctuations about a black brane solution of the gravitational theory, holographically dual to a thermal state of the boundary QFT. Writing the fluctuation of a field \(\f_a\) as \(\f_a \to \f_a + \d\f_a (t,\mathbf{x},z)\), we consider fluctuations of a single Fourier mode for each field,\footnote{Different Fourier modes decouple due to the translational invariance of the background solution.} schematically
\begin{equation} \label{eq:single_fourier_mode}
    \d\f_a(t,\mathbf{x},z) = e^{-i \w t + i \mathbf{k}\cdot \mathbf{x}} \d\f_a(\w,\mathbf{k},z),
\end{equation}
for some fixed frequency \(\w\) and momentum \(\mathbf{k}\). We seek solutions to the linearised equations of motion for the Fourier modes which are:
\begin{enumerate}
    \item Normalisable at the boundary. Each Fourier mode will have a near-boundary expansion like~\eqref{eq:generic_scalar_expansion} (with different values of \(\D\)). We impose that the leading coefficient at the boundary vanishes. Holographically, we are keeping the sources for all operators fixed.
    \item Ingoing at the horizon. Near the horizon, there are two independent solutions to the equations of motion for the fluctuations, corresponding to waves moving into or out of the horizon. We choose boundary conditions such that the solution is purely ingoing. Holographically, this means that the quasinormal modes correspond to poles of the retarded --- as opposed to advanced --- Green's functions.
\end{enumerate}
Solutions satisfying these boundary conditions generically exist only for a discrete set of frequencies, at a given momentum. These frequencies are the quasinormal modes.

To compute the retarded Green's functions \(G_{ab}\) themselves, the method is as follows~\cite{Son:2002sd,Herzog:2002pc,Kaminski:2009dh}. We consider fluctuations of the fields \(\f_a\) as above, writing the Fourier transformed fluctuations as
\begin{equation} \label{eq:real_time_correlator_fourier_transform}
    \d \f_a(t, \mathbf{x}, z) = \int \frac{ \diff^{d}k}{(2\pi)^d} e^{i k \cdot x} F_{ab}(k,z) z^{d-\D_b} \d\f_{(0),b}(k),
\end{equation}
where \(\D_b\) is the dimension of the operator dual to \(\f_b\), \(k = (\w, \mathbf{k})\), and \(F_{ab}(k,0) = \d_{ab}\). We impose ingoing boundary conditions on \(F_{ab}\) at the horizon. By construction, \(\d\f_{(0),a}(k)\) is the Fourier transform of the perturbation in the source, \(\d\f_{(0),a}(x)\).

Substituting~\eqref{eq:real_time_correlator_fourier_transform} into the linearised action for the fluctuations, one finds that when the equations of motion are satisfied, the action may be written in the form
\begin{equation} \label{eq:minkowski_correlator_action}
    S = \int \frac{ \diff^{d}k}{(2\pi)^d} \le[\d\f_{(0)a}(-k) \mathcal{F}_{ab} (k,z) \d\f_{(0)b}(k)\ri]_{z=0}^{z=z_H},
\end{equation}
where \(z_H\) is the location of the black brane's horizon, and \(\mathcal{F}_{ab}\) depends quadratically on \(F_{ab}\). The precise form of \(\mathcal{F}_{ab}\) is determined by the action of the gravitational theory. The retarded two-point functions are given by
\begin{equation} \label{eq:minkowski_correlator_prescription}
    G_{ab} = - 2 \mathcal{F}_{ab}(k, z=0).
\end{equation}

Notice that this is not what one would obtain by naive functional differentiation of~\eqref{eq:minkowski_correlator_action} with respect to the boundary values \(\f_{(0),a}(k)\). However, ref.~\cite{Herzog:2002pc} showed that~\eqref{eq:minkowski_correlator_prescription} may be derived using functional derivatives in a more careful approach that holographically implements the Schwinger-Keldysh formalism.

\subsection{Wilson loops}
\label{sec:wilson_loops}

In the duality between \(\cN = 4\) \(\SU(N)\) SYM and type IIB string theory on \(\ads[5] \times \sph[5]\), a Wilson line in the fundamental representation is holographically dual to a string which reaches the boundary of \ads[5]~\cite{Maldacena:1998im,Rey:1998ik,Drukker:1999zq}.

\begin{figure}
    \begin{subfigure}{0.5\textwidth}
        \begin{center}
        \includegraphics[scale=0.9]{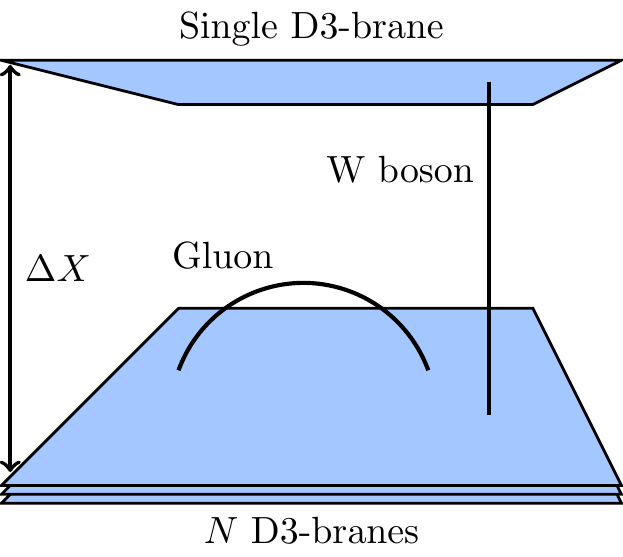}
        \caption{}
        \end{center}
    \end{subfigure}
    \begin{subfigure}{0.5\textwidth}
        \begin{center}
        \includegraphics[scale=0.9]{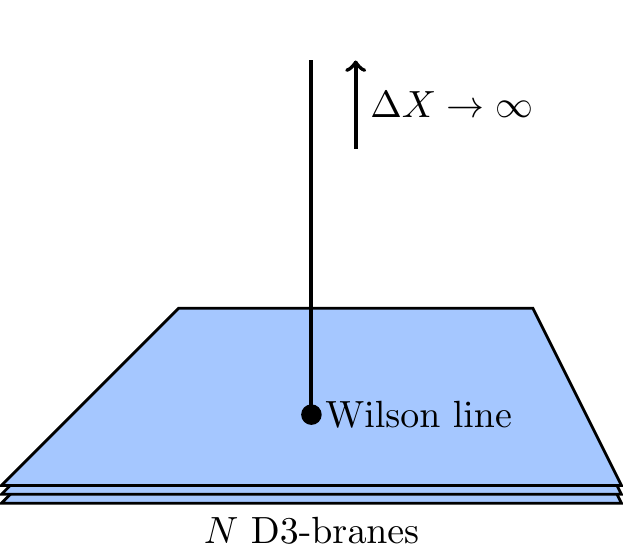}
        \caption{}
        \end{center}
    \end{subfigure}
    \caption[Schematic illustration of Wilson loops in gauge/gravity duality.]{
        \textbf{(a):} Cartoon of the holographic description of the spontaneous symmetry breaking pattern \(\SU(N+1) \to \SU(N) \times \U(1)\), in \(\cN=4\) SYM. The symmetry breaking is caused by non-zero separation \(\D X\) between one of the D3-branes and the other \(N\). Open strings which begin and end on the stack of \(N\) D3-branes correspond to the ``gluons'', the massless gauge bosons of the \(\SU(N)\) factor. Open strings stretched between the separated branes correspond to ``W bosons'', the gauge bosons which acquire a mass proportional to \(\D X\) due to spontaneous symmetry breaking.
        \textbf{(b):} In the limit \(\D X \to \infty\), the W bosons become infinitely massive. From the point of view of the \(\SU(N)\) factor, which decouples from the \(\U(1)\) in this limit, a W boson insertion looks like infinitely massive quarks, i.e. a Wilson loop. Note that we have suppressed the time direction in the figure.
    }
    \label{fig:wilson_line_cartoon}
\end{figure}
To see why, let us sketch the argument presented in ref.~\cite{Maldacena:1998im} and illustrated in figure~\ref{fig:wilson_line_cartoon}. Consider a stack of \(N+1\) D3-branes, holographically dual to \(\cN=4\) SYM with gauge group \(\SU(N+1)\). If we now separate one of the branes from the others by a distance \(\D X^i\), this spontaneously breaks the gauge group to \(\SU(N) \times \U(1)\). The W bosons correspond to strings stretched between the \(\SU(N)\) and \(\U(1)\) branes. They have mass proportional to \(|\D X|\), the minimum length of such strings, and transform in the fundamental representation of \(\SU(N)\).

In the limit \(|\D X| \to \infty\), the \(\SU(N)\) and \(\U(1)\) factors decouple. The W boson strings reach the boundary of the \ads[5] region near the horizon of the \(\SU(N)\) branes. The W bosons themselves become infinitely massive as \(|\D X| \to \infty\). Recalling the interpretation of the Wilson loop as the phase factor of a heavy charged particle travelling around the loop, it is natural to expect that Wilson loops are holographically related to strings ending on the boundary of \ads.

In fact, such a string is holographically dual to a Wilson loop operator modified by a term depending on the \(\cN=4\) SYM scalar fields \(\f^i\). The operator is~\cite{Maldacena:1998im}
\begin{equation} \label{eq:bps_wilson_loop}
    \mathcal{W}(\mathcal{C}) = \frac{1}{N} \tr \mathcal{P} \exp \le[
        i \oint \diff s \le( A_\m \frac{\diff x^\m}{\diff s} +  \f^i \q^i \le|\frac{\diff x}{\diff s}\ri|\ri)
    \ri],
\end{equation}
where \(\mathcal{C} \equiv x(s)\) is the contour defining the Wilson loop, \(\mathcal{P}\) denotes path ordering, and \(\q^i \equiv \D X^i/|\D X|\). The trace is taken in the fundamental representation. If the integration contour is a straight line or a circle, the operator~\eqref{eq:bps_wilson_loop} is known as the half-BPS Wilson loop operator, since in either case it commutes with eight independent linear combinations of the fermionic symmetry generators of \(\cN = 4\) SYM, due to the additional \(\f^i\)-dependent term~\cite{Drukker:1999zq,Bianchi:2002gz}.

Half-BPS Wilson loops in representations other than the fundamental are dual to configurations of multiple strings, one string for each box in the representation's Young tableau. However, when the number of strings is of order \(N\), interactions between the strings become important, and the appropriate holographic dual is instead a configuration of D3- and D5-branes~\cite{Drukker:2005kx,Gomis:2006sb,Gomis:2006im}, interpreted as a bound state of strings.

\section{Probe branes}
\label{sec:probe_branes}

In this section we review the concept of probe branes, which feature in the holographic models studied in chapters~\ref{chap:zero_sound} and \ref{chap:probe_m5}. Early examples of probe branes in gauge/gravity duality appear in refs.~\cite{Karch:2000gx,Karch:2002sh,Skenderis:2002vf}.

Consider a \(D\)-dimensional gravitational theory, into which we embed a \(p\)-brane, i.e. a \((p+1)\)-dimensional dynamical object. The full action is
\begin{equation}
    S_\mathrm{grav} = S_\mathrm{bulk} + S_\mathrm{brane},
\end{equation}
where \(S_\mathrm{bulk}\) is the action for the \(D\)-dimensional theory, and \(S_\mathrm{brane}\) is the action for the brane. We assume that \(S_\mathrm{brane}\) is proportional to a tension \(\mathcal{T}_p\), and therefore so is the contribution of the probe brane  to the stress tensor \(\Q_{MN}\) of the gravitational theory. The probe limit is that of small tension in units of Newton's constant,
\begin{equation} \label{eq:probe_limit_general}
    \t_p \equiv \mathcal{T}_p \gn L^{p + 3 - D} \ll 1. 
\end{equation}
In this limit, the contribution of the brane to the stress tensor is small compared to the curvature terms in Einstein's equations, and may be neglected to a first approximation. One solves the equations of motion which follow from \(S_\mathrm{bulk}\) to obtain a background solution, into which the probe brane is embedded. This will generate a series of corrections to the bulk fields, which we denote collectively as \(g\),
\begin{equation}
    g = g^{(0)} + g^{(1)} + g^{(2)} + \dots \;,
\end{equation}
where \(g^{(n)} \propto \t_p^n\), and \(g^{(0)}\) is the background solution. These corrections are known as the back-reaction of the brane.

We similarly expand other physical quantities as a series in the tension, for example we write the on-shell action as
\begin{equation}
    S^\star = S^{(0)} + S^{(1)} + S^{(2)} + \dots \; ,
\end{equation}
where the \(S^{(n)} \propto \t_p^n\). The probe approximation provides a simple way to compute \(S^{(1)}\), the leading order correction to the on-shell action of the background solution, which in turn gives the leading order correction to the generating functional of the dual QFT using~\eqref{eq:duality_statement}. Let us write the on-shell action on the solution \(g\) as \(S^\star[g]\). Then, expanding in \(\t_p\),
\begin{equation} \label{eq:probe_correction_action}
    S^\star[g] = S_\mathrm{bulk}^\star[g^{(0)}] + \underbrace{\int \diff^D x
      \le. \frac{\d S_\mathrm{bulk}}{\d g} \ri|_{g^{(0)}} g^{(1)} + S_\mathrm{brane}[g^{(0)}]
    }_{S^{(1)}} + \dots \;.
\end{equation}
The functional derivatives \(\d S_\mathrm{bulk}/\d g\) give the equations of motion for the bulk theory, which vanish when evaluated on the background solution \(g^{(0)}\). Hence
\begin{equation}
    S^{(1)} = S_\mathrm{brane}[g^{(0)}],
\end{equation}
so the leading order correction to the on-shell action is determined by evaluating the action of the probe brane on the background solution. This provides a great simplification. However, care must be taken as not every quantity may reliably computed in this way. For example, computing the energy density from the probe brane's stress tensor yields the wrong answer at non-zero temperature~\cite{Karch:2008uy}.

The presence of a probe brane in a holographic model often indicates the presence of matter fields in the boundary QFT. To illustrate this, consider the D3-D7 system of ref.~\cite{Karch:2002sh}.\footnote{See ref.~\cite{Erdmenger:2007cm} for a detailed review of this system.} The background solution is \(\ads[5] \times \sph[5]\), generated by a stack of \(N\) D3-branes, to which we add a stack of \(N_f\) coincident D7-branes described by the bosonic action~\cite{Becker:2007zj}
\begin{equation} \label{eq:d7_action}
    S_\mathrm{brane} = - \mathcal{T}_7 \int_\Sigma \diff^8 \xi \sqrt{-\det ( g + 2 \pi \a' F) } + 2 \pi^2 \a'^2 \mathcal{T}_7 \int_\Sigma  P[C_4] \wedge F \wedge F,
\end{equation}
where \(P\) denotes the pullback of a bulk supergravity field, \(g \equiv P[G]\), \(F\) is the field strength for a \(\U(1)\) gauge field, and \(\xi\) are coordinates on the brane world volume \(\Sigma\). The tension is \(\mathcal{T}_7 = N_f/((2\pi)^7 g_s \a'^4)\), from which one finds that the probe limit~\eqref{eq:probe_limit_general} is \(N_f \ll N\),\footnote{In string units, Newton's constant is given by \(\gn = 8 \pi^6 g_s^2 \a'^4\). Note that in the holographic limit we took \(g_s \to 0 \) with \(g_s N \gg 1\).} i.e. the number of fundamental flavours is very small compared to the rank of the gauge group.

Consider the \(\ads[5] \times \sph[5]\) metric~\eqref{eq:d3_brane_near_horizon} with the metric on the \sph[5] written as
\begin{equation}
    \diff s_{\sph[5]}^2 = \diff \q^2 + \sin^2 \q \diff \f^2 + \cos^2 \q \diff s_{\sph[3]}^2.
\end{equation}
Ref.~\cite{Karch:2002sh} derived a family of embeddings by employing a static gauge in which the D7-branes are parameterised by the \(x^\m\) coordinates, the coordinates on the \sph[3], and \(z \equiv L^2/r\), taking an ansatz where \(F=0\), \(\f\) is constant, and \(\q = \q(z)\). With these choices, the D7-brane action~\eqref{eq:d7_action} becomes
\begin{equation}
    S_\mathrm{brane} = -2 \pi^2 L^8 T_7 \int \diff^4 x \diff z \frac{\cos^3 \q}{z^5} \sqrt{1 + z^2 \q'^2},
\end{equation}
where \(\q' \equiv \p_z \q\) and we have integrated over the wrapped \sph[3]. The Euler-Lagrange equation for \(\q\) has the solution \(\sin \q = M z\). At the boundary, the D7-brane fills \ads[5] and wraps an equatorial \(\sph[3] \subset \sph[5]\). Moving into the bulk, the radius of the wrapped \sph[3] shrinks, eventually vanishing at \(z = 1/M\). This is the maximal extent of the D7-brane into the bulk. Expanding the solution for small \(z\), one finds \(\q = M z + M^3 z^3/6 + \cO(M^5z^5)\), indicating that \(\q\) is dual to a scalar operator \(\cO\) of dimension \(\D = 3\) (with source \(\propto M\) of dimension \(4-\D=1\)). Holographic renormalisation reveals that \(\vev{\cO} = 0\) for any value of \(M\)~\cite{Karch:2005ms}. 

\begin{figure}
\begin{center}
    \includegraphics{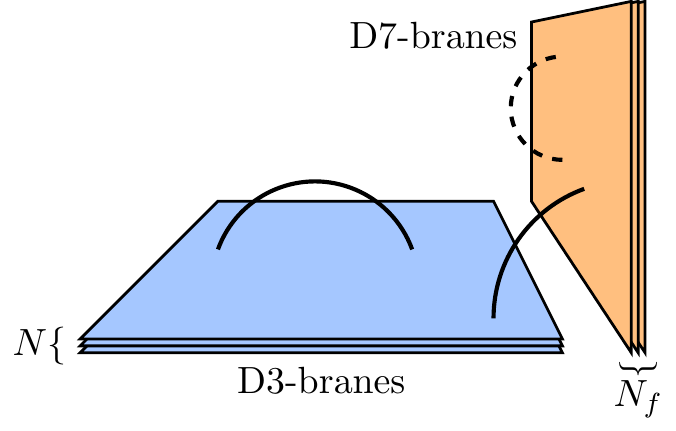}
    \caption[Cartoon of the D3-D7 system.]{Cartoon of the D3-D7 system. The curves represent strings, which can end on the branes. Strings which stretch between the D3- and D7-brane stacks give rise to fields which transform under the (anti-) fundamental of \(\SU(N)\), and under the (anti-) fundamental of a \(\U(N_f)\) global flavour symmetry, with masses determined by the closest approach of the two stacks.}
    \label{fig:d3_d7}
\end{center}
\end{figure}

The holographic dual of the D3-D7 system can be understood as follows. As discussed above, the holographic dual of a stack of D3-branes is \(\SU(N)\) \(\cN = 4\) SYM, with the fields of the gauge multiplet intuitively arising from strings beginning and ending on the D3-branes (commonly called 3-3 strings). The presence of the D7-branes allows for two new types of string: 3-7 strings, with one endpoint on each stack, and 7-7 strings, with both endpoints on the D7-branes. In principle one would expect the 7-7 strings to give rise to a \(\U(N_f)\) gauge theory, but a consequence of the supergravity limit is that the 't Hooft coupling for this theory vanishes, so the 7-7 strings decouple and may be ignored. The 3-7 strings therefore transform in the fundamental representations of a \(\U(N_f)\) global symmetry and the \(\SU(N)\) gauge symmetry. Analysis of the preserved supersymmetry shows that the open string states form hypermultiplets of four-dimensional \(\cN = 2\) supersymmetry. The different sorts of strings are sketched in figure~\ref{fig:d3_d7}.

If the D3- and D7-branes are separated along some transverse direction, then the 3-7 strings have a minimum length. This gives a non-zero mass to the \(\cN=2\) hypermultiplet, given by the separation multiplied by the string tension. The separation, and therefore the mass, is proportional to the integration constant \(M\) in the solution described above. The operator \(\cO\), dual to \(\q\), is a linear combination of bilinears of the fermion and scalar fields that compose the \(\cN=2\) hypermultiplet.

\chapter{Entanglement entropy}
\label{chap:entanglement_entropy}

In chapters~\ref{chap:entanglement_density} and~\ref{chap:probe_m5} we will use gauge/gravity duality to calculate entanglement entropy in a variety of systems. We now review the definition of entanglement entropy, and how it is calculated holographically. For further details, see refs.~\cite{Nishioka:2009un,Rangamani:2016dms}.

\section{Definition and properties}

The state of a quantum system may be specified by an operator called the density matrix, \(\dMatrix\). A system in a state~\(\ket{\y}\) is said to be in a pure state. The density matrix in this case is \(\dMatrix = \ket{\y}\bra{\y}\). However, the state of a physical system is not normally known with absolute certainty. Given a basis \(\{\ket{n}\}\), if we assign a classical probability \(p_n\) that the system is in the state \(\ket{n}\), the density matrix is~\(\dMatrix = \sum_n p_n \ket{n}\bra{n}\)~\cite{binney2013physics}. When the density matrix may not be expressed as \(\ket{\y}\bra{\y}\), this is called an impure or mixed state. The expectation value of an operator \(\cO\) in a state with density matrix \(\dMatrix\) is
\begin{equation}
    \vev{\cO} = \tr \le( \dMatrix \cO \ri) = \sum_n p_n \bra{n} \cO \ket{n}.
\end{equation}

Suppose the Hilbert space \(\cH\) of a system may be decomposed into a tensor product structure, \(\cH  = \cA \otimes \bar{\cA} \). We will refer to the subspace \(\cA\) as the entangling region. The reduced density matrix on \(\cA\), \(\dMatrix_\cA\), is defined by tracing out the states in its complement
\begin{equation}
    \dMatrix_\cA \equiv \tr_{\bar{\cA}} \dMatrix \equiv \sum_{\ket{b} \in \bar{\cA}} \bra{b} \dMatrix \ket{b}.
\end{equation}
The entanglement entropy \(\see\) of \(\cA\) is the von Neumann entropy associated to \(\dMatrix_\cA\),
\begin{equation} \label{eq:ee_definition}
    \see[\cA] \equiv - \tr_\cA \le( \dMatrix_\cA \ln \dMatrix_\cA \ri),
\end{equation}
see ref.~\cite{Nishioka:2009un}, for example. If the total density matrix \(\dMatrix\) describes a pure state, then the entanglement entropy satisfies \(\see[\cA] = \see[\bar{\cA}]\).

Entanglement entropy is a quantitative measure of the entanglement between the entangling region and its complement. For example, consider the case when \(\cA\) and \(\bar{\cA}\) are single qubits, each with possible states \(\ket{0}\) and \(\ket{1}\). If the system is in the state
\begin{equation} \label{eq:two_spin_half_state}
    \ket{\y} = \cos\q \ket{0}_\cA \ket{1}_{\bar{\cA}} + \sin\q \ket{1}_\cA \ket{0}_{\bar{\cA}},
\end{equation}
then the entanglement entropy is 
\(
    \see[\cA] = - \cos^2\q \ln \le( \cos^2\q \ri) - \sin^2 \q \ln \le( \sin^2\q \ri)
\)
~\cite{hongliulectures}. This vanishes when \(\q = 0\) or \(\pi/2\), corresponding to when~\eqref{eq:two_spin_half_state} is a product state. The entanglement entropy is maximal for \(\q = \pi/4\), in which case~\eqref{eq:two_spin_half_state} is the maximally entangled state \(\ket{\y} = \le( \ket{0}_\cA \ket{1}_{\bar{\cA}} + \ket{1}_\cA  \ket{0}_{\bar{\cA}} \ri)/\sqrt{2} \).

In a QFT, the Hilbert space may be taken as the set of states on a given Cauchy surface, roughly speaking a spatial slice at a single instant in time. It is therefore natural to divide the Hilbert space geometrically, by choosing \(\cA\) to be a spatial subregion on the Cauchy surface. We will assume that the Hilbert space factorises into the tensor product structure \(\cH = \cA \otimes \bar{\cA}\) necessary for the definition of entanglement entropy.\footnote{This assumption is not always true, see, for example, the discussion in ref.~\cite{Pretko:2018yxl}.}

\begin{figure}
\begin{subfigure}{0.5\textwidth}
    \includegraphics{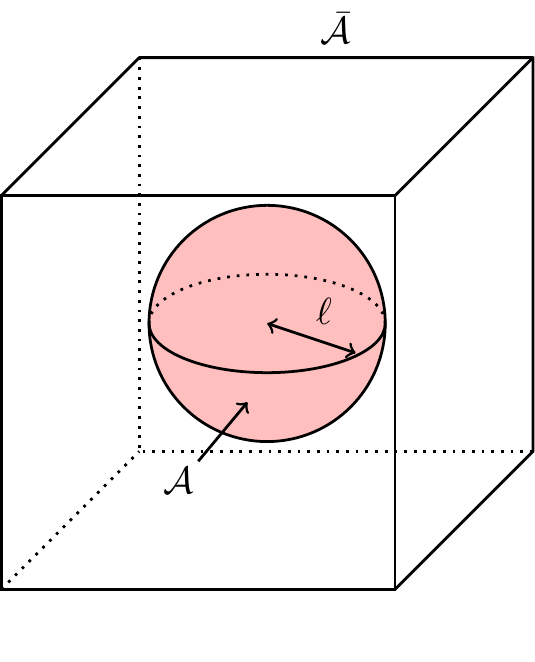}
    \caption{The sphere geometry.}
\end{subfigure}
\begin{subfigure}{0.5\textwidth}
    \includegraphics{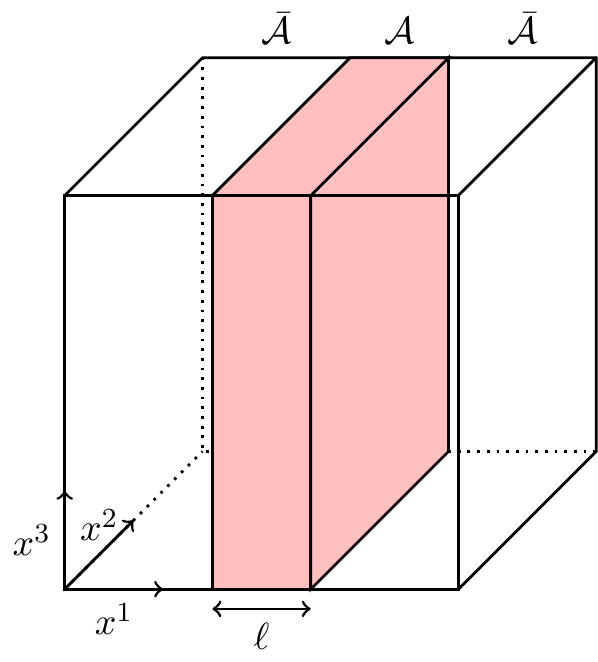}
    \caption{The strip geometry.}
\end{subfigure}
\caption[Sketches of the two entangling region geometries used in this thesis.]{The two geometries we will use for the entangling region. In the sphere geometry (left), \(\cA\) is a ball of radius \(\ell\). In the strip geometry (right), \(\cA\) is the region between two parallel planes, separated by a distance \(\ell\).}
\label{fig:entangling_region_geometries}
\end{figure}

Throughout this thesis we will use two different geometries for the subregion \(\cA\), which we will call the sphere and the strip. In the sphere geometry, \(\cA\) is the interior of a sphere of radius \(\ell\). In the strip geometry, \(\cA\) is the region between two parallel planes, separated by a distance \(\ell\) along a direction we denote \(x^1\). Choosing coordinates for \(d\)-dimensional flat space such that the metric is
\begin{equation} \label{eq:minkowski_metric}
    \diff s^2 = - \diff t^2 + \diff x^i \diff x^i,
\end{equation}
we take \(t=0\) as the Cauchy surface. The sphere and strip geometries are
\begin{align}
    \text{Sphere:}& \quad \cA = \le\{ x^i ~ | ~ x^i x^i \leq \ell^2 \ri\},
    \nonumber \\  
    \text{Strip:}& \quad \cA = \le\{x^i ~ | -\ell/2 \leq x^1 \leq \ell/2  \ri\}.
\end{align}
Both geometries are sketched in figure~\ref{fig:entangling_region_geometries}. We have chosen to work with the sphere and strip geometries because they are invariant under subgroups of the \(\SO(d-1,1)\) isometries of Minkowski space, simplifying calculations. 

Entanglement entropy in QFT is UV divergent, due to contributions from near the boundary of \(\cA\). We will regularise this divergence with a short distance cutoff \(\e\). The leading order divergence is typically proportional to the area of the boundary \(\p\cA\) of the entangling region~\cite{Bombelli:1986rw,Srednicki:1993im},
\begin{equation} \label{eq:area_law}
    \see \propto \frac{\mathrm{Area}[\p\cA]}{\e^{d-2}} + \dots \; ,
\end{equation}
where the ellipsis indicates terms which are subleading as \(\e \to 0\). When \(d=2\), there is instead a logarithmic divergence; the entanglement entropy of a connected subregion of length \(\ell\) in a 2D CFT of central charge \(c\) takes the universal form~\cite{Holzhey:1994we,Calabrese:2004eu}
\begin{equation}
    \see =  \frac{c}{3} \ln \le( \frac{\ell}{\e} \ri) + \cO(\e^0).
\end{equation}

The logarithm of \(\dMatrix_\cA\) appearing in~\eqref{eq:ee_definition} makes it difficult to calculate the entanglement entropy directly. However, computation of the logarithm can be avoided by a method known as the replica trick, which works as follows~\cite{Holzhey:1994we,Calabrese:2004eu}. Let us define the \(q\)th R\'enyi entropy~\cite{renyi1}
\begin{equation} \label{eq:renyi_definition}
    S_q \equiv \frac{1}{1-q} \ln \tr \dMatrix_\cA^q.
\end{equation}
In the limit \(q \to 1\), this reduces to the entanglement entropy~\eqref{eq:ee_definition}.\footnote{Other interesting limits of the R\'enyi entropies are \(q\to0\), which measures the number of non-zero eigenvalues of the reduced density matrix, and \(q\to\infty\), which measures the largest eigenvalue. See \cite{Hung:2011nu} for a detailed discussion on these limits in QFTs.} The trace appearing in \eqref{eq:renyi_definition} may be evaluated for integer \(q\) by performing a path integral on a manifold consisting of \(q\) copies of the original manifold, sewn together along the entangling region \(\cA\). One then analytically continues to non-integer \(q\) and takes the limit \(q\to1\) to obtain the entanglement entropy.

Of course, evaluating the path integral on the replicated manifold is not trivial, and will be intractable for strongly coupled theories. As we review in the next section, gauge/gravity duality provides a relatively simple alternative way to calculate entanglement entropy in holographic QFTs.

\section{Holographic calculation of entanglement entropy}

\subsection{The Ryu-Takayanagi prescription}

\begin{figure}
    \begin{center}
        \includegraphics{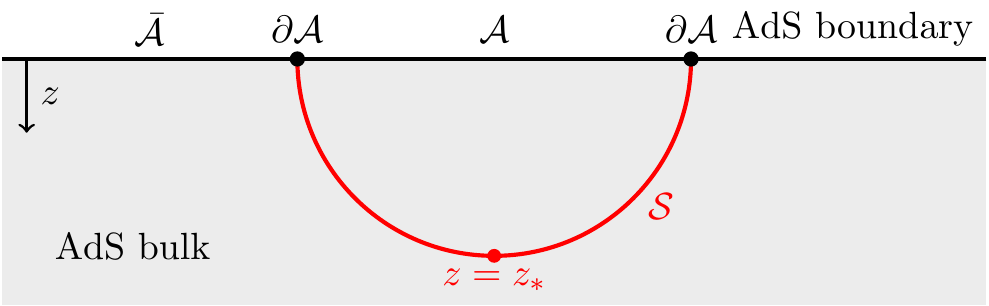}
    \end{center}
    \caption[The Ryu-Takayanagi prescription for holographic entanglement entropy.]{The entanglement entropy of a subregion \(\cA\) on the boundary of \ads\ is proportional to the area of a minimal area surface \(\cS\) homologous to \(\cA\) and anchored at \(\p\cA\), but extending into the bulk of \ads. We denote the maximal extension of the minimal surface into the bulk as \(z = z_*\).}
    \label{fig:rt_cartoon}
\end{figure}

The holographic dual of entanglement entropy in time-independent systems was proposed by Ryu and Takayanagi (RT) \cite{Ryu:2006bv,Ryu:2006ef}. Thinking of the QFT as living on the boundary of \ads, a partition into subspaces \(\cA\) and \(\bar\cA\) is a partition of a constant time slice of the boundary. To compute the entanglement entropy, one must find the bulk surface \(\cS\) of minimal area which is homologous to \(\cA\) and shares the same boundary, \(\p \cS  = \p \cA\). The entanglement entropy of \(\cA\) is proportional to the area of this surface,
\begin{equation} \label{eq:ryu_takayanagi_formula}
    \see[\cA] = \frac{\mathrm{Area}[\cS]}{4 \gn}.
\end{equation}
This formula is reminiscent of the Bekenstein-Hawking formula~\eqref{eq:bh_entropy} for the thermodynamic entropy of a black hole, with the area of the event horizon replaced by the area of the minimal surface \(\cS\). The various surfaces in this prescription are sketched in figure~\ref{fig:rt_cartoon}.

The RT prescription was proved by Lewkowycz and Maldacena (LM)~\cite{Lewkowycz:2013nqa}, assuming the validity of an analytic continuation which was used to implement the replica trick holographically, as described in section~\ref{sec:generalised_gravitational_entropy}. The generalisation of the RT prescription to time-dependent systems was proposed in ref.~\cite{Hubeny:2007xt}, and derived by a generalisation of the LM construction in ref.~\cite{Dong:2016hjy}.

The entanglement entropies for the sphere and strip geometries in a \(d\)-dimensional CFT holographically dual to \ads[d+1]  were computed in~\cite{Ryu:2006bv,Ryu:2006ef}. For the strip geometry, the result is
\begin{equation} \label{eq:pure_ads_strip_ee}
    \see^\mathrm{strip} = \frac{L^{d-1} \vol(\mathbb{R}^{d-2})}{4 (d-2) \gn} \le\{
        \frac{2}{\e^{d-2}} - \le[\frac{2 \sqrt {\pi} \G \le( \frac{d}{2d-2} \ri)}{\G \le( \frac{1}{2d-2} \ri)} \ri]^{d-1} \frac{1}{\ell^{d-2}}
    \ri\} + \dots \; ,
\end{equation}
where the ellipsis denotes terms which vanish as \(\e \to 0\). The factor of \(\vol(\mathbb{R}^{d-2})\) is the infinite surface area of one of the planar boundaries of the strip, which requires regularisation. The entanglement entropy for the sphere geometry is
\begin{equation} \label{eq:pure_ads_sphere_ee}
    \see^\mathrm{sphere} =
    \begin{cases}
    \displaystyle
        \frac{L^{d-1} \vol(\sph[d-2])}{4 \gn} \le[
            \sum_{m=1}^{d/2} p_m \le( \frac{\ell}{\e} \ri)^{d-2m} +  p_L \ln \le( \frac{\ell}{\e} \ri) + \dots
    \ri],~&d~\mathrm{even},
    \\[2em]
    \displaystyle
    \frac{L^{d-1} \vol(\sph[d-2])}{4 \gn} \le[
        \sum_{m=1}^{(d-1)/2} p_m \le( \frac{\ell}{\e} \ri)^{d-2m} +  \tilde{p}_{d/2} + \dots
    \ri],&d~\mathrm{odd},
    \end{cases} 
\end{equation}
where the ellipses denote terms which vanish as \(\e \to 0\), and
\begin{align} \label{eq:pure_ads_sphere_ee_coefficients}
    p_{m < d/2} &=  \frac{(-1)^{m + 1}}{d-2m} \frac{ \Gamma\le(\frac{d-1}{2}\ri)}{ \Gamma \le( m \ri)\Gamma \le( \frac{d + 1 -2m}{2} \ri) },
    &
    p_L & = - \frac{(-1)^{d/2}}{\sqrt{\pi}} \frac{ \Gamma\le( \frac{d-1}{2} \ri)}{ \Gamma \le( \frac{d}{2} \ri)},
    \nonumber \\
    p_{d/2} &= - \frac{(-1)^{d/2}}{2 \sqrt{\pi}} \frac{\G \le( \frac{d-1}{2} \ri)}{\G \le( \frac{d}{2} \ri) } \le[ \y(d/2) + \g_E + 2 \ln 2 \ri],
    &
    \tilde{p}_{d/2} &= - \frac{(-1)^{d/2} \sqrt{\pi} }{2} \frac{\G \le( \frac{d-1}{2} \ri)}{\G \le( \frac{d}{2} \ri) },
\end{align}
where \(\y\) is the digamma function, and \(\g_E\) is the Euler-Mascheroni constant. We note that the leading small-\(\e\) divergence in the entanglement entropies for the strip and the sphere may be written in the unified form
\begin{equation}
    \see = \frac{L^{d-1}}{4 (d-2) \gn} \frac{\mathrm{Area[\p \cA]}}{\e^{d-2}} + \dots \; ,
\end{equation}
providing an example of the area law~\eqref{eq:area_law}.

\subsection{The Casini-Huerta-Myers method}
\label{sec:casini_huerta_myers}

For a holographic CFT, there is an alternate method for calculating the entanglement entropy of a spherical geometry, due to Casini, Huerta, and Myers (CHM) \cite{Casini:2011kv}. This method was developed as a step toward proving the RT prescription, and is also a useful calculational tool. The prescription is as follows.

\begin{figure}
    \begin{center}
    \includegraphics{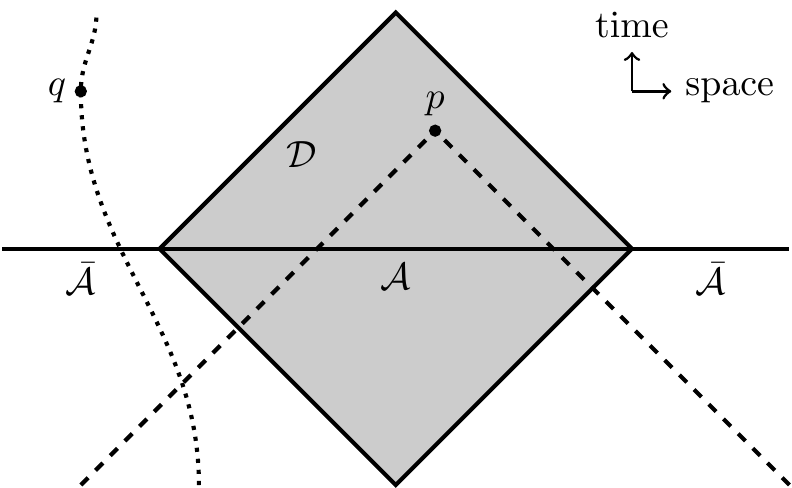}
    \end{center}
    \caption[Diagram of the causal development \(\cD\) of a spatial region.]{The grey diamond is the causal development \(\cD\) of the region \(\cA\). The dashed lines show the past light cone of a point \(p\) in \(\cD\). Clearly any causal curve that passes through \(p\) must intersect \(\cA\). The point \(q\) is not in \(\cD\) since there exist causal curves which pass through \(q\) but do not intersect \(\cA\), the dotted line being an example.}
    \label{fig:causal_development}
\end{figure}

The modular Hamiltonian \(H\) is a Hermitian operator, defined as
\begin{equation}
    H \equiv - \ln \dMatrix_\cA,
\end{equation}
which generates a symmetry of the subsystem \(\cA\),\footnote{The identities~\eqref{eq:modular_flow} and~\eqref{eq:modular_flow_periodicity} follow from the observation that \(\dMatrix_\cA = e^{-H}\) commutes with \(e^{-i H s}\).}
\begin{equation} \label{eq:modular_flow}
    \tr \le( \dMatrix_\cA  e^{-i H s} \cO e^{i H s} \ri) = \tr (\dMatrix_\cA \cO),
\end{equation}
where \(\cO\) is any operator defined on \(\cA\), and \(s \in \mathbb{R}\). Causality requires that the algebra of operators inside the causal development \(\cD\) of \(\cA\) is closed under such transformations \cite{Casini:2011kv}. The causal development of \(\cA\) is the set of spacetime points \(p\) such that any causal path through \(p\) necessarily intersects \(\cA\), see figure~\ref{fig:causal_development}.
The causal development for a spherical entangling region of radius \(\ell\) at \(t=0\) is the region satisfying
\begin{equation} \label{eq:sphere_causal_development}
   x^\pm \equiv  r \pm t \leq \ell,
\end{equation}
where we use a spherical coordinate system with radius \(r\). Under the modular flow~\eqref{eq:modular_flow}, correlation functions obey the periodicity relation
\begin{equation} \label{eq:modular_flow_periodicity}
    \tr \le[ \dMatrix_\cA \cO_1(s + i) \cO_2(s') \ri] = \tr \le[ \dMatrix_\cA \cO_2(s') \cO_1(s) \ri],
\end{equation}
where \(\cO_i(s) \equiv e^{-i H s} \cO_i e^{i H s}\).

In a CFT, for a spherical entangling region the modular Hamiltonian is a local operator. In this special case, the flow generated by \(e^{-i H s}\) is \(x^\pm \to x^\pm(s)\) with all other coordinates invariant, where~\cite{Casini:2011kv}
\begin{equation} \label{eq:modular_flow_2}
    x^{\pm}(s) = \ell \frac{(\ell + x^\pm) - e^{\mp 2\pi s} (  \ell - x^\pm)}{(\ell + x^\pm) + e^{\mp 2\pi s} (  \ell - x^\pm)}.
\end{equation}
The causal development~\eqref{eq:sphere_causal_development} is manifestly closed under this flow.

The coordinate transformation
\begin{equation} \label{eq:flat_to_hyperbolic_coordinate_transformation}
        t = \frac{\ell \sinh \t}{\cosh u + \cosh \t},
        \quad
        r = \frac{\ell \sinh u}{\cosh u + \cosh \t}.
\end{equation}
with \(\t \in (-\infty,\infty)\) and \(u \in [0,\infty)\) turns the Minkowski metric~\eqref{eq:minkowski_metric} (with \(x^i x^i = r^2\)) into the form
\begin{equation}
    \diff s^2 = \frac{\ell^2}{(\cosh u + \cosh \t)^2} \le(- \diff \t^2 + \diff u^2 + \sinh^2 u \, \diff s_{\sph[d-2]}^2 \ri).
\end{equation}
Removing the prefactor with a Weyl transformation turns this into the metric of \(\mathbb{R} \times \mathbb{H}^{d-1}\), where \(\mathbb{H}^{d-1}\) is \(d\)-dimensional hyperbolic space of unit radius. The right hand sides of~\eqref{eq:flat_to_hyperbolic_coordinate_transformation} imply \(x^\pm = \ell \tanh \le(\frac{u \pm \t}{2} \ri) \leq \ell\), so the hyperbolic space coordinates cover the causal development \(\cD\) of the spherical entangling region.

This conformal transformation to hyperbolic space maps the modular flow~\eqref{eq:modular_flow_2} to time translation, \(\t \to \t + 2\pi s\).  The periodicity relation~\eqref{eq:modular_flow_periodicity} shows that hyperbolic space correlation functions are periodic under imaginary time translations, \(\t \sim \t  + 2\pi i \), which is the Kubo-Martin-Schwinger condition of statistical mechanics (see ref.~\cite{Bellac:2011kqa}, for example). In hyperbolic space the CFT is therefore in a thermal state, at inverse temperature \(\b = 2\pi\), implying that the reduced density matrix may be written as
\begin{equation} \label{eq:chm_reduced_density_matrix}
    \dMatrix_\cA = \frac{1}{Z} U^{-1} e^{- 2\pi H_\t} U,
\end{equation}
where \(Z \equiv \tr  e^{-2\pi H_\t}\), \(H_\t\) is the Hamiltonian generating translations in \(\t\), and \(U\) is the unitary operator implementing the conformal transformation to hyperbolic space on states in \(\cA\).

Substituting the reduced density matrix~\eqref{eq:chm_reduced_density_matrix} into the entanglement entropy~\eqref{eq:ee_definition}, all factors of \(U\) cancel,
\begin{equation} \label{eq:entanglement_entropy_hyperbolic_space}
    \see = \frac{2\pi}{Z} \tr \le(H_\t e^{-2\pi H_\t} \ri) + \ln Z.
\end{equation}
The right hand side of~\eqref{eq:entanglement_entropy_hyperbolic_space} is precisely the thermodynamic entropy in the hyperbolic space at \(\b = 2\pi\), since \(\tr \le(H_\t e^{-2\pi H_\t} \ri)/Z\) is the energy and \(-\ln Z/2\pi\) is the free energy. We thus arrive at a key result of ref.~\cite{Casini:2011kv}: the entanglement entropy of a spherical region \(\cA\) in a CFT is equal to the thermodynamic entropy of that CFT on hyperbolic space at \(\b = 2\pi\).

In holography, the conformal transformation to hyperbolic space is implemented by a bulk diffeomorphism and a change in defining function. Starting from \ads[d+1] in flat slicing, with metric~\eqref{eq:ads_metric}, we perform the coordinate transformation
\begin{equation} \label{eq:hyperbolic_slicing_diffeomorphism}
    t = \O^{-1} \ell \sqrt{v^2 - 1} \sinh \t,
    \quad
    z = \O^{-1} \ell,
    \quad
    r = \O^{-1} \ell v \sinh u,
\end{equation}
where \(t=x^0\), \(r = \sqrt{x^i x^i}\), and  \(\O = v \cosh u + \sqrt{v^2 - 1} \cosh \t\). The new coordinates take values \(v \in [1,\infty)\), \(\t \in (-\infty, \infty)\) and \(u \in [0,\infty)\). The metric becomes
\begin{equation} \label{eq:hyperbolic_slicing_metric}
    \diff s^2 = L^2 \le( \frac{\diff v^2}{f(v)} - f(v) \diff \t^2 + v^2 \diff u^2 + v^2 \sinh^2 u \, \diff s_{\sph[d-2]}^2 \ri),
\end{equation}
where \(f(v) = v^2 - 1\). The boundary of \ads\ is approached by sending \(v \to \infty\) with \(\t\) and \(u\) fixed. This limit reduces the bulk diffeomorphism~\eqref{eq:hyperbolic_slicing_diffeomorphism} to the transformation~\eqref{eq:flat_to_hyperbolic_coordinate_transformation}, as desired. If we take the natural defining function in these coordinates, \(1/Lv\), then the boundary metric is precisely that of \(\mathbb{R} \times \mathbb{H}^{d-1}\).

The boundary of \ads\ may also be approached by fixing \(v\) and \(\t\), and sending \(u \to \infty\), which sends \((t,r) \to (0,\ell)\), or by fixing \(v\) and \(u\), and sending \(\t \to \pm\infty\), which sends \((t,r) \to (\pm \ell, 0)\). These are the corners of the causal development \(\cD\).

\begin{figure}
    \begin{subfigure}{0.6\textwidth}
        \includegraphics[width=\textwidth, trim={1cm 0 1cm 0}, clip]{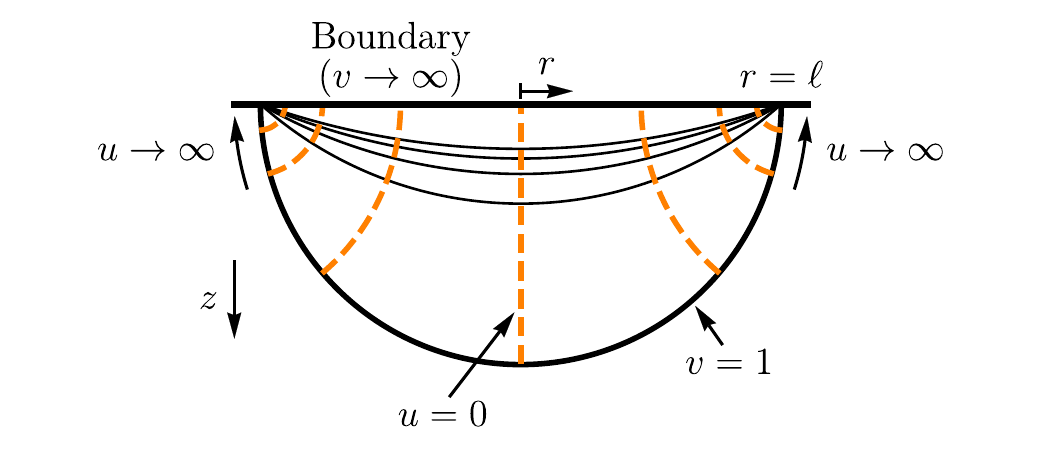}
    \end{subfigure}
    \begin{subfigure}{0.4\textwidth}
        \includegraphics[width=\textwidth]{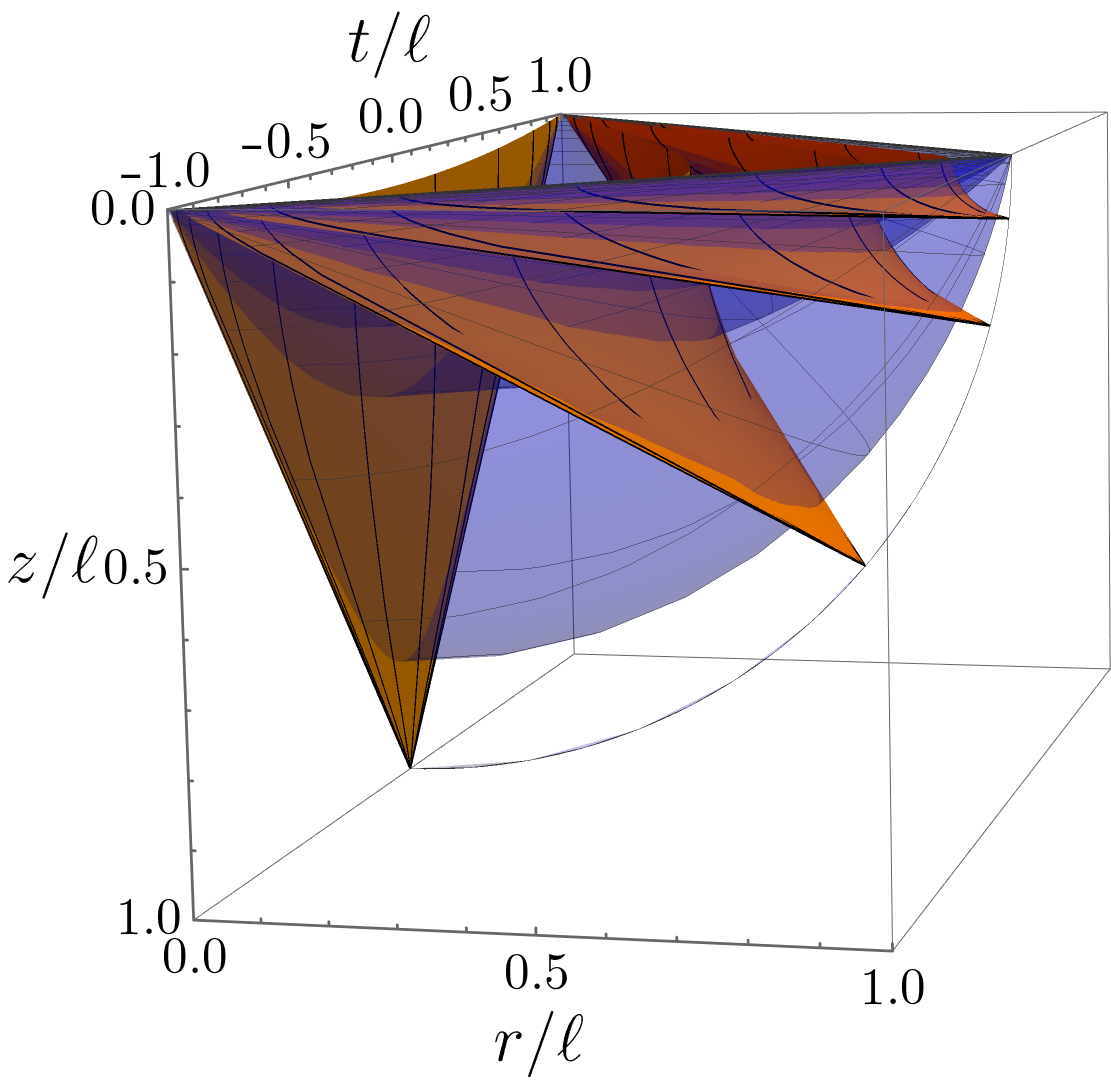}
    \end{subfigure}
    \caption[Cartoon of the hyperbolic slicing of \ads.]{How hyperbolic slices in \ads\ relate to flat slicing. The left shows the \((r,z)\) plane at \(\t = 0\), so \(t = 0\) by~\eqref{eq:hyperbolic_slicing_diffeomorphism}. The solid black curves are the hyperbolic slices (curves of constant \(v\)), the thickest being the horizon at \(v = 1\).  As \(v \to \infty\) the slices tend to the boundary of \ads. The dashed orange lines are curves of constant \(u\). The orange surfaces in the figure on the right are the slices of constant \(u\) in the space parameterised by \((t,r,z)\). The black curves on each slice are curves of constant \(\t\). The transparent blue surfaces are constant \(v\) hyperbolic slices. The black line which passes through the tip of each of the orange slices is the horizon at \(v=1\).}
    \label{fig:hyperbolic_slicing}
\end{figure}

The coordinate system~\eqref{eq:hyperbolic_slicing_metric} does not cover all of the Poincar\'e patch of \ads. The horizon at \(v = 1\) is part of the boundary of the covered coordinate patch. From~\eqref{eq:hyperbolic_slicing_diffeomorphism}, we see that in terms of the flat slicing coordinates the horizon is the surface \(r^2 + z^2 = \ell^2\) at \(t = 0\), which is precisely the RT surface for the sphere geometry~\cite{Ryu:2006bv,Ryu:2006ef}. The coordinate patch covered by the hyperbolic slicing is the region known as the entanglement wedge~\cite{Headrick:2014cta}, which in general is defined as the causal development of the region bounded by the RT surface at \(t = 0\). For the sphere geometry in pure \ads, this region is \(\sqrt{r^2 + z^2} \pm t \leq \ell\). Figure~\ref{fig:hyperbolic_slicing} illustrates how the hyperbolic slicing covers this coordinate patch.

As discussed above, the entanglement entropy of the spherical region in the boundary theory is given by the thermodynamic entropy in hyperbolic space. Holographically, the thermodynamic entropy is proportional to the area of the horizon at \(v=1\), according to the Bekenstein-Hawking formula~\eqref{eq:bh_entropy}. Since the horizon coincides with the RT surface, this proves the Ryu-Takayanagi prescription~\eqref{eq:ryu_takayanagi_formula} for the special case of the sphere geometry in a CFT.

The map to hyperbolic slicing also provides a convenient method for calculating the R\'enyi entropies \(S_q\), as follows~\cite{2011arXiv1102.2098B,Hung:2011nu}. Substituting the reduced density matrix~\eqref{eq:chm_reduced_density_matrix} into the R\'enyi entropy~\eqref{eq:renyi_definition} one finds again that all factors of \(U\) cancel,
\begin{equation}
    S_q = \frac{1}{1-q} \le[
        \ln \le(\tr e^{- 2\pi q H_\t}\ri) - q \ln \le( \tr e^{- 2\pi H_\t} \ri)
    \ri].
\end{equation}
Identifying \(\tr e^{- 2\pi q H_\t} = Z( 2\pi q)\) as the thermal partition function of the CFT in hyperbolic space at inverse temperature \(\b = 2\pi q\), we can rewrite the R\'enyi entropy as
\begin{equation} \label{eq:chm_renyi}
    S_q = \frac{2 \pi q}{1 - q} \le[ F (2\pi) - F(2\pi q) \ri],
\end{equation}
where \(F(\b) = - \b^{-1} \ln Z(\b)\) is the free energy in hyperbolic space at inverse temperature \(\b\).

To compute~\eqref{eq:chm_renyi} holographically, we require the geometry holographically dual to the CFT in hyperbolic space at arbitrary temperature. For the cases of interest in this thesis, the appropriate metric is given by~\eqref{eq:hyperbolic_slicing_metric}, with the more general metric function
\begin{equation} \label{eq:hyperbolic_metric_function}
    f(v) = v^2 - 1 - \frac{v_H^{d} - v_H^{d-2}}{v^{d-2}}.
\end{equation}
The metric~\eqref{eq:hyperbolic_slicing_metric} with this \(f(v)\) remains a solution the vacuum Einstein equations with negative cosmological constant. The horizon is now at a position \(v= v_H\), related to the inverse Hawking temperature \(\b\) by
\begin{equation} \label{eq:hyperbolic_horizon_position}
    v_H  = \frac{1}{d \b} \le(2\pi + \sqrt{ 4 \pi^2 + d(d-2) \b^2} \ri).
\end{equation}
Using the holographic dictionary in table~\ref{tab:holographic_dictionary}, the free energies appearing in~\eqref{eq:chm_renyi} are given by the on-shell action of this solution.

\subsubsection{Casini-Huerta-Myers and probe branes}

Consider the computation of entanglement entropy in a holographic model containing a probe brane. We expand the entanglement entropy in powers of the dimensionless tension \(\t_p\),
\begin{equation}
    \see = \see^{(0)} + \see^{(1)} + \dots \;,
\end{equation}
where \(\see^{(n)} \propto \t_p^n\). The first term, \(\see^{(0)}\), is the entanglement entropy computed in the absence of the probe brane. 

Suppose we wish to compute the leading correction, \(\see^{(1)}\). The RT prescription~\eqref{eq:ryu_takayanagi_formula} implies that \(\see^{(1)}\) is proportional to the leading order change in the area of the minimal surface \(\cS\) in the presence of the brane. Since the area of \(\cS\) is determined by the metric, naively one cannot compute \(\see^{(1)}\) without calculating the back-reaction of the brane --- often a difficult problem.

One way to compute the entanglement entropy without directly computing the back-reaction of the brane was found by the authors of ref.~\cite{Chang:2013mca}, who showed that \(\see^{(1)}\) is equal to a certain double integral over the world volume of the RT surface at \(\t_p=0\) and the world volume of the probe brane. However, this integral is not easy to compute, in general.

For probe branes dual to planar, conformal defects, the CHM method provides a convenient way to compute \(\see^{(1)}\) in the sphere geometry, without computing back-reaction or a double integral~\cite{Jensen:2013lxa}. The restriction that the defect is planar and conformal is necessary for the generalisation of the CHM proof in the presence of the defect. Holographically, these restrictions mean that the probe brane has an \ads[p+1] world volume inside \ads[d+1], and that the brane's world-volume fields do not depend on the radial coordinate in this \ads[p+1] subspace.

If these conditions hold, then the R\'enyi entropy contribution from the probe brane may be computed from~\eqref{eq:chm_renyi}. Expanding this formula in powers of the brane tension, and equating the \(\cO(\t_p)\) terms, we find
\begin{equation} \label{eq:probe_brane_renyi}
    S_q^{(1)} = \frac{2\pi q}{1-q} \le[ F^{(1)} (2\pi) - F^{(1)} (2\pi q) \ri],
\end{equation}
where as usual the superscripts denote powers of \(\t_p\). The right hand side depends only on the leading order contribution of the brane to the free energy, which we argued in section~\ref{sec:probe_branes} may usually be computed in the probe limit without needing the back-reaction of the brane. One may then compute the entanglement entropy by taking the \(q \to 1\) limit of~\eqref{eq:probe_brane_renyi}, or equivalently
\begin{equation} \label{eq:probe_brane_entropy}
    \see^{(1)} = \b^2 \le.\frac{\p F^{(1)}}{\p \b} \ri|_{\b = 2\pi}.
\end{equation}

While useful, this method suffers from the restriction that the probe brane must preserve defect conformal symmetry. In the next section we discuss how the method may be adapted to relax this condition~\cite{Karch:2014ufa}.

\subsection{Generalised gravitational entropy}
\label{sec:generalised_gravitational_entropy}

Lewkowycz and Maldacena have developed a quantity called generalised gravitational entropy~\cite{Lewkowycz:2013nqa}. Consider a gravitational theory on a Euclidean-signature manifold with a boundary. We suppose that the boundary geometry has a direction which is topologically a circle, parameterised by a coordinate \(\t \sim \t + 2\pi\), and that the boundary conditions respect this periodicity.

Now increase the period of the circle to \(\t \sim \t + 2\pi q\) where \(q \in \mathbb{Z}\), with the boundary conditions respecting the original periodicity \(\t \sim \t + 2\pi\). Let \(Z(q)\) be the generating functional for the solution with \(\t \sim \t + 2\pi q\). The generalised gravitational entropy is defined by
\begin{equation} \label{eq:generalised_gravitational_entropy}
    S_\mathrm{G} = \lim_{q \to 1} q \p_q \le[I_\mathrm{grav}^\star(q) - q I_\mathrm{grav}^\star(1) \ri],
\end{equation}
where it is assumed that it is possible to continue \(I_\mathrm{grav}^\star\) to non-integer \(q\).

Specialising to static \aads\ spacetime, let us assume there is a circle that wraps the boundary \(\p\cA\) of a subregion \(\cA\) of the \aads\ boundary. In this case, increasing the period by a factor \(q\) is effectively a holographic implementation of the replica trick, replacing the original boundary of \aads\ with \(q\) copies of itself sewn together along \(\cA\). The generating functional is therefore related to the reduced density matrix by \(Z(q) = \tr \dMatrix_\cA^q\), and the generalised gravitational entropy~\eqref{eq:generalised_gravitational_entropy} is equal to the entanglement entropy \(\see\) of the subregion \(\cA\) in the dual QFT.

In order to evaluate~\eqref{eq:generalised_gravitational_entropy}, we require a prescription to perform the continuation to non-integer \(q\). A convenient one is as follows. For integer \(q\), we demanded that the boundary conditions were invariant under \(\t \to \t + 2\pi\), implying the existence of a \(\mathbb{Z}_q\) replica symmetry under cyclic permutations of the \(q\) arcs of length \(2\pi\) which make up the \(\t\) circle. Assuming this replica symmetry is unbroken, we may therefore write \(\ln Z(q) = q [\ln Z(q)]_{2\pi}\), where \([\cdot]_{2\pi}\)  indicates that \(\t\) is integrated only over the range \([0,2\pi]\). If we take this to be true also for non-integer \(q\), then we may rewrite~\eqref{eq:generalised_gravitational_entropy} as
\begin{equation} \label{eq:generalised_gravitational_entropy_alt}
    \see = \lim_{q \to 1} q^2 \p_q [I_\mathrm{grav}^\star(q)]_{2\pi},
\end{equation}
where we have taken the saddle-point approximation \(Z(q) = \exp[-I_\mathrm{grav.}^\star(q)]\). It was shown in ref.~\cite{Lewkowycz:2013nqa} that~\eqref{eq:generalised_gravitational_entropy} reduces to the area of a minimal surface which intersects the \aads\ boundary at \(\p\cA\). This amounts to a proof of the RT prescription, subject to the validity of the continuation to non-integer \(q\).

Generalised gravitational entropy was adapted to probe branes in ref.~\cite{Karch:2014ufa}. Since the formula~\eqref{eq:generalised_gravitational_entropy} depends only on the on-shell action, we expect that the leading non-trivial term in the probe limit may be calculated without computing the back-reaction of the brane. Expanding in powers of the tension, we find
\begin{equation} \label{eq:probe_generalised_gravitational_entropy}
    \see^{(1)} = \lim_{q \to 1} q^2 \p_q [I_\mathrm{brane}(q)]_{2\pi}
\end{equation}

\begin{figure}
    \begin{center}
    \includegraphics[width=0.5\textwidth]{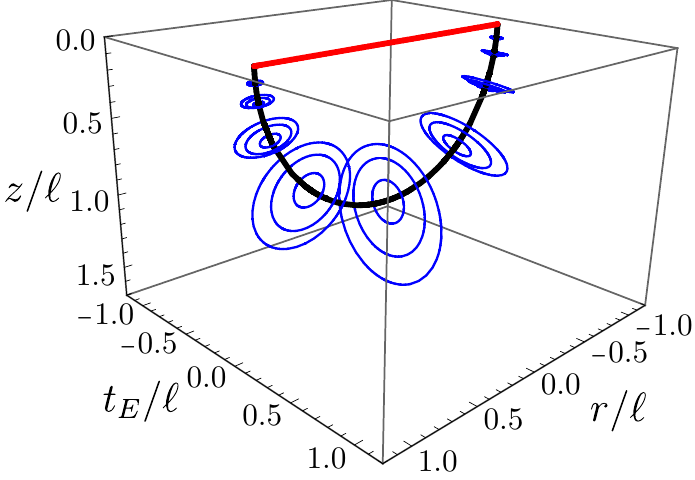}
    \caption[\ads\ in flat and hyperbolic slicing in Euclidean signature.]{The relationship between flat and hyperbolic slicing of \ads\ in Euclidean signature, with \(t_E\) the Euclidean time. The red line is the entangling region \(\cA\), the black line is the horizon at \(v=1\), or equivalently the Ryu-Takayanagi surface. The blue circles are curves of constant \(u\) and \(v\), parameterised by \(\t\), which degenerate at the horizon.}
    \label{fig:hyperbolic_slicing_euclidean}
    \end{center}
\end{figure}
To compute \eqref{eq:probe_generalised_gravitational_entropy} we need to know the gravitational solution at arbitrary \(q\). For a probe brane embedded in pure \ads\ with  spherical entangling region, this solution is provided by the map to hyperbolic slicing~\eqref{eq:hyperbolic_slicing_diffeomorphism}, Wick rotated to Euclidean signature \(\t \to i \t\)~\cite{Karch:2014ufa}. In Euclidean signature, the hyperbolic slicing covers the whole of local \ads. The Euclidean \(\t\) is periodic, \(\t \sim \t + 2\pi\), with the \(\t\) circles winding the horizon as illustrated in figure~\ref{fig:hyperbolic_slicing_euclidean}. To increase the period to \(\b = 2\pi q\), one changes the metric function \(f(v)\) to that given in~\eqref{eq:hyperbolic_metric_function}, with \(v_H\) given by~\eqref{eq:hyperbolic_horizon_position}. The entanglement entropy then becomes
\begin{equation} \label{eq:probe_generalised_gravitational_entropy_beta}
    \see^{(1)} = 2\pi \lim_{\b \to 2\pi} \p_\b [I^\star_\mathrm{brane}(\b)]_{2\pi}.
\end{equation}

We will make use of this method in chapter~\ref{chap:probe_m5}. In several of the examples in that chapter, we will only know the brane solution at \(\b = 2\pi\). We cannot then analytically compute the on-shell action \(I^\star_\mathrm{brane}\) as a function of \(\b\). However, we may still calculate the entanglement entropy using~\eqref{eq:probe_generalised_gravitational_entropy_beta} by making use the observation of ref.~\cite{Kumar:2017vjv} that the first variation of the action with respect to the fields vanishes when the fields satisfy the classical equations of motion, and therefore we may choose to take the embedding on-shell only after performing the derivative in \eqref{eq:probe_generalised_gravitational_entropy_beta}.

As a concrete example, consider a probe \(p\)-brane with a Dirac-Nambu-Goto action
\begin{equation} \label{eq:example_brane_action}
    I_\mathrm{brane} = \mathcal{T}_p \int \diff \t \int \diff^{p} \xi \sqrt{-\det {G_{MN} \p_a X^M \p_b X^N}}
\end{equation}
with tension \(\mathcal{T}_p\), where the background space has metric \(G\), and \(X\) specifies the embedding of the brane. In~\eqref{eq:example_brane_action} we have chosen one of the coordinates on the brane to be \(\t\), with \(\xi\) denoting the remaining world volume coordinates. In this example, the derivative with respect to inverse temperature appearing in~\eqref{eq:probe_generalised_gravitational_entropy_beta} is
\begin{equation}
    \p_\b [I_\mathrm{brane}^\star]_{2\pi} = \int_0^{2\pi} \diff \t \int \diff^{p}\xi \le[
       \le(\frac{\d I_\mathrm{brane}}{\d G_{MN}}\ri)^\star \p_\b G_{MN} + \le(\frac{\d I_\mathrm{brane}}{\d X^M} \ri)^\star \p_\b X^M 
    \ri].
\end{equation}
 Since the action is on shell, the functional derivative with respect to \(X^M\) vanishes by the equations of motion. Hence, the only contribution to the \(\b\) derivative comes from the temperature dependence of the background metric, or more generally the full set of background fields. As a result, we may use the \(\b = 2\pi\) form of the \(X^M\), or more generally all world volume fields, throughout the calculation of the generalised gravitational entropy~\eqref{eq:probe_generalised_gravitational_entropy_beta}.

\part{Research} \label{part:research}

\chapter{Entanglement density}
\label{chap:entanglement_density}

\section{Introduction}
\label{intro}

The scaling of entanglement entropy with subregion size \(\ell\) provides a diagnostic tool for characterising the state of matter. For example, in \(d\)-dimensional spacetime, entanglement entropy receives contributions \(\propto N_G \ln \ell^{d-2}\) from \(N_G\) Goldstone bosons~\cite{Metlitski:2011pr}, \(\propto \ell^{d-2} \ln \ell\) from a Fermi surface~\cite{Wolf:2006zzb,2006PhRvL..96j0503G,Swingle:2009bf,Swingle:2010yi}, or independent of \(\ell\) from topologically-ordered degrees of freedom~\cite{2006PhRvL..96k0404K,2006PhRvL..96k0405L,2008PhRvB..78o5120C,2011PhRvB..84s5120G}.

The goal of this chapter is to use gauge/gravity duality to study the dependence of entanglement entropy on \(\ell\) in holographic CFTs with a variety of deformations.\footnote{The research in this chapter was conducted in collaboration with Nikola I. Gushterov (in addition to my supervisor, Andy O'Bannon), and was also submitted as part of his PhD thesis at the University of Oxford. My primary contributions were the calculation of the numerical results that are presented in this chapter and the next, and the analytic calculations in appendix~\ref{app:entanglement_density}.} We use holography since it provides a simple way to calculate entanglement entropy in interacting QFTs, allowing us to study a wide range of deformations. 

\subsection{Definition of entanglement density}
\label{sec:entanglement_density}

As discussed in chapter~\ref{chap:entanglement_entropy}, entanglement entropy in QFT is generally UV divergent. In order to obtain a finite quantity characterising our deformed CFTs, we will subtract the entanglement entropy \(\see^\mathrm{CFT}\) of the undeformed CFT.  In order 
to highlight the qualitative features of the subtracted entanglement entropy, we find it useful to divide by the volume \(V\) of the entangling region, defining the \textit{entanglement density}\footnote{A different quantity, also called entanglement density, was defined in refs.~\cite{Nozaki:2013wia,Bhattacharya:2014vja} as a second variation of the entanglement entropy with respect to the boundary of the entangling region.} 
\begin{equation} \label{eq:eddef}
	\sigma\equiv\frac{\see-\see^{\mathrm{CFT}}}{V}.
\end{equation}

We could also have chosen to remove the UV divergences using covariant counterterms~\cite{Taylor:2016aoi,Taylor:2017zzo}. Our subtraction is motivated by the definition of entanglement temperature~\cite{Bhattacharya:2012mi,Blanco:2013joa}. Under a small change in the state of a QFT, the change in the entanglement entropy of a spherical subregion, or a strip in a holographic QFT, is proportional to the change in the energy contained in the entangling region~\cite{Bhattacharya:2012mi,Blanco:2013joa}. If the initial and final states are translationally invariant, the change in energy is \(V \d \vev{T_{tt}}\), where \(\d \vev{T_{tt}}\) is the small change in the expectation value of the stress tensor, so the change in entanglement entropy is
\begin{equation}
	\d \see = \tent^{-1} V \d\vev{T_{tt}},
\end{equation}
where by definition the proportionality coefficient \(\tent\) is the entanglement temperature. For a small perturbation in the state, the entanglement density~\eqref{eq:eddef} reduces to \(\tent^{-1} \d\vev{T_{tt}}\), but it is also well defined for large changes in the state, and for changes in the Hamiltonian of the QFT.

Our goal is to characterise deformed CFTs using the dependence of the entanglement density on subregion size \(\ell\). We will consider \(d\)-dimensional QFTs with $d\geq 3$, and use both the strip and sphere geometries for the subregions. In section~\ref{general} we derive formulae for the holographic entanglement density in an asymptotically \ads[d+1] metric of a general form that encompasses all but one of our later examples. In the subsequent sections we numerically solve for the entanglement density as a function of \(\ell\).

For both the sphere and the strip, \(\tent \propto \ell^{-1}\). Hence, we find that \(\s \propto \ell \vev{T_{tt}}\) at small \(\ell\) for deformations of the state, such as non-zero temperature. For deformations of the Hamiltonian, the small-\(\ell\) behaviour of the entanglement density is determined by the dimension $\Delta$ of the perturbing operator, as we discuss in section~\ref{rg}.

As $\ell \to \infty$ compared to all other scales, the leading behaviour of the subtracted entanglement entropy is\footnote{Some deformations can also introduce a term $\propto A \ln A$ in the subtracted entanglement entropy. One example is a chemical potential $\mu$ in a free fermion CFT, which produces a Fermi surface~\cite{Wolf:2006zzb,2006PhRvL..96j0503G,Swingle:2009bf,Swingle:2010yi} resulting in a logarithmic term in the entanglement entropy as mentioned above. For discussions about the conditions under which such ``area law violation'' can occur, see for example ref.~\cite{Swingle:2011np}.}
\begin{equation}
\label{eq:EElargeL}
\see - \see^\mathrm{CFT} = s  V + \a A + \dots,
\end{equation}
where $s$ is the thermodynamic entropy density (which vanishes in some of our examples), \(A\) is the surface area of the entangling region, $\a$ is a dimensionful constant, and the ellipsis represents terms subleading in \(1/\ell\).

The leading volume term $\propto V$ in~\eqref{eq:EElargeL} is expected for mixed states, such as thermal states. In such cases, when $\ell \to \infty$ the subregion becomes the entire system, and the reduced density matrix \(\dMatrix_\cA\) becomes the total density matrix \(\dMatrix\). Hence the entanglement entropy becomes equal to the thermodynamic entropy. In holography, this occurs because the RT surface lies along a horizon at large \(\ell\)~\cite{Hubeny:2012ry,Liu:2013una}, so that after subtracting UV divergences, the dominant contribution to its area comes from the area of the horizon. For a sphere, $V \propto \ell^{d-1}$ while for the strip, $V \propto\vol\left(\mathbb{R}^{d-2}\right) \ell$, where $\vol\left(\mathbb{R}^{d-2}\right)$ is the regularised area of one of the planar boundaries of the strip.

For Lorentz-invariant RG flows to a $d$-dimensional CFT in the IR, $\a$ obeys a weak monotonicity theorem, called the \textit{area theorem}~\cite{Casini:2012ei,Casini:2016udt}, which implies that \(\a \leq 0\). The area theorem has been proven in $d=3$ using strong subadditivity~\cite{Casini:2012ei} and for a sphere in $d \geq 3$ using positivity of relative entropy~\cite{Casini:2016udt}.\footnote{Strong subadditivity is the inequality \(\see[\cA] + \see[\cB] \geq \see[\cA \cup \cB] + \see[\cA \cap \cB]\), for subspaces \(\cA\) and \(\cB\). The relative entropy between two density matrices \(\r_1\) and \(\r_2\) is \(S(\r_1|\r_2) = \tr (\r_1 \ln \r_1 - \r_1 \ln \r_0)\). Positivity of relative entropy is \(S(\r_1 | \r_2) \geq 0\).} Roughly speaking, strong subadditivity is holographically dual to the null energy condition~\cite{Headrick:2007km,Wall:2012uf}. All of our holographic examples will obey the null energy condition.\footnote{A different monotonicity theorem for the entanglement entropy in holographic RG flows appears in refs.~\cite{Ryu:2006ef,Myers:2012ed}, also following from the null energy condition.}

As discussed in section~\ref{sec:weyl_anomaly}, quantities satisfying monotonicity theorems typically count degrees of freedom. It is thus natural to ask whether \(\a\) counts degrees of freedom in any precise sense. Further, one may wonder whether the monotonicity extends to other types of deformations, such as finite temperature or chemical potential, or to non-Lorentz invariant RG flows~\cite{Swingle:2013zla}. We will use entanglement density to study some of these issues in holographic systems. 

From~\eqref{eq:EElargeL}, we find that when \(\ell\) is much larger than all other scales, the entanglement density satisfies
\begin{equation}
\label{eq:EDlargeL}
\s = s + \a \frac{A}{V} + \ldots \; .
\end{equation}
For both the sphere and strip $A/V \propto 1/\ell$. In section~\ref{general} we use techniques from refs.~\cite{Liu:2012eea,Liu:2013una,Kundu:2016dyk} to show that for a holographic QFT dual to a geometry with a horizon, the entanglement density indeed takes the form~\eqref{eq:EDlargeL} for both the sphere and the strip, with the same coefficient \(\a\) for both geometries.

Equation~\eqref{eq:EDlargeL} shows that we can easily read off the sign of $\a$ from the large-\(\ell\) behaviour of the entanglement density. In section~\ref{general} we find a formula for the coefficient of the $1/\ell$ correction as an integral over bulk metric components, which typically must be performed numerically, making it particularly easy to extract the sign of \(\a\).

\section{General Analysis}
\label{general}

Except in section~\ref{soliton}, we will use the symmetries of translations in time and translations and rotations in space to write the metric of \aads[d+1] in the form
\begin{equation} \label{eq:aads_metric}
	\diff s^{2} = G_{MN} \diff X^M \diff X^N = \frac{L^{2}}{z^{2}}\left[-f(z) \diff t^{2}+ \d_{ij} \diff x^i \diff x^j +\frac{\diff z^{2}}{g(z)}\right].
\end{equation}
As $z \to 0$, the metric~\eqref{eq:aads_metric} asymptotically approaches the metric~\eqref{eq:ads_metric} of \ads[d+1] with radius \(L\). In several of our examples, the \(z\to0\) asymptotics of the functions in~\eqref{eq:aads_metric} are
\begin{equation}	\label{eq:fgasymp}
	f(z) = 1 - m z^d + \ldots,
	\qquad
	g(z) = 1 - m z^d + \ldots,
\end{equation}
where $m$ is a constant and the ellipses represent powers of $z$ that vanish faster than those shown as $z \to 0$. The energy density of the dual QFT is then~\cite{Balasubramanian:1999re,deHaro:2000xn}
\begin{equation} \label{eq:energy_density}
	\langle T_{tt} \rangle=\frac{(d-1)L^{d-1}}{16\pi \gn}\, m.
\end{equation}

The exceptions to the behaviour~\eqref{eq:fgasymp} are the holographic RG flows studied in sections~\ref{rg} and~\ref{hyper}. Each RG flow will be triggered by a non-zero source or vacuum expectation value (VEV) of a relevant operator, and there may be smaller powers of \(z\) appearing in the near boundary expansions~\eqref{eq:fgasymp}, depending on the dimension of the operator.

In some of our examples there is a horizon at \(z=z_H > 0\), such that \(f(z_H) = 0\). As discussed in~\ref{sec:thermodynamics}, the temperature \(T\) and thermodynamic entropy density \(s\) are determined by the horizon's Hawking temperature and Bekenstein-Hawking entropy, respectively. In the coordinate system~\eqref{eq:aads_metric}, one finds
\begin{equation}
	\label{eq:thermo_entropy}
	T = \frac{\sqrt{f'(z_H)g'(z_H)}}{4\pi}, \qquad s = \frac{L^{d-1}}{4 \gn}\frac{1}{z_H^{d-1}}.
\end{equation}

In the next two subsections we derive formulae for the entanglement density in the sphere and strip geometries, for the holographic duals of spacetimes of the form~\eqref{eq:aads_metric}, using the Ryu-Takayanagi prescription. The results will be integrals over the metric function \(g(z)\). The details of the different theories giving rise to our different examples will enter into the entanglement density through the form of this function.

\subsection{Strip geometry}

Recall that in the strip geometry the entangling region is enclosed by parallel planes at \(x^1 = \pm \ell/2\). To reduce notation we will drop the superscript on \(x^1\) from now on. Using the translational and rotational symmetry of the strip we can parameterise the minimal surface as $x(z)$. As depicted in figure~\ref{fig:rt_cartoon}, the minimal surface will have some maximal extension into the bulk at some \(z = z_*\). By symmetry, \(x(z_*) = 0\), since the strip is invariant under \(x \to -x\). The area functional is then
\begin{equation} \label{eq:strip_area_functional}
		\mathrm{Area}[\cS] = 2 L^{d-1} \vol(\mathbb{R}^{d-2}) \int_\e^{z_*} \diff z \frac{1}{z^{d-1}} \sqrt{\frac{1}{g(z)} + x'(z)^2 },
\end{equation}
where \(\e\) is a short-distance cutoff. The factor of \(\vol(\mathbb{R}^{d-2})\) arises from the integrals over the \(x^i\) directions with \(i > 1\). This is an infrared divergence which will cancel from the entanglement density.

Since the area~\eqref{eq:strip_area_functional} depends only on $x'(z)$, there is a conserved quantity, the value of which is fixed in terms of \(z_*\) by requiring that \(\lim_{z\to z_*} x'(z) = \infty\). Solving for \(x'(z)\), one finds
\begin{equation}
		x'(z) = \pm \le( \frac{z}{z_*} \ri)^{d-1} \frac{1}{\sqrt{1 - (z/z_*)^{2d-2}}} \frac{1}{\sqrt{g(z)}} .
\end{equation}

Substituting this into the area functional and applying the RT formula~\eqref{eq:ryu_takayanagi_formula}, we obtain the entanglement entropy
\begin{equation} \label{eq:strip_area}
	\see^\mathrm{strip} = \frac{L^{d-1} \vol(\mathbb{R}^{d-2})}{2\gn} \int_\e^{z_*} \frac{\diff z}{z^{d-1}} \frac{1}{\sqrt{1 - (z/z_*)^{2d-2}}} \frac{1}{\sqrt{g(z)}}.
\end{equation}
The turning point \(z_*\) is implicitly determined by the width of the strip through
\begin{equation} \label{eq:ldef}
	\ell = 2 \int_0^{z_*} \diff z \, x'(z) = 2 \int_0^{z_*} \diff z\le( \frac{z}{z_*} \ri)^{d-1} \frac{1}{\sqrt{1 - (z/z_*)^{2d-2}}} \frac{1}{\sqrt{g(z)}},
\end{equation}
where we have used~\eqref{eq:strip_area}. For \ads[d+1], where $g(z)=1$, the integrals in~\eqref{eq:strip_area} and~\eqref{eq:ldef} may be performed exactly, leading to the result given in~\eqref{eq:pure_ads_strip_ee}. For non-trivial \(g(z)\) we perform the integrals numerically.

To evaluate the entanglement density~\eqref{eq:eddef}, we must subtract the pure \ads[d+1] result~\eqref{eq:pure_ads_strip_ee} and divide by the volume of the strip, \(V = \ell \, \vol(\mathbb{R}^{d-2})\). To avoid numerical imprecision caused by the subtraction of two UV-divergent quantities, we rewrite the term depending on \(\e\) in~\eqref{eq:pure_ads_strip_ee} as an integral over \(z\), yielding
\begin{multline} \label{eq:strip_sigma}
		\s_\mathrm{strip} = \frac{L^{d-1}}{2 \gn \ell} \Biggl\{
				\int_\e^{z_*} \frac{\diff z}{z^{d-1}} \le( \frac{1}{\sqrt{1 - (z/z_*)^{2d-2} } }  \frac{1}{\sqrt{g(z)}} - 1 \ri)
				\\
				- \frac{1}{(d-2) z_*^{d-2}} + \frac{1}{2(d-2)} \le[\frac{2 \sqrt{\pi} \G\le( \frac{d}{2d-2} \ri) }{\G \le( \frac{1}{2d-2} \ri)} \ri]^{d-1} \frac{1}{\ell^{d-2}}
		\Biggr\}.
\end{multline}
For all the cases that we study, the integrand on the first line of~\eqref{eq:strip_sigma} remains finite as \(\e \to 0\), so is well-adapted to numerical computation.

\subsection{Sphere geometry}

For the sphere subregion we first write $\d_{ij} \diff x^i \diff x^j = \diff r^2 + r^2 \diff s_{\sph[d-2]}^2$, where \(r = \sqrt{x^i x^i}\)  and $\diff s_{\sph[d-2]}^2$ is the metric of a round, unit-radius \((d-2)\)-sphere. We use rotational symmetry to parameterise the minimal surface as \(r(z)\). The resulting area functional is
\begin{equation}
\label{eq:sphere_area}
\mathrm{Area}[\cS] = L^{d-1}\mathrm{Vol}(\sph[d-2]) \int_{\e}^{z_{*}} \diff z  \frac{r(z)^{d-2}}{z^{d-1}}\sqrt{\frac{1}{g(z)} + r'(z)^{2}},
\end{equation}
where \(\mathrm{Vol}(\sph[d-2])\) is the volume of \sph[d-2]. Extremising the area leads to a non-linear second order ordinary differential equation for $r(z)$,
\begin{equation} \label{eq:sphere_equation_of_motion}
		r''(z) - \le[\frac{d-2}{g(z)r(z)} + \frac{(d-1)r'(z)}{z}  \ri] \le[ 1 + g(z) r'(z)^2 \ri] + \frac{g'(z)r'(z)}{2 g(z)} = 0.
\end{equation}
For \ads[d+1], where \(g(z) = 1\), the exact solution of~\eqref{eq:sphere_equation_of_motion} is \(r(z) = \sqrt{\ell^2 - z^2}\), and the area may be evaluated explicitly to obtain~\eqref{eq:pure_ads_sphere_ee}. For non-trivial \(g(z)\) we solve this equation numerically, and substitute the result into the area integral~\eqref{eq:sphere_area}. The integral is then also evaluated numerically to obtain the entanglement entropy.

We compute the entanglement density~\eqref{eq:eddef} by subtracting the \ads\ result and dividing by the volume of the entangling region, \(V = \ell^{d-1} \vol(\sph[d-2])/d\),
\begin{equation} \label{eq:sphere_sigma}
	\s_\mathrm{sphere} = \frac{d L^{d-1}}{4 \ell^{d-1} \gn} \int_{\e}^{z_{*}} \diff z  \frac{r(z)^{d-2}}{z^{d-1}}\sqrt{\frac{1}{g(z)} + r'(z)^{2}} - \frac{d}{4 \gn \ell^{d-1} \vol(\sph[d-2])} \see^{\mathrm{CFT},\,\mathrm{sphere}},
\end{equation}
where \(r(z)\) is the solution to the equation of motion~\eqref{eq:sphere_equation_of_motion}, and \(\see^{\mathrm{CFT},\,\mathrm{sphere}}\) is the pure \ads\ result~\eqref{eq:pure_ads_sphere_ee}. As for the strip, we rewrite the divergent terms in the pure \ads\ result as integrals over \(z\),  and combine them with the integral in~\eqref{eq:sphere_sigma} to obtain a UV-finite integral suitable for numerical evaluation. The explicit form of the subtracted integrand is cumbersome, so we do not reproduce it here.

\subsection{Asymptotic behaviour of the entanglement density}
\label{sec:entanglement_asymptotics}

\subsubsection{Small subregions}

In this section we derive results for the behaviour of the entanglement density for subregion size \(\ell\) small compared to all other scales in theory apart from the UV cutoff. We follow the method of ref.~\cite{Bhattacharya:2012mi}.

\begin{figure}
	\begin{center}
		\includegraphics{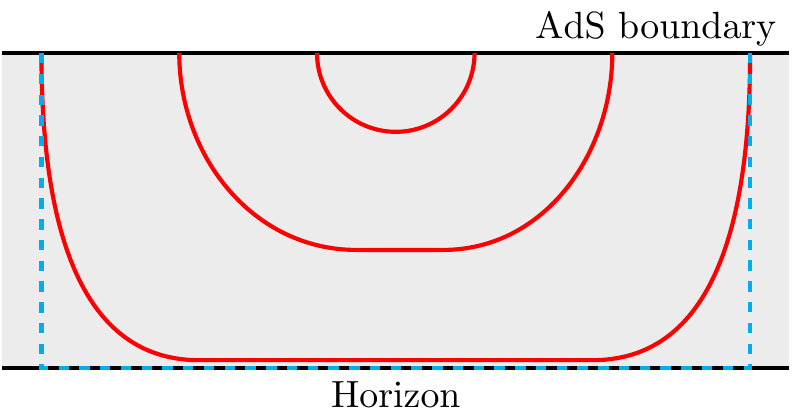}
	\end{center}
	\caption[Ryu-Takayanagi surfaces in the presence of a horizon.]{Schematic depiction of how the Ryu-Takayanagi (RT) surface changes shape with the subregion size in the presence of a planar horizon. The grey region is the bulk of \ads, with the boundary and the horizon represented by the black lines at the top and bottom, respectively. The red curves are typical shapes of the RT surface for different subregion sizes. For small subregions, the RT surface does not probe deep into the bulk, and so is insensitive to the presence of the horizon. For large subregions, part of the RT surface almost lies flat along the horizon. The dashed blue line is a surface that drops straight from the boundary to the horizon. In our examples, we find that such a surface is an extremum, but not a global minimum of the area functional.}
	\label{fig:minimal_surface_horizon}
\end{figure}

Suppose the metric function \(g(z)\) takes the form
\begin{equation} \label{eq:fgasymp_general}
		g(z) \approx 1 + a z^b + \dots \;
\end{equation}
at small \(z\), where \(b > 0\) and the ellipsis denotes terms with larger powers of \(z\). When the entangling region is small, the RT surface will remain close to the boundary rather than probing deep into the bulk, as depicted in figure~\ref{fig:minimal_surface_horizon}. Therefore, in this subsection we assume that \(a z_*^b \ll 1\) for sufficiently small \(\ell\), where \(z_*\) is the maximal extension of the RT surface into the bulk.

Expanding~\eqref{eq:ldef} and~\eqref{eq:strip_sigma} for small values of \(a z^b\), we find
\begin{subequations}
	\begin{align}
		\ell &= 
				 \varrho z_* 
				- \frac{
					\varrho' a z_*^{b+1}
				}{
					2(b+1)
				} 
			+ \dots \; ,
			\label{eq:strip_width_small_l_expansion}
			\\
			\s_\mathrm{strip} &= \frac{L^{d-1}}{4 (d-2) \gn \ell} \le(
				\frac{\varrho^{d-1}}{\ell^{d-2}} -  \frac{\varrho}{z_*^{d-2}} - 
				\frac{d-2}{2(b+2-d)} \varrho' a z_*^{b+2-d}
			\ri)  + \dots \;,
			\label{eq:strip_sigma_small_l_expansion}
	\end{align}
\end{subequations}
where the ellipses denote terms of higher order in \(a z_*^b\), and 
\begin{equation} \label{eq:small_strip_coefficients}
	\varrho \equiv \frac{
			2\sqrt{\pi} \G\le( \frac{d}{2d-2} \ri)
	 }{
			\G \le( \frac{1}{2d-2} \ri)
	 },
	 \qquad
	 \varrho' \equiv \frac{
			2 \sqrt{\pi} \G\le( \frac{d+b}{2d-2} \ri)
 		}{
			\G \le( \frac{b+1}{2d-2} \ri)
 		}.
\end{equation}

Inverting the expansion appearing in~\eqref{eq:strip_width_small_l_expansion}, we find
\begin{equation}
		z_* = 
			\frac{\ell}{\varrho} + \frac{\varrho' a \ell^{b+1}}{2 (b+1) \varrho^{b+2}} + \dots\;,
\end{equation}
which may be substituted into the entanglement density~\eqref{eq:strip_sigma_small_l_expansion} to give the strip entanglement density at small \(\ell\),
\begin{equation} \label{eq:strip_sigma_small_l_expansion_final}
		\s_\mathrm{strip} = -\frac{L^{d-1}(d-1) \varrho'}{8 \gn (b+1)(b+2-d) \varrho^{b+2-d}} a \ell^{b + 1 - d} + \ldots \; .
\end{equation}
If we take the special case \(a = -m\) and \(b = d\), as in~\eqref{eq:fgasymp}, we find
\begin{equation} \label{eq:strip_flee}
		\s_\mathrm{strip} = \frac{L^{d-1}}{32 \sqrt{\pi} (d+1) \gn} \frac{
			\G \le( \frac{d}{d-1} \ri)
			\G^2 \le( \frac{1}{2d-2} \ri)
		}{
			\G \le( \frac{d+1}{2d-2} \ri)
			\G^2 \le( \frac{d}{2d-2} \ri)
		} m \ell + \dots \; .
\end{equation}

In principle, one could find the analogous asymptotic behaviour of the entanglement density in the sphere geometry by solving the equation of motion~\eqref{eq:sphere_equation_of_motion} and area integral~\eqref{eq:sphere_area} perturbatively in \(a \ell^b\). However, we have not been able to find a perturbative solution to~\eqref{eq:sphere_equation_of_motion} for arbitrary \(d\) and \(b\). For \(b = d\), the solution is given in ref.~\cite{Bhattacharya:2012mi}, as is the corresponding entanglement entropy. The resulting entanglement density is
\begin{equation} \label{eq:sphere_flee}
	\s_\mathrm{sphere} = \frac{L^{d-1} (d-1)}{8 (d+1) \gn} m \ell + \dots \;,
\end{equation}
where \(m = -a\).

The metric function \(g(z)\) has the asymptotic form~\eqref{eq:fgasymp} for deformations by a finite temperature or chemical potential, in which case the results~\eqref{eq:strip_flee} and~\eqref{eq:sphere_flee} are special cases of the first law of entanglement~\cite{Bhattacharya:2012mi}. The entanglement densities satisfy \(\s \approx \tent^{-1} \vev{T_{tt}}\), where  the entanglement temperature is
\begin{equation}
\label{eq:striptent}
\tent^\mathrm{strip} = \frac{2 (d^2-1) \G\left(\frac{d+1}{2d-2}\right) \G^2 \left(\frac{d}{2d-2}\right)}{\sqrt{\pi} \G\left(\frac{1}{d-1}\right) \G^2 \left(\frac{1}{2d-2}\right)^2  \ell },
\quad
\tent^\mathrm{sphere} = \frac{d+1}{2\pi \ell}
\end{equation}
for the strip and sphere respectively.

\subsubsection{Large subregions}

For large subregions, in the presence of a planar horizon in the bulk of \aads, the RT surface will typically have a component that almost lies flat along the horizon, as illustrated in figure~\ref{fig:minimal_surface_horizon}. This provides the dominant contribution to the entanglement density as \(\ell \to \infty\), so in this limit the entanglement density tends to the thermodynamic entropy density \(s\), determined from~\eqref{eq:bh_entropy}.

Computing the leading order correction to this, we find
\begin{subequations} \label{eq:sigma_large_l}
\begin{align}
	\s_{\mathrm{strip}} &= s \le[1 + \frac{2 z_H}{\ell} C(z_H) \ri] + \cO\le(\frac{z_H^2}{\ell^2} \ri),
	\label{eq:strip_sigma_large_l}
	\\
	\s_\mathrm{sphere} &= s \le[1 + \frac{(d-1) z_H}{\ell} C(z_H) \ri] + \cO\le(\frac{z_H^2}{\ell^2} \ri),
	\label{eq:sphere_sigma_large_l}
\end{align}
\end{subequations}
for the sphere and the strip, respectively, where 
\begin{equation} \label{eq:large_l_coefficient_integral}
	C(z_H) \equiv - \frac{1}{d-2} + \int_0^{1} \diff u \frac{1}{u^{d-1}} \le( \sqrt{\frac{1 - u^{2d-2} }{ g(z_H u) } } - 1 \ri).
\end{equation}
The result for the sphere, \eqref{eq:sphere_sigma_large_l}, is derived in appendix~\ref{app:entanglement_density_large_subregions} using the method of matched expansions developed in ref.~\cite{Liu:2013una}. The proof of \eqref{eq:strip_sigma_large_l} was carried out by my collaborator, Nikola I. Gushterov, and is detailed in ref.~\cite{Gushterov:2017vnr}.

We may bring~\eqref{eq:sigma_large_l} into a unified form by noting that the ratio of the surface area \(A\) to volume \(V\) of the entangling region is \(A/V = 2/\ell\) for the strip and \(A/V = (d-1)/\ell\) for the sphere. We may therefore rewrite~\eqref{eq:sigma_large_l} as
\begin{equation} \label{eq:sigma_large_l_area}
		\s = s \le[ 1 + \frac{A}{V} z_H C(z_H) \ri] + \cO\le(\frac{z_H^2}{\ell^2} \ri),
\end{equation}
which is valid for both the strip and the sphere. The first correction to the extensive behaviour \(\s \approx s\) at large \(\ell\) is proportional to the surface area of the entangling region, with coefficient determined by~\eqref{eq:large_l_coefficient_integral}.

Comparing to~\eqref{eq:EDlargeL}, we see that the area law coefficient is given by \(\a = s z_H C(z_H)\). Since \(s\) and \(z_H\) are non-negative, the area theorem would imply that \(C(z_H) \leq 0\). However, the area theorem only applies to Lorentz-invariant RG flows, and therefore does not apply to the holographic duals of geometries with horizons. Indeed, in many of our examples we find that \(C(z_H) > 0\). We will refer to this as area theorem violation, although strictly speaking the area theorem does not apply to these examples.

\section{Lorentz invariant RG flows}
\label{rg}

We now turn to our first example, Lorentz invariant holographic RG flows to an IR fixed point. We consider a bulk action\footnote{In this chapter we will need the explicit form of the boundary terms in \(S_\mathrm{grav}\), so we drop them from our expressions for notational simplicity.}
\begin{equation}
\label{eq:holo_rg_action}
		\sgrav = \frac{1}{16 \pi \gn} \int \diff^{d+1}x \sqrt{-G} \left[R -  \frac{1}{2} \partial_M \phi \partial^M \phi - V(\phi) \right],
\end{equation}
where $R$ is the Ricci scalar and $\phi$ is a real scalar field with potential $V(\phi)$. We want solutions to the equations of motion derived from $\sgrav$ that describe Lorentz-invariant RG flows between CFTs, driven by the scalar operator $\mathcal{O}$ holographically dual to $\phi$. We thus assume $V(\phi)$ has at least two stationary points, at which the equations of motion are solved by pure \ads[d+1], with radius of curvature $L$ given by
\begin{equation} \label{eq:rg_flow_curvature}
8 \pi \gn \left . V(\f) \right |_{\textrm{stationary}}= - \frac{d(d-1)}{2 L^2},
\end{equation}
where we assume the potential is negative at the stationary points. In RG flow solutions \(\f\) interpolates between two stationary points, beginning from a maximum of \(V(\f)\) in the UV (\(z = 0\)), and tending to a minimum in the IR (\(z\to\infty\)). The metric interpolates from \ads\ with radius \(L\), to \ads\ with radius \(L_\mathrm{IR}\). Since the potential is less negative in the UV than in the IR, we find from~\eqref{eq:rg_flow_curvature} that \(L > L_\mathrm{IR}\). This inequality is guaranteed by the holographic $c$-theorem~\cite{Freedman:1999gp}.

To find RG flow solutions, we make an ansatz
\begin{equation} \label{eq:rg_flow_metric}
\diff s^2 = G_{MN} \diff x^M \diff x^N=\frac{L^2}{z^2} \left[ - \diff t^2 + \diff \vec{x}^2 + \frac{\diff z^2}{g(z)} \right],
\quad
\phi=\phi(z),
\end{equation}
with $0 \leq z < \infty$. This metric is of the~\eqref{eq:aads_metric} with \(f(z) = 1\). Following ref.~\cite{Freedman:1999gp}, if we introduce a function \(W(\f)\) satisfying
\begin{equation}
\label{eq:VW}
V(\phi) = \frac{1}{16 \pi \gn} \left(\partial_{\phi}W\right)^2 - \frac{1}{2} \frac{d}{d-1} W^2,
\end{equation}
then the equations of motion derived from \(\sgrav\) are solved by any solution to~\cite{Freedman:1999gp,Ammon:2015wua}
\begin{equation}
\label{eq:superpotential}
\phi' = \frac{d-1}{8 \pi \gn} \frac{1}{z W} \, \partial_{\phi}W, \qquad g(z) = \frac{8 \pi \gn}{(d-1)^2} \, L^2 \, W^2.
\end{equation}
We therefore only need to solve the first-order equation~\eqref{eq:superpotential}. In fact, for our purposes, we can construct bottom-up holographic models of RG flows by choosing $g(z)$, which then determines $W$ and hence $\phi(z)$ via~\eqref{eq:superpotential}, which in turn is guaranteed to solve the equations of motion for the corresponding potential $V(\phi)$ in~\eqref{eq:VW}.

The metric function $g(z)$ obeys several constraints which will restrict our choice. Equation~\eqref{eq:superpotential} implies
\begin{equation}
\label{eq:rg_flow_g_condition_1}
\phi'(z)^2 = \frac{d-1}{16 \pi \gn} \frac{g'(z)}{z g(z)},
\end{equation}
so that $g'(z)\geq 0$, since \(\f(z)\) is real and by assumption $g(z)>0$. The null energy condition also requires $g'(z)\geq 0$, so any solution of~\eqref{eq:superpotential} is guaranteed to obey the null energy condition. We require that $\mathcal{O}$ is relevant, so its scaling dimension is $\D<d$, and unitary, so $\Delta \geq \frac{d-2}{2}$. We will actually impose a stricter upper bound, $\Delta \leq (d+2)/2$. Larger values of \(\D\) introduce additional UV divergences in the entanglement entropy, which the subtraction $\see - \see^\mathrm{CFT}$ does not cancel~\cite{Liu:2013una,Casini:2016udt}. As discussed in section~\ref{sec:holographic_dictionary}, the asymptotic behaviour of the scalar field is then \(\f(z) = \f_{(0)} z^{\D_-}+\dots\), where  \(\D_-=\mathrm{Min}(d-\D,\D)\). The coefficient \(\f_{(0)}\) is proportional either to the source of \(\cO\) ($\D_-=d-\D$) or to the VEV \(\vev{\cO}\) ($\D_-=\D$). The range of allowed \(\D_-\) values is
\begin{equation} \label{eq:rg_delta_range}
		\frac{d}{2} - 1 \leq \D_- \leq \frac{d}{2} + 1,
\end{equation}
independent of whether \(\f_{(0)}\) corresponds to a source or VEV.

From~\eqref{eq:rg_flow_g_condition_1}, the asymptotic expansion of \(g(z)\) is
\begin{equation}
\label{eq:rg_flow_g_condition_2}
g(z) = 1 + \frac{8 \pi \gn \D_-}{d-1} \phi_{(0)}^2 z^{2\D_{-}} + \dots \; ,
\end{equation}
where the ellipsis denotes terms with higher powers of $z$. This has the form of~\eqref{eq:fgasymp_general}, with \(a = 8 \pi \gn \D_- \phi_{(0)}^2/ (d-1)\) and \(b = 2 \D_-\). Substituting these values into~\eqref{eq:strip_sigma_small_l_expansion_final}, we find the small-\(\ell\) behaviour of the entanglement density for the strip
\begin{equation} \label{eq:rgsmallL}
		\s_\mathrm{strip} = - \frac{2 \pi^{3/2} \D_- \f_{(0)}^2 L^{d-1} \ell^{2\D_- + 1-d}}{(2\D_- + 1)(2 \D_- + 2 - d)}
		\frac{
				\G \le( \frac{2 \D_- + d}{2d-2} \ri)
		}{
			\G \le( \frac{2 \D_- + 1}{2d-2} \ri)
		} \le[
			\frac{
						\G \le( \frac{1}{2d-2} \ri)
				}{
						2 \sqrt{\pi} \G \le( \frac{d}{2d-2} \ri)
				}
		\ri]^{2\D_- + 2 - d}  + \dots \; .
\end{equation}
Note that the leading order term is always negative over the allowed range of \(\D_-\)~\eqref{eq:rg_delta_range}.

Entanglement entropy in holographic RG flows has been studied in detail before, for example in refs.~\cite{Albash:2011nq,Myers:2012ed,Liu:2012eea,Liu:2013una,Taylor:2017zzo}, so we focus only on a few cases that illustrate some possible behaviour of the entanglement density as a function of $\ell$. We restrict to $d=4$ and choose
\begin{subnumcases}{\label{eq:rgchoices} g(z)=} 1 + \tanh^4(\m z), \label{eq:rg_flow_single_peak} \\  1 + \tanh^4(\m z)  + \frac{3}{2}\tanh(\m z - 2) \tanh^5(\m z),\label{eq:holo_rg_double_trough} \\ 1 + \tanh^4(\m z) + \frac{20(\m z - 1)^2 + 1}{(\m z - 1)^2 + 1} \left[1 + \tanh(\m z) \right]\tanh^4(\m z),\label{eq:holo_rg_transition} \\ 1 + \tanh^3(\m z), \label{eq:rg_flow_constant} \\ 1 + \tanh^{7/2}(\m z), \label{eq:rg_flow_fractional_power}
\end{subnumcases}
where in each case $\m$ is a constant of mass dimension \([\m]=1\), which may be related to $\phi_{(0)}$ by~\eqref{eq:rg_flow_g_condition_2}. Table~\ref{tab:rgchoices} summarizes some properties of our choices of $g(z)$. The second column in the table is $L_{\textrm{IR}}$, the value of the $\ads[5]$ radius at $z \to \infty$, determined by the value of $\lim_{z \to \infty} g(z)$. The third column shows the leading powers of $z$ in the asymptotics of \(g(z)\), which via~\eqref{eq:rg_flow_g_condition_2} determine $\Delta_-$, listed in the fourth column, with the corresponding $\Delta$ in the fifth column. The sixth column indicates whether the RG flow is driven by a source or vacuum expectation value (VEV) of the scalar operator \(\cO\), as determined by the analysis of refs.~\cite{Bianchi:2001kw,Bianchi:2001de,Papadimitriou:2004rz}.

\begin{table}
\centering
\begin{tabular}[c]{c|c|c|c|c|c}
$g(z)$ & $L_{\textrm{IR}}$ & Asymptotics & $\Delta_-$ & $\Delta$ & Flow driven by \\ \hline
\eqref{eq:rg_flow_single_peak} & $L/\sqrt{2}$ & $1+\left(\m z\right)^4+\ldots$ & $2$ & $2$ & VEV \\ 
\eqref{eq:holo_rg_double_trough} & $L/\sqrt{7/2}$ & $1+\left(\m z\right)^4+\ldots$ & $2$ & $2$ & VEV \\ 
\eqref{eq:holo_rg_transition} & $L/\sqrt{42}$ & $1+\frac{23}{2}\left(\m z\right)^4+\ldots$ & $2$ & $2$ & VEV \\ 
\eqref{eq:rg_flow_constant} & $L/\sqrt{2}$ & $1+\left(\m z\right)^3+\ldots$ & $3/2$ & $5/2$ & Source \\
\eqref{eq:rg_flow_fractional_power} & $L/\sqrt{2}$ & $1+\left(\m z\right)^{7/2}+\ldots$ & $7/4$ & $9/4$ & Source \\ 
\end{tabular}
\caption[Summary of key properties of different choices for holographic RG flows.]{\label{tab:rgchoices} Summary of properties of our choices of $g(z)$ in~\eqref{eq:rgchoices}, as discussed in the text.}
\end{table}

\begin{figure}
	\begin{subfigure}{0.5\textwidth}
		\includegraphics[width=\textwidth]{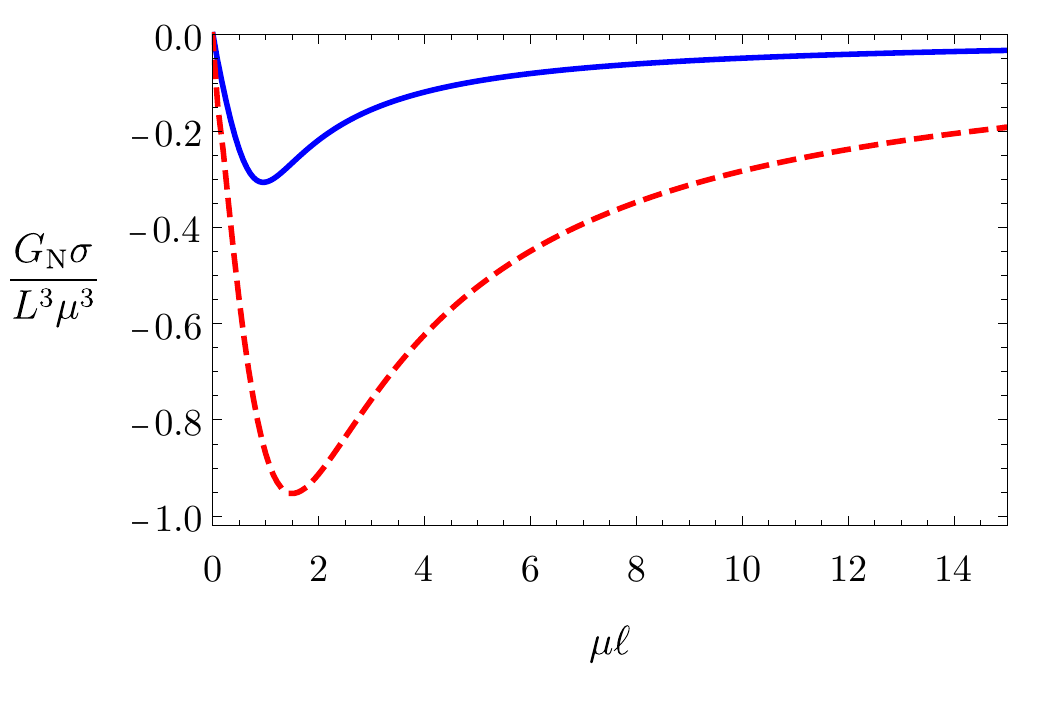}
		\caption{Eq.~\eqref{eq:rg_flow_single_peak}}
		\label{fig:holo_rg_single_peak}
	\end{subfigure}
	\begin{subfigure}{0.5\textwidth}
		\includegraphics[width=\textwidth]{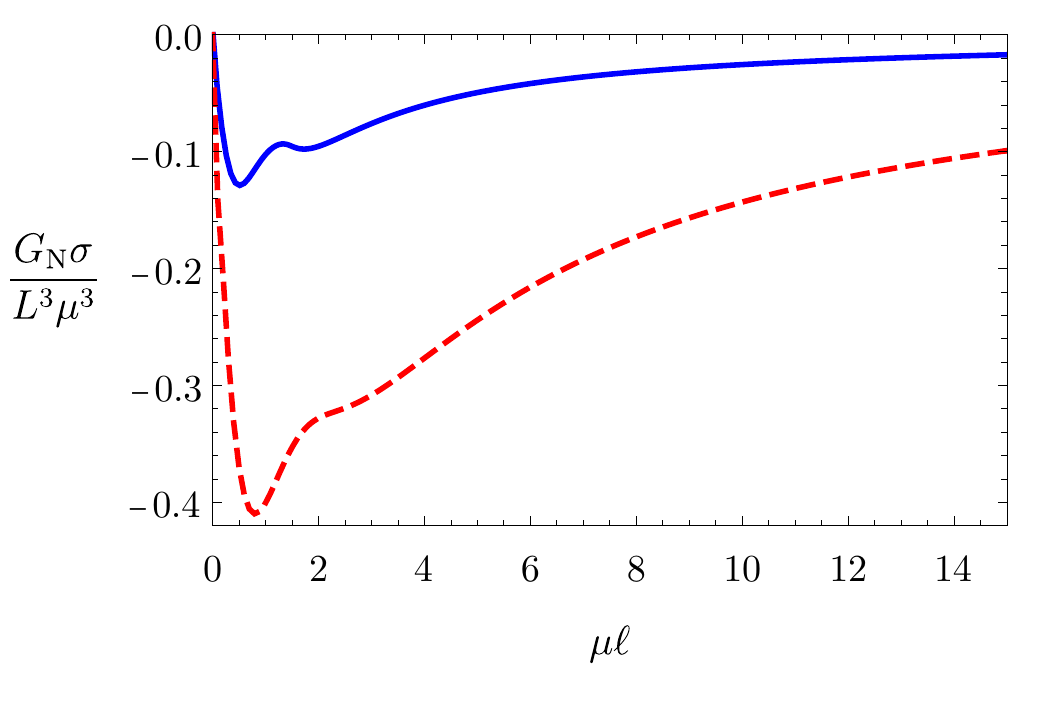}
		\caption{Eq.~\eqref{eq:holo_rg_double_trough}}
		\label{fig:holo_rg_double_peak}
	\end{subfigure}
	\begin{subfigure}{0.5\textwidth}
		\includegraphics[width=\textwidth]{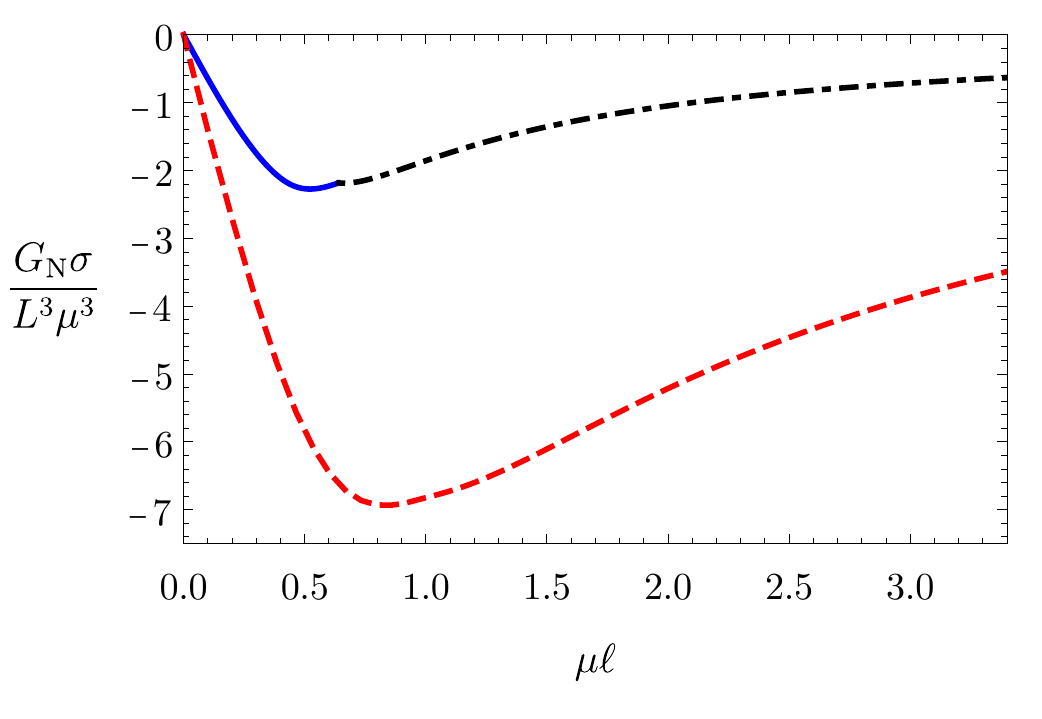}
		\caption{Eq.~\eqref{eq:holo_rg_transition}}
		\label{fig:holo_rg_transition}
	\end{subfigure}
	\begin{subfigure}{0.5\textwidth}
		\includegraphics[width=\textwidth]{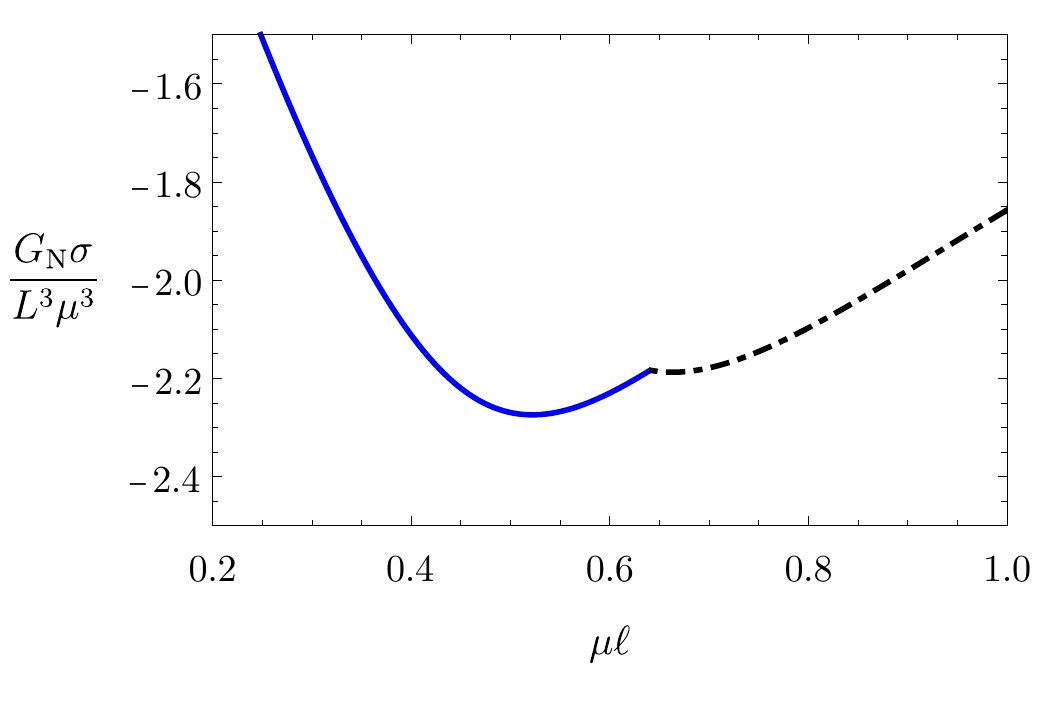}
		\caption{Eq.~\eqref{eq:holo_rg_transition}, close-up}
		\label{fig:holo_rg_transition_close}
	\end{subfigure}
	\begin{subfigure}{0.5\textwidth}
		\includegraphics[width=\textwidth]{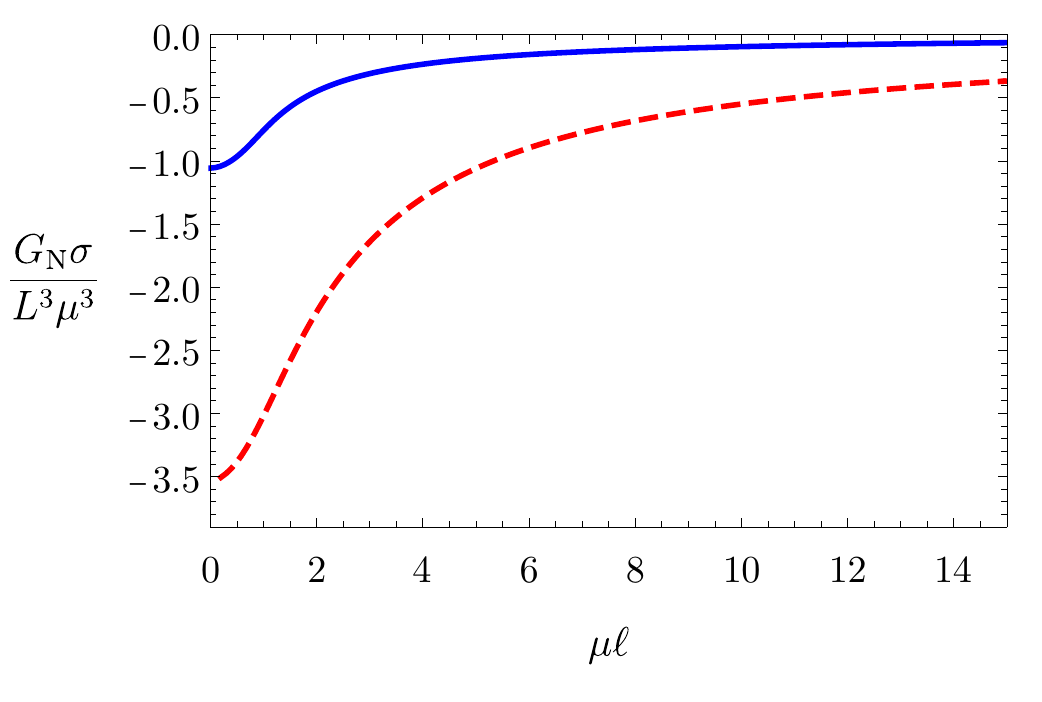}
		\caption{Eq.~\eqref{eq:rg_flow_constant}}
		\label{fig:holo_rg_constant}
	\end{subfigure}
	\begin{subfigure}{0.5\textwidth}
		\includegraphics[width=\textwidth]{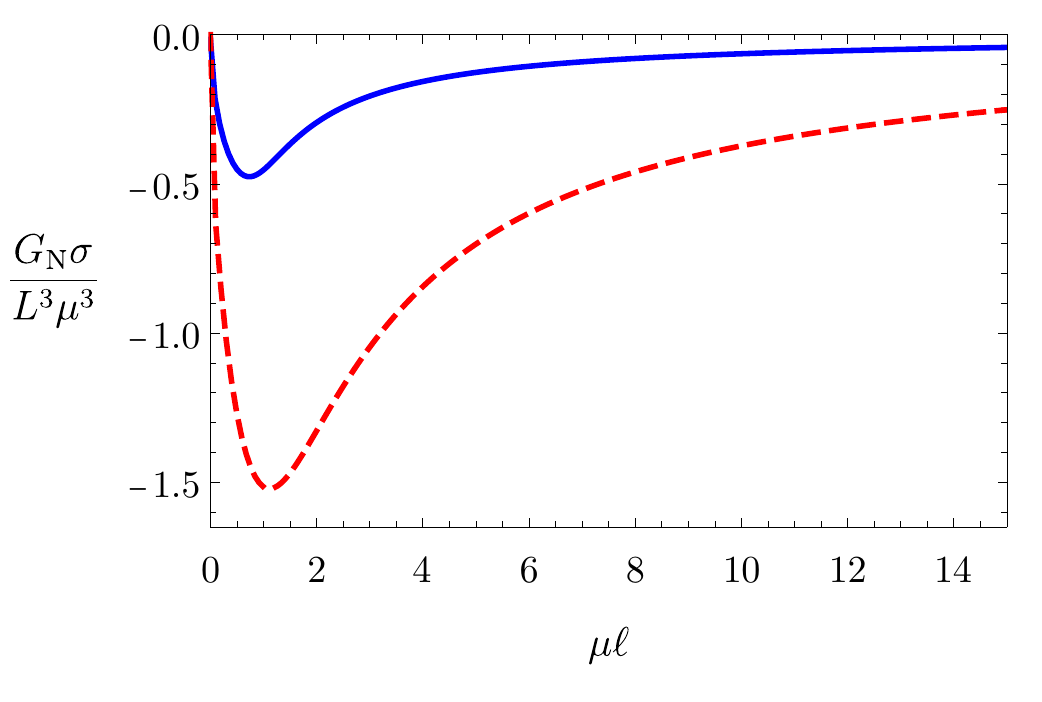}
		\caption{Eq.~\eqref{eq:rg_flow_fractional_power}}
		\label{fig:holo_rg_fractional_power}
	\end{subfigure}
	\caption[Entanglement density of holographic RG flows.]{\label{fig:rged} The entanglement density, $\s$ as a function of $\m\ell$ for RG flows between holographic CFTs in $d=4$. In each plot, the solid blue line is for the strip and the dashed red line is for the sphere. The caption to each plot indicates the $g(z)$ we chose from~\eqref{eq:rgchoices}. For the $g(z)$ in~\eqref{eq:holo_rg_transition}, there are three extremal surfaces for the strip geometry (see figure~\ref{fig:rgphasetrans}). The strip entanglement entropy undergoes a ``first-order phase transition'' at $\m\ell\approx0.65$, where the surface of minimal area changes. This leads to the kink in the entanglement density shown in (c) and (d), where the solid blue and dot-dashed black curves meet.}
\end{figure}
%
Figure~\ref{fig:rged} shows our numerical results for \(\s\) as a function of $\ell$, for each of the flows determined by~\eqref{eq:rgchoices}. We plot $\s$ in units of $\m^3 L^3/\gn$, where $L^3/\gn$ is proportional the Weyl anomaly of the UV CFT~\cite{Henningson:1998ey,Henningson:1998gx}. In all cases, $\s<0$ for all $\m \ell$, with $\s \to 0^-$ as $\m \ell \to \infty$, as required by the area theorem.

Figure~\ref{fig:holo_rg_single_peak} shows the simplest behaviour, for the $g(z)$ in~\eqref{eq:rg_flow_single_peak}, in which $\sigma \propto - \ell$ at small $\ell$, and then a single minimum appears before $\s \to 0^-$ as $\m\ell \to \infty$, for both the strip and sphere. Figure~\ref{fig:holo_rg_double_peak}, for the $g(z)$ in~\eqref{eq:holo_rg_double_trough}, is similar, but with a second, local minimum, and an intermediate local maximum, at intermediate $\ell$, for the strip. The sphere exhibits only a single local minimum.

\begin{figure}[t]
	\begin{subfigure}{0.5\textwidth}
		\includegraphics[width=\textwidth]{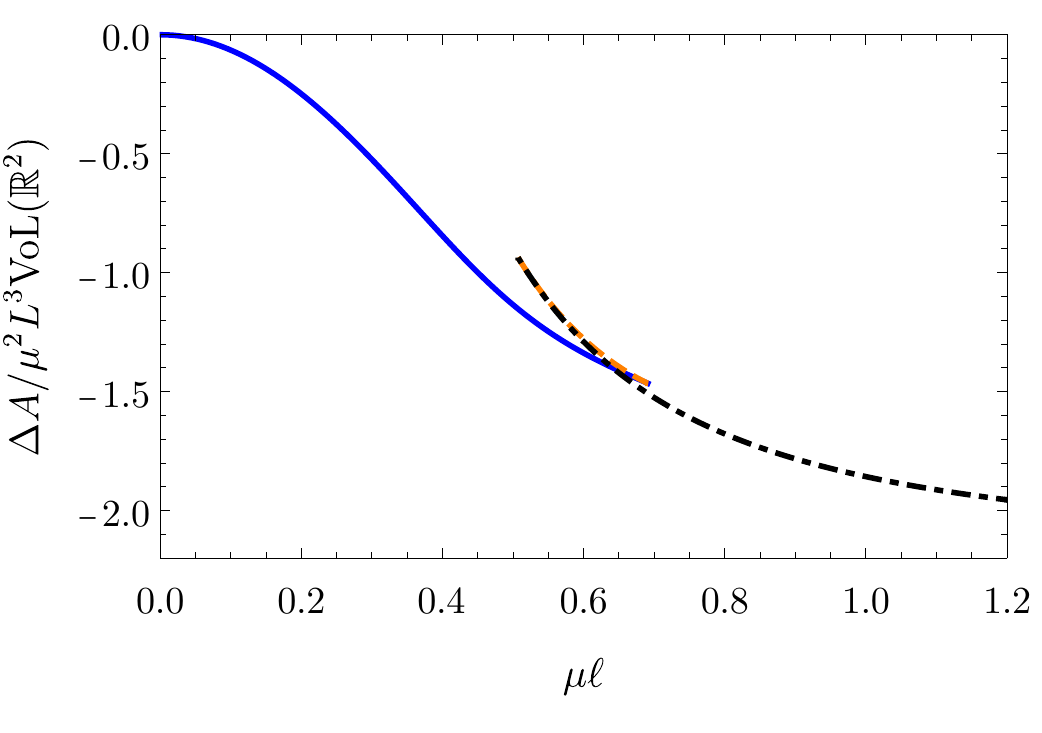}
	\end{subfigure}
	\begin{subfigure}{0.5\textwidth}
		\includegraphics[width=\textwidth]{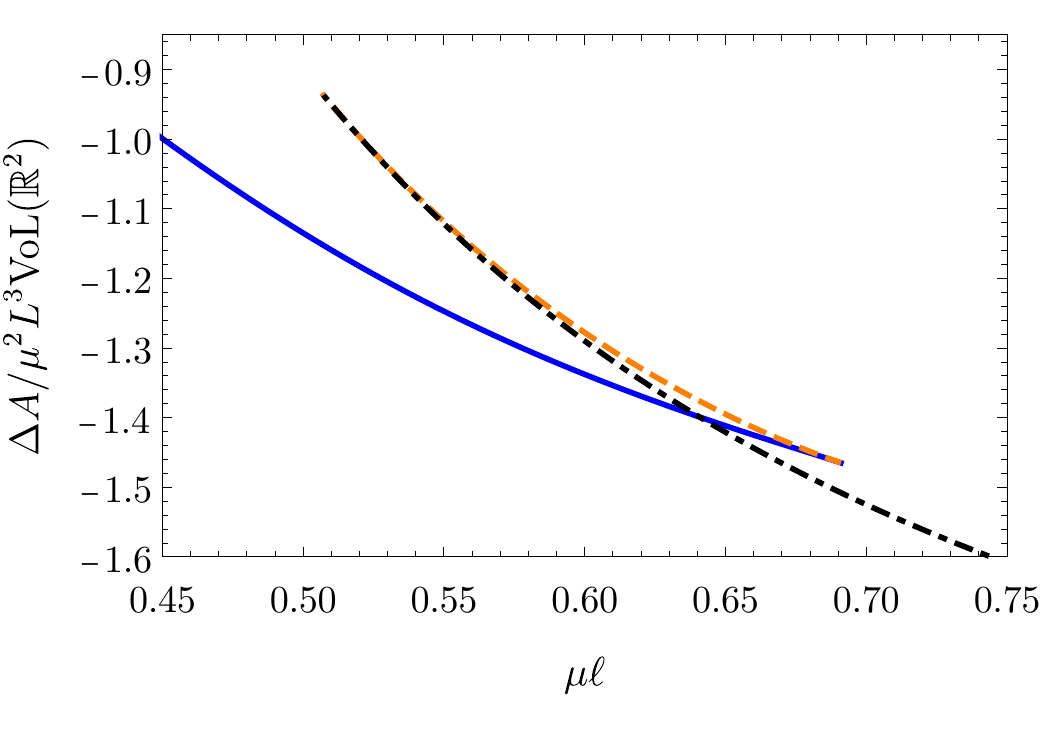}
	\end{subfigure}
    \caption[Transition in minimal area for a holographic RG flow.]{
		The difference in area \(\D A\) between extremal surfaces in the geometry	corresponding to \(g(z)\) in~\eqref{eq:holo_rg_transition} and the minimal surface in pure \ads[5] with the same \(\ell\). For \(0.51 \lesssim \m\ell \lesssim 0.69\) there are three extremal surfaces, indicated by solid blue,  dashed orange, and dot-dashed black lines. Of these, the solid blue line has the smallest area for \(\m\ell \lesssim 0.65\), while the dot-dashed black line has the smallest area for larger values of \(\m\ell\), leading to a ``first-order phase transition'' in the entanglement entropy. The right-hand plot shows a close-up of this transition.}\label{fig:rgphasetrans}
\end{figure}

For the $g(z)$ in~\eqref{eq:holo_rg_transition}, three extremal surfaces exist for the strip over the range \(0.51 \lesssim \m\ell \lesssim 0.69\). Figure~\ref{fig:rgphasetrans} shows the difference in area, $\D A$, between each of these three surfaces and the minimal surface in \ads[5] with the same $\ell$, indicating a transition from one to the other as the global minimum of the area functional. We find that this transition occurs at $\m\ell\approx 0.65$. The entanglement density for the strip therefore exhibits a kink at the critical $\ell$, shown in figures~\ref{fig:holo_rg_transition} and~\ref{fig:holo_rg_transition_close}. In contrast, no transition occurs for the sphere, and \(\s\) for the sphere does not exhibit a kink.

The $g(z)$ in~\eqref{eq:rg_flow_constant} yields $\D_-=3/2$, so~\eqref{eq:rgsmallL} implies \(\s\) tends to a negative constant at \(\ell = 0\). As $\ell$ increases, \(\s\) monotonically increases to zero as shown in figure~\ref{fig:holo_rg_constant}. The $g(z)$ in~\eqref{eq:rg_flow_fractional_power} yields $\D_-=7/4$, hence~\eqref{eq:rgsmallL} implies $\s \propto - \ell^{1/2}$ at small $\ell$. Aside from the fractional power of $\ell$, the entanglement density for this \(g(z)\) behaves similar to that for the \(g(z)\) in~\eqref{eq:rg_flow_single_peak}, with a single global minimum before $\s \to 0^-$ as \(\m\ell \to \infty\).

In summary, the entanglement density can exhibit a variety of behaviour as a function of $\ell$, depending on details of the RG flow. However, $\s$ often exhibits a unique \textit{global} minimum, which by dimensional analysis must be at an $\ell \propto 1/\m$. This value of $\ell$ provides a candidate quantity to characterize and compare RG flows. For example, it may provide a precise definition of the crossover scale from the UV to IR.

\section{AdS-Schwarzschild}
\label{adssc}

In this section we consider a bulk action
\begin{equation}
\label{eq:adsscaction}
\sgrav = \frac{1}{16 \pi \gn} \int \diff^{d+1} x \sqrt{-G} \left[ R + \frac{d(d-1)}{L^2} \right],
\end{equation}
i.e. the Einstein-Hilbert action with a negative cosmological constant. The corresponding equations of motion admit the $(d+1)$-dimensional \ads[d+1]-Schwarzschild black brane solution, of the form in~\eqref{eq:aads_metric} with
\begin{equation}
\label{eq:schwarzschild_metric}
f(z) = g(z) = 1 - m z^d.
\end{equation}
There is a horizon at $z_H = m^{-1/d}$, with $\vev{T_{tt}}$, $T$, and $s$ given by equations~\eqref{eq:energy_density} and~\eqref{eq:thermo_entropy}.  The numerical results for the entanglement entropy of a strip in \ads-Schwarzschild presented in this section have been reproduced analytically in ref.~\cite{Erdmenger:2017pfh}.

\begin{figure}
	\centering
	\includegraphics[width=0.5\textwidth]{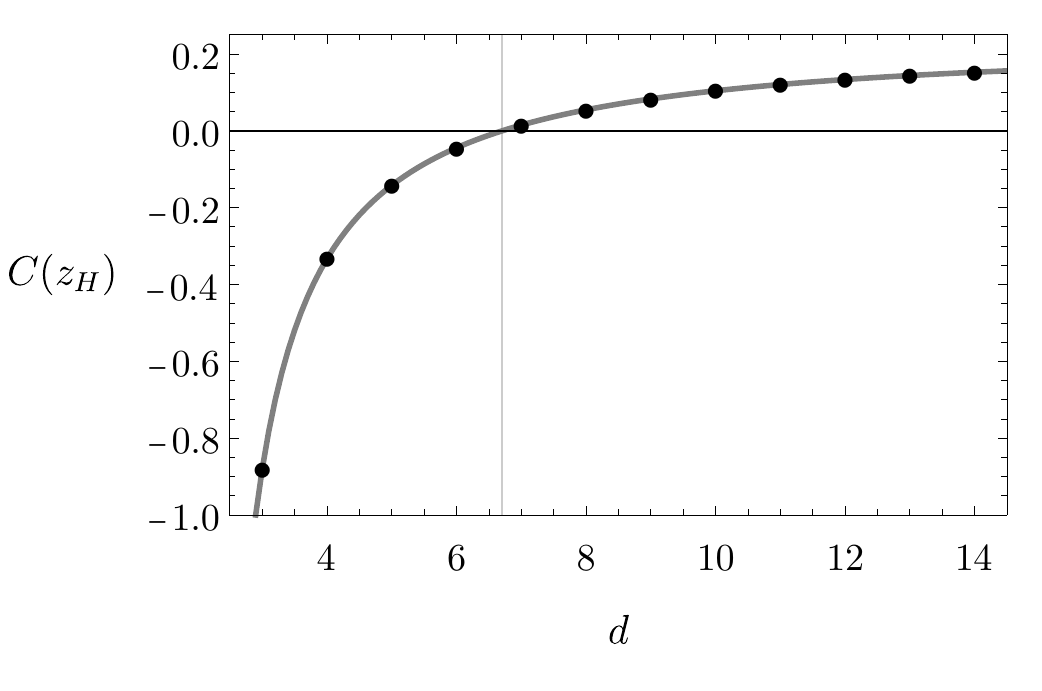}
	\caption[The coefficient controlling the large-\(\ell\) behaviour of the entanglement density for \ads-Schwarzschild.]{The dimensionless coefficient $C(z_H)$ for \ads[d+1]-Schwarzschild. The sign of \(C(z_H)\) controls whether \(\s \to s\) from above or below at large \(\ell\), see~\eqref{eq:sigma_large_l}. The black dots show the value of \(C(z_H)\) for integer \(d\), while the grey curve is an interpolation obtained by letting \(d\) take non-integer values in~\eqref{eq:large_l_coefficient_integral}. At $d=3$, $C(z_H)\approx -0.88$, and $C(z_H)$ then increases monotonically with $d$, reaching zero at $\dcrit \approx 6.7$, indicated by the vertical line.}\label{fig:adssc_cplot}
\end{figure}

The entanglement density for \(T \ell \ll 1\) is given by the first law of entanglement entropy~\cite{Bhattacharya:2012mi}, \eqref{eq:strip_flee} and~\eqref{eq:sphere_flee} for the strip and the sphere, respectively. Note that in particular \(\s \propto \ell\) with a positive proportionality coefficient, in contrast to the entanglement density for RG flows. As \(T\ell\to \infty\), the entanglement density tends to the thermodynamic entropy density \(s\), according to~\eqref{eq:sigma_large_l_area}. Whether this occurs from above or below depends on the sign of the coefficient \(C(z_H)\), given in~\eqref{eq:large_l_coefficient_integral}. Figure~\ref{fig:adssc_cplot} shows the value of \(C(z_H)\) as a function of \(d\). We find that \(\s \to s\) from below (\(\s \to s^-\))  for \(d \leq 6\), while \(\s \to s\) from above (\(\s \to s^+\)) for \(d \geq 7\),  signalling area theorem violation.\footnote{In appendix~\ref{app:monotonicity_of_c_for_ads_schwarzschild} we show that \(C(z_H)\) for \ads-Schwarzschild monotonically increases with \(d\), and therefore remains positive for all values of \(d\) larger than those shown in figure~\ref{fig:adssc_cplot}.}

\begin{figure}
	\begin{subfigure}{0.5\textwidth}
		\includegraphics[width=\textwidth]{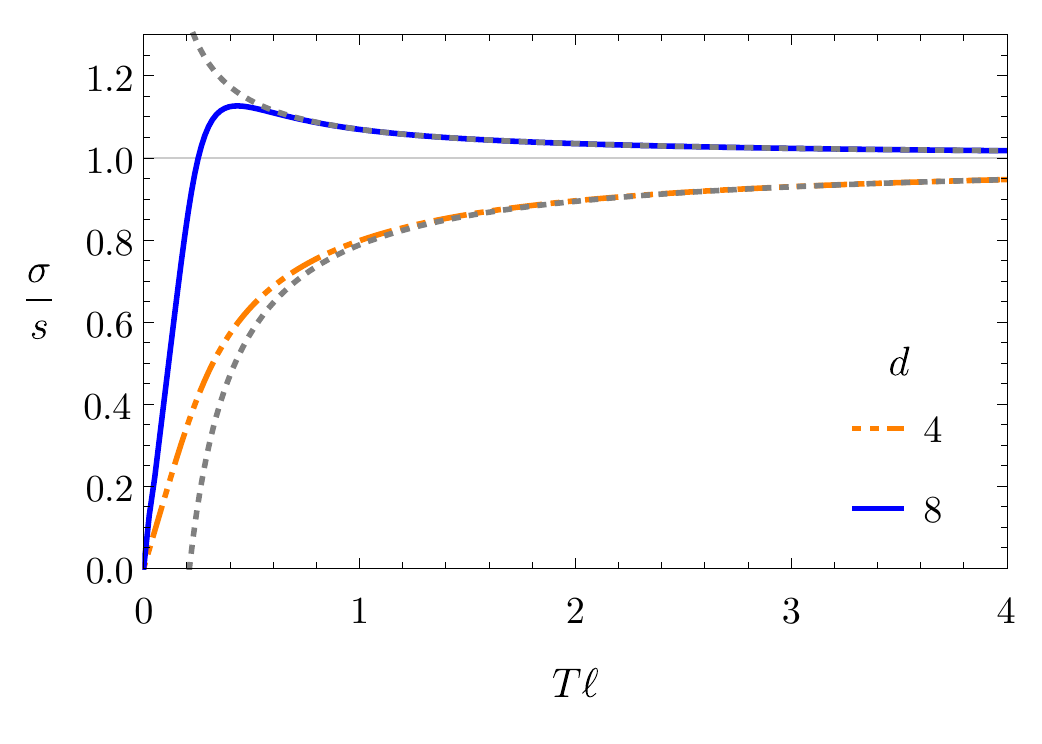}
		\caption{Strip}
		\label{fig:adsscd48_strip}
	\end{subfigure}
	\begin{subfigure}{0.5\textwidth}
		\includegraphics[width=\textwidth]{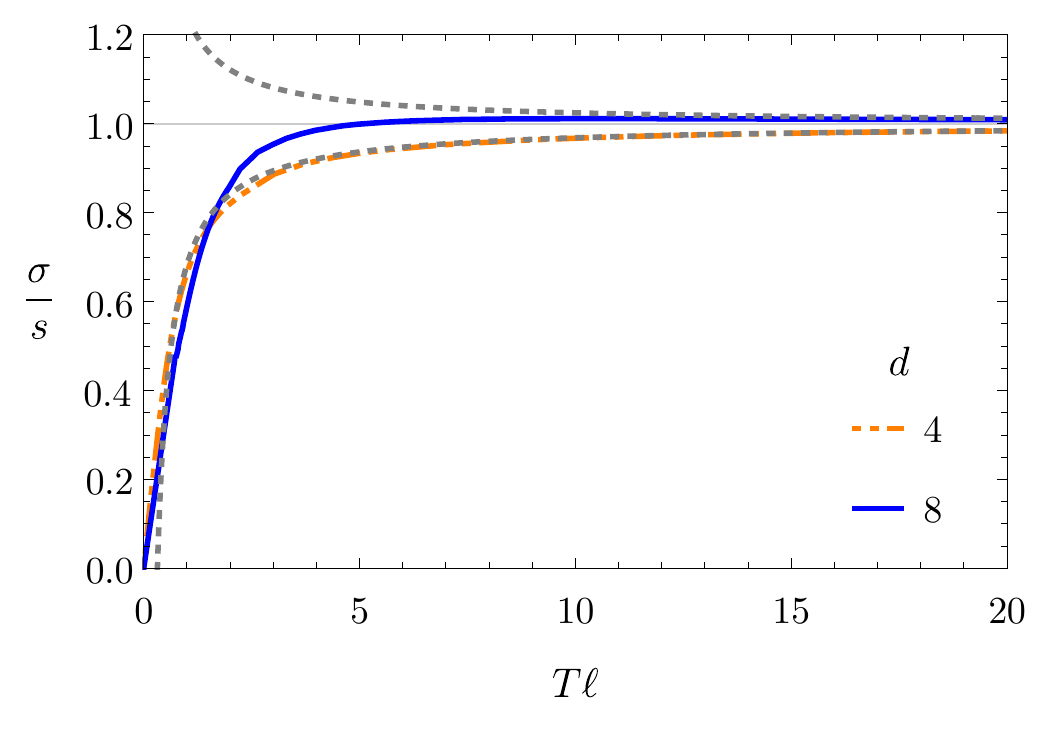}
		\caption{Sphere}
		\label{fig:adsscd48_sphere}
	\end{subfigure}
    \begin{subfigure}[b]{0.5\textwidth}
        \centering
        \includegraphics[width=\textwidth]{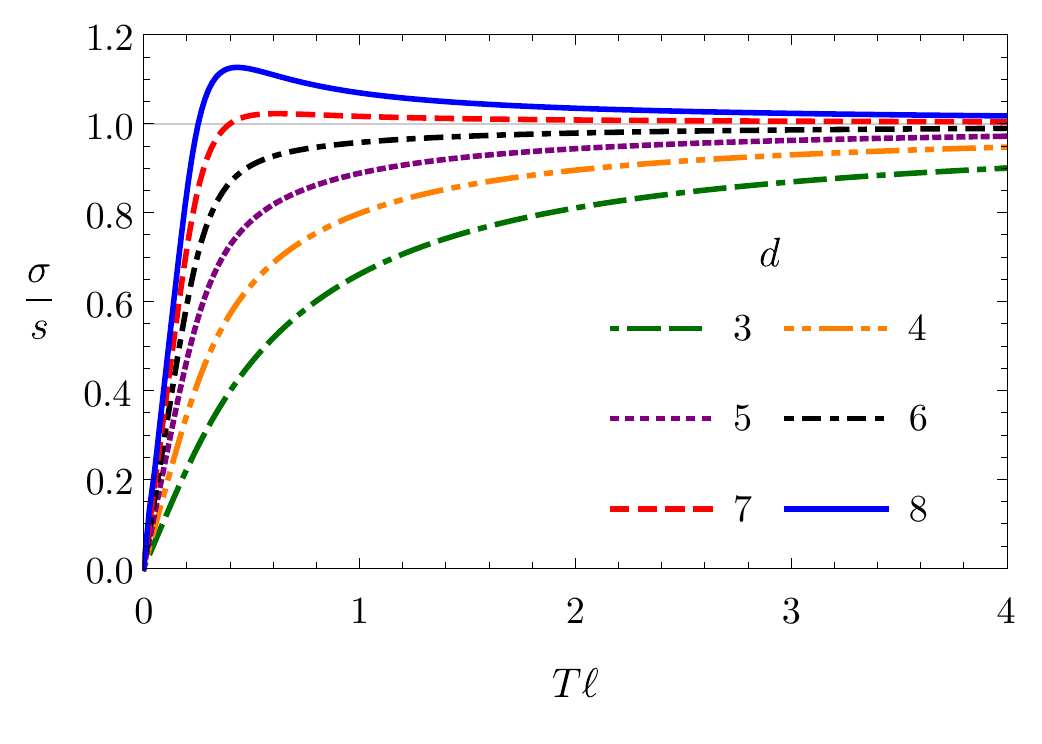}
				\caption{Strip}
				\label{fig:schwarzschild_dimensions_strip}
    \end{subfigure}
    \begin{subfigure}[b]{0.5\textwidth}
        \centering
        \includegraphics[width=\textwidth]{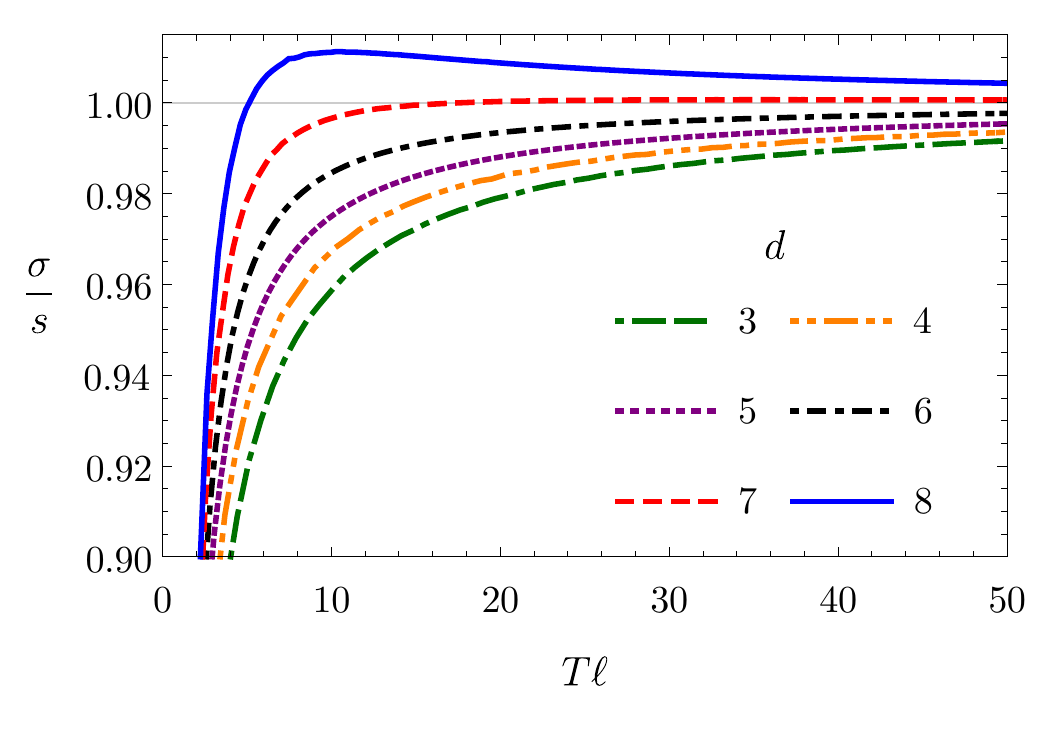}
				\caption{Sphere}
				\label{fig:schwarzschild_dimensions_sphere}
    \end{subfigure}
		\caption[Entanglement density for \ads-Schwarzschild.]{
			The entanglement density, $\s$, in units of entropy density $s$, versus $T\ell$ for \ads-Schwarzschild, for the strip (left column) and the sphere (right column). The top row shows numerical results for \(d = 4\) (dot-dashed orange) and \(d = 8\) (sold blue), along with the large-\(\ell\) approximation~\eqref{eq:sigma_large_l} in dotted grey. The bottom row shows numerical results for \(d = 3,4,\dots,8\). When \(d \leq 6\), \(\s/s\) increases monotonically with \(T\ell\), approaching one from below as \(T\ell \to \infty\). When \(d \geq 7\), \(\s/s\) exhibits a global maximum at intermediate \(\ell\), before approaching one from above as \(T\ell \to \infty\).
		}
		\label{fig:schwarzschild_dimensions}
\end{figure}

For example, figures~\ref{fig:adsscd48_strip} and~\ref{fig:adsscd48_sphere} show our numerical results for $\s/s$ as a function of $T\ell$ for \ads[d+1]-Schwarzschild in $d=4$ and $8$. In all cases we find $\s/s \propto \ell$ at small $T\ell$, as expected. For $d=4$ and for both the strip and sphere, we find $\s/s$ increases monotonically and $\s \to s^-$ as $T\ell \to \infty$, whereas for $d=8$, $\s/s$ rises to a global maximum at an $\ell$ that by dimensional analysis must be $\propto 1/T$, and then $\s \to s^+$ as $T\ell \to \infty$. The dotted lines in figures~\ref{fig:adsscd48_strip} and~\ref{fig:adsscd48_sphere} show the large-\(\ell\) approximation~\eqref{eq:sigma_large_l}. Figures~\ref{fig:schwarzschild_dimensions_strip} and \ref{fig:schwarzschild_dimensions_sphere} show the entanglement density for all values of \(d\) between three and eight. When \(d \leq 6\) we find that the entanglement density increases monotonically with \(\ell\), while for \(d \geq 7\) the entanglement density exhibits a global maximum at intermediate values of \(\ell\).
%

This pattern extends to CFTs at non-zero $T$ in $d=2$, where the entanglement entropy for an interval of length $\ell$ is known exactly~\cite{Calabrese:2004eu}. Given $d=2<\dcrit$, we expect $\s \to s^-$ as $T\ell \to \infty$. The result of ref.~\cite{Calabrese:2004eu} leads to
\begin{equation}
\s = \frac{c}{3 \ell} \ln\left[\frac{\sinh\left(\pi T \ell \right)}{\pi T \ell}\right]\nonumber = \frac{c}{3} \pi T - \frac{c}{3} \frac{\ln\left(2 \pi T \ell\right)}{\ell} + \mathcal{O}\left(e^{-2\pi T \ell}/\ell\right), \nonumber
\end{equation}
where $c$ is the central charge, and in the second equality we performed the $1/\ell$ expansion. The first term in the expansion is the thermodynamic entropy density \(s\)~\cite{Cardy:1986ie}. The second term is negative, so  $\s \to s^-$ as $T\ell \to \infty$.

In the $d\to\infty$ limit AdS-Schwarzschild is dual to an RG flow from a $(d+1)$-dimensional UV CFT to a $(0+1)$-dimensional scale-invariant theory in the IR, which is clearly only possible when Lorentz symmetry is broken. In the limit $d \to \infty$, the near-horizon geometry of the AdS-Schwarzschild black brane becomes $SL(2,\mathbb{R})/U(1) \times \mathbb{R}^{d-1}$, with \(\mathbb{R}^{d-1}\) corresponding to the QFT spatial directions \(\vec{x}\)~\cite{Emparan:2013moa,Emparan:2013xia}. The \(SL(2,\mathbb{R})\) factor implies that linearized fluctuations in the near-horizon region exhibit scale invariance in $t$ and $z$ but not $\vec{x}$~\cite{Emparan:2013xia,Castro:2010fd}. This suggests that area-theorem violation may be associated to an IR fixed point with different scaling to the UV. We find area theorem violation in some, but not all, of our other examples with different scaling in the IR.

\section{\ads-Reissner-Nordstr\"om}
\label{adsrn} 

In this section we consider the bulk Einstein-Maxwell action
\begin{equation}
\sgrav = \frac{1}{16 \pi \gn} \int \diff^{d+1} x  \sqrt{-G} \left[R + \frac{d(d-1)}{L^2} - L^2 F_{MN}F^{MN}\right],
\end{equation}
where $F_{MN}=\partial_M A_N - \partial_N A_M$ is the field strength for a gauge field $A_M$, dual to a conserved $\U(1)$ current. The equations of motion admit the $(d+1)$-dimensional \ads-Reissner-Nordstr\"om black brane solution~\cite{Chamblin:1999tk}, with metric of the form~\eqref{eq:aads_metric}, with 
\begin{equation} \label{eq:ads_rn_metric_factors}
f(z) = g(z)=1-mz^{d}+q^{2}z^{2d-2}.
\end{equation}
The black brane has charge density proportional to \(q\). The solution has a horizon at \(z = z_H\), the smallest positive root of $g(z_H)=0$, related to \(m\) by \(m = (1  +q^2 z_H^{2d-2})/z_H^d\). The gauge field's only non-zero component is
\begin{equation}
A_t = \mu \left(1 - \frac{z^{d-2}}{z_H^{d-2}}\right), \qquad \mu = \sqrt{\frac{d-1}{2(d-2)}} z_H^{d-2} q.
\end{equation}
\ads-Reissner-Nordstr\"om is dual to a CFT with non-zero chemical potential $\mu$ and charge density proportional to $q$. The temperature is given by
\begin{equation}
T = \frac{d}{4\pi z_H} \left(1-\frac{d-2}{d} q^2 z_H^{2d-2}\right).
\end{equation}
Requiring $T\geq 0$ implies $q^2 \leq \frac{d}{d-2} z_H^{-2(d-1)}$. In the limit where $q$ saturates the upper bound, and therefore $T=0$, an extremal horizon is present. There is non-zero entropy density~\eqref{eq:thermo_entropy}, even at \(T=0\), and also non-zero energy density~\eqref{eq:energy_density}.

The CFT states are parameterised by $T/\mu$, which determines $\vev{T_{tt}}$ and $q$. When $T/\mu \gg 1$, AdS-Reissner-Nordstr\"om approaches AdS-Schwarzschild. When $T=0$, \(f(z) = g(z)\) has a double zero at the horizon, so that near the horizon the metric~\eqref{eq:aads_metric} is approximately
\begin{equation} \label{eq:extremal_rn_near_horizon_metric}
		\diff s^2  \approx \frac{L^2}{z_H^2} \le[-\frac{1}{2} f''(z_H) (z_H-z)^2 \diff t^2 + \frac{2}{f''(z_H) (z_H - z)^2} \diff z^2 \ri] + \frac{L^2}{z_H^2} \d_{ij} \diff x^i \diff x^j.
\end{equation}
The factor in brackets is the metric of \ads[2],\footnote{To put this metric into the form we have been using for \ads, define a new coordinate \(\tilde{z} = 2/f''(z_H)(z_H-z)\).} so the near-horizon geometry of extremal \ads-Reissner-Nordstr\"om is \(\ads[2]\times\mathbb{R}^{d-2}\), with \ads[2] of radius $L/\sqrt{d(d-1)}$~\cite{Faulkner:2009wj}. The holographic dual of extremal \ads-Reissner-Nordstr\"om is known as a semi-local quantum liquid state~\cite{Iqbal:2011in}, describing an RG flow from a $d$-dimensional UV CFT to a $(0+1)$-dimensional scale invariant theory in the IR. 

\begin{figure}
	\centering
	 \includegraphics[width=0.5\textwidth]{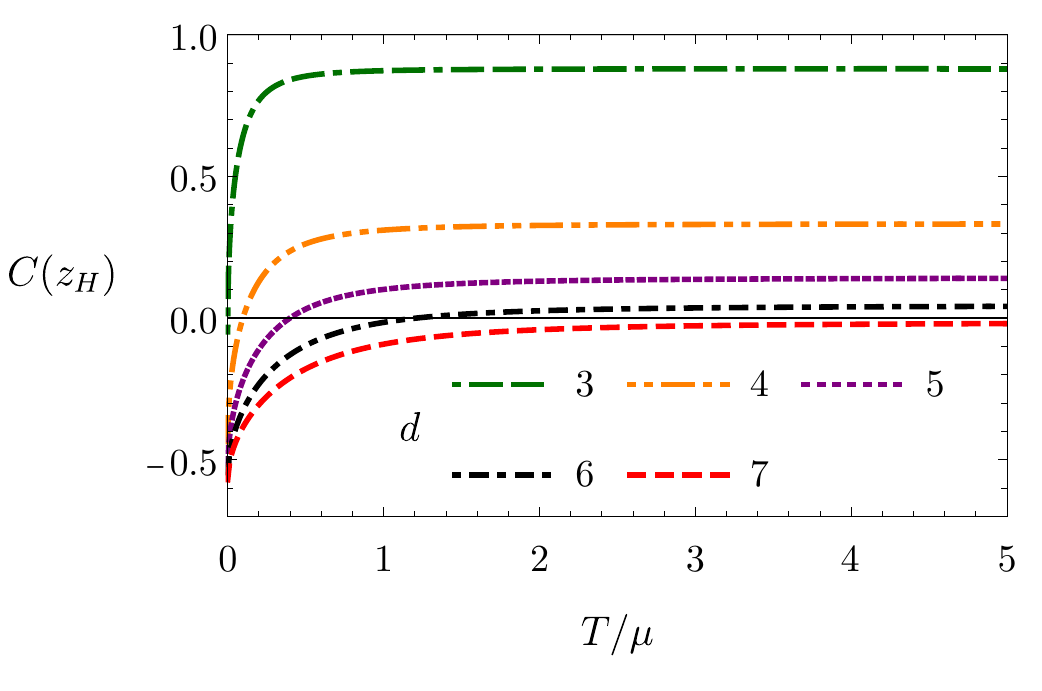}
	 \caption[The coefficient controlling the large-\(\ell\) behaviour of the entanglement density for \ads-Reissner-Nordstr\"om.]{The dimensionless coefficient $C(z_H)$ for \ads[d+1]-Reissner-Nordstr\"om, as a function of \(T/\m\). The sign of \(C(z_H)\) controls whether \(\s \to s\) from above or below at large \(\ell\), see~\eqref{eq:sigma_large_l}. We find that \(C(z_H) \geq 0\) at \(T/\m = 0\) for all values of \(d\) that we have checked. As \(T/\m \to \infty\), \(C(z_H)\) tends to the value for \ads[d+1]-Schwarzschild, which is negative for \(d \leq 6\) and positive for \(d \geq 7\). For \(T/\m = 0\), \(C(z_H)\) is positive for all \(d\).}
	 \label{fig:rnc}
\end{figure}

Since \(g(z)\) for \ads-Reissner-Nordstr\"om takes the form~\eqref{eq:fgasymp}, the small-\(\ell\) entanglement density is linear in \(\ell\) for both the strip and sphere geometries. As \(\ell \to \infty\) the entanglement density tends to the thermodynamic entropy density, with subleading behaviour given by~\eqref{eq:sigma_large_l}. The coefficient \(C(z_H)\) which determines whether the \(\s \to s\) from above or below is plotted as a function of \(T/\m\) in figure~\ref{fig:rnc}. We find that \(C(z_H)\) is positive for \(T/\m = 0\), for all \(d\). For large \(T/\m\), \(C(z_H)\) tends to the value for \ads-Schwarzschild, so is negative for \(d \leq 6\) and positive for \(d \geq 7\).
	
\begin{figure}
	\begin{subfigure}{0.5\textwidth}
		\includegraphics[width=\textwidth]{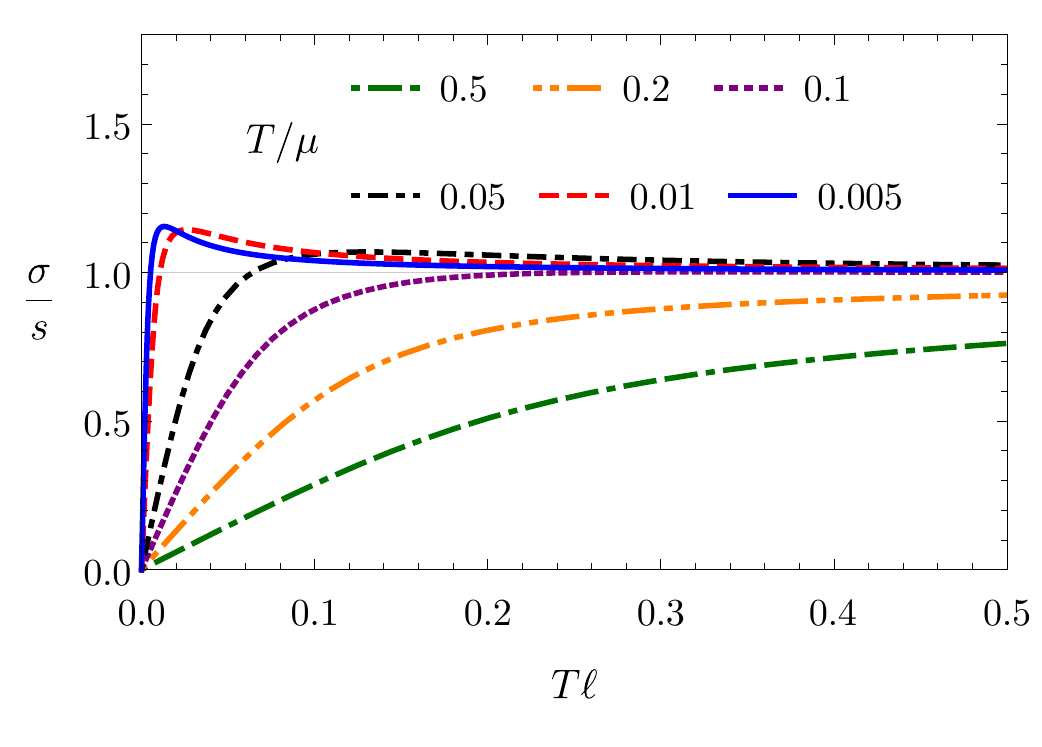}
		\caption{Strip, \(d = 4\).}
		\label{fig:adsrnd4}
	\end{subfigure}
	\begin{subfigure}{0.5\textwidth}
		\includegraphics[width=\textwidth]{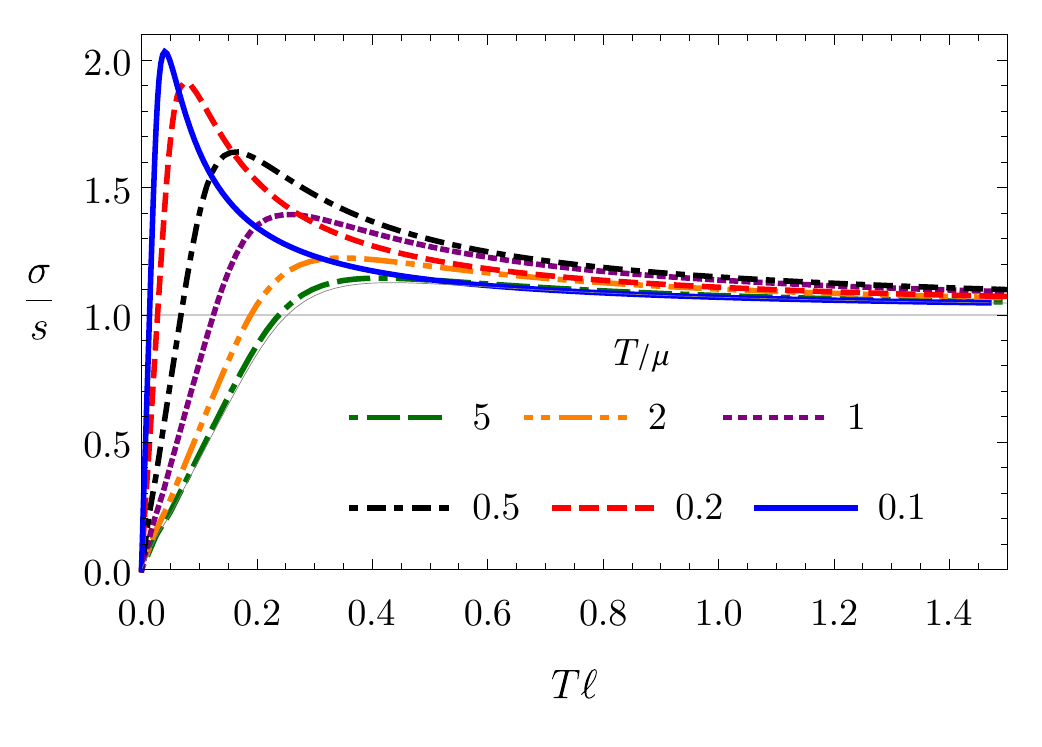}
		\caption{Strip, \(d = 8\).}
		\label{fig:adsrnd8}
	\end{subfigure}
	\begin{subfigure}{0.5\textwidth}
		\includegraphics[width=\textwidth]{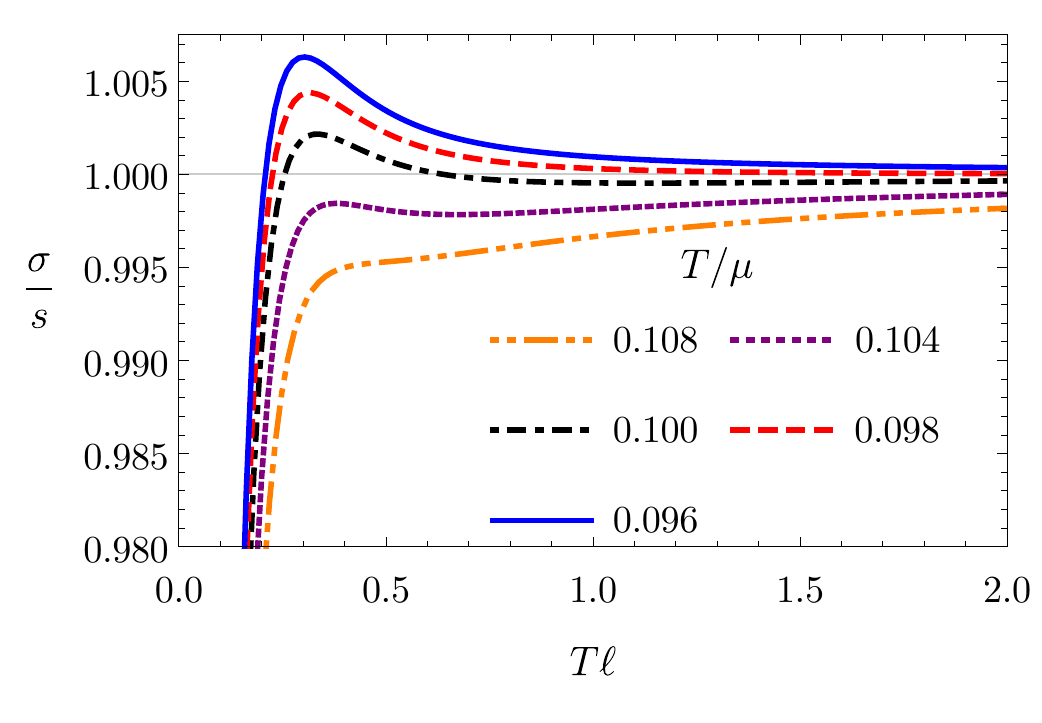}
		\caption{Strip, \(d=4\).}
		\label{fig:adsrnd4_close}
	\end{subfigure}
	\begin{subfigure}{0.5\textwidth}
		\includegraphics[width=\textwidth]{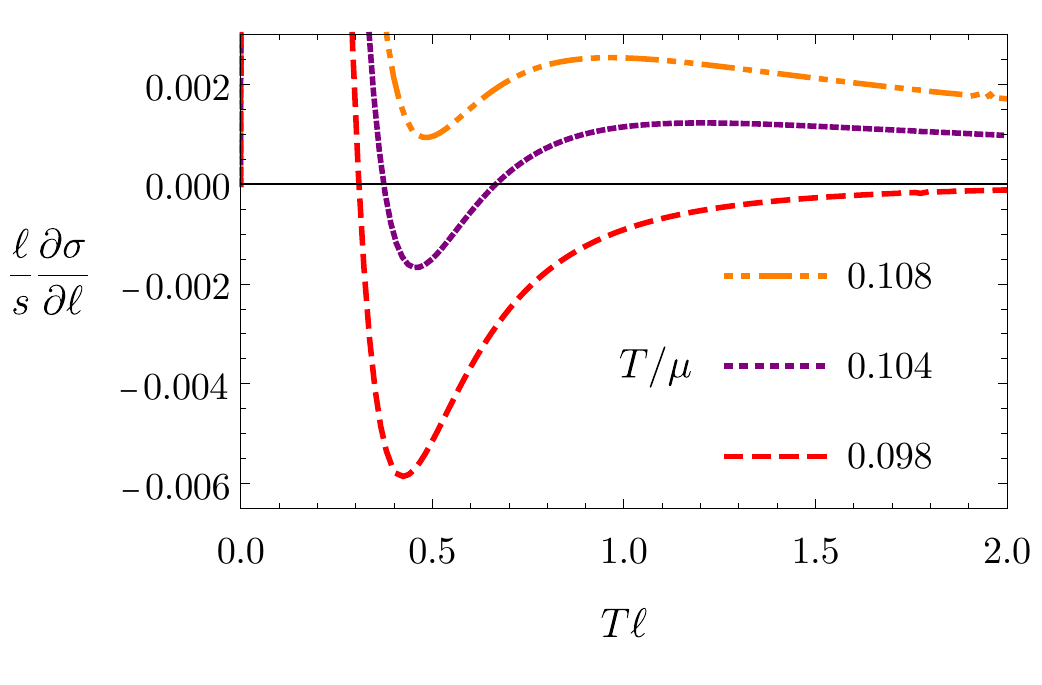}
		\caption{Strip, \(d=4\).}
		\label{fig:adsrnd4_close_deriv}
	\end{subfigure}
	\begin{subfigure}{0.5\textwidth}
		\includegraphics[width=\textwidth]{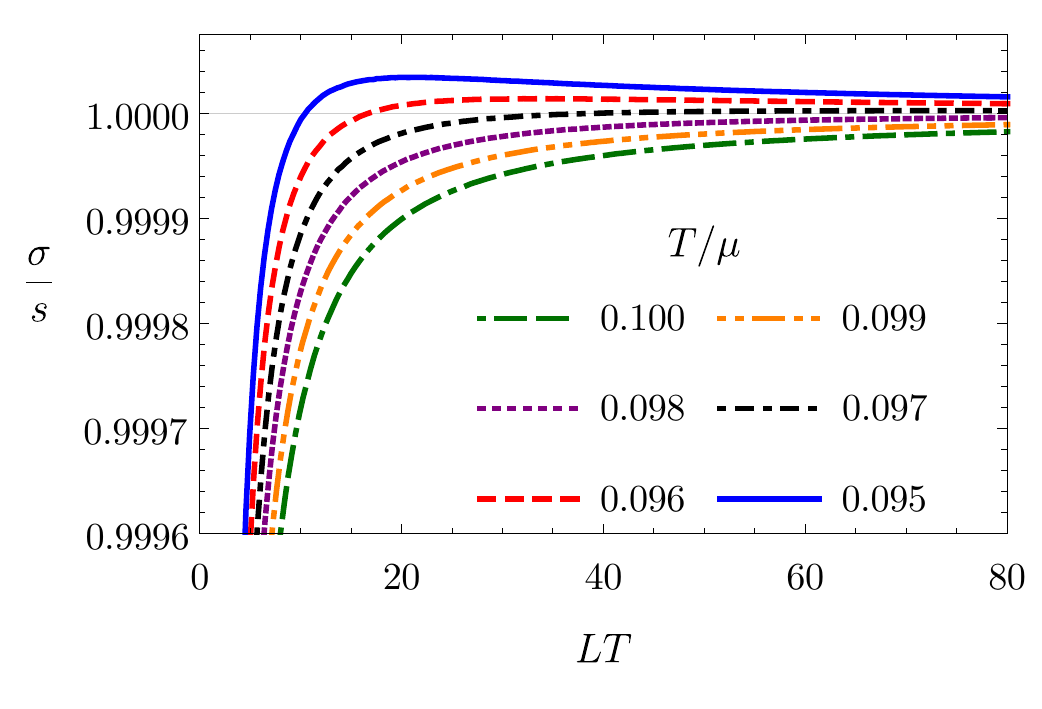}
		\caption{Sphere, \(d=4\).}
		\label{fig:adsrnd4_sphere_close}
	\end{subfigure}
	\begin{subfigure}{0.5\textwidth}
		\includegraphics[width=\textwidth]{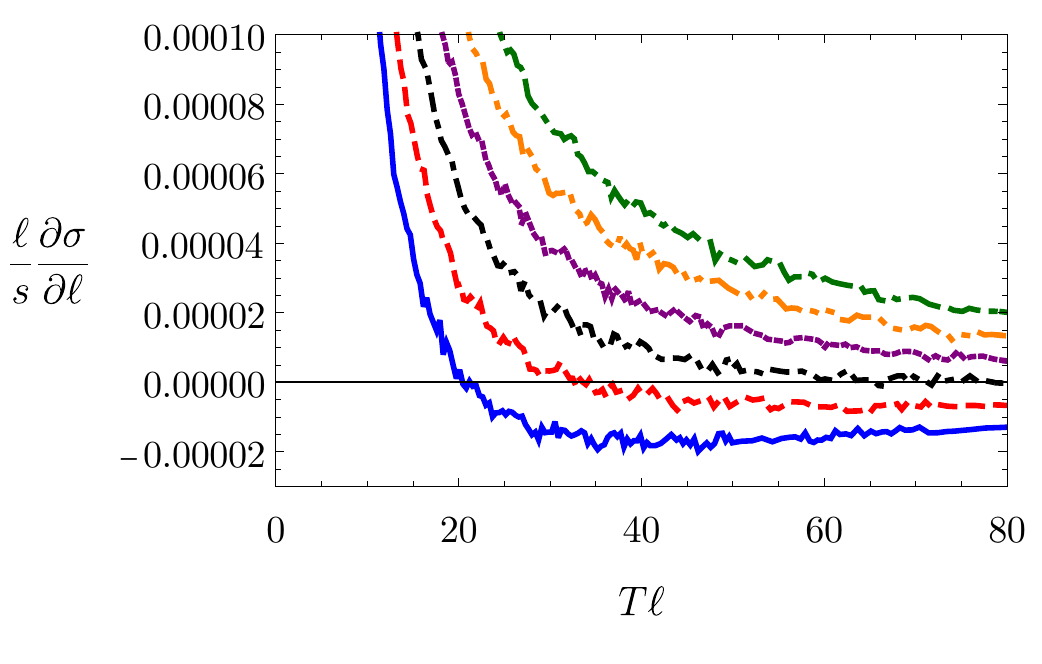}
		\caption{Sphere, \(d=4\).}
		\label{fig:adsrnd4_sphere_close_deriv}
	\end{subfigure}
	 \caption[Entanglement density in \ads-Reissner-Nordstr\"om.]{\textbf{(a,\,b):} The entanglement density as a function of $T\ell$ for the strip in \ads[d+1]-Reissner-Nordstr\"om with \(d=4\) and \(8\), for different values of $T/\mu$. The results for the sphere are qualitatively similar. \textbf{(c):} Close-up of the transition from \(\s \to s^+\) to \(\s \to s^-\) for the strip in \(d =4\). For a small range of \(T/\m\), \(\s\) exhibits a maximum as a function of \(\ell\), but approaches \(s\) from below as \(\ell \to \infty\). \textbf{(d):} Derivative of \(\s/s\) with respect to \(\ln\ell\), illustrating the disappearance of the local maximum. The colour coding is the same as in (c). \textbf{(e,\,f):} Entanglement density and its derivative for the sphere in \(d=4\). No local maximum is observed, within the numerical precision we have achieved.}
	 \label{fig:adsrnd4_8}
\end{figure}

\begin{figure}
\begin{subfigure}{0.5\textwidth}
\includegraphics[width=\textwidth]{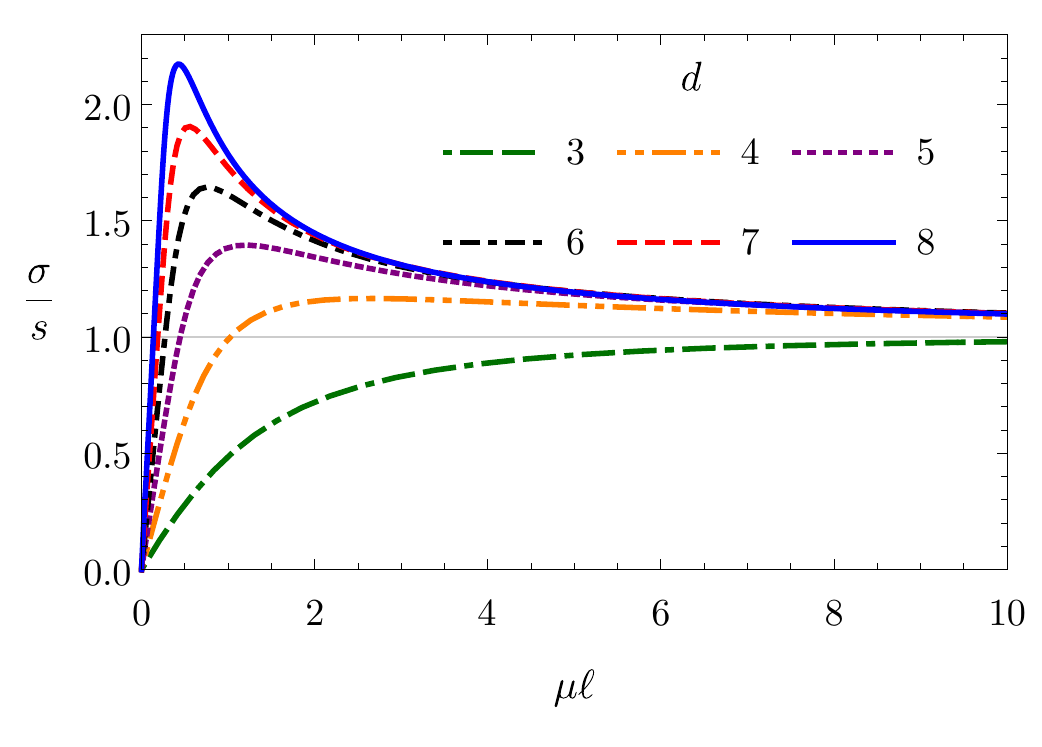}
\caption{Strip.}
\end{subfigure}
\begin{subfigure}{0.5\textwidth}
	\includegraphics[width=\textwidth]{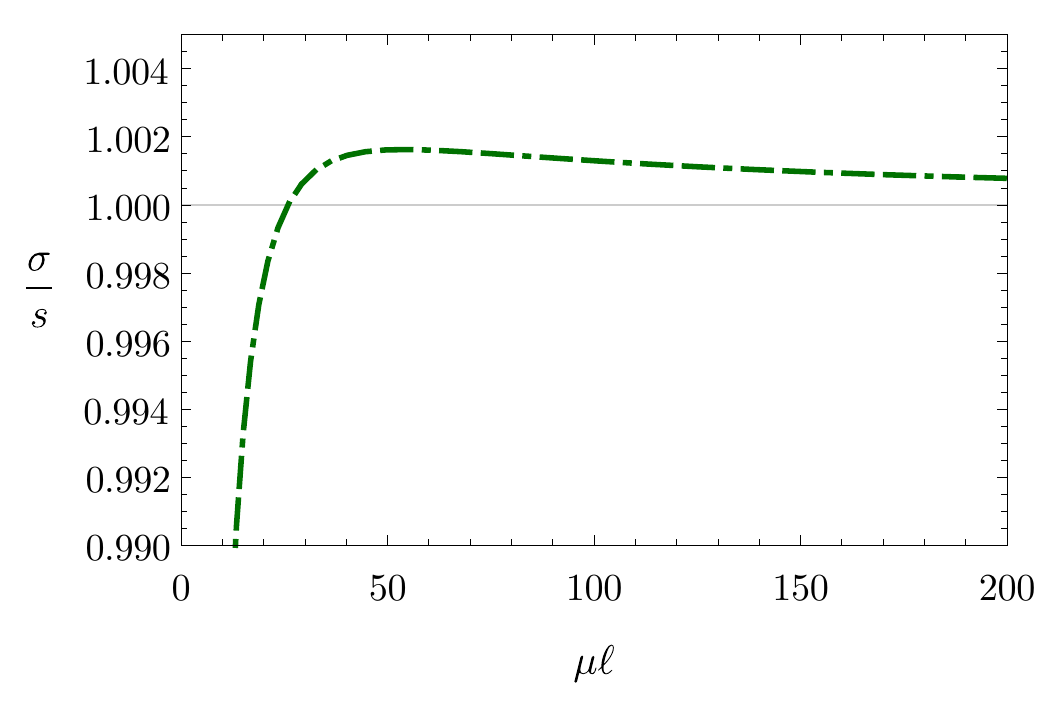}
	\caption{Strip.}
\end{subfigure}
\begin{subfigure}{0.5\textwidth}
	\includegraphics[width=\textwidth]{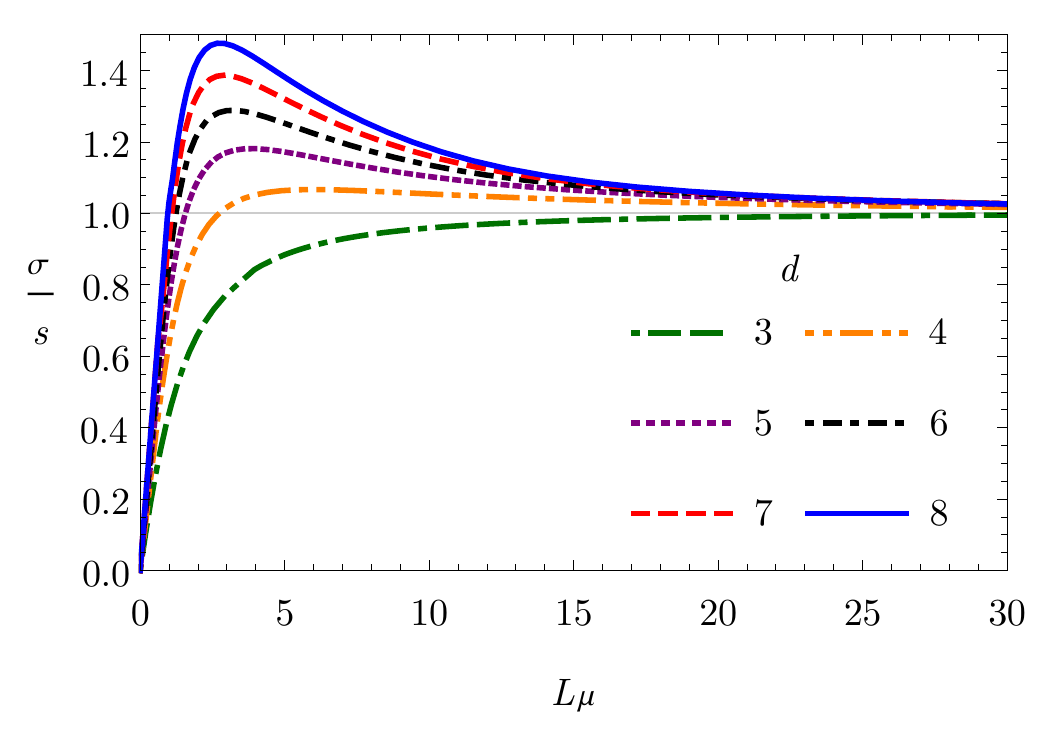}
	\caption{Sphere.}
\end{subfigure}
\begin{subfigure}{0.5\textwidth}
	\includegraphics[width=\textwidth]{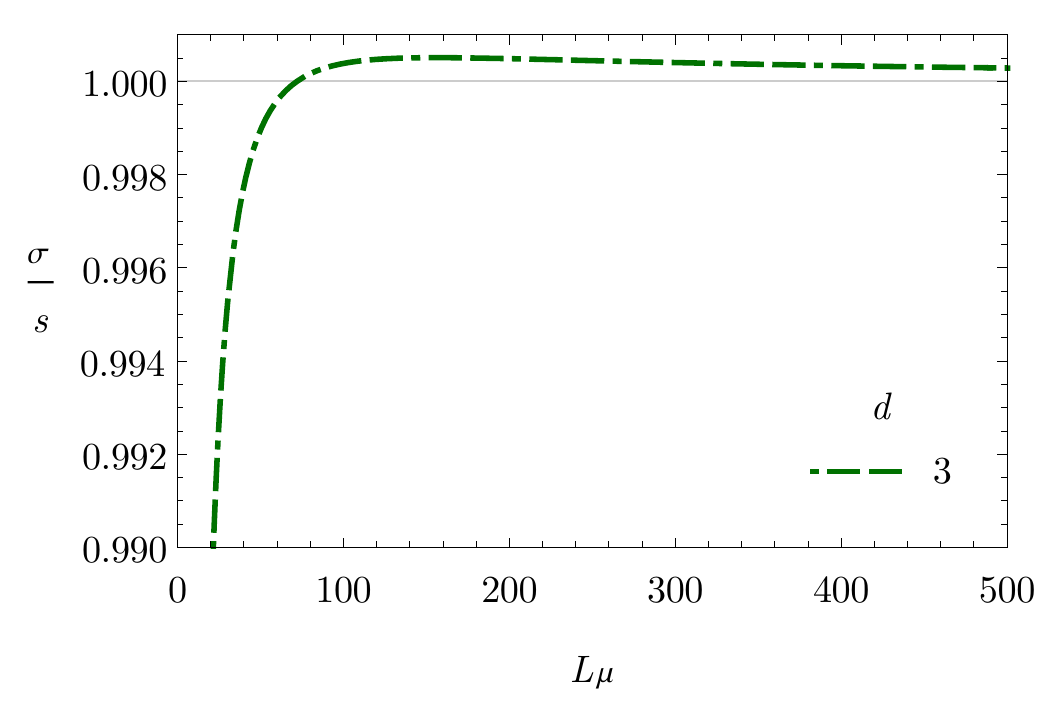}
	\caption{Sphere.}
\end{subfigure}
\caption[Entanglement density in extremal \ads-Reissner-Nordstrom.]{The entanglement density for the strip (top row) and sphere (bottom row) as a function of \(\m\ell\), for extremal \ads[d+1]-Reissner-Nordstr\"om, for various values of \(d\). For all \(d\), the entanglement density exhibits a global maximum, where \(\s/s > 1\). For \(d=3\) the maximum occurs at very large values of \(\m\ell\), as plotted in the right column.}
\label{fig:adsrnextremal}
\end{figure}

For example, figures~\ref{fig:adsrnd4} and~\ref{fig:adsrnd8} show the entanglement density as a function of $T\ell$ for the strip in \ads[d+1]-Reissner-Nordstr\"om for \(d = 4\) and \(8\) respectively, for different values of $T/\mu$.\footnote{The results for the strip in \ads-Reissner-Nordstr\"om with \(d=4\) (figures~\ref{fig:adsrnd4} and~\ref{fig:adsrnd4_close}) were submitted in partial fulfillment of my MPhys degree at the University of Oxford. The same is true of the results for the strip and the sphere in extremal \ads-Reissner-Nordstr\"om with \(d=4\) (the orange dot-dashed curves in figure~\ref{fig:adsrnextremal}).} We find $\s/s \propto T\ell$ at small $\ell$ for all $T/\mu$, as expected. For large \(T/\m\), the entanglement density reproduces that of \ads[d+1]-Schwarzschild; in particular it is monotonic in \(T\ell\) for \(d=4\), and exhibits a global maximum for \(d=8\). As \(T/\m\) is lowered, a global maximum appears also for \(d=4\), consistent with \(C(z_H)\) as plotted in figure~\ref{fig:rnc}. The sphere shows the same qualitative behaviour.

As plotted in figure~\ref{fig:adsrnd4_close}, for the strip in \(d=4\) the transition between \(\s/s \to 1^\pm\) at large \(\ell\) occurs in three stages. At high temperature, \(\s\) monotonically approaches \(s\) from below.  As we lower the temperature, at $T/\mu\approx 0.107$ a local minimum and maximum appear, with $\s/s<1$ for all $T\ell$. Then, at $T/\mu \approx 0.102$, the maximum rises above $\s/s=1$, becoming a global maximum, but a local minimum persists at $\s/s<1$, and then $\s/s\to 1^-$ as $T\ell \to \infty$. Finally, at $T/\mu \approx 0.098$, a transition occurs from $\s/s\to1^-$ to $\s/s\to1^+$ as $T\ell \to \infty$, and the local minimum disappears. Figure~\ref{fig:adsrnd4_close_deriv} shows the logarithmic derivative $\frac{\ell}{s} \frac{\partial \s}{\partial \ell}$, which clearly has no zero for $T/\mu> 0.107$, indicating $\s/s$ is monotonic in $T\ell$, then develops two zeroes for $0.107 > T/\mu > 0.102$, indicating a local minimum and maximum in $\s/s$, and then develops a single zero for $T/\mu < 0.098$, indicating a global maximum in $\s/s$. For the sphere, on the other hand, to within our numerical precision we find no value of \(T/\m\) for which \(\s\) exhibits a local maximum with \(\s/s < 1\), as shown in figures~\ref{fig:adsrnd4_sphere_close} and~\ref{fig:adsrnd4_sphere_close_deriv}.

We find qualitatively similar behaviour for the strip in all $d\leq6$: at some $(T/\mu)_1$ a local minimum and maximum appear, but $\s/s$ remains below one for all $T\ell$; at some $(T/\mu)_2<(T/\mu)_1$ a global maximum emerges, but still $\s/s\to1^-$ for $T\ell\to\infty$ (in other words \(C(z_H) < 0\)); and finally at some $(T/\mu)_3<(T/\mu)_2$ the transition occurs to $\s/s\to1^+$ as $T\ell \to \infty$ (so that \(C(z_H) > 0\)). Our numerical estimates of $(T/\mu)_1$, $(T/\mu)_2$, and $(T/\mu)_3$ are listed in table~\ref{tab:adsrn}. For \(d \geq 7 \) the global maximum with \(\s/s > 1\) exists for all \(T/\m\).
\begin{table}
\centering
\begin{tabular}[c]{ c|c c c }
$d$ & $(T/\mu)_1$ & $(T/\mu)_2$ & $(T/\mu)_3$ \\ \hline
$3$ & $6.343 \times 10^{-4}$ & $4.858 \times 10^{-4}$ & $2.967 \times 10^{-4}$ \\
$4$ & $0.107$ & $0.102$ & $0.098$  \\
$5$ & $0.407$ & $0.403$ & $0.399$ \\
$6$ & $1.219$ & $1.215$ & $1.213$ \\
\end{tabular}
\caption[Critical values of the ratio of temperature to chemical potential for the strip entanglement density in \ads-Reissner-Nordstr\"om.]{For the strip in \ads[d+1]-Reissner-Nordstr\"om with $d\leq6$, as $(T/\mu)$ decreases: at $(T/\mu)_1$ a local minimum and maximum appear in $\s/s$ as a function of $T\ell$; at $(T/\mu)_2$ the local maximum becomes a global maximum, but a local minimum remains, and $\s/s<1$ for all $T\ell$; and then at $(T/\mu)_3$ the global maximum rises above one, and the transition occurs to $\s/s\to1^+$ as $T\ell \to \infty$. }
\end{table}
\label{tab:adsrn}

In all cases above, the transition between $\s/s\to 1^{\pm}$ as $T\ell \to \infty$ indicates area theorem violation. Since \(C(z_H) > 0\) at sufficiently small \(T/\m\) for all \(d\), the area theorem is violated in extremal \ads-Reissner-Nordstr\"om for all \(d\). Figure~\ref{fig:adsrnextremal} shows $\s/s$ versus $\mu\ell$ in extremal \ads-Reissner-Nordstr\"om for various values of \(d\), illustrating that in all cases $\s/s$ indeed has a global maximum and $\s/s\to1^+$ as $\mu\ell\to\infty$.

In summary, in AdS-Reissner-Nordstr\"om for either $d\geq7$ at any $T/\mu$, or for any $d$ and sufficiently small $T/\mu$, we find a global maximum in $\s/s$, and $\s/s \to 1^+$ as $\ell \to \infty$, indicating area theorem violation. In other words, as we dial a parameter towards a limiting value in which an IR fixed point appears with different scaling from the UV CFT ($d \to \infty$ or $T/\mu \to 0$), we find area theorem violation.

\section{Conformal-to-Hyperscaling-Violating RG flows}
\label{hyper}

In this section we consider the bulk action
\begin{equation} \label{eq:hyperscaling_violating_action}
\sgrav = \frac{1}{16 \pi \gn}\int \diff^{d+1} x \sqrt{-G} \left[R - 2 \left(\partial \F\right)^2 - V(\Phi)- \frac{Z(\F)}{4} F^2 - \frac{\tilde{Z}(\F)}{4} \tilde{F}^2\right],
\end{equation}
where $\Phi$ is a real scalar field with potential $V(\Phi)$, $F_{MN}$ and $\tilde{F}_{MN}$ are field strengths for two $\U(1)$ gauge fields $A_M$ and $\tilde{A}_M$, respectively, and $Z(\Phi)$ and $\tilde{Z}(\Phi)$ are real functions of $\Phi$. The scalar field $\Phi$ is dual to a scalar operator $\cO$, while $A_M$ and $\tilde{A}_M$ are dual to conserved $\U(1)$ currents. Ref.~\cite{Lucas:2014sba} constructed solutions to the equations of motion that follow from~\eqref{eq:hyperscaling_violating_action}, with a metric of the form
\begin{equation}
\label{eq:sachdev_metric}
\diff s^2 = \frac{L^2}{z^2} \left[- a(z)b(z) \diff t^2 + \d_{ij} \diff x^i \diff x^j+\frac{a(z)}{b(z)} \diff z^2 \right],
\end{equation}
with real functions $a(z)$ and $b(z)$. This metric is of the form in~\eqref{eq:aads_metric} with $f(z)=a(z) b(z)$ and $g(z) = b(z)/a(z)$. If $b(z_H)=0$ then a horizon exists at $z=z_H$, with $\left< T_{tt} \right>$, $T$, and $s$ given by~\eqref{eq:energy_density} and~\eqref{eq:thermo_entropy}. The solutions of ref.~\cite{Lucas:2014sba} also include non-zero $\Phi(z)$, $F_{zt}(z)$, and $\tilde{F}_{zt}(z)$, with all other components of $F_{MN}$ and $\tilde{F}_{MN}$ vanishing.

Ref.~\cite{Lucas:2014sba} found a family of black brane solutions by splitting $b(z)$ as
\begin{equation}
\label{eq:bsplit}
b(z) = b_0(z) + \eta^2 \, b_2(z),
\end{equation}
where $b_0(z)$ and $b_2(z)$ are independent of the temperature \(T\) or chemical potential \(\m\) dual to \(A_M\), but the real parameter $\eta$ depends on $T/\m$, with \(\h = 0\) at \(T/\m=0\). Rescaling one of the gauge fields \(\tilde A_M \to \h \tilde A_M\), the equations of motion may be simplified by separating terms by powers of $\eta$, and solved by freely choosing two functions in the solution, which then determine all other functions and the corresponding potentials $V(\Phi)$, $Z(\Phi)$, and $\tilde{Z}(\Phi)$, leaving only a choice of boundary conditions. Following ref.~\cite{Lucas:2014sba}, we choose $b_2(z)$ and $F_{zt}(z)$, and obtain $a(z)$ by solving, from the equations of motion,
\begin{equation}
\label{eq:aeq}
\frac{\partial}{\partial\hat{z}}\left[\frac{a(\hat{z}) b_2(\hat{z})}{\hat{z}^{2(d-1)}}\right] = \hat{c} \, \frac{a(\hat{z})}{\hat{z}^{d-1}},
\end{equation}
with constant $\hat{c}$, and then obtain $b_0(z)$ by solving,
\begin{equation}
	\label{eq:b0eq}
	\frac{\p}{\p \hat{z}} \le(
		\frac{1}{\hat{z}^{d-1} a(\hat{z})} \frac{\p}{\p \hat{z}} \le[ a(\hat{z}) b_0(\hat{z})\ri]	\ri)  = - 2 \hat{F}_{zt},
\end{equation}
where $\hat{z}$ and $\hat{F}_{zt}$ are defined by the re-scalings
\begin{equation}
	z = Q^{-\frac{1}{d-1}} L^{\frac{d-2}{d-1}}\left(8\pi \gn\right)^{\frac{1}{1-d}} \hat{z},
	\qquad
	F_{zt} = Q^{\frac{1}{d-1}} L^{\frac{1}{d-1}}\left(8\pi \gn\right)^{\frac{1}{d-1}} \hat{F}_{zt},
\end{equation}
and \(Q\) is the charge density dual to \(A_M\), $Q \equiv - \delta S_{\textrm{grav}}/\delta F_{zt}$.

Specifying $b_2(z)$ and $F_{zt}(z)$ and solving~\eqref{eq:aeq} and~\eqref{eq:b0eq} with the above boundary conditions completely determines the metric, and therefore the holographic entanglement entropy. There are further equations which fix the remaining fields and potentials, for a detailed discussion of which we refer to ref.~\cite{Lucas:2014sba}. These additional equations imply that $\tilde{F}_{tz}(z)$ and $\Phi(z)$ are generically non-zero, indicating that the dual theory has non-zero chemical potential and charge density for the second $\U(1)$, and also $\vev{\cO} \neq 0$ and possibly a non-zero source for $\cO$. For the family of solutions under consideration, all of these quantities are determined by \(T/\m\).

We will solve~\eqref{eq:aeq} and~\eqref{eq:b0eq} numerically. We focus on the three solutions presented in ref.~\cite{Lucas:2014sba}, which at $T=0$ have no horizon, and describe domain walls from \aads[d+1] as $z \to 0$ to a hyperscaling-violating geometry as $z \to \infty$. We impose boundary conditions
\begin{equation}
\label{eq:sachdev_small_z}
a(z = 0) = b_0(z = 0) = 1 ,
\end{equation}
and at leading order $b_2(z) \propto - z^d$. If we choose
\begin{equation}
\hat{c} = d-2 + \frac{(d-1)(\zeta-1)-\theta}{d-1-\theta},
\end{equation}
then when $z \to \infty$ we find the following scalings
\begin{equation}
\label{eq:hvscalings}
	a(z) \sim z^{-\left[(d-1)(\z-1) - \q\right]/(d-\q-1)},
	\quad
	b_0(z) \sim z^{-\left[(d-1)(\z-1) + \q\right]/(d-\q-1)},
	\quad
	b_2(z) \sim - z^d.
\end{equation} 
The solutions we study all have \(\F \propto z^{d-2}\) at small \(z\), so \(\D_- = d-2\) in the notation of section~\ref{rg}. From~\eqref{eq:rgsmallL} we therefore find \(\s \propto \ell^{d-3}\) at small \(\ell\).

When \(\h = 0\) (corresponding to \(T=0\) in the dual QFT) the asymptotic form of the metric as \(z \to \infty\) is
\begin{equation} \label{eq:hv_metric}
	\diff s^2 \approx \frac{L^2}{z^2} \le[- z^{-2 (d-1)(\z-1)/(d-\q-1)} \diff t^2 +  \d_{ij} \diff x^i \diff x^j + z^{2\q/(d-\q-1)} \diff z^2\ri].
\end{equation}
When \(\h \neq 0\) there is instead a horizon at some \(z = z_H\). The entanglement density at \(T\ell \gg 1\) is given by~\eqref{eq:sigma_large_l}. When $\zeta \to \infty$ with $-\theta/\zeta$ fixed, the metric~\eqref{eq:hv_metric} becomes conformal to $\ads[2] \times \mathbb{R}^{d-1}$, with no horizon~\cite{Hartnoll:2012wm}.

Under a Lifshitz scaling, $t \to \lambda^{\zeta} t$, $\vec{x} \to \lambda \vec{x}$, $z \to \lambda^{(d-\q-1)/(d-1)} z$, the metric~\eqref{eq:hv_metric} rescales as $\diff s^2 \rightarrow \lambda^{2\q/(d-1)} \diff s^2$. For pure \ads, the holographic dual to a CFT, \(\q=0\) and \(\z = 1\). When \(\q = 0\) but \(\z \neq 1\), the holographic dual is referred to as a Lifshitz theory. Non-zero \(\q\) indicates hyperscaling violation for \(\q \neq 0\)~\cite{Huijse:2011ef,Dong:2012se}. Hyperscaling refers to the naive scaling of the free energy according to its mass dimension, \(\mathcal{F} \to \l^{-\z} \mathcal{F}\) under the Lifshitz scaling \(t \to \l^\z t\) and \(\vec{x} \to \l \vec{x}\). With hyperscaling violation, the free energy instead scales as \(\mathcal{F} \to \l^{\q - \z} \mathcal{F}\).

Hyperscaling violation is rare in non-holographic models, although Fermi liquids have \(\q = d-2\)~\cite{zaanen_liu_sun_schalm_2015}. Strange metals --- such as cuprate high-temperature superconductors --- have an optical conductivity with unusual high-temperature frequency dependence, which suggests that hyperscaling is violated in these materials~\cite{Patel:2015dda,Eberlein:2016jlt}. Hyperscaling-violating holographic models with \(\q = d-2\) are believed to possess \textit{hidden Fermi surfaces}, corresponding to fermions in non-trivial representations of the gauge group~\cite{Huijse:2011ef}. Hidden Fermi surfaces are difficult to probe directly in gauge/gravity duality, which relates only gauge-invariant observables in the two theories. However,  entanglement entropy in other holographic models with \(\q = d-2\) exhibits logarithmic violation of the area law, characteristic of a Fermi surface~\cite{Ogawa:2011bz,Huijse:2011ef}. We find that the same is true in the model of ref.~\cite{Lucas:2014sba}.

In this section we only compute the entanglement density for the strip geometry. Since we solve for \(a(z)\) and \(b_0(z)\) numerically, it is computationally intensive obtaining satisfactory numerical precision when solving for the RT surface in the sphere geometry and evaluating its area. The strip is simpler since we do not need to solve for the embedding of the RT surface. As discussed in section~\ref{sec:entanglement_asymptotics}, the large-\(\ell\) behaviour of the entanglement density is the same for the strip and the sphere, so it is enough to compute the entanglement density for the strip to determine whether the area theorem is violated.

We will now show results for the entanglement entropy for the three choices of \(b_2\) and \(F_{zt}\) given in ref.~\cite{Lucas:2014sba}. We first consider a \(d=3\) solution specified by
\begin{equation}
\label{eq:sachdev1}
b_2 = - \hat{z}^3 \, \frac{9\hat{z}^2+20\hat{z}+80}{9\hat{z}^2+20\hat{z}+40}, \qquad \hat{F}_{zt} = - \left(1+0.891\,\hat{z}\right)^{-4}.
\end{equation}
At $T=0$ this is a domain wall from \ads[4] to a hyperscaling-violating geometry with $\zeta = 2$ and $\theta=-2$.

\begin{figure}
	\begin{subfigure}{0.5\textwidth}
		\includegraphics[width=\textwidth]{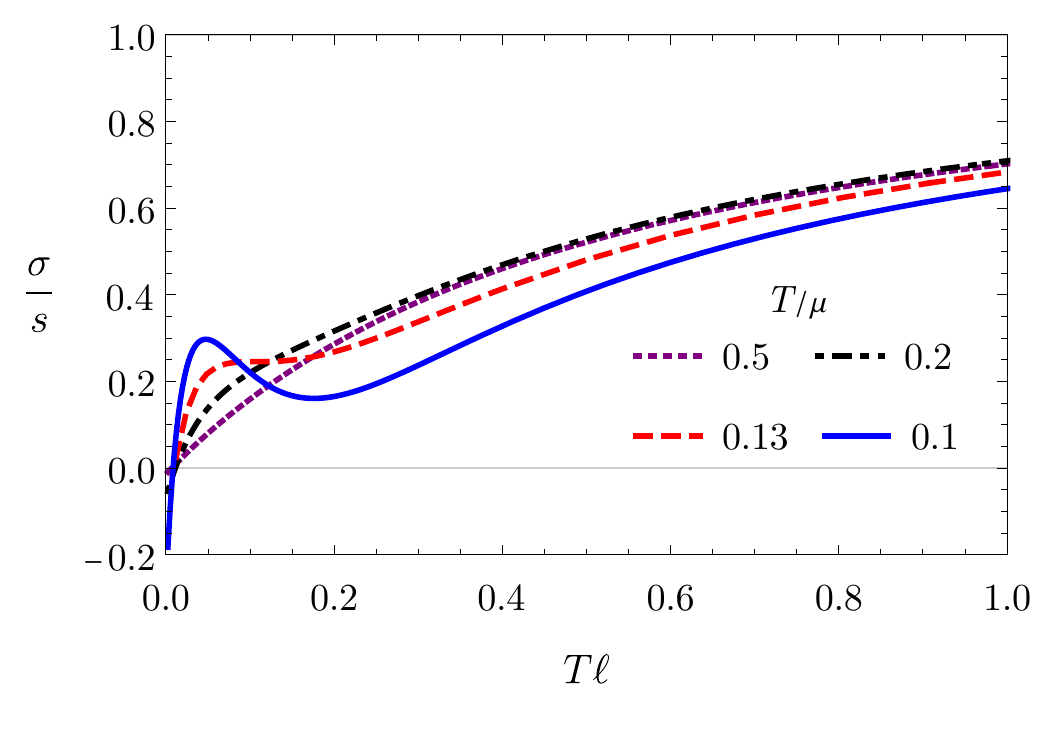}
		\caption{High temperature.}
		\label{fig:sachdev1_high_temperature}
	\end{subfigure}
	\begin{subfigure}{0.5\textwidth}
		\includegraphics[width=\textwidth]{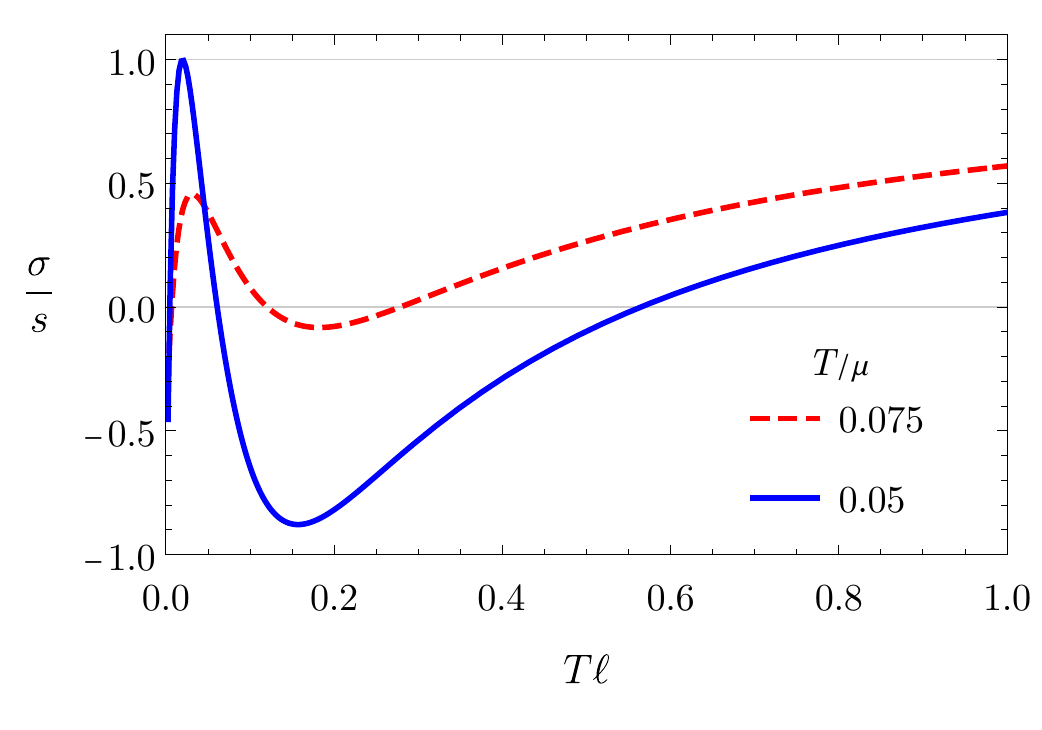}
		\caption{Low temperature.}
		\label{fig:sachdev1_low_temperature}
	\end{subfigure}
	\begin{subfigure}{0.5\textwidth}
		\includegraphics[width=\textwidth]{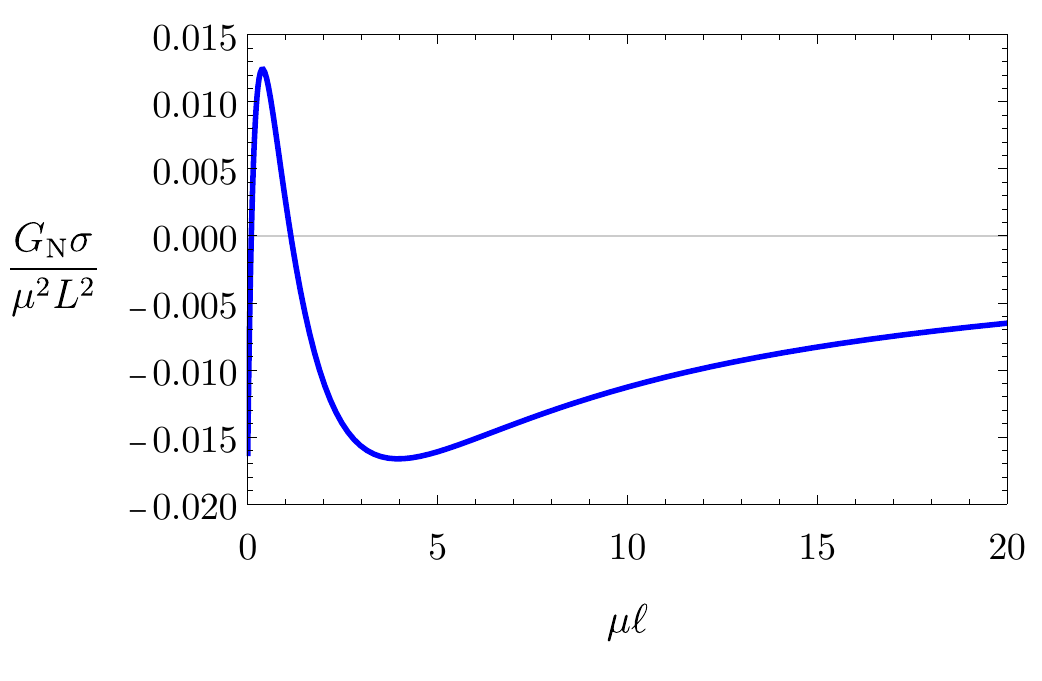}
		\caption{Zero temperature \(T = 0\).}
		\label{fig:sachdev1_zero_temperature}
	\end{subfigure}
	\begin{subfigure}{0.5\textwidth}
		\includegraphics[width=\textwidth]{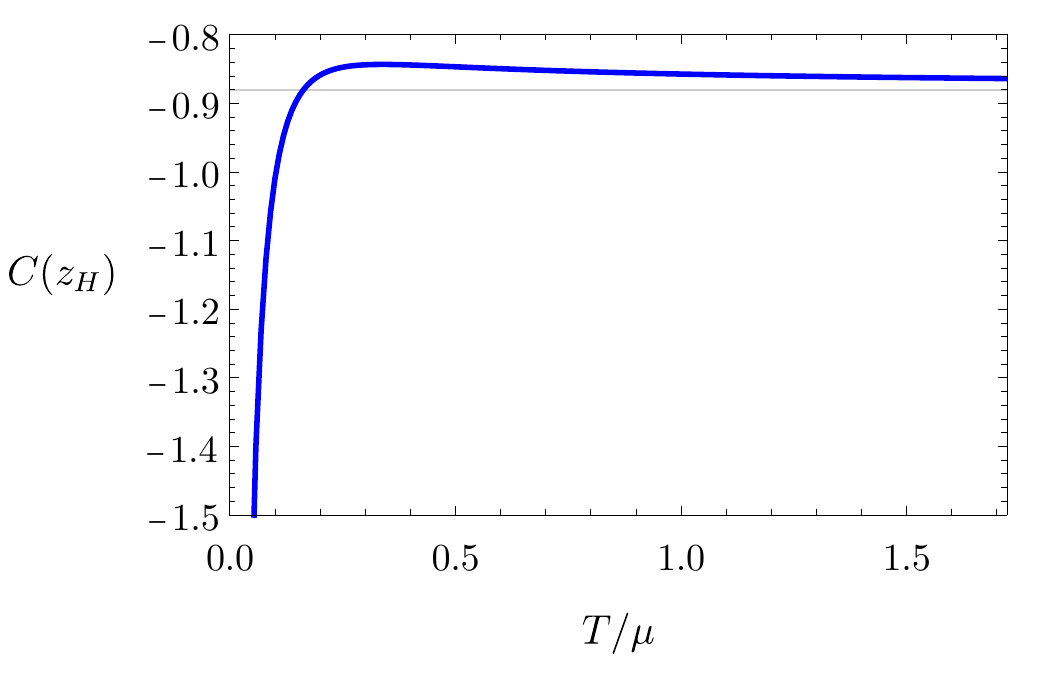}
		\caption{Large-\(\ell\) behaviour.}
		\label{fig:sachdev1_c}
	\end{subfigure}
	\caption[Entanglement density for hyperscaling-violating RG flows with $d=3$, $\zeta=2$, and $\theta=-2$.]{\textbf{(a,\,b,\,c):} The entanglement density for the strip geometry in the solution corresponding to equation~\eqref{eq:sachdev1} ($d=3$, $\zeta=2$, $\theta=-2$). Figures (a) and (b) show the entanglement density at non-zero temperature (the distinction between high and low temperature is an arbitrary choice made for clarity of presentation). A local maximum and minimum appear when $T/\mu \lesssim 0.130$. Figure (c) is the entanglement density for \(T=0\), corresponding to a domain wall between a CFT in the UV and a hyperscaling-violating IR. \textbf{(d):} The coefficient \(C(z_H)\), the sign of which controls whether \(\s \to s\) from above or below, according to~\eqref{eq:sigma_large_l}. We find $C(z_H)<0$ for all $T/\mu$, indicating that the area theorem is obeyed. As \(T/\m \to \infty\), \(C(z_H)\) approaches its \ads[4]-Schwarzschild value, indicated by the horizontal line.}
	\label{fig:sachdev1}
\end{figure}
Figures~\ref{fig:sachdev1_high_temperature} and~\ref{fig:sachdev1_low_temperature} show the entanglement density versus $T\ell$ for the strip in this solution for several non-zero values of $T/\mu$. For all $T/\mu$, we find that \(\s/s\) tends to a negative constant value as \(\ell \to 0\). At large \(T/\m\), the entanglement density tends to that of \ads[4]-Schwarzschild, which is monotonic in \(T\ell\).  Lowering \(T/\m\), when $T/\mu\approx0.130$ a local maximum and minimum appear at intermediate $T\ell$.  As $T/\mu$ is decreased further, the maximum grows in height, and the minimum increases in depth. Figure~\ref{fig:sachdev1_zero_temperature} shows the entanglement density at \(T=0\). We find that the entanglement density tends to the entropy density from below as $\ell \to \infty$ for all $T/\mu$; figure~\ref{fig:sachdev1_c} shows $C(z_H)$, which is negative for all $T/\mu$.

This example shows that an IR fixed point with non-relativistic scaling does not require violation of the area theorem. If there is indeed a connection between the area theorem and scaling in the IR, then the connection only goes one way. In other words, it is possible that area theorem violation implies non-relativistic scaling in the IR, but the reverse statement is not true.

We next consider the \(d=4\) solution
\begin{equation}
\label{eq:sachdev2}
b_2 = - \hat{z}^4 \, \frac{\hat{z}^2+12}{\hat{z}^2+6}, \qquad \hat{F}_{zt} = - \hat{z}\left(1+0.852\,\hat{z}^2\right)^{-3},
\end{equation}
which at \(T=0\) has $\zeta \to \infty$ with $\theta/\zeta=-3$.

\begin{figure}
	\begin{subfigure}{0.5\textwidth}
		\includegraphics[width=\textwidth]{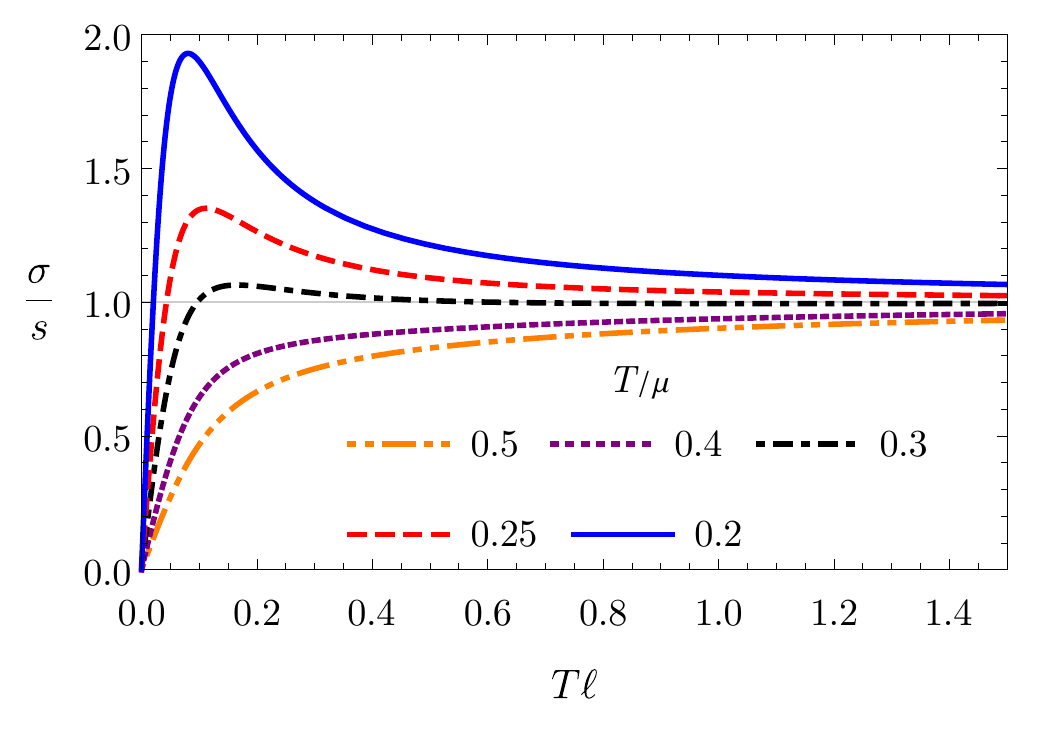}
		\caption{Non-zero temperature.}
		\label{fig:sachdev2_finite_temperature}
	\end{subfigure}
	\begin{subfigure}{0.5\textwidth}
		\includegraphics[width=\textwidth]{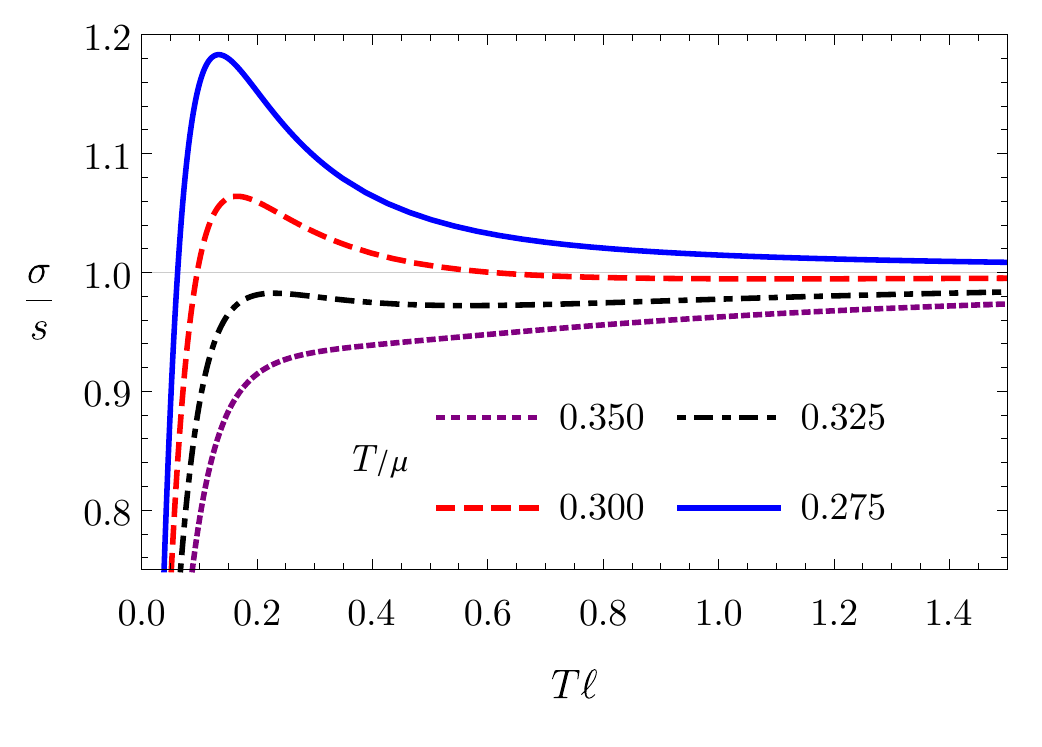}
		\caption{Non-zero temperature.}
		\label{fig:sachdev2_finite_temperature_close}
	\end{subfigure}
	\begin{subfigure}{0.5\textwidth}
		\includegraphics[width=\textwidth]{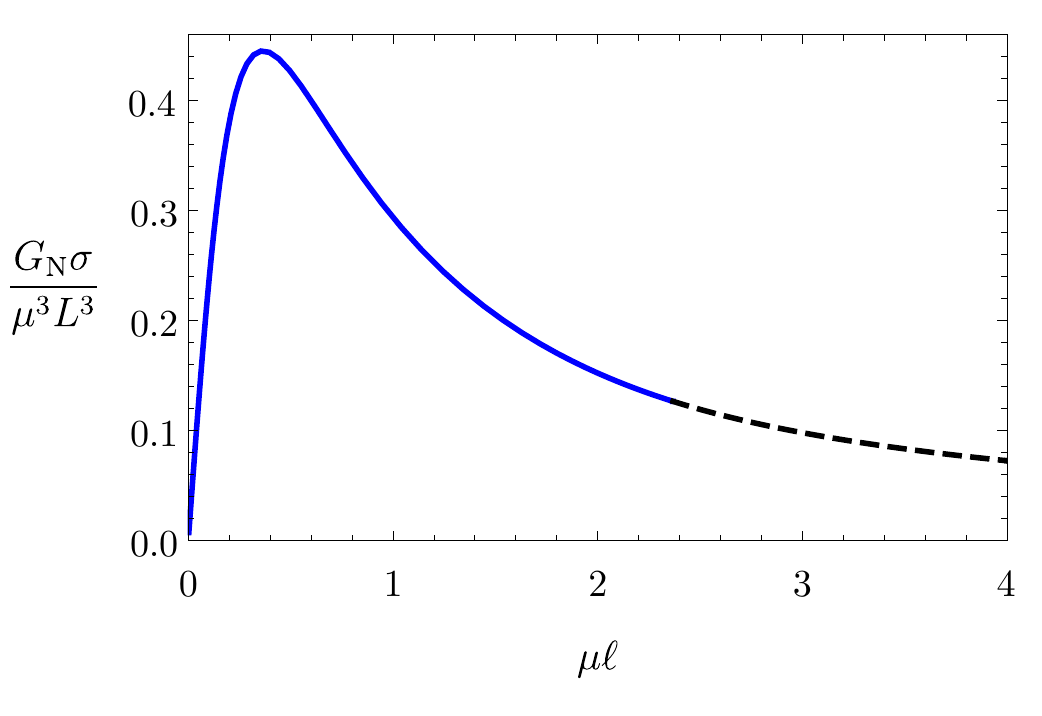}
		\caption{Zero temperature.}
		\label{fig:sachdev2_zero_temperature}
	\end{subfigure}
	\begin{subfigure}{0.5\textwidth}
		\includegraphics[width=\textwidth]{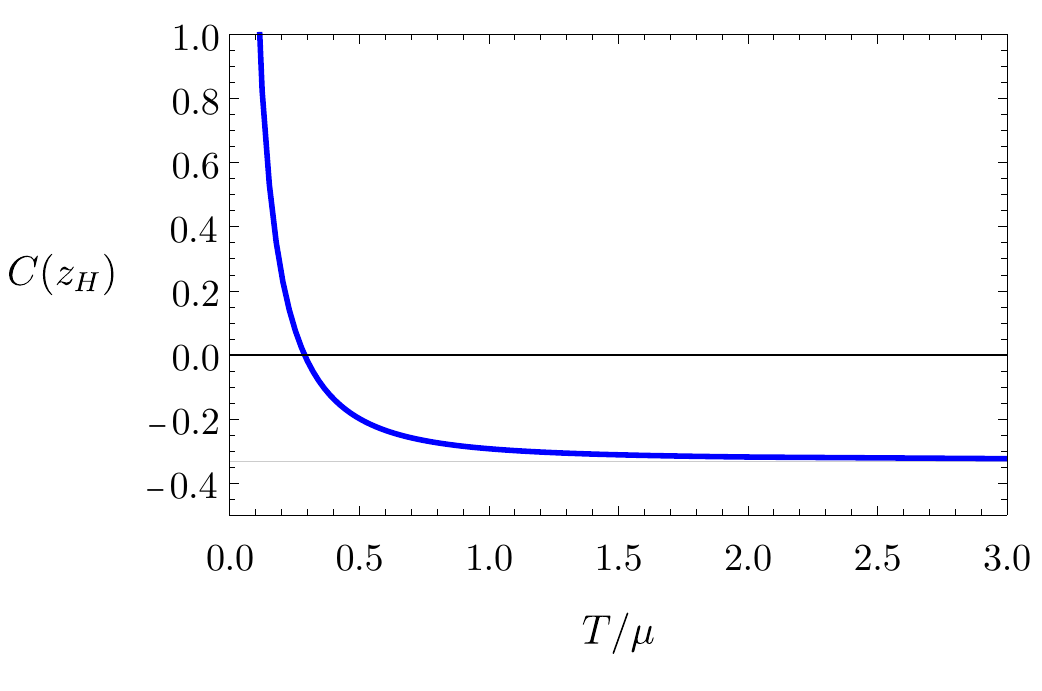}
		\caption{Large-\(\ell\) behaviour.}
		\label{fig:sachdev2_c}
	\end{subfigure}
	\caption[Entanglement density for hyperscaling-violating RG flows with $d=4$, $\zeta=\infty$, and $\theta/\zeta = -3$.]{
		\textbf{(a,\,b,\,c):} The entanglement density for the strip geometry in the solution corresponding to equation~\eqref{eq:sachdev1} ($d=4$, $\zeta\to \infty$, $\theta/\zeta=-3$). Figures (a) and (b) show the entanglement density at non-zero temperature. At large \(T/\m\), \(\s \to s^-\) monotonically as \(T\ell \to \infty\), while at small \(T/\m\) the entanglement density exhibits a global maximum, and then \(\s \to s^+\) as \(T\ell \to \infty\). Figure (b) focuses on the transition between these behaviours. Figure (c) is the entanglement density for \(T=0\). A phase transition occurs in the entanglement entropy at \(\m\ell \approx 2.37\). For small \(\m\ell\) the entanglement entropy is given by the area of a connected surface, indicated by the blue line in figure (c). For large \(\m\ell\) the entanglement entropy is given by the area of a disconnected surface, denoted by the dashed black line. As \(\m\ell \to \infty\), \(\s \to 0\) from above, indicating area theorem violation. \textbf{(d):} The coefficient \(C(z_H)\), the sign of which controls whether \(\s \to s\) from above or below, according to equation~\eqref{eq:sigma_large_l}. We find \(C(z_H) < 0\) for \(T/\m \gtrsim 0.289\), while \(C(z_H) > 0\) for smaller \(T/\m\). As \(T/\m \to \infty\), \(C(z_H)\) approaches its \ads[5]-Schwarzschild value, indicated by the horizontal line.}
	\label{fig:sachdev2}
\end{figure}
%
Figures~\ref{fig:sachdev2_finite_temperature} and~\ref{fig:sachdev2_finite_temperature_close} show the entanglement density as a function of $T\ell$ for the strip in this solution, for several values of $T/\mu$. For all $T/\mu$, we find $\s/s \propto \ell$ at small $T\ell$. At sufficiently large $T/\mu$, \(\s/s\) increases monotonically with \(T\ell\), so in particular $\s/s \to 1^-$ as $T\ell \to \infty$. As we decrease $T/\mu$ we find a transition very similar to that of \ads-Reissner-Nordstr\"om with $d \leq 6$, discussed in section~\ref{adsrn}. At some $(T/\mu)_1$, a local minimum and maximum appear, with $\s/s$ remaining below one for all $T\ell$. Then, at some $(T/\mu)_2$, the local maximum rises above one to become a global maximum, but still $\s/s\to1^-$ for $T\ell\to\infty$. Finally, at some $(T/\mu)_3$, the local minimum disappears and $\s/s\to1^+$ as $T\ell \to \infty$. Our numerical results for $(T/\mu)_1$, $(T/\mu)_2$, and $(T/\mu)_3$ are listed in table~\ref{tab:hyper}.

Figure~\ref{fig:sachdev2_zero_temperature} shows our numerical results for the entanglement density at \(T/\m = 0\). As mentioned above, for a solution such as this, with $\zeta \to \infty$, when $T/\mu=0$ the $z \to \infty$ geometry is conformal to $\ads[2] \times \mathbb{R}^{d-2}$, with no horizon. As for other geometries conformal to \(\ads[2]\times \mathbb{R}^{d-2}\), there are two different extremal surfaces~\cite{Kulaxizi:2012gy,Erdmenger:2013rca}. One is a smooth, connected surface, with area given by~\eqref{eq:strip_area}. The other is a disconnected surface, consisting of two sheets at \(x(z) = \pm \ell/2\). The connected surface only exists for \(\m \ell \lesssim 2.37\). When it exists, it always has smaller area than the disconnected surface, and therefore determines the entanglement entropy (the blue curve in figure~\ref{fig:sachdev2_zero_temperature}). For \(\m \ell \gtrsim 2.37\), the entanglement entropy is determined by the disconnected surface (the dashed black curve in figure~\ref{fig:sachdev2_zero_temperature}).

Figure~\ref{fig:sachdev2_c} shows $C(z_H)$ versus $T/\mu$. As expected from our results for the entanglement density, for $T/\mu<(T/\mu)_3$ we find \(C(z_H) < 0\).

\begin{table}
\centering
\begin{tabular}[c]{c c c|c c c}
$d$ & $\zeta$ & $\theta$ & $(T/\mu)_1$ & $(T/\mu)_2$ & $(T/\mu)_3$ \\ \hline
$4$ & $\infty$ & $-3\zeta$ & $0.336$ & $0.319$ & $0.289$ \\
$3$ & $3$ & $1$ & $0.0629$ & $0.0516$ & $0.0334$  \\
\end{tabular}
\caption[Critical values of the ratio of temperature to chemical potential for the strip entanglement density in \ads-Reissner-Nordstr\"om.]{Critical values of \(T/\m\) for the strip entanglement entropy in the hyperscaling-violating geometries determined by~\eqref{eq:sachdev2} ($d=4$, $\zeta=\infty$, $\theta=-3\zeta$) and~\eqref{eq:sachdev3} ($d=3$, $\zeta=3$, $\theta=1$). When \(T/\m\) is large, the entanglement density tends to the entropy density monotonically, with \(\s/s \to 1^-\) as \(T\ell \to \infty\). Decreasing \(T/\m\), at $(T/\mu)_1$ a local minimum and maximum appear in $\s/s$ as a function of $T\ell$, at $(T/\mu)_2$ the local maximum becomes a global maximum, and then at $(T/\mu)_3$ the local minimum disappears, and $\s/s\to1^+$ as $T\ell \to \infty$. }
\label{tab:hyper}
\end{table}

The final solution we consider has $d=3$ and
\begin{equation}
\label{eq:sachdev3}
b_2 = - \hat{z}^3 \, \frac{\hat{z}^2+12\hat{z}+288}{\hat{z}^2+12\hat{z}+72}, \qquad \hat{F}_{zt} = - \left(1+0.891\,\hat{z}\right)^{-4}.
\end{equation}
At $T=0$, the solution is a domain wall from \ads[4] to a hyperscaling-violating geometry with $\zeta =3$ and $\theta=1$.

\begin{figure}
	\begin{subfigure}{0.5\textwidth}
		\includegraphics[width=\textwidth]{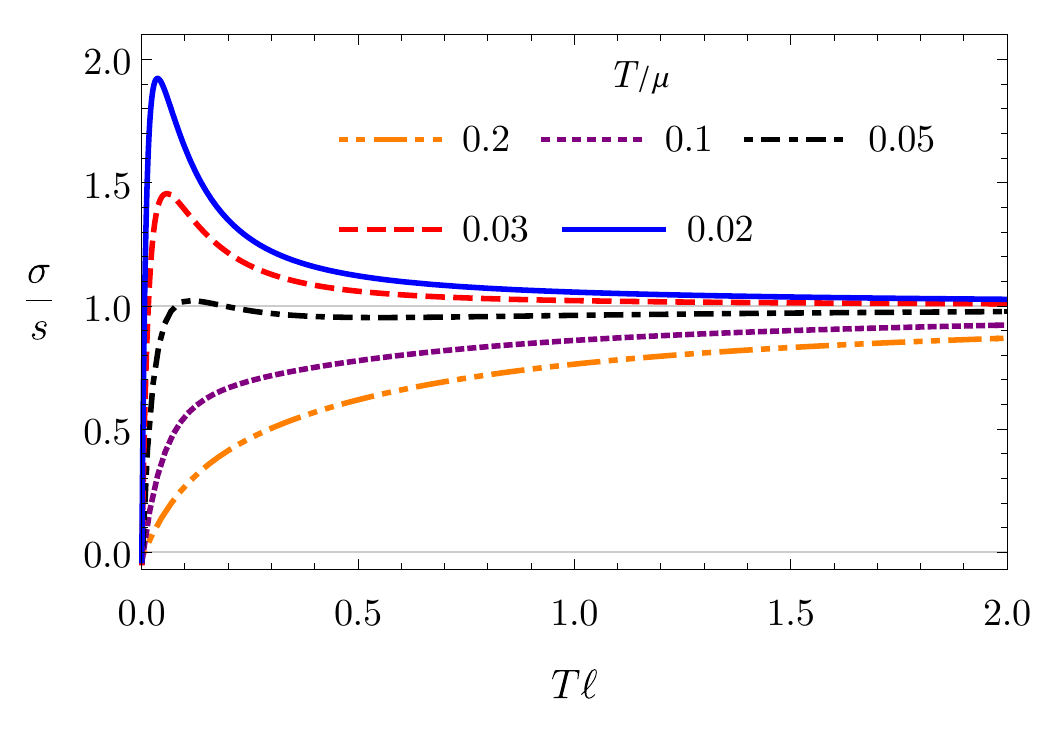}
		\caption{Non-zero temperature.}
		\label{fig:sachdev3_finite_temperature}
	\end{subfigure}
	\begin{subfigure}{0.5\textwidth}
		\includegraphics[width=\textwidth]{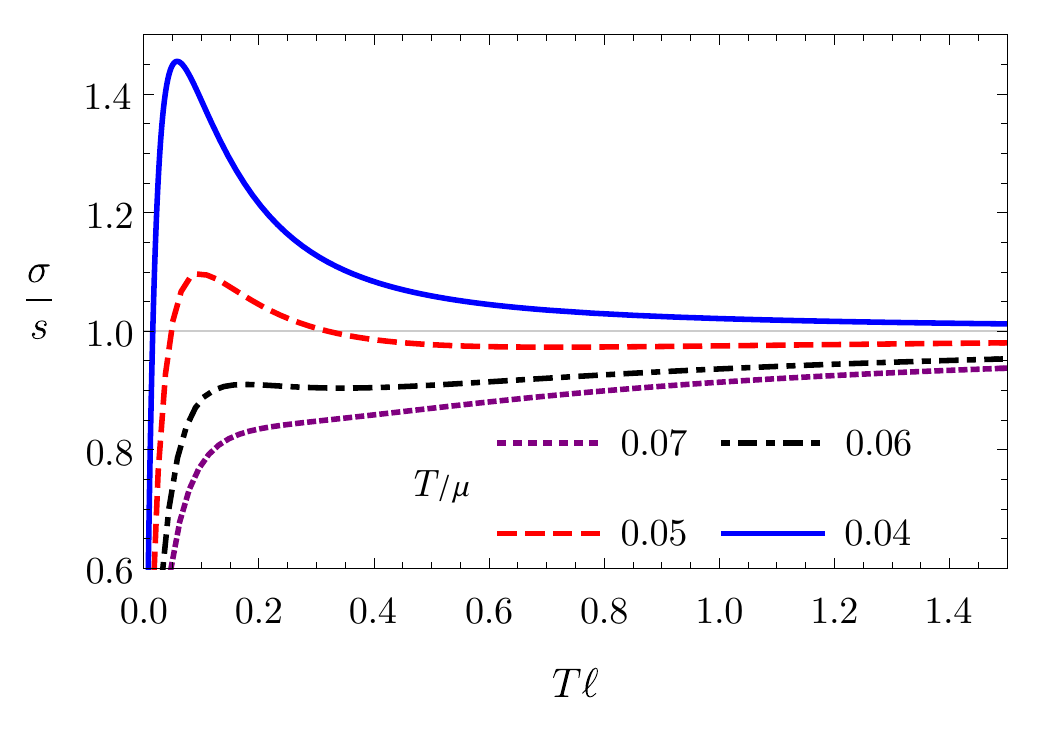}
		\caption{Non-zero temperature.}
		\label{fig:sachdev3_finite_temperature_close}
	\end{subfigure}
	\begin{subfigure}{0.5\textwidth}
		\includegraphics[width=\textwidth]{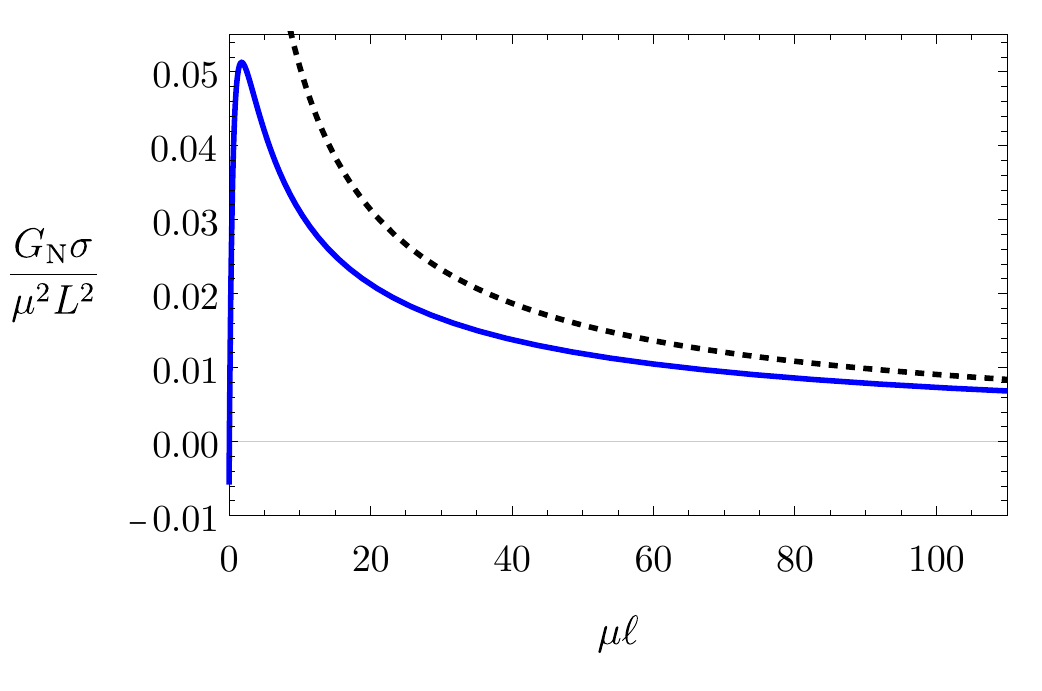}
		\caption{Zero temperature.}
		\label{fig:sachdev3_zero_temperature}
	\end{subfigure}
	\begin{subfigure}{0.5\textwidth}
		\includegraphics[width=\textwidth]{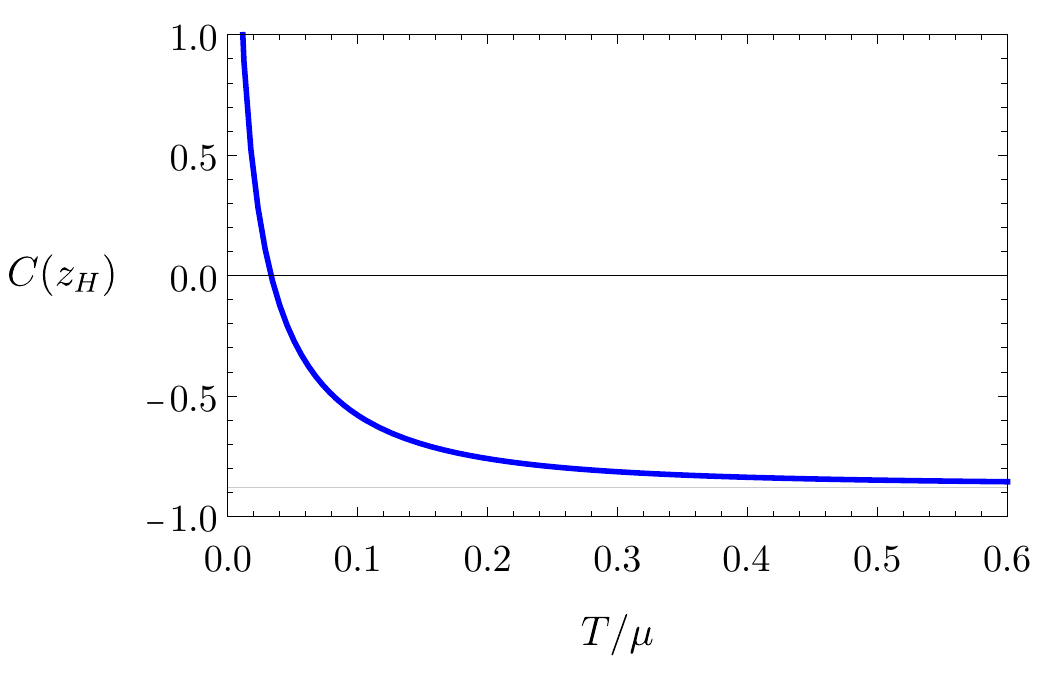}
		\caption{Large-\(\ell\) behaviour.}
		\label{fig:sachdev3_c}
	\end{subfigure}
	\caption[Entanglement density for hyperscaling-violating RG flows with $d=4$, $\zeta=3$, and $\theta = 1$.]{
		\textbf{(a,\,b):} The entanglement density for the strip geometry in the solution corresponding to equation~\eqref{eq:sachdev1} ($d=3$, $\zeta=3$, $\theta=1$) at non-zero temperature. At large \(T/\m\), \(\s \to s^-\) monotonically as \(T\ell \to \infty\), while at small \(T/\m\) the entanglement density exhibits a global maximum, and then \(\s \to s^+\) as \(T\ell \to \infty\). Figure (b) focuses on the transition between these behaviours. \textbf{(c):} The entanglement density for \(T=0\). As \(\m\ell \to \infty\), \(\s \to 0\) from above, indicating (logarithmic in this case) area theorem violation. The dotted black line is the large-width approximation~\eqref{eq:arealawviolation}. \textbf{(d):} The coefficient \(C(z_H)\), the sign of which controls whether \(\s \to s\) from above or below, according to equation~\eqref{eq:sigma_large_l}. We find \(C(z_H) < 0\) for \(T/\m \gtrsim 0.0334\), while \(C(z_H) > 0\) for smaller \(T/\m\). As \(T/\m \to \infty\), \(C(z_H)\) approaches its \ads[4]-Schwarzschild value, indicated by the horizontal line.}
	\label{fig:sachdev3}
\end{figure}
Figures~\ref{fig:sachdev3_finite_temperature} and~\ref{fig:sachdev3_finite_temperature_close} show our results for the entanglement density in this solution with \(T/\m > 0\). At high temperature we find that \(\s/s \to 1^-\) monotonically. Lowering $T/\mu$, we find similar behaviour to the previous solution, with a local minimum and maximum emerging at \((T/\m)_1\), the maximum becoming global at \((T/\m)_2\), and finally \(\s/s \to 1^-\) for \(T/\m \leq (T/\m)_3\), indicating area theorem violation at low temperatures. The values of these critical values of \((T/\m)\) are listed in table~\ref{tab:hyper}. Figure~\ref{fig:sachdev3_zero_temperature} shows the entanglement density as a function of \(\m\ell\) at $T/\mu=0$. We find $\s \to 0^+$ as $\mu\ell \to \infty$, indicating that the area theorem violation persists at zero temperature.

At \(T/\m = 0\), the IR hyperscaling violation exponent is \(\q = d-2\).  We therefore expect the dual QFT to contain a hidden Fermi surface, which should produce a logarithmic term in the entanglement density at large \(\ell\). To see the origin of the logarithm, we rewrite the entanglement density for the strip as~\cite{Gushterov:2017vnr}
\begin{equation} \label{eq:hyper_entanglement_density}
	\s_\mathrm{strip} = \frac{L^{d-1}}{4 \gn} \le[\frac{1}{z_*^{d-1}} + \frac{2C(z_*)}{z_*^{d-2} \ell} + \frac{\varrho^{d-1}}{(d-2)} \frac{1}{\ell^{d-1}} \ri],
\end{equation}
where \(\varrho\) was defined in~\eqref{eq:small_strip_coefficients} and \(C(z_*)\) is the integral~\eqref{eq:large_l_coefficient_integral}, with the horizon position \(z_H\) replaced by \(z_*\), the maximal extension of the extremal surface into the bulk. At \(T/\m = 0\), as \(z \to \infty\)  one finds $g(z)=b(z)/a(z) \sim z^{-2\q/(d-\q-1)}$ for general \(d\), \(\z\), and \(\q\). Substituting this into \(C(z_*)\), with \(\q = (d-2)\) we find an integral \(\int \diff z/z\) at large \(z_*\), producing a logarithm. We detail this calculation in appendix~\ref{app:hyperscaling_large_width}. The result is
\begin{equation}
\label{eq:arealawviolation}
\sigma \approx \frac{\m^2 L^2}{\gn} \le[ 
	0.174 \frac{\ln \left(\m \ell\right)}{\m\ell}
	 + \frac{0.106}{\m\ell} 
	 + \mathcal{O}\le(\m^{-2}\ell^{-2}\ri) \ri].
\end{equation}
This is the logarithmic violation of the area law expected due to a Fermi surface~\cite{Gioev:2006zz,Swingle:2009bf}, as found in other holographic models with \(\q = d-2\)~\cite{Ogawa:2011bz,Huijse:2011ef}.

\section{AdS soliton}
\label{soliton}

In this section we study the \ads\ soliton solution to the Einstein-Hilbert bulk action~\eqref{eq:adsscaction}~\cite{Witten:1998zw,Horowitz:1998ha,Klebanov:2007ws,Nishioka:2009zj}, obtained from \ads-Schwarzschild by double Wick rotation. The metric is
\begin{equation}
\label{eq:ads_soliton_metric}
\diff s^2 = \frac{L^2}{z^2} \left[ - \diff t^2 + \d_{ij} \diff x^i \diff x^j + g(z) \diff \chi^2 + \frac{\diff z^2}{g(z)}\right],
\end{equation}
where $g(z)=1 - z^d/z_0^d$, the coordinate $\chi$ is compact, $\chi \sim \chi + 4 \pi z_0 /d$, and $x^i$ represents $(d-2)$ non-compact spatial directions. There is a hard wall at $z=z_0$, indicating that the dual field theory has a mass gap and confinement~\cite{Witten:1998zw,Klebanov:2007ws}. The \ads\ soliton has $T=0$, $s=0$, and a negative Casimir energy
\begin{equation}
\label{eq:casimir}
\vev{T_{tt}}=- \frac{L^{d-1}}{16 \pi \gn z_0^d}.
\end{equation}

Since the \ads\ soliton metric~\eqref{eq:ads_soliton_metric} is not of the form~\eqref{eq:aads_metric}, the results of section~\ref{general} do not apply. However, the minimal area calculations generalise straightforwardly~\cite{Klebanov:2007ws,Bueno:2016rma}. As our entangling region, we take a strip of width $\ell$, with planar boundaries separated along a non-compact direction \(x\). The entangling region therefore wraps around $\chi$. We do not consider the sphere geometry, since compactifying one of the spatial directions breaks rotational symmetry, rendering the sphere more complicated to study than in the previous sections.

As shown in ref.~\cite{Klebanov:2007ws}, multiple extremal surfaces exist. For any $\ell$, there are disconnected extremal surfaces with \(x'(z) = 0\). The disconnected surfaces drop straight from the boundary of \ads\ to the hard wall, analogous to the dashed blue surface in figure~\ref{fig:minimal_surface_horizon}, but with the horizon replaced by a hard wall. The area of a disconnected surface is
\begin{equation}
\label{eq:ads_soliton_disconnected_area}
\mathrm{Area}^\mathrm{strip}_\mathrm{discon.} = L^{d-1}\mathrm{Vol}(\mathbb{R}^{d-3})\frac{8\pi z_0}{d(d-2)} \left(\frac{1}{\e^{d-2}} - \frac{1}{z_0^{d-2}}
\right).
\end{equation}
For sufficiently small $\ell$, connected extremal surfaces also exist, which extend into the bulk up to some maximal $z = z_*$, like the red surfaces in figure~\ref{fig:minimal_surface_horizon}. The turning point is related to the strip width by
\begin{equation}
\label{eq:ads_soliton_strip_width}
\ell = 2 z_* \int_0^1 \diff u \sqrt{\frac{g(z_*)}{g(z_* u)}} \frac{u^{d-1}}{\sqrt{g(z_* u) - g(z_*) u^{2(d-1)}}},
\end{equation}
and the area of such a surface is
\begin{align}
\label{eq:ads_soliton_connected_area}
\mathrm{Area}^\mathrm{strip}_\mathrm{con.} = L^{d-1} \mathrm{Vol}(\mathbb{R}^{d-3}) \frac{8\pi z_0}{d}\Biggl[&
\frac{1}{d-2} \left(\frac{1}{\e^{d-2}} - \frac{1}{z_*^{d-2}} \right)
\nonumber \\ & + \frac{1}{z_*^{d-2}}\int_{0}^1 \diff u \frac{1}{u^{d-1}} \left( \sqrt{\frac{g(z_* u)}{g(z_* u) - g(z_*) u^{2(d-1)}}} - 1\right) \Biggr].
\end{align}

\begin{figure}
\begin{subfigure}{0.5\textwidth}
	\includegraphics[width=\textwidth]{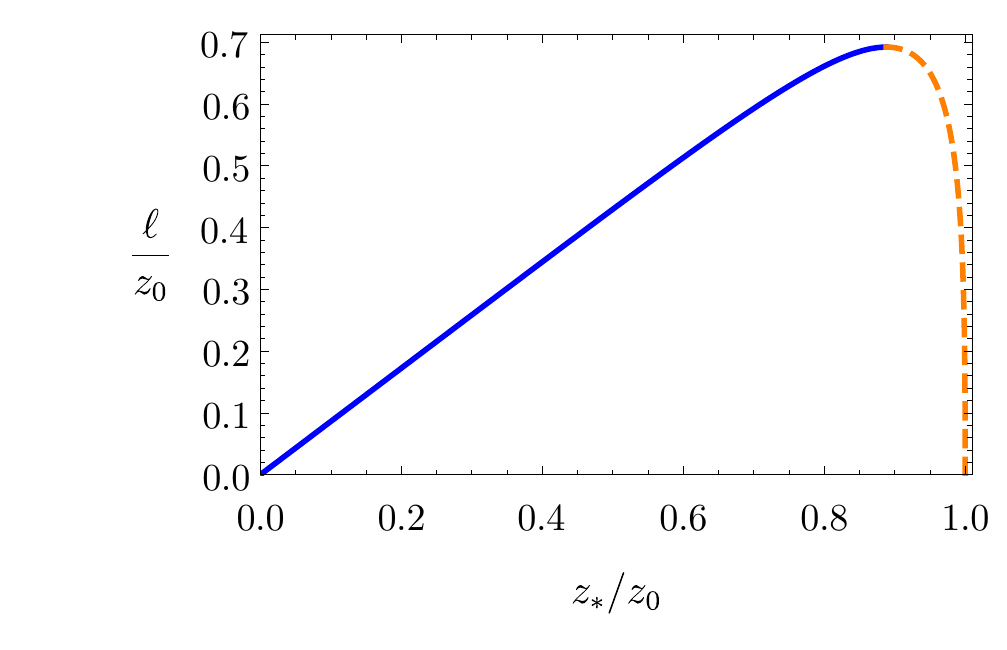}
	\caption{Widths of connected surfaces.}
	\label{fig:soliton_width}
\end{subfigure}
\begin{subfigure}{0.5\textwidth}
	\includegraphics[width=\textwidth]{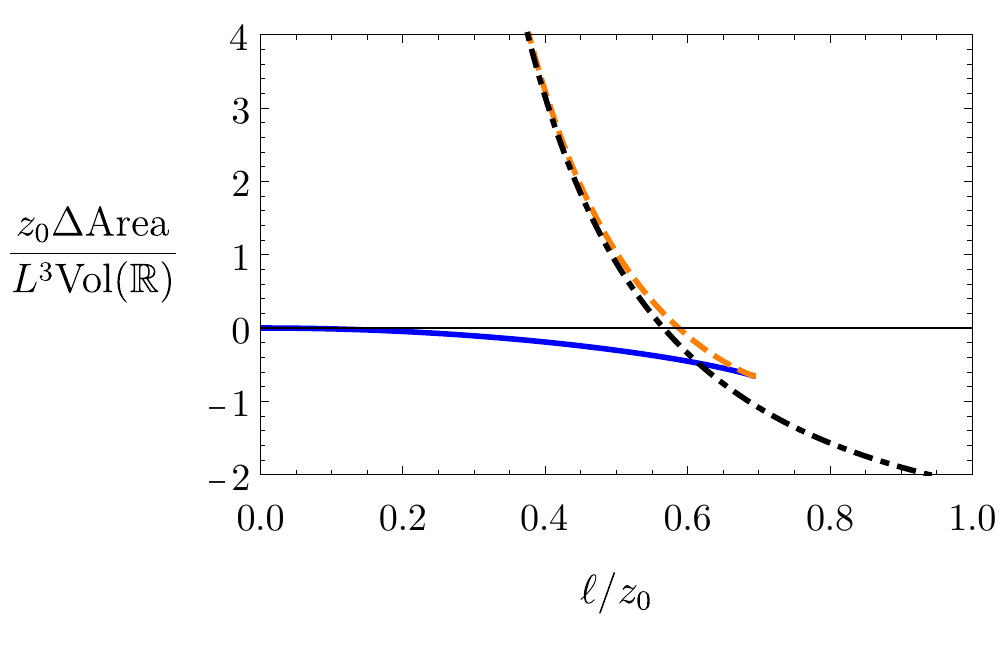}
	\caption{Areas of extremal surfaces.}
	\label{fig:solitonarea}
\end{subfigure}
\begin{center}
\begin{subfigure}{0.5\textwidth}
	\includegraphics[width=\textwidth]{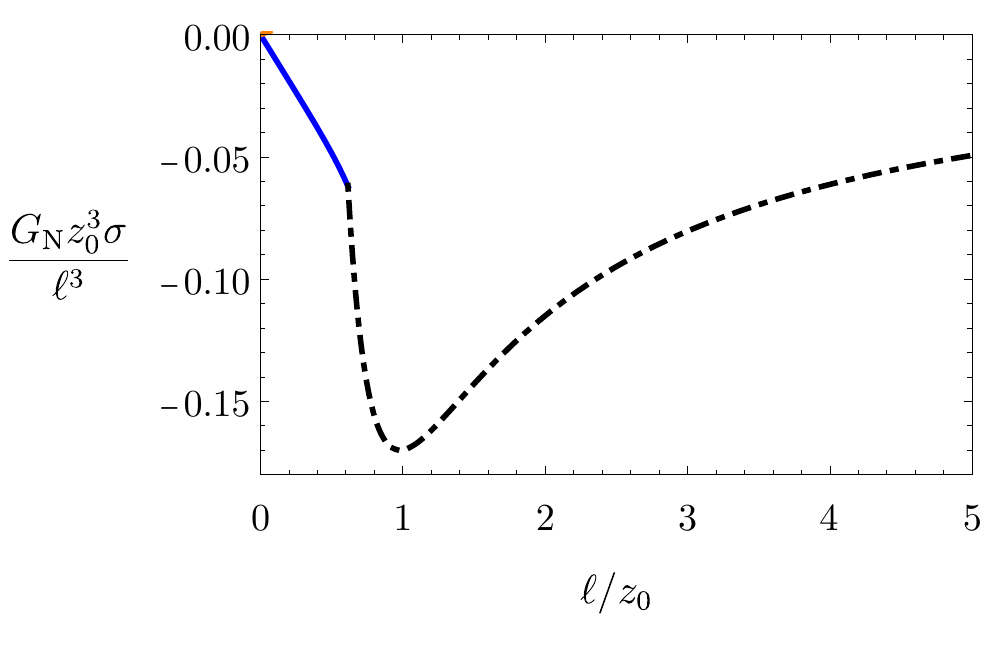}
	\caption{Entanglement density.}
	\label{fig:solitonED}
\end{subfigure}
\end{center}
\caption[Entanglement density and properties of the extremal surfaces for the strip in the \ads\ soliton geometry.]{
		\textbf{(a):} The relationship between the strip widths and turning points for the two classes of connected extremal surfaces in the \(d=4\) \ads\ soliton geometry. Connected surfaces only exist for \(\ell \lesssim 0.7 z_0\). \textbf{(b):} The difference in area between each of the extremal surfaces in the \ads[4] soliton geometry and the surface with the same \(\ell\) in compactified \ads[4]. The dot-dashed black curve is the disconnected surface, while the solid blue and dashed orange curves are the connected surfaces, with the same colour coding as in (a). \textbf{(c):} The entanglement density of the strip in the \ads[4] soliton geometry. The colour-coding is the same as in (b), and shows which surface determines the entanglement entropy. The results shown in this figure are qualitatively similar for other values of \(d\).
}
\label{fig:soliton}
\end{figure}

Connected surfaces exist only for values of \(\ell\) that can be obtained from~\eqref{eq:ads_soliton_strip_width} with positive \(z_*\). Numerically, one finds that connected surfaces exist in \(d=4\) for \(\ell \lesssim 0.7 z_0\). We plot \(\ell\) as a function of \(z_0\) for \(d=4\) in figure~\ref{fig:soliton_width}. For a given \(\ell \lesssim 0.7 z_0\), there are two solutions for \(z_*\), and therefore two connected extremal surfaces~\cite{Klebanov:2007ws}. 

To define the entanglement density in \ads\ soliton, we take a slightly different background subtraction to our previous examples, since the compact spatial direction of \ads\ soliton changes the UV divergence of the holographic entanglement entropy compared to \ads[d+1]. The factor of \(\vol(\mathbb{R}^{d-2})\) in the \ads[d+1] strip entanglement entropy~\eqref{eq:pure_ads_sphere_ee} arises from the area of the planar boundaries of the strip. Since the strip we consider wraps the compact spatial direction \(\chi\), the planar boundaries instead have area \((4\pi z_0/d) \times \vol(\mathbb{R}^{d-3})\), changing the coefficient of the UV divergent \(\e^{-(d-2)}\) term in the entanglement entropy. For a detailed discussion of the divergence of \(\see\) in the \ads\ soliton and regularisation schemes, see ref.~\cite{Bueno:2016rma}.

We will instead remove the UV divergence by subtracting the strip entanglement entropy in \ads[d+1] with a compact direction of length $4 \pi z_0/d$, which we will refer to as compactified \ads[d+1]. The compactified \ads[d+1] metric is locally identical to~\eqref{eq:ads_metric}, but produces divergences in the holographic entanglement entropy identical to those in the \ads\ soliton.

A key caveat is that compactified \ads[d+1] has a conical singularity at \(z=\infty\)~\cite{Gibbons:1998th}. The singularity could affect the behaviour of the entanglement density as $\ell\to\infty$, since in this limit the RT surface probes deep into the bulk. We have compared our subtraction to renormalisation via covariant counterterms~\cite{Taylor:2016aoi,Taylor:2017zzo}, and found no difference at large $\ell$. The counterterms for the strip only remove the divergent term proportional to \(\e^{-(d-2)}\), and so the renormalised entanglement entropy differs from the result of our subtraction only by a term $\propto 1/\ell^{d-2}$, which primarily affects the behaviour at small \(\ell\). In particular, this means that our subtraction is sufficient to determine whether the area theorem is violated.

Figure~\ref{fig:solitonarea} shows the subtracted areas of each of the extremal surfaces as functions of \(\ell\) for \(d=4\), with the solid blue lines and dashed orange lines representing the two connected surfaces, and the dot-dashed black line representing the disconnected surface. The blue connected surface area has smallest area for \(\ell \lesssim 0.615 z_0\), while the disconnected surface has smallest area for \(\ell \gtrsim 0.615 z_0\)~\cite{Bueno:2016rma}. There is therefore a ``phase transition'' in the entanglement entropy at \(\ell = 0.615 z_0\).

Applying our subtraction to the area of the connected surfaces~\eqref{eq:ads_soliton_connected_area} gives a formula for the entanglement density for $\ell$ below the transition,
\begin{equation}
\label{eq:solitonsmallL}
\s=\frac{L^{d-1}}{4 \gn}\left[ 
	\frac{\sqrt{g(z_*)}}{z_*^{d-1}}
	+ \frac{2\hat{C}(z_*)}{z_*^{d-2} \ell}
	+ \frac{ \varrho^{d-1}}{(d-2)\ell^{d-1}} \right],
\end{equation}
where \(\varrho\) was defined in~\eqref{eq:small_strip_coefficients} and $\hat{C}(z_*)$ is defined in analogy to the integral in~\eqref{eq:large_l_coefficient_integral},
\begin{equation}
\hat{C}(z_*) = - \frac{1}{d-2} + \int_0^1 \frac{\diff u}{u^{d-1}} \left( \sqrt{1 - \frac{g(z_*)}{g(z_*u)} u^{2(d-1)}} - 1\right).
\end{equation}
The area of the disconnected surfaces may be computed analytically, and we find that for \(\ell\) above the transition the entanglement density is given by
\begin{equation}
\label{eq:solitonlargeL}
\s=\frac{L^{d-1}}{4 (d-2) \gn} \left[\frac{\varrho^{d-1}}{\ell^{d-1}}  -  \frac{2}{z_0^{d-2} \ell}\right].
\end{equation}
Figure~\ref{fig:solitonED} shows  the entanglement density as a function of $\ell/z_0$ for \(d=4\). We have checked numerically that the qualitative behaviour of $\s$ is the same as that in figure~\ref{fig:solitonED} up to $d=40$.

At small \(\ell\), we find that the entanglement density is proportional to \(\ell\). Expanding~\eqref{eq:ads_soliton_strip_width} and~\eqref{eq:solitonsmallL} for small \(z_*\), in a similar manner to the analysis in section~\ref{sec:entanglement_asymptotics}, we find
\begin{equation} \label{eq:ads_soliton_small_l}
	\s = 2 \vev{T_{tt}} \tent^{-1},
\end{equation}
at small \(\ell\), where \(\tent\) is the entanglement temperature for the strip, given in~\eqref{eq:striptent}. The difference between \ads\ soliton and compactified \ads\ is not a change of state, so we have no reason to expect the first law of entanglement to apply. Instead,~\eqref{eq:ads_soliton_small_l} shows that the small-\(\ell\) entanglement density is twice what one would find from the first law.

As $\ell$ increases, the entanglement density decreases until the transition from connected to disconnected minimal surface in the bulk, at $\ell/z_0 \approx 0.615$ in \(d=4\).  For larger values of $\ell$, the entanglement density is given by the analytic expression~\eqref{eq:solitonlargeL}. There is a global minimum at
\begin{equation}
\ell = 2 (d-1)^{\frac{1}{d-2}} \pi^{\frac{d-1}{2(d-2)}} \left[
	\frac{\Gamma\left(\frac{d}{2d-2)}\right)}{\Gamma\left(\frac{1}{2d-2)}\right)}\right]^{\frac{d-1}{d-2}} z_0.
\end{equation}

Since the \ads\ soliton solution is not Lorentz-invariant, the proofs of the area theorem in refs.~\cite{Casini:2012ei,Casini:2016udt} do not apply. However, as \(\ell \to \infty\) we find that the entanglement density tends to zero from below. This is consistent with the interpretation of the coefficient of the entanglement entropy area term as counting degrees of freedom; the hard wall of the \ads\ soliton geometry implies that there are no degrees of freedom in the deep IR of the dual QFT. It would be interesting to determine whether the area theorem is satisfied in other holographic systems with a mass gap, to test the interpretation of the area term as counting degrees of freedom.

\section{Discussion}

We have computed the entanglement density of holographic CFTs with a variety of deformations, paying particular attention to its behaviour for large subregions. For Lorentz invariant RG flows, the coefficient \(\a\) of the area contribution to entanglement entropy satisfies a monotonicity theorem~\cite{Casini:2012ei,Casini:2016udt} which ensures \(\a \leq 0\), suggesting that the area term in entanglement entropy may count degrees of freedom.

In section~\ref{rg}, we found \(\a < 0\) in holographic examples of Lorentz invariant RG flows, as required by the area theorem. We also found \(\a < 0\) for the \ads\ soliton solution in section~\ref{soliton}, despite the fact that the proof of the area theorem does not hold for this system. The \ads\ soliton is dual to a QFT with a mass gap, so \(\a < 0\) is consistent with the idea that \(\a\) counts degrees of freedom.

However, in sections~\ref{adssc},~\ref{adsrn}, and~\ref{hyper} we found many examples with \(\a > 0\). These examples all had non-zero temperature or chemical potential, or both, and included examples of RG flows from a Lorentz-invariant UV to a hyperscaling-violating IR. Such area theorem violation occurred at or near regimes where the IR exhibited different scaling to the UV. Area theorem violation has also been observed at low temperatures in holographic systems with broken translational invariance~\cite{Gushterov:2017vnr}, in which the dual geometry had a near-horizon \ads[2] factor at zero temperature, similar to \ads-Reissner-Nordstr\"om. This suggests that the large-subregion behaviour of the entanglement entropy may be a probe of new scaling in the IR. However, one of the examples of RG flows to a hyperscaling-violating IR (with \(d = 3\), \(\z = 2\), and \(\q = -2\)) did not exhibit area theorem violation. Hence, it cannot always be true that different scaling in the IR implies area theorem violation.

There are a number of avenues for future research. For example, it would be interesting to study more examples of Lorentz-invariant to hyperscaling-violating RG flows, with different values of \((d,\,\q,\,\z)\). One could then look for correlation between the values of these parameters and area theorem violation. In addition, computing the entanglement entropy for different examples of RG flows to an IR with a given \((d,\,\q,\,\z)\) would provide a test of whether area theorem violation is determined only by these parameters, or is model-dependent. As a final example, it would be useful to look for area theorem violation in non-holographic models. This would provide a crucial test of whether the results presented in this chapter may be applied to real physical systems.

\chapter[Holographic zero sound]{Holographic zero sound\footnote{As with chapter~\ref{chap:entanglement_density}, the research in this chapter was conducted in collaboration with Nikola I. Gushterov and submitted as part of his PhD thesis at the University of Oxford. My primary contribution was the calculation of the numerical results that are presented in this chapter.}}
\label{chap:zero_sound}

\section{Introduction and background}
\label{bg}

\subsection{Hydrodynamics and sound waves}

Hydrodynamics is an effective theory, describing the long-wavelength excitations of systems near thermal equilibrium. Here, ``long-wavelength'' means in comparison to the mean free path of the quasiparticles in the system. The degrees of freedom in hydrodynamics are the conserved currents. A key assumption of hydrodynamics is that these conserved currents are functions only of a small set of hydrodynamic variables: a local temperature, a local fluid four-velocity, and local chemical potentials for any internal global symmetries. Since hydrodynamics deals with long-wavelength phenomena, one writes the components of the conserved currents as a series of terms with increasing number of derivatives of the hydrodynamic variables.\footnote{See ref.~\cite{Kovtun:2012rj} for details.}

Substituting this expansion into the conservation equations, one can derive plane wave solutions describing long-wavelength fluctuations around thermal equilibrium. For example, conservation of the stress tensor in a translationally invariant system implies the existence  of sound waves, with dispersion
\begin{equation} \label{eq:sound_dispersion}
		\w = \pm v_s k - i \G_s k^2 + \cO(k^3/T^3),
\end{equation}
where \(\w\) is the angular frequency of the sound wave, \(k = |\mathbf{k}|\) is the momentum (i.e. the wavenumber), and \(T\) is temperature. The speed \(v_s\) and attenuation constant \(\G_s\) are given by
\begin{equation} \label{eq:hydro_sound_attenuation}
		v_s^2 = \frac{\p p}{\p \ve},
		\quad
		\G_s = \frac{2(d-2)\h + (d-1)\z}{2(d-1)(\ve + p)},
\end{equation}
where \(\ve\) is the energy density, \(p\) is the pressure, and \(\h\) and \(\z\) are the shear and bulk viscosities of the fluid, respectively. The sound modes appear as poles in the longitudinal\footnote{Using rotational invariance to orient the momentum \(\mathbf{k}\) along \(x^1\) direction, the longitudinal Green's functions are two-point functions of \(T^{00}\), \(T^{01}\), \(T^{1 1}\), and \(\sum_{a\geq 2} T^{a a}\)~\cite{Policastro:2002tn}. These are the components of \(T^{\m\n}\) invariant under rotations around the \(x^1\) axis.} retarded Green's functions of the  stress tensor.

If the system possesses a \(\U(1)\) global symmetry, there will also be a hydrodynamic mode corresponding to diffusion of the associated conserved charge, with dispersion relation
\begin{equation} \label{eq:diffusion_dispersion}
		\w = - i D k^2 + \cO(k^3/T^3),
\end{equation}
for some diffusion constant \(D\). The charge diffusion mode is a pole in the longitudinal\footnote{Two-point functions of \(J^0\) and \(J^1\).} retarded Green's functions of the conserved current associated to the \(\U(1)\) symmetry.

\subsection{Fermi liquids and zero sound}

In this section we provide a brief review of elements of Fermi liquids, in particular the phenomenon of zero sound. Zero sound was predicted in ref.~\cite{Landau} and observed in ref.~\cite{PhysRevLett.17.74}. Our discussion will follow that of ref.~\cite{Pines}.

A quantum liquid is an interacting fluid in which quantum properties, such as the Pauli exclusion principle, play an important role. A Fermi liquid is a quantum liquid of fermionic quasiparticles. Quantum effects typically become important when the thermal de Broglie wavelength is of the order of the average distance between particles, requiring low temperatures or large densities. Most materials freeze into solids at temperatures larger than this, so fermi liquids are relatively rare in nature. In fact, the only known example of a true Fermi liquid is helium-3.\footnote{Quantum effects become important in helium-3 at temperatures of around \SI{4}{K}~\cite{Pines}. At still colder temperatures of around \SI{3}{mK}, helium-3 undergoes a transition to a superfluid phase~\cite{Osheroff:1972zz,1972PhRvL..29..920O}, which is not a fermi liquid.} However, in many metals the electron sea is well approximated as a fermi liquid~\cite{Pines}.

The ground state of a gas of free fermions possesses a Fermi surface. For a fixed number of particles, the low-energy excited states are reached by promoting a particle from just inside the Fermi surface to just outside. The ground state of a Fermi liquid also possesses a Fermi surface. However, promoting a particle out of the Fermi surface no longer produces an energy eigenstate, due to the interactions between particles. A key assumption of the theory of Fermi liquids is that the low-lying excited states can instead be described by quasiparticles of well-defined momentum, particles dressed by their interactions.

The distribution function, which gives the number of quasiparticles per unit 
momentum at a given position and time, obeys the Boltzmann equation. At low temperatures, the frequency of hard collisions between quasiparticles becomes small. The term in the Boltzmann equation which accounts for these collisions may then be neglected, in which case one can derive plane wave solutions with dispersion~\cite{Landau}\footnote{In the Fermi liquid literature, the dispersion of zero sound is usually given as \(k(\w)\), with real \(\w\) and complex \(k\). This may be straightforwardly obtained by inverting the expansion in~\eqref{eq:zero_sound_dispersion}, yielding \(k = \pm \w/v - i \G \w^2/v^2 + \cO(\w^3/\ve_F^3)\).}
\begin{equation} \label{eq:zero_sound_dispersion}
		\w = \pm v k - i \G k^2 + \cO(k^3/\ve_F^3),
\end{equation}
where \(\w\) is the angular frequency of the wave, \(k\) is the absolute value of the momentum, and \(\ve_F\) is the Fermi energy. The speed \(v\) and attenuation constant \(\G\) depend on the form of the interactions between quasiparticles. The dispersion relation~\eqref{eq:zero_sound_dispersion} is of the same form as the dispersion of sound waves~\eqref{eq:sound_dispersion}, although with different speed \(v\) and attenuation \(\G\). This excitation is called \textit{zero sound}, and arises from fluctuations in the shape of the Fermi surface. Experimental observation of zero sound in helium-3 was reported in ref.~\cite{PhysRevLett.17.74}.

\begin{figure}
	\begin{center}
	\includegraphics[width=0.6\textwidth]{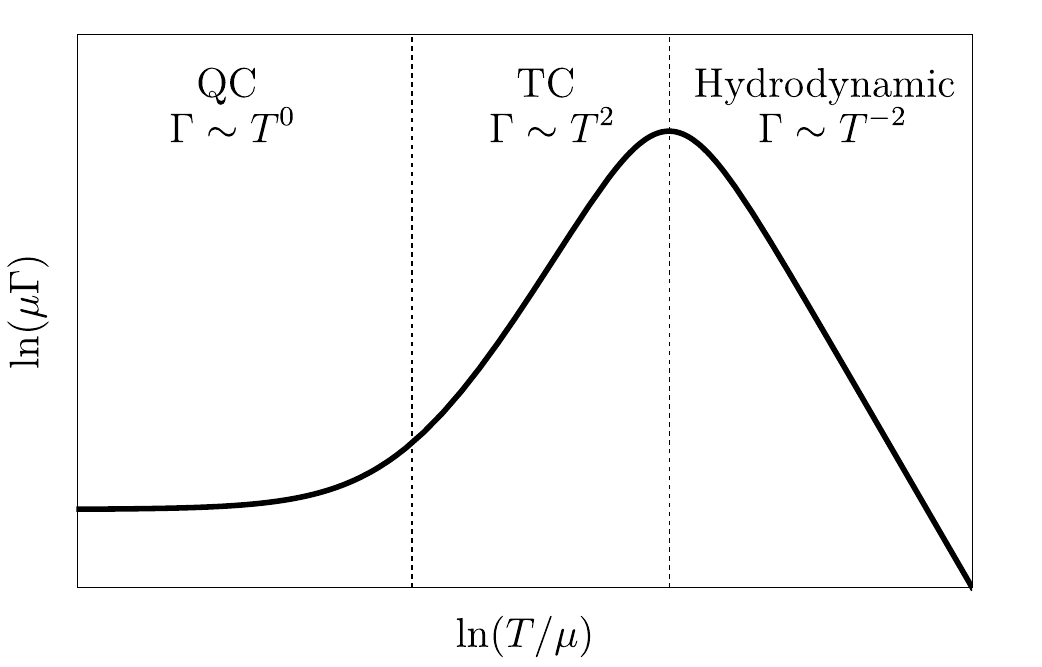}
	\caption[Scaling of the attenuation of zero sound with temperature.]{\label{fig:attenuation_cartoon} Schematic depiction of the attenuation of zero sound in Fermi liquids as a function of temperature \(T\), for fixed frequency \(\w\) and momentum \(k\) in units of chemical potential \(\m\). The two vertical dashed black lines represent $\pi T /\mu = \omega/\mu$ (left) and $\sqrt{\omega/\mu}$ (right). In the quantum collisionless (QC) regime $\Gamma \propto T^0$, in the thermal collisionless (TC) regime $\Gamma \propto T^2$, and in the hydrodynamic regime $\Gamma \propto T^{-2}$. A maximum appears between the thermal collisionless and hydrodynamic regimes, which is conventionally taken to define the collisionless-to-hydrodynamic crossover.}
	\end{center}
	\end{figure}
%
At zero temperature, the attenuation \(\G\) takes a non-zero value due to decay into quasiparticle-quasihole pairs (known as multipair decay)~\cite{Pines}. As the \(T\) is raised, the attenuation passes through several regimes~\cite{Landau,Khalatnikov,Pines}, as sketched in figure~\ref{fig:attenuation_cartoon}. First, for \(T\) small compared to \(\w\), collisions between quasiparticles are infrequent, and the attenuation remains dominated by the multipair decay. As a result, \(\G\) is approximately independent of temperature. This is known as the quantum collisionless regime.

When \(T\) becomes large compared to \(\w\), collisions between thermally excited quasiparticles become frequent enough to provide the dominant contribution to zero sound attenuation. The attenuation is proportional to the collision rate, \(\G \propto T^2\). This regime is known as the thermal collisionless regime.

As the temperature is raised further, the approximations used in the Landau Fermi liquid theory break down. However, the collisions become frequent enough to ensure local thermodynamic equilibrium, so hydrodynamics should be a valid effective theory. There will therefore be a hydrodynamic sound mode~\eqref{eq:sound_dispersion}, with attenuation~\eqref{eq:hydro_sound_attenuation}. The shear viscosity of a Fermi liquid satisfies \(\h/(\ve + p) \propto T^{-2}\)~\cite{Pines} and \(\h \gg \z\)~\cite{Abrikosov_1959}, so that \eqref{eq:hydro_sound_attenuation} gives \(\G \propto T^{-2}\). As observed experimentally~\cite{PhysRevLett.17.74}, zero sound smoothly becomes hydrodynamic sound as the temperature is raised, leading to a maximum between the \(\G \propto T^{2}\) and \(\G \propto T^{-2}\) behaviour of the thermal collisionless and hydrodynamic regimes, respectively.

\subsection{Holographic zero sound}
\label{sec:hzs_background}

Longitudinal modes with sound-like dispersion~\eqref{eq:sound_dispersion} have been observed in a wide range of holographic models of compressible quantum matter~\cite{Karch:2008fa,Kulaxizi:2008kv,Kulaxizi:2008jx,Kim:2008bv,Karch:2009zz,Kaminski:2009dh,Edalati:2010pn,HoyosBadajoz:2010kd,Nickel:2010pr,Lee:2010ez,Bergman:2011rf,Ammon:2011hz,Davison:2011ek,Davison:2011uk,Jokela:2012vn,Goykhman:2012vy,Brattan:2012nb,Jokela:2012se,Pang:2013ypa,Dey:2013vja,Edalati:2013tma,Brattan:2013wya,Davison:2013uha,DiNunno:2014bxa,Jokela:2015aha,Itsios:2016ffv,Jokela:2016nsv,Hartnoll:2016apf,Roychowdhury:2017oed,Chen:2017dsy}. These are models of systems with non-zero chemical potential \(\m\), and corresponding charge density \(\vev{J^t}\), with non-vanishing compressibility \(\diff \vev{J^t}/\diff \m\). By quantum we mean $T=0$, so that quantum, rather than thermal, effects determine the ground state~\cite{Sachdev:2011wg}. We will refer to such zero-temperature sound modes as \textit{holographic zero sound} (HZS). The purpose of the adjective ``holographic'' is to distinguish these modes from Fermi liquid zero sound, since as discussed further below, HZS appears in holographic models in which the dual QFT is not a Fermi liquid.

HZS modes have been found in two classes of holographic models. The first class is probe brane models~\cite{Karch:2008fa,Kulaxizi:2008kv,Kulaxizi:2008jx,Kim:2008bv,Karch:2009zz,Kaminski:2009dh,HoyosBadajoz:2010kd,Nickel:2010pr,Lee:2010ez,Bergman:2011rf,Ammon:2011hz,Davison:2011ek,Jokela:2012vn,Goykhman:2012vy,Brattan:2012nb,Jokela:2012se,Pang:2013ypa,Dey:2013vja,Edalati:2013tma,Brattan:2013wya,DiNunno:2014bxa,Jokela:2015aha,Itsios:2016ffv,Jokela:2016nsv,Hartnoll:2016apf,Roychowdhury:2017oed,Chen:2017dsy}, consisting of gravity coupled to a \(\U(1)\) gauge field through a Dirac-Born-Infeld (DBI) action,
\begin{equation}
\label{sdbi}
\sgrav = \frac{1}{16\pi \gn} \int \diff^{d+1} x \, \sqrt{- \det G} \left ( R + \frac{d(d-1)}{L_0^2}\right)  - \mathcal{T} \int \diff^{d+1}x \, \sqrt{-\textrm{det}\left(g+ \alpha F\right)}.
\end{equation}
with tension \(\mathcal{T}\), and a coupling constant \(\a\) of mass dimension \([\a] = -2\). These models employ the probe limit discussed in section~\ref{sec:probe_branes}, taking $\gn \mathcal{T} L_0^2 \ll 1$ to leading non-trivial order. We specialize to spacetime-filling branes~\cite{Tarrio:2013tta}, which is why the second integral in~\eqref{sdbi} is over all $(d+1)$ bulk dimensions, although defect branes (which are not spacetime-filling) can also give rise to HZS modes~\cite{Karch:2008fa,Karch:2009zz}. In field theory terms, the probe limit corresponds to charged degrees of freedom making up only a small proportion of the total number of degrees of freedom, as discussed in section~\ref{sec:probe_branes}.

The second class of models exhibiting HZS is Einstein-Maxwell theory~\cite{Edalati:2010pn,Davison:2011uk},
\begin{equation} \label{eq:einstein_maxwell_action}
		\sgrav = \frac{1}{16 \pi \gn} \int \diff^{d+1}x \,\sqrt{-\det{G}} \le( R + \frac{d(d-1)}{L^2} - \frac{1}{4e^2} F^2 \ri),
\end{equation}
with coupling constant \(e\), possibly coupled to an uncharged scalar field~\cite{Davison:2013uha}. These models  do not take a probe limit, so the gauge field back-reacts on the metric. In field theory terms, in back-reacted models the charged fields comprise a non-negligible fraction of the total number of degrees of freedom. 

In both classes of models, sound modes appear in zero-temperature solutions in which \(G\) and \(F\) depend only on \(z\) (in a suitably chosen coordinate system), and the only non-zero component of the \(\U(1)\) field strength is an electric field in the holographic radial direction, giving rise to non-zero chemical potential and charge density in the dual QFT. For example, in Einstein-Maxwell theory HZS modes appear in the extremal \ads[d+1]-Reissner-Nordstr\"om charged black brane solution~\cite{Edalati:2010pn,Davison:2011uk}.

There are three examples of compressible quantum matter in traditional condensed matter physics, Fermi liquids, solids, and superfluids~\cite{Landau,PhysRevLett.17.74,Pines,LP1,LP2,Negele,Sachdev:2011wg}, each of which supports a low temperature sound mode. Fermi liquids support zero sound, due to fluctuations of their Fermi surface, while solids and superfluids support phonons due to their spontaneous breaking of translational and \(\U(1)\) particle number symmetries, respectively.

There are holographic systems supporting HZS that do not fall into any of these three categories, making the physical origin of HZS unclear. HZS modes appear in holographic systems which have unbroken translational and \(\U(1)\) global symmetries, so HZS cannot be a phonon. In addition, the effective description of these holographic systems differs from Landau Fermi liquid theory in key respects.

Probe brane models do not exhibit a Fermi surface~\cite{Karch:2008fa,Kulaxizi:2008kv,Kulaxizi:2008jx,Kim:2008bv,Karch:2009zz,Kaminski:2009dh,HoyosBadajoz:2010kd,Nickel:2010pr,Lee:2010ez,Bergman:2011rf,Ammon:2011hz,Davison:2011ek,Jokela:2012vn,Goykhman:2012vy,Brattan:2012nb,Jokela:2012se,Pang:2013ypa,Dey:2013vja,Edalati:2013tma,Brattan:2013wya,DiNunno:2014bxa,Jokela:2015aha,Itsios:2016ffv,Jokela:2016nsv,Hartnoll:2016apf,Roychowdhury:2017oed,Chen:2017dsy}. The effective description of these models is that of the hydrodynamics of a weakly conserved current~\cite{Chen:2017dsy}. Einstein-Maxwell models can have a Fermi surface~\cite{Liu:2009dm,Cubrovic:2009ye,Faulkner:2009wj}, but violate Luttinger's theorem: the Fermi surface volume is smaller than $\vev{J^t}$ by powers of $N$~\cite{Edalati:2010pn,Davison:2011uk,Davison:2013uha,Hartnoll:2016apf}. As discussed in chapter~\ref{chap:entanglement_density}, the near horizon \ads[2] of extremal \ads[d+1]-Reissner-Nordstr\"om indicates that that the low-energy effective description is a semi-local quantum liquid~\cite{Iqbal:2011in}. This leads to branch cuts in the retarded Green's functions of \(T^{\m\n}\) and \(J^\m\)~\cite{Edalati:2010hk,Edalati:2010pn}, which are not present in Fermi liquid two-point functions.

\begin{figure}
		\begin{subfigure}{0.5\textwidth}
			\includegraphics[width=\textwidth]{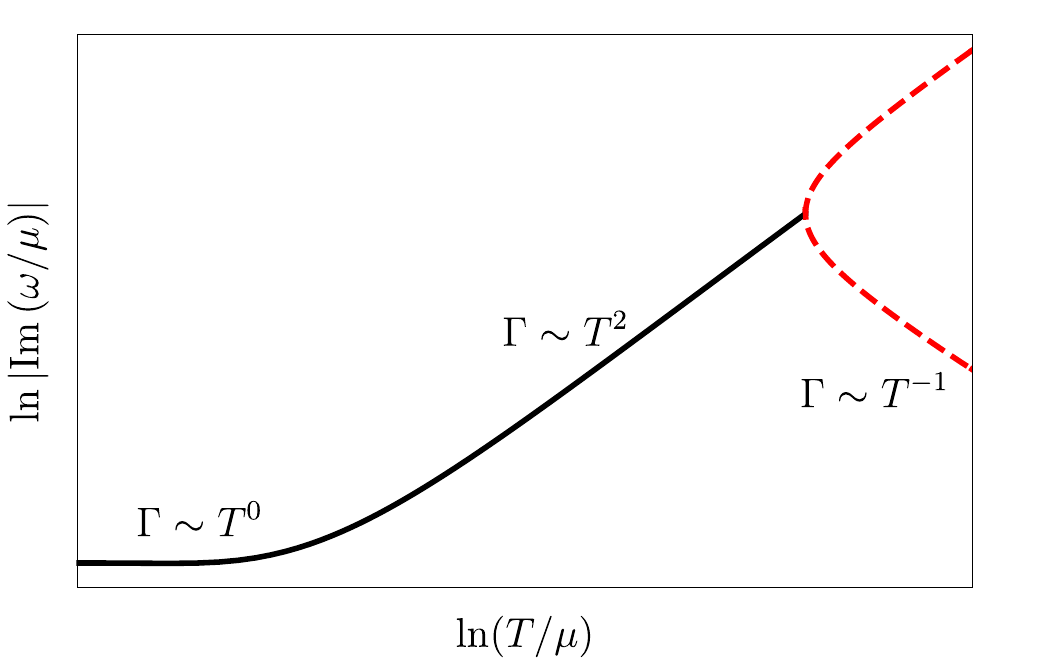}
			\caption{Probe brane.} \label{fig:probe_attenuation}
		\end{subfigure}
		\begin{subfigure}{0.5\textwidth}
			\includegraphics[width=\textwidth]{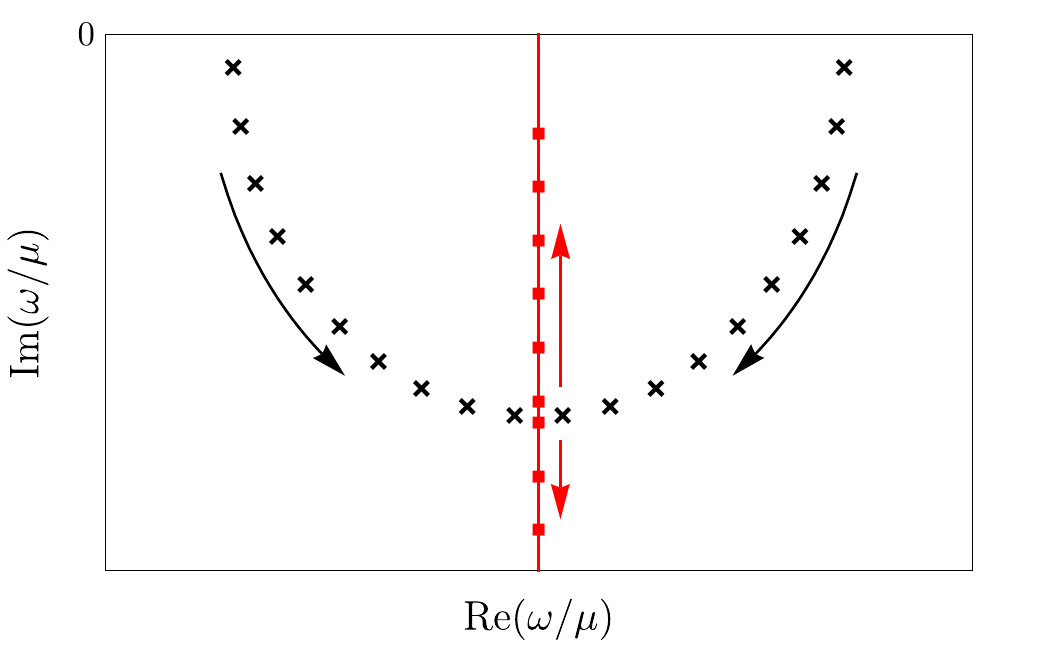}
			\caption{\label{fig:probe_cartoon} Probe brane.}
		\end{subfigure}
		\begin{subfigure}{0.5\textwidth}
			\includegraphics[width=\textwidth]{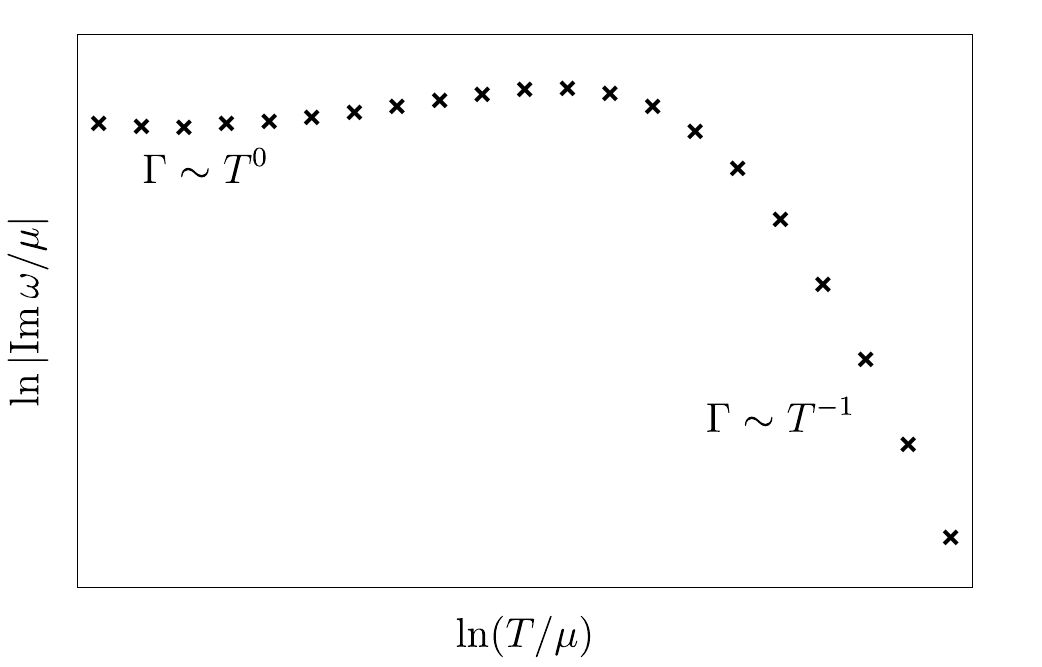}
			\caption{\label{fig:rn_attenuation} Einstein-Maxwell.}
		\end{subfigure}
		\begin{subfigure}{0.5\textwidth}
			\includegraphics[width=\textwidth]{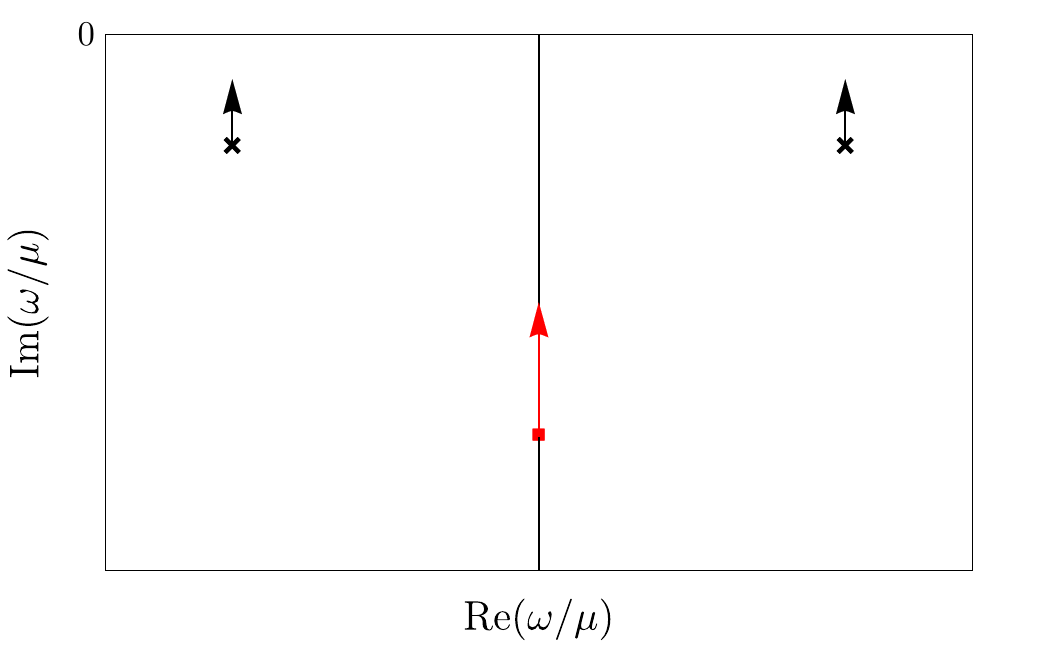}
			\caption{\label{fig:rn_cartoon} Einstein-Maxwell.}
		\end{subfigure}
		\caption[Sound-channel quasinormal modes for probe brane and Einstein-Maxwell models, in the complex frequency plane.]{
			\textbf{(a):} Typical temperature dependence of the imaginary part of poles in longitudinal \(J^\m\) two-point functions, in probe brane models~\cite{Davison:2011ek}. The solid black line is the attenuation of HZS, which also has non-zero real part. The dashed red lines correspond to purely imaginary poles, one of which (the branch with \(\G \sim T^{-1}\)) becomes the hydrodynamic charge diffusion pole at large \(T/\m\). \textbf{(b):} Schematic depiction of the longitudinal poles in probe brane models. The plot shows the locations of the poles in the complex frequency plane, for fixed momentum in units of the chemical potential. Arrows show how the poles move as the temperature is increased. At zero temperature there are two HZS poles (black crosses). As the temperature is increased, the poles move in approximate semicircles until they collide on the imaginary axis, producing two purely imaginary poles (red squares). As the temperature is increased still further, one of the imaginary poles moves towards the real axis, eventually becoming the charge diffusion pole, with \(\G \sim T^{-1}\). \textbf{(c):} Typical form for the imaginary part of HZS in Einstein-Maxwell models, as a function of temperature~\cite{Davison:2011uk}. \textbf{(d):} Schematic depiction of poles in the longitudinal $J^{\mu}$ and $T^{\mu\nu}$ Green's functions, in Einstein-Maxwell models. At low temperatures there are three poles close to the real axis: a pair with non-zero real part (the HZS poles), and a purely imaginary pole. As the temperature increases, the real part of the frequencies stay constant, while their imaginary parts mostly become less negative. At high temperature, the purely imaginary pole becomes charge diffusion.
		}
		\label{fig:cartoons}
\end{figure}

Given these differences, HZS exhibits some remarkable similarities to Fermi liquid zero sound. The attenuation of HZS in probe brane models behaves identically to zero sound in the quantum and thermal collisionless regimes~\cite{Davison:2011ek}, as sketched in figure~\ref{fig:probe_attenuation}. At very low temperatures, \(\G \sim T^0\) analogous to the quantum collisionless regime in a Fermi liquids. At intermediate temperature, \(\G \sim T^2\) like the thermal collisionless regime. However, the probe limit decouples the bulk \(\U(1)\) gauge field from the metric, with the result that the HZS pole appears only in Green's functions of $J^{\mu}$, and not those of $T^{\mu\nu}$. As a result, HZS cannot possibly crossover to hydrodynamic sound at large temperatures. Instead, at large temperatures HZS crosses over to diffusion, with dispersion given by~\eqref{eq:diffusion_dispersion}. As a result, the sound attenuation exhibits no maximum.

In probe brane models, the crossover to hydrodynamics is instead defined by the motion of the poles with \(T/\m\) in the complex frequency plane~\cite{Davison:2011ek}, sketched in figure~\ref{fig:probe_cartoon}. At zero temperature, the poles closest to the real axis are the zero sound poles (the highest black crosses in the figure). As the temperature is raised, the poles move in an approximate semicircle toward the imaginary axis, eventually colliding, and forming two purely imaginary poles (the red squares in the figure). One of the imaginary poles moves deeper into the complex plane, while the other moves up towards the real axis, becoming charge diffusion at high temperatures. The critical value of \(T/\m\) at which the poles collide is taken to define the hydrodynamic crossover~\cite{Davison:2011ek}.

In Einstein-Maxwell models, the gauge field back-reacts on the metric. Hence, HZS modes in these models are poles of both the \(J^\m\) and \(T^{\m\n}\) two-point functions. As the temperature is raised, HZS smoothly evolves into the sound pole found in \ads-Schwarzschild~\cite{Herzog:2003ke,Kovtun:2005ev}, with attenuation proportional to \(T^{-1}\), as required by hydrodynamics. Figure~\ref{fig:rn_attenuation} shows the typical form of the HZS attenuation as a function of temperature~\cite{Davison:2011uk}. At low temperatures, the sound attenuation is approximately temperature-independent, as in the quantum collisionless regime in Fermi liquids. However, there is no range of temperatures for which \(\G \sim T^2\). There is only a small, broad, maximum in the sound attenuation before the hydrodynamic regime, so the Fermi liquid definition of the hydrodynamic crossover is impractical.

The motion of the poles in the complex plane is qualitatively much simpler for Einstein-Maxwell than for probe branes~\cite{Edalati:2010pn,Davison:2011uk}. At zero temperature, there are HZS poles, and a branch cut along the imaginary axis. At small non-zero temperature, the branch cut splits into a discrete set of purely imaginary poles. Since the gauge field back-reacts on the metric, all of these poles appear in both the \(J^\m\) and \(T^{\m\n}\) Green's functions. As the temperature is increased, HZS smoothly evolves into hydrodynamic sound, while the purely imaginary pole closest to the real axis evolves into charge diffusion. Since there is no pole collision for Einstein-Maxwell models, the probe brane definition of the hydrodynamic crossover may not be applied.

Instead, ref.~\cite{Davison:2011uk} defined the crossover using the spectral function for the charge density, which we denote $\rho_{qq}$.  At low \(T/\m\), \(\r_{qq}\) is dominated by a peak due to HZS modes. As $T/\mu$ increases, this peak is suppressed, and a peak produced by the charge diffusion pole rises. The temperature at which the charge diffusion peak first becomes taller than the sound peak is taken to define the crossover~\cite{Davison:2011uk}. No crossover is apparent in the energy density spectral function, which we denote $\rho_{\ve\ve}$, which is always dominated by the sound pole.

The ubiquity of HZS in holographic models naturally raises the question of whether sound modes are present more generally in compressible quantum matter. The Fermi liquid and holographic results provide three possible definitions for the crossover temperature: the maximum in sound attenuation, the collision of poles on the imaginary axis, and the transfer of dominance in $\rho_{qq}$ from the sound peak to the charge diffusion peak. Another natural question is how common each of these behaviours is, and whether a ``universal'' definition of the hydrodynamic regime exists, applicable to all of the cases above, and more generally to all quantum compressible matter. 

\section{The Model}
\label{model}

In order to gain a better understanding of the nature of zero sound in compressible quantum matter, we will study the fate of zero sound and the hydrodynamic crossover when we relax the probe limit in the holographic model~\eqref{sdbi}. For concreteness, and ease of comparison to the \ads[4]-Reissner-Nordstr\"om results presented in ref.~\cite{Davison:2011uk}, we will set \(d=3\) for the remainder of this chapter.

It will be convenient at this stage to reparameterise the model slightly. First, we note that when \(F=0\) in~\eqref{sdbi}, the tension of the brane provides a contribution to the cosmological constant. As a result, the asymptotic \ads-radius on the gravity side of our holographic model will be \(L\), given by
\begin{equation}
		L^2 \equiv \frac{L_0^2}{1 - 8 \pi \gn \mathcal{T} L_0^2/3}.
\end{equation}
We will study the dependence of HZS on the dimensionless tension
\begin{equation} \label{taudef}
		\t \equiv 8 \pi \gn \mathcal{T} L^2,
\end{equation}
which determines the back-reaction of the brane. Note that the probe limit discussed in section~\ref{sec:probe_branes} is \(\t \ll 1\). In top-down probe brane models, such as the D3-D7 system discussed in section~\ref{sec:probe_branes}, $\tau$ would measure the ratio of the number of degrees of freedom charged under the \(\U(1)\) global symmetry to the total number of degrees of freedom. We also define the dimensionless coupling
\begin{equation}
\label{talphadef}
\talpha \equiv \alpha/L^2,
\end{equation}
which controls the strength of non-linear $F_{MN}$ self-interactions.

In terms of these parameters, the bulk action~\eqref{sdbi} becomes
\begin{multline} \label{eq:hzs_model}
		\sgrav = \frac{1}{16 \pi \gn} \int \diff^4 x \sqrt{-\det G} \le( R + \frac{6}{L^2} \ri)
		\\
		+
		\frac{\t}{8 \pi \gn L^2} \int \diff^4 x \sqrt{-\det G} \le( 1 - \sqrt{
			- \det \le(\id + \talpha L^2 G^{-1} F \ri)
		} \ri),
\end{multline}
where \(\id\) is the four-dimensional identity matrix. When \(\talpha \ll 1\), we may expand the final square root up to terms quadratic in \(\talpha\), in which case we recover the Einstein-Maxwell action~\eqref{eq:einstein_maxwell_action}.\footnote{Strictly it is the combination \(\talpha L^2 G^{-1} F\) which must be small to make this expansion. However, \(L^2 G^{-1} F\) remains \(\cO(1)\) when we take \(\talpha \to 0\) on the solution that we study, so it is enough to take \(\talpha\) to be small.} The combination \(\t\talpha^2\) must be kept fixed when taking this limit, so that the back-reaction of the gauge field on the metric remains finite.

Note that $\talpha$ appears in the action~\eqref{eq:hzs_model} only as the prefactor of \(F\). One may therefore eliminate \(\talpha\) by a redefinition of the gauge field, so that changing \(\talpha\) is equivalent to changing the chemical potential in the solutions we study. We will retain \(\talpha\) in this section primarily as a bookkeeping parameter to ease comparison with the Einstein-Maxwell models.

The equations of motion arising from the action~\eqref{eq:hzs_model} admit a charged black brane solution, dual to a CFT at non-zero temperature \(T\) and chemical potential \(\m\)~\cite{Fernando:2003tz,Dey:2004yt,Cai:2004eh,Pal:2012zn,Tarrio:2013tta}. We will choose coordinates in which the metric and gauge field strength take the form
\begin{gather}
	\diff s^2 = G_{MN}\diff x^M \diff x^N = \frac{L^2}{z^2} \left[\frac{\diff z^2}{f(z)} - f(z) \diff t^2 + \diff x^2 + \diff y^2\right],
	\nonumber \\
	F_{tz} = - F_{zt} = \frac{Q/z_H^2}{\sqrt{1+ \tilde{\alpha}^2 Q^2 z^4/z_H^4}},
	\label{bgsol}
\end{gather}
where \(Q\) is a dimensionless integration constant and 
\begin{multline}
	f(z) \equiv  1- \frac{z^3}{z_H^3} + \frac{\tau}{3} \left[1 - \frac{z^3}{z_H^3} + {}_2F_1\left(-\frac{1}{2},-\frac{3}{4};\frac{1}{4};-\tilde{\alpha}^2 Q^2\right) \frac{z^3}{z_H^3}\right]
	\\ -\frac{\t}{3} \, {}_2F_1\left(-\frac{1}{2},-\frac{3}{4};\frac{1}{4};\tilde{\alpha}^2 Q^2\frac{z^4}{z_H^4}\right),
	\label{eq:hzs_metric_function}
\end{multline}
where \(_2 F_1\) is the hypergeometric function. The asymptotic \ads\ boundary is at \(z=0\), while the horizon of the black brane is at \(z = z_H > 0\). The temperature and chemical potential in the CFT are given by
\begin{subequations} \label{eq:hzs_t_mu}
\begin{align}
\label{teq}
T &= \frac{|f'(z_H)|}{4 \pi} = \frac{3+\tau\left(1-\sqrt{1+\tilde{\alpha}^2 Q^2}\right)}{4\pi z_H},
\\
\label{mueq}
\mu &=  \int_0^{z_H} \diff z \, F_{tz} = \frac{Q}{z_H} \,_2F_1\left(\frac{1}{2},\frac{1}{4};\frac{5}{4};-\tilde{\alpha}^2 Q^2\right),
\end{align}
\end{subequations}
where $f'(z) \equiv \diff f(z)/\diff z$. The parameter \(Q\) is therefore implicitly determined by the dimensionless ratio \(T/\m\), through
\begin{equation}
\label{tmu}
\frac{T}{\mu} = \frac{3+\tau\left(1-\sqrt{1+\tilde{\alpha}^2 Q^2}\right)}{4\pi Q \,_2F_1\left(\frac{1}{2},\frac{1}{4};\frac{5}{4};-\tilde{\alpha}^2 Q^2\right)}.
\end{equation}

The Bekenstein-Hawking entropy~\eqref{eq:bh_entropy} is
\begin{equation}
\label{entropy}
s = \frac{L^2}{4 \gn} \frac{1}{z_H^2} = \frac{L^2}{4 \gn} \left(\frac{4 \pi T}{3}\right)^2 \left[1+\frac{\tau}{3}\left(1-\sqrt{1+\talpha^2 Q^2}\right)\right]^{-2}.
\end{equation}
The solution~\eqref{bgsol} is of the form~\eqref{eq:aads_metric} with \(f(z) = g(z)\) having a near-boundary expansion of the form~\eqref{eq:fgasymp}. The energy density is therefore given by~\eqref{eq:energy_density},
\begin{equation}
	\label{energy}
	\ve \equiv \vev{T_{tt}} =  \frac{L^2}{8\pi \gn}\left(\frac{4 \pi T}{3}\right)^3 \, \frac{1+ \frac{\tau}{3} \left [ 1 -  \,_2F_1\left(-\frac{1}{2},-\frac{3}{4};\frac{1}{4};-\tilde{\alpha}^2 Q^2\right)\right]}{\left[1+\frac{\tau}{3}\left(1-\sqrt{1+\talpha^2 Q^2}\right)\right]^3},
\end{equation}
where we have expanded \(f(z)\) in small \(z\) to determine the coefficient \(m\) defined in~\eqref{eq:fgasymp}. The charge density $\vev{J^t}$ may be obtained from the standard relation \(\varepsilon + p = s\,T + \mu \vev{J^t}\), which yields
\begin{equation}
\label{jt}
\vev{J^t} = 
\frac{L^2}{8\pi \gn} 
\left(\frac{4 \pi T}{3}\right)^2
 \frac{ \tau \talpha^2  Q}{\left[1+\frac{\tau}{3}\left(1-\sqrt{1+\talpha^2 Q^2}\right)\right]^2}= \frac{\tau \talpha^2 Q  s}{2\pi}.
\end{equation}

The trace of the stress tensor vanishes, since the curvature invariants in~\eqref{eq:weyl_anomaly} vanish individually. The pressure is therefore \(p = \ve/2\) (since the CFT is three-dimensional). From~\eqref{eq:hydro_sound_attenuation}, the speed of sound in the hydrodynamic regime is  then $v_s^2 = 1/2$. We note that for both \ads[4]-Reissner-Nordstr\"om and probe branes in \ads[4]-Schwarzschild, the speed of HZS is also $v^2 = 1/2$~\cite{Karch:2008fa,Edalati:2010pn,Davison:2011ek,Davison:2011uk}. In a Fermi liquid, the speeds of hydrodynamic and zero sound coincide in the limit of infinite quasiparticle interaction strength~\cite{Pines}.

The solution in~\eqref{bgsol} admits an extremal limit, in which \(Q = Q_\mathrm{ext}\) is given by the solution to \(T = 0\) using~\eqref{teq},
\begin{equation}
\label{qext}
Q_\mathrm{ext}^2 = \frac{1}{\tau \tilde{\alpha}^2} \left( 6 + \frac{9}{\tau}\right).
\end{equation}
In this limit, \(f(z)\) has a double zero at \(z = z_H\), as for extremal \ads-Reissner-Nordstr\"om, so the near horizon geometry takes the \(\ads[2]\times \mathbb{R}^2\) form~\eqref{eq:extremal_rn_near_horizon_metric}. The \ads[2] radius is
\begin{equation} \label{ads2radius}
		L_{\ads[2]} = \sqrt{\frac{3 + \t}{9 + 6 \t}} L.
\end{equation}

\subsection{Quasinormal modes and Green's functions}

We will study linearised fluctuations of the metric and gauge field about the solution~\eqref{bgsol}, making use of the formalism of \cite{Kovtun:2005ev}. Let us write the fluctuations as
\begin{equation}
	G_{MN}(z) \to G_{MN}(z) + \d G_{MN}(z,t,\mathbf{x}),
	\quad
	A_M(z) \to A_M(z) + \d A_M(z,t,\mathbf{x}),
\end{equation}
where \(\mathbf{x} = (x,y)\). Substituting these into the equations of motion following from the action~\eqref{eq:hzs_model}, and expanding to leading non-trivial order in the fluctuations, one finds a set of coupled, linear PDEs for \(\d G_{MN}\) and \(\d A_M\). We now Fourier transform with respect to the boundary coordinates,
\begin{equation}
	\d G_{MN}(z,t,\mathbf{x}) = \int \frac{\diff \w \diff^2 \mathbf{k}}{(2\pi)^3} e^{- i \w t + i \mathbf{k}\cdot \mathbf{x}} \d G_{MN} (z,\w,\mathbf{k}),
\end{equation}
and similar for \(\d A_M\). Since the background~\eqref{bgsol} is invariant under translations in \(t\) and \(\mathbf{x}\), the Fourier modes with different frequencies or momenta decouple in the equations of motion. 

As discussed in section~\ref{sec:quasinormal_modes}, for a given momentum \(\mathbf{k}\), the quasinormal modes are the values of \(\w\) for which there exists a solution with \(\d G_{MN}(z,\w,\mathbf{k})\) and \(\d A_{M}(z,\w,\mathbf{k})\) normalisable at the boundary, and ingoing at the horizon. By rotational invariance in the \(x\)--\(y\) plane, these frequencies can depend on the momentum only through its absolute value \(k \equiv |\mathbf{k}|\). Without loss of generality, we may therefore take \(\mathbf{k}\) to be oriented along the \(x\) direction. Non-zero momentum breaks the \(\SO(2)\) rotational symmetry into a \(\mathbb{Z}_2\) subgroup, under which \(y \to -y\). The fluctuations may be classified into even (longitudinal) and odd (transverse) representations of the residual \(\mathbb{Z}_2\), where even fluctuations are invariant under \(y \to -y\), whereas odd fluctuations change sign. The classification is:
\begin{align}
	&\text{Even:} & & \d G_{zz},\,\d G_{tt},\,\d G_{xx},\,\d G_{yy},\,\d G_{zt},\,\d G_{zx},\,\d G_{tx},\, \d A_z,\, \d A_t, \, \d A_x.
	\nonumber \\
	&\text{Odd:} &  & \d G_{zy}, \, \d G_{ty}, \, \d G_{xy}, \, \d A_y.
\end{align}
The two representations decouple, by symmetry, so may be considered separately. We will restrict to the even channel, which also call the sound channel since it contains the sound modes. The quasinormal modes and Green's functions in the odd channel were computed in ref.~\cite{Gushterov:2018nht}.

The ten sound-channel fluctuations satisfy ten coupled ordinary differential equations. The problem of extracting the quasinormal modes from these equations may be simplified by defining the gauge-invariant linear combinations~\cite{Kovtun:2005ev,Davison:2011uk,Davison:2013bxa,Tarrio:2013tta},
\begin{align}
	Z_1 & \equiv k  \delta A_t + \omega \delta A_x + \frac{1}{2} k  z F_{zt} \, \delta {G^y}_{y}, 
	\nonumber \\
	Z_2 & \equiv - k^2 f \delta {G^t}_{t} + \omega^2  \delta {G^x}_{x} + 2 \w k \delta {G^x}_{t} + \left(-\w^2+ k^2 f - \frac{1}{2} k^2 z f'\right) \delta {G^y}_{y},
	\label{zdef}
\end{align}
where indices on \(\d G\) have been raised using the unperturbed metric~\eqref{bgsol}. Substituting these for \(\d A_t\) and \({\d G^t}_t\) in the equations of motion, one finds that \(Z_{1,2}\) decouple from the remaining fluctuations. They satisfy two coupled second-order ordinary differential equations, of the form
\begin{align}
Z_1''  + A_1 Z_1' + A_2 Z_2' + A_3 Z_1 + A_4 Z_2 &= 0, 
\nonumber \\
Z_2''  + B_1 Z_1' + B_2 Z_2' + B_3 Z_1 + B_4 Z_2 &= 0,
\label{eq:fluctuation_eoms}
\end{align}
where \(A_a\) and \(B_a\) are functions of \(z\), whose form is given in appendix~\ref{app:zero_sound_eom}.
The asymptotic expansions of $Z_1$ and $Z_2$ near the boundary, obtained by solving~\eqref{eq:fluctuation_eoms} order-by-order at small \(z\), are
\begin{align}
	Z_1 &= Z^{(0)}_1 + Z^{(1)}_1 z + \mathcal{O}(z^2),
	\nonumber \\
	Z_2 &= Z^{(0)}_2 - \frac{1}{2} Z^{(0)}_2(k^2-\omega^2) z^2 + Z^{(3)}_2 z^3 + \mathcal{O}(z^4),
	\label{eq:hzs_Z_near_boundary}
\end{align}
where \(Z_{1}^{(0)}\), \(Z_1^{(1)}\), \(Z_2^{(0)}\), and \(Z_2^{(3)}\) are determined by the boundary conditions. The values of $A_{\mu}$ and ${G^{\mu}}_{\nu}$ at the boundary are sources for $J^{\mu}$ and  ${T_{\mu}}^{\nu}$, respectively. Since $\lim_{z \to 0} \left(z F_{zt}\right)= \lim_{z \to 0} \left(z f' \right)= 0$, from~\eqref{zdef} we identify \(Z_1^{(0)}\) as a linear combination of the sources for \(J^t\) and \(J^x\), while \(Z_2^{(0)}\) is a linear combination of the sources for \({T_t}^t\), \({T_x}^x\), \({T_x}^t\), and \({T_y}^y\).

The expansions of $Z_1$ and $Z_2$ near the horizon are
\begin{align}
	Z_1 &= c_1^\mathrm{in} (z_H-z)^{-i\omega/4 \pi T} \zeta^\mathrm{in}_1(z)+ c_1^\mathrm{out} (z_H-z)^{i\omega/4 \pi T} \zeta^\mathrm{out}_1(z),
	\nonumber \\
	Z_2 &= c_2^\mathrm{in} (z_H-z)^{-1-i\omega/4 \pi T}  \zeta^\mathrm{in}_2(z)+ c_2^\mathrm{out} (z_H-z)^{-1+i\omega/4 \pi T} \zeta^\mathrm{out}_2(z),
\end{align}
where \(c_{1,2}^\mathrm{in,\,out}\) are constants, determined by the boundary conditions, and \(\z_{1,2}^\mathrm{in,\,out}(z)\) are functions which are regular at \(z=z_H\), with \(\z_{1,2}^\mathrm{in,\,out} (z_H) = 1\). The quasinormal modes are those frequencies for which there exists a solution to~\eqref{eq:fluctuation_eoms} satisfying \(Z_{1,2}^{(0)} = c_{1,2}^\mathrm{out} =0\).

The coefficients \(B_1\) and \(B_3\) in~\eqref{eq:fluctuation_eoms} are proportional to \(\t\). Thus, in the probe limit \(Z_1\) drops out of the second equation in~\eqref{eq:fluctuation_eoms}. One may then consistently set \(Z_2 = 0\), and solve the remaining equation of motion for \(Z_1\). The quasinormal modes one finds in this manner are poles of the sound-channel Green's functions involving \(J^\m\). Similarly, the poles of the sound-channel \(T^{\m\n}\) Green's functions are the quasinormal modes found by solving the second equation in~\eqref{eq:fluctuation_eoms} for \(Z_2\). Away from the probe limit, the quasinormal modes are poles of the Green's functions of \textit{both} the \(J^\m\) and \(T^{\m\n}\) Green's functions.

We will also compute the spectral functions for the charge and energy densities. These are defined as
\begin{align}
	\r_{qq}(\w,k) &\equiv i \le[
		G_{J^t J^t} (\w,k) - G_{J^t J^t} (\w,k)^*
	\ri],
	\nonumber \\
	\r_{\ve\ve}(\w,k) &\equiv i \le[
		G_{T^{tt} T^{tt}} (\w,k) - G_{T^{tt} T^{tt}} (\w,k)^*
	\ri],
	\label{eq:hzs_spectral_functions}
\end{align}
respectively, where \(G_{\cO\cQ}\) denotes the retarded Green's function of operators \(\cO\) and \(\cQ\). To compute the quasinormal modes and spectral functions numerically, we use the shooting method of ref.~\cite{Kaminski:2009dh}, described in detail in appendix~\ref{app:zero_sound_numerics}.

\section{Numerical Results}
\label{numerical}

We now present results for the frequencies of poles in the sound-channel retarded two-point functions of \(J^\m\) and $T^{\mu\nu}$, in the state holographically dual to the solution~\eqref{bgsol}. We will use \(G_{JJ}\) to collectively denote all of the sound-channel retarded Green's functions of \(J^\m\), and similarly \(G_{TT}\) for the \(T^{\m\n}\) Green's functions. We will focus on the sound and diffusion poles, and the crossover to hydrodynamic behaviour at large temperatures. We will also compute spectral functions~\eqref{eq:hzs_spectral_functions}.

As discussed in section~\ref{adsrn}, in extremal \ads[4]-Reissner-Nordstr\"om the near-horizon $\ads[2] \times \mathbb{R}^2$ indicates that the dual CFT state is a semi-local quantum liquid~\cite{Iqbal:2011in}. This leads to branch cuts along the imaginary frequency axis in \(G_{JJ}\) and \(G_{TT}\) at \(T=0\)~\cite{Edalati:2010hk,Edalati:2010pn}. At non-zero temperature, this branch cut splits into a discrete set of purely imaginary poles. We expect the same physics to occur for the back-reacted brane model, due to the near-horizon \ads[2] factor. However, for numerical stability of our shooting method we will always restrict to non-zero temperature, so we will not see direct evidence for a branch cut. To obtain accurate results at \(T=0\), we would need to use a different numerical method, such as Leaver's matrix method, which was used for extremal \ads[4]-Reissner-Nordstr\"om in refs.~\cite{Edalati:2010hk,Edalati:2010pn}.

We present our numerical results for the poles in the Green's functions in section~\ref{sec:poles}, and for the spectral functions in section~\ref{sec:spectral_functions}. In section~\ref{sec:soundatt} we analyse the attenuation of the sound poles in detail.

\subsection{Poles and Dispersion Relations}
\label{sec:poles}

\subsubsection{Changing back-reaction}

We will begin in the probe limit, \(\t = 0\), and then move through successively larger values of \(\t\), seeing how the behaviour of the poles in the Green's functions changes with each step. 

In the probe limit, the metric and gauge field decouple, so the poles of the \(G_{JJ}\) and \(G_{TT}\) are distinct. At \(T=0\), the metric is that of \ads[4] and the \(G_{TT}\) are fixed by conformal invariance. They have no poles, but branch points at \(\w = \pm k\) and \(\w = \infty\)~\cite{DiFrancesco:1997nk}. When \(T\neq0\), the metric becomes that of \ads[4]-Schwarzschild, and \(G_{TT}\) develops poles. HZS appears as poles of the \(G_{JJ}\) Green's functions at \(T=0\)~\cite{Karch:2008fa,Karch:2009zz}. 

\begin{figure}
	\begin{center}
		\(\talpha = 1\), \(k/\m = 10^{-2}\)
	\end{center}
		\begin{subfigure}{0.5\textwidth}
			\includegraphics[width=\textwidth]{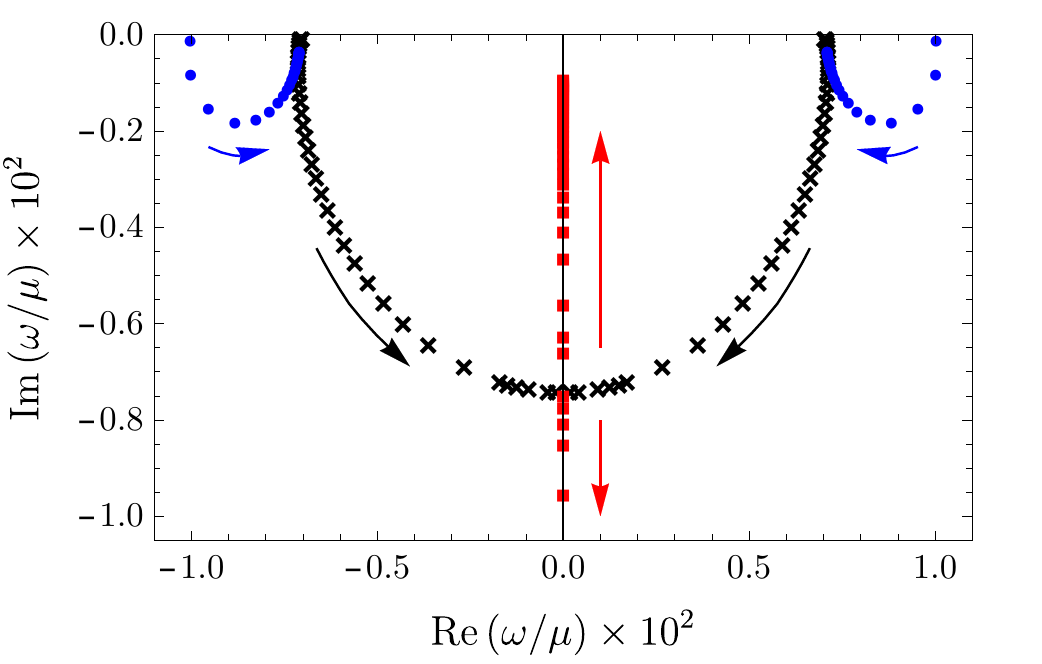}
			\caption{
				Probe limit: \(\t = 0\).
			}
			\label{plane_probe}
		\end{subfigure}
		\begin{subfigure}{0.5\textwidth}
			\includegraphics[width=\textwidth]{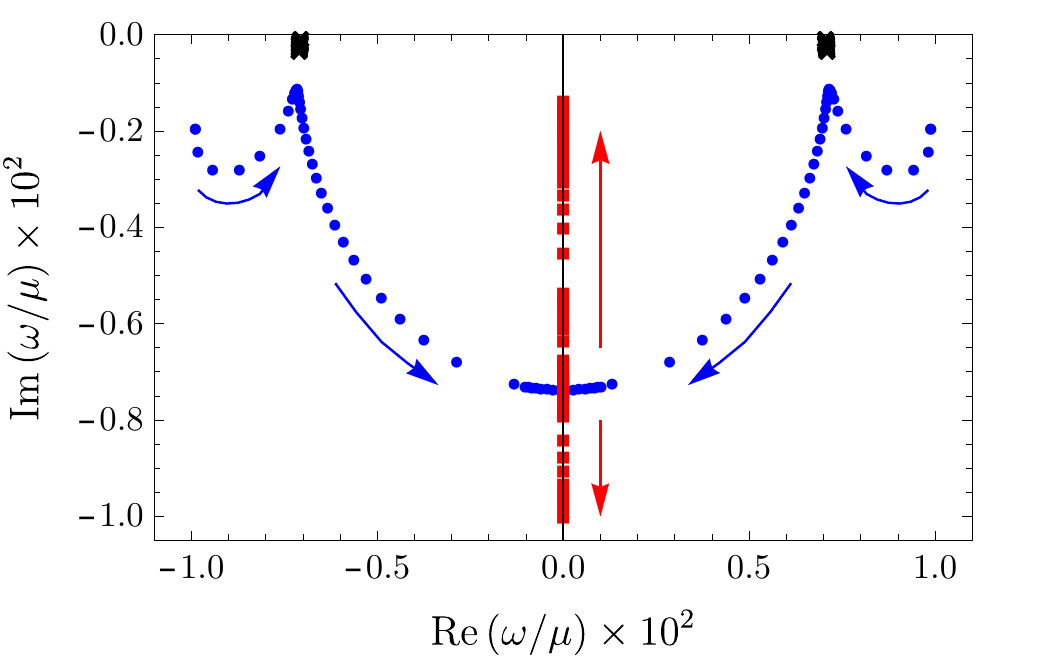}
			\caption{
				\(\t = 10^{-4}\).
			}
			\label{plane_10Em4}
		\end{subfigure}
		\begin{subfigure}{0.5\textwidth}
			\includegraphics[width=\textwidth]{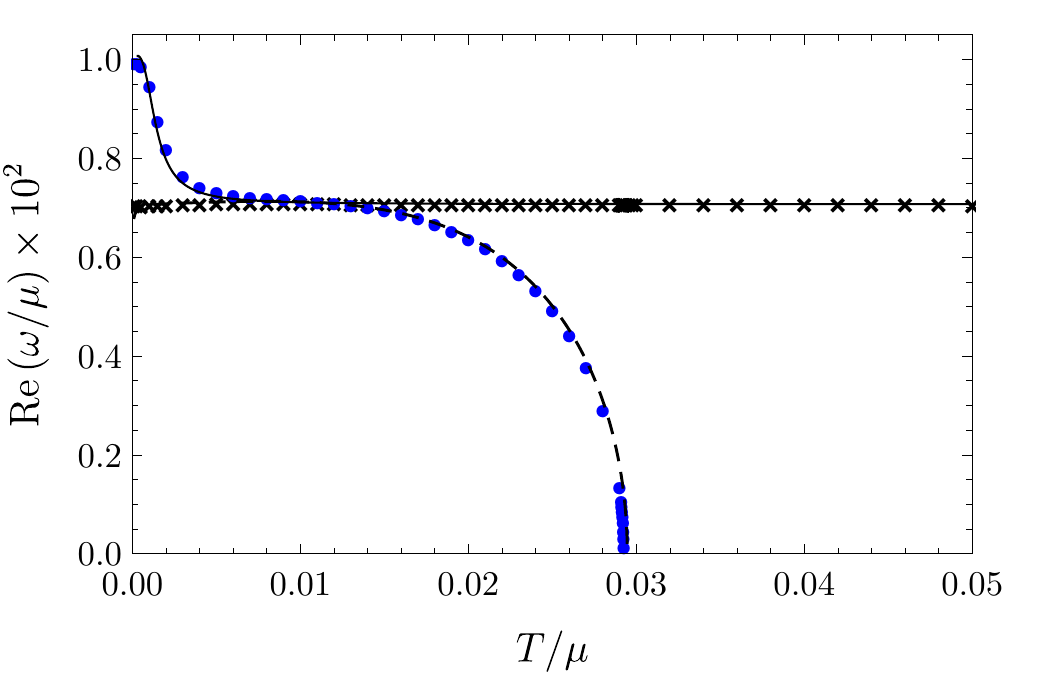}
			\caption{\(\t = 10^{-4}\), real part.}
			\label{realimsmalltaua}
			\end{subfigure}
			\begin{subfigure}{0.5\textwidth}
			\includegraphics[width=\linewidth]{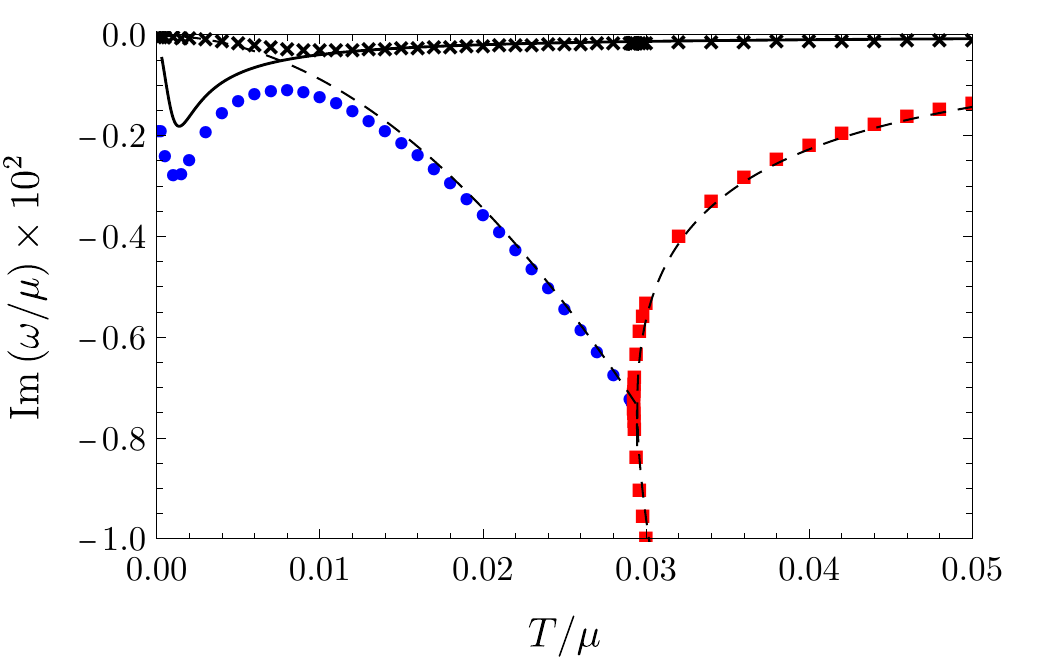}
			\caption{\(\t = 10^{-4}\), imaginary part.}
			\label{realimsmalltaub}
			\end{subfigure}
		\caption[Poles in Green's functions of \(J^\m\) and \(T^{\m\n}\) as a function of temperature in the holographic dual to a spacetime filling brane, at small back-reaction.]{\label{planesmalltau}
			Poles of \(G_{JJ}\) and \(G_{TT}\), as a function of temperature $T/\mu$, with $\talpha=1$ and $k/\mu=10^{-2}$.  In the top two plots, the arrows indicate the movement of poles as $T/\mu$ increases. \textbf{(a):} The poles in the complex frequency plane in the probe limit, $\tau=0$, for temperatures in the range $5\times10^{-4}\leq T/\mu\leq0.1$. At $T/\mu=5\times10^{-4}$ we find four poles, two in \(G_{TT}\), with \(\Re \w \approx \pm k\) (blue dots), and two in \(G_{JJ}\), with dispersion well approximated by the HZS dispersion~\eqref{eq:zero_sound_dispersion} (black crosses). As $T/\mu$ increases, the blue dots move to smaller real part, eventually becoming hydrodynamic sound poles with dispersion~\eqref{eq:sound_dispersion}. The black crosses move toward the imaginary axis in an approximate semicircle, eventually colliding on the imaginary axis. At higher temperatures, the poles in \(G_{JJ}\) are purely imaginary (red squares). One of these moves up and becomes the charge diffusion pole.
			\textbf{(b):} Poles for $\tau=10^{-4}$, for $10^{-4}\leq T/\mu \leq 0.05$. Since \(\t \neq 0\), all poles are poles of both \(G_{JJ}\) and \(G_{TT}\). At $T/\mu=10^{-4}$ we find HZS poles (black crosses), and poles with larger real part (blue dots), similar to \(\t=0\). However, as \(T/\m\) is increased, HZS now smoothly evolves into hydrodynamic sound, and it is the blue dots which collide on the imaginary axis to eventually form charge diffusion.
			\textbf{(c,\,d):} The real and imaginary parts of the pole frequencies as a function of temperature, for \(\t=10^{-4}\). The colour coding and shapes are the same as in figure (b). The solid and dashed black lines are the probe limit results for the poles in \(G_{TT}\) and \(G_{JJ}\), respectively.
		}
\end{figure}

Figure~\ref{plane_probe} shows numerical results for the motion of the poles of \(G_{JJ}\) and \(G_{TT}\) in the complex frequency plane as we increase the temperature. Our results are qualitatively the same as the \(d=4\) results of refs.~\cite{Policastro:2002tn,Kovtun:2005ev} for $G_{TT}$ and refs.~\cite{Davison:2011ek,Chen:2017dsy} for \(G_{JJ}\). At low temperature, $T/\mu=5\times10^{-4}$, we find four poles, two in $G_{TT}$ at $\Re \omega \approx \pm k$~\cite{Kovtun:2005ev}, denoted by blue dots in figure~\ref{plane_probe}, and two in $G_{JJ}$ with dispersion well approximated by the zero sound dispersion~\eqref{eq:zero_sound_dispersion}, with \(v \approx 1/\sqrt{2}\)~\cite{Davison:2011ek,Chen:2017dsy}, denoted by black crosses in the figure.

As \(T/\m\) is increased, the blue dots move to smaller real part, eventually becoming the hydrodynamic sound modes at large \(T/\m\), with dispersion~\eqref{eq:sound_dispersion}. The black crosses move in an approximate semicircle, eventually colliding on the imaginary axis at \(T/\m \approx 0.033\). As the temperature is increased further, the poles in \(G_{JJ}\) become purely imaginary (red squares in the figure). One of these poles moves deeper into the complex plane as the temperature is raised, while the other pole moves towards the real axis, eventually becoming hydrodynamic charge diffusion, with dispersion~\eqref{eq:diffusion_dispersion}.

We now introduce small back-reaction, setting \(\t = 10^{-4}\).\footnote{We also studied the motion of the poles for \(\t=10^{-5}\), finding qualitatively similar behaviour to \(\t =10^{-4}\).}  Figure~\ref{plane_10Em4} shows our numerical results for the motion of the poles in the complex frequency plane, for $10^{-4}\leq T/\mu \leq 0.05$, while figures~\ref{realimsmalltaua} and~\ref{realimsmalltaub} show the real and imaginary parts of \(\w/\m\) as a function of \(T/\m\).

At the lowest temperature we studied, $T/\mu=10^{-4}$, we again find four poles, two with \(\Re \w \approx \pm k\), again denoted by blue dots, and two HZS poles, again denoted by black crosses.  As \(T/\m\) is increased, there is a significant qualitative difference to the probe limit: the black crosses no longer collide. Instead, they stay at approximately constant real part and small imaginary part. When \(T/\m\) becomes large, the black crosses become hydrodynamic sound.

It is now the blue dots which are responsible for forming charge diffusion. At low temperatures they execute motion similar to in the probe limit, until they reach a closest approach to the black crosses. As the temperature is raised further, the blue dots trace out an approximate semicircle in the complex plane, similar to the poles of \(G_{JJ}\) in the probe limit. The blue dots eventually collide on the imaginary axis, at \(T/\m \approx 0.029\), forming two purely imaginary poles, one of which becomes charge diffusion. It is therefore still possible to define a precise moment of crossover in the same way as the probe limit~\cite{Davison:2011ek}, when the two poles collide on the imaginary axis and produce the charge diffusion pole.

\begin{figure}[t]
	\begin{center}
		\(\t = 10^{-4}\), \(\talpha = 1\), \(T/\m = 10^{-2}\)
	\end{center}
	\begin{subfigure}{0.5\textwidth}
		\includegraphics[width=\textwidth]{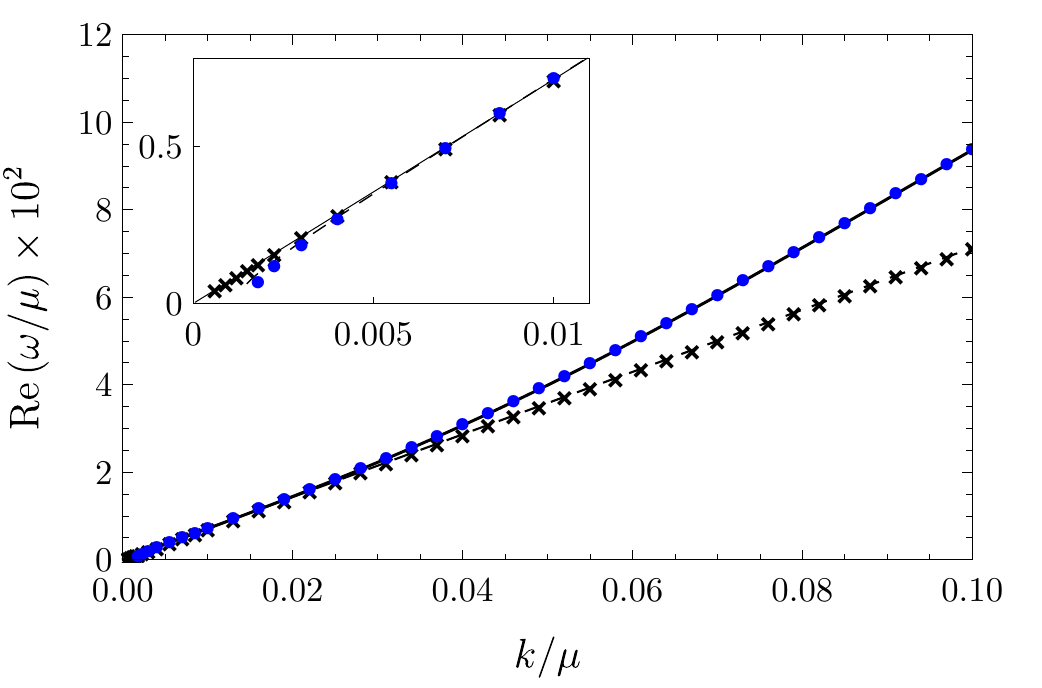}
		\caption{Real part.}
	\end{subfigure}
	\begin{subfigure}{0.5\textwidth}
		\includegraphics[width=\textwidth]{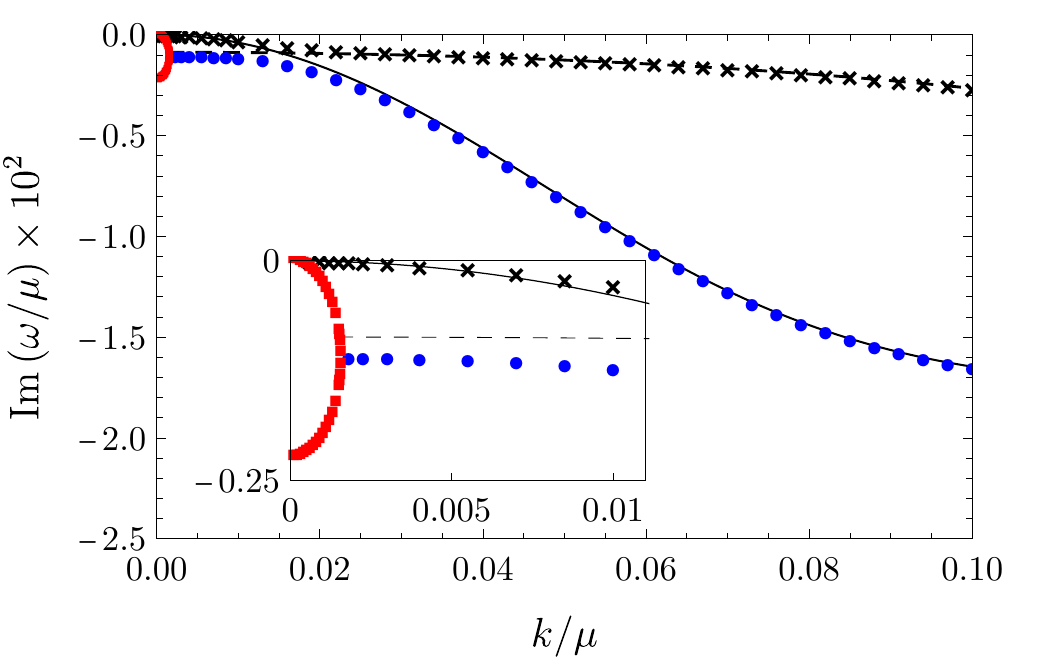}
		\caption{Imaginary part.}
		\label{fig:dispersionsmalltau_imaginary}
	\end{subfigure}
	\caption[Sample dispersion relations for the poles in Green's functions of \(J^\m\) and \(T^{\m\n}\) in the holographic dual to a spacetime filling brane.]{\label{fig:dispersionsmalltau} Dispersion relations of the four highest poles for $\tau=10^{-4}$, $\talpha=1$, $T/\mu = 10^{-2}$, and momenta in the range $10^{-4} \leq k/\mu \leq 0.1$. The plots show the real (left) and imaginary (right) parts of \(\w/\m\), as a function of \(k/\m\), with insets detailing the behaviour at small \(k/\m\). The solid and dashed black lines show the poles in $G_{TT}$ and $G_{JJ}$ in the probe limit, respectively. The colour coding is consistent with figure~\ref{planesmalltau}. The black crosses follow the probe HZS dispersion for large $k/\mu$, and the hydrodynamic sound dispersion for small $k/\mu$. At large $k/\mu$, the blue dots have the dispersion of the poles in $G_{TT}$, with $\Re\left(\omega\right) = \pm k$, but for \(k/\m \approx 0.02\)--\(0.03\), they satisfy $\Re \left(\omega\right) = \pm k/\sqrt{2}$. For $k/\mu \lesssim 0.02$, the real part of the blue dots rapidly decreases, reaching $\Re \left(\omega\right) = 0$ around $k/\mu \approx 2\times 10^{-3}$. At lower momenta, the blue dots are replaced by purely imaginary poles, indicated by the red squares.}
\end{figure}

The identification of poles as HZS, hydrodynamic sound, or charge diffusion is made using their dispersion relations. For example, figure~\ref{fig:dispersionsmalltau} shows dispersion relations for $\tau=10^{-4}$, $\talpha=1$, $T/\mu=10^{-2}$, and momenta in the range $10^{-4} \leq k/\mu \leq 0.1$. The colour coding is the same as in the \(\t=10^{-4}\) plots in figure~\ref{planesmalltau}.

The black crosses, which exist for all values of \(k/\m\), have sound-like dispersion \(\w = \pm k/\sqrt{2} - i \G k^2 + \cO(k^3)\). From the inset in figure~\ref{fig:dispersionsmalltau_imaginary}, we observe that \(\G\) is well approximated by the attenuation of the hydrodynamic sound pole in the holographic dual to \ads[4]-Schwarzschild. In fact, as detailed in section~\ref{sec:soundatt}, for any \(\t \neq 0\) we find that the hydrodynamic prediction~\eqref{eq:hydro_sound_attenuation} for the attenuation of sound works well even for \(T/\m \ll 1\).\footnote{The same phenomenon has been observed in other holographic models~\cite{Davison:2013bxa,Davison:2013uha}.} The distinction between HZS and hydrodynamic sound at non-zero back-reaction is therefore a matter of terminology. We will continue to refer to sound modes at \(T/\m \ll 1\) as HZS, to emphasise the fact that sound modes survive to low temperatures in these models.

At sufficiently large momenta, we also observe poles with \(\Re \w \approx \pm k\), indicated by blue dots in the figure. Lowering the momentum, around \(k/\m \approx 0.02\)--\(0.03\) the dispersion becomes \(\Re \w \approx \pm k/\sqrt{2}\), similar to the sound modes. However, as the momentum is lowered still further, we find that the blue dots collide on the imaginary axis, forming two purely imaginary poles, indicated by red squares. One of these poles moves to zero frequency as \(k \to 0\), and is identified with hydrodynamic charge diffusion.

\begin{figure}
	\begin{center}
		\(\t = 10^{-3}\), \(\talpha = 1\), \(k/\m = 10^{-2}\)
	\end{center}
	\begin{subfigure}{0.5\textwidth}
		\includegraphics[width=\textwidth]{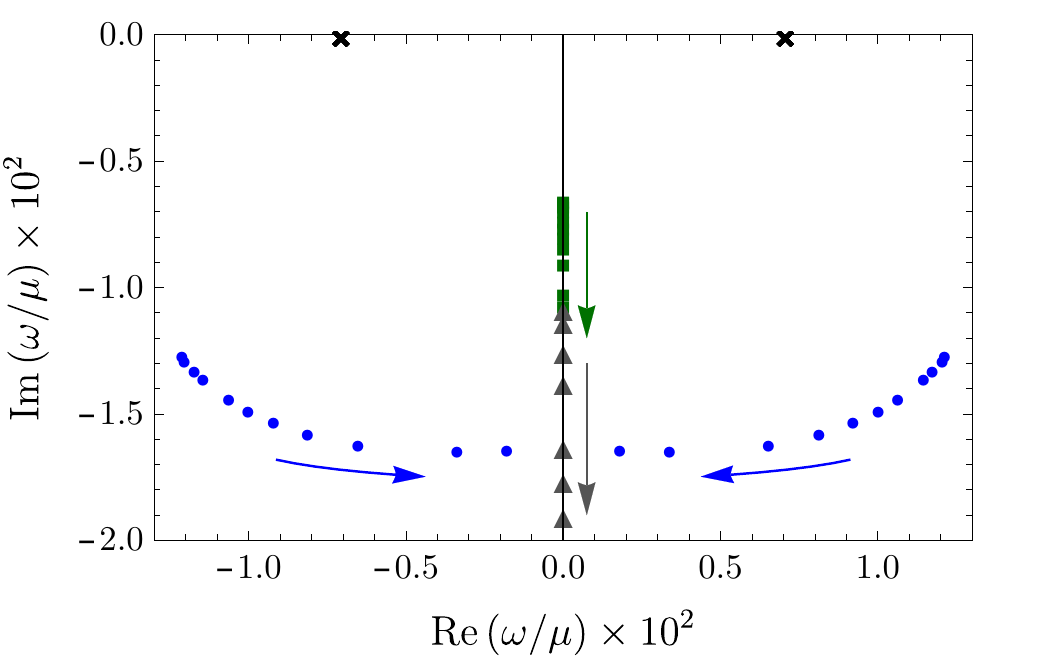}
		\caption{$T/\mu=1.25\times10^{-3}$ to $2.23\times10^{-3}$.}
		\label{plane_tau_0p001_low_T}
	\end{subfigure}
	\begin{subfigure}{0.5\textwidth}
		\includegraphics[width=\textwidth]{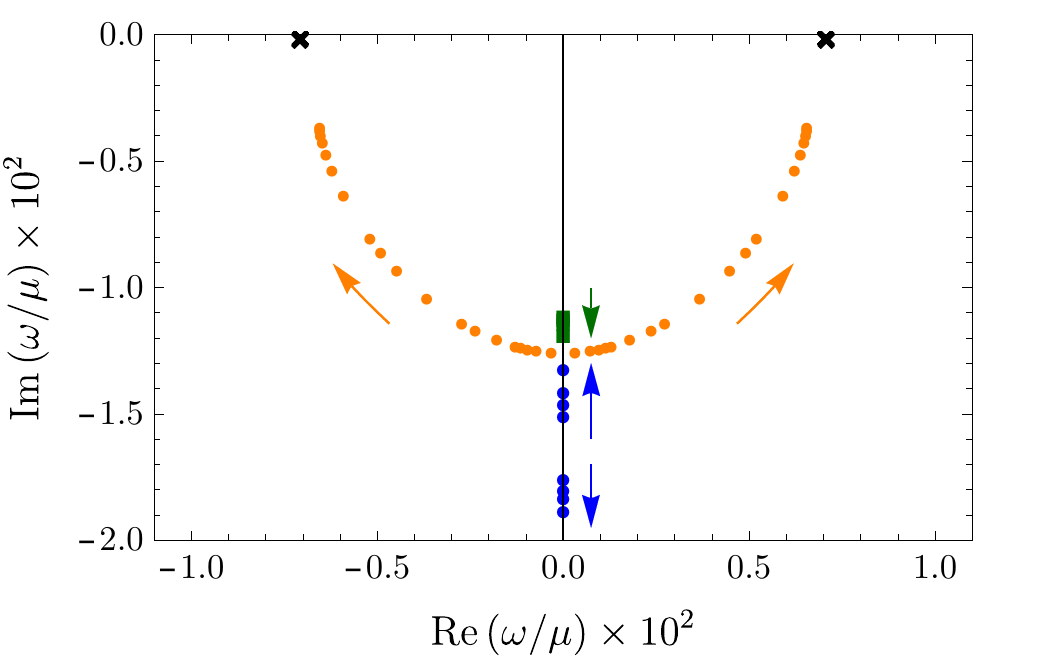}
		\caption{$T/\mu = 2.23\times10^{-3}$ to $10^{-2}$.}
		\label{plane_tau_0p001_mid_T}
	\end{subfigure}
	\begin{center}
	\begin{subfigure}{0.5\textwidth}
		\includegraphics[width=\textwidth]{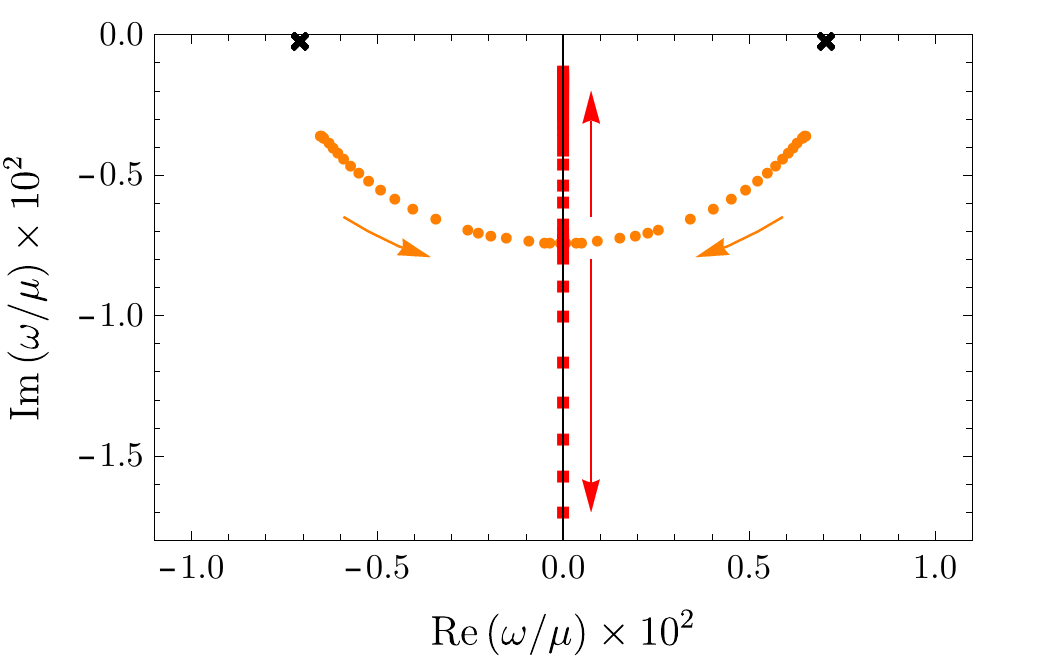}
	\caption{\(T/\mu = 0.011\) to \(0.05\).}
	\label{plane_tau_0p001_high_T}
	\end{subfigure}
	\end{center}
	\begin{subfigure}{0.5\textwidth}
		\includegraphics[width=\textwidth]{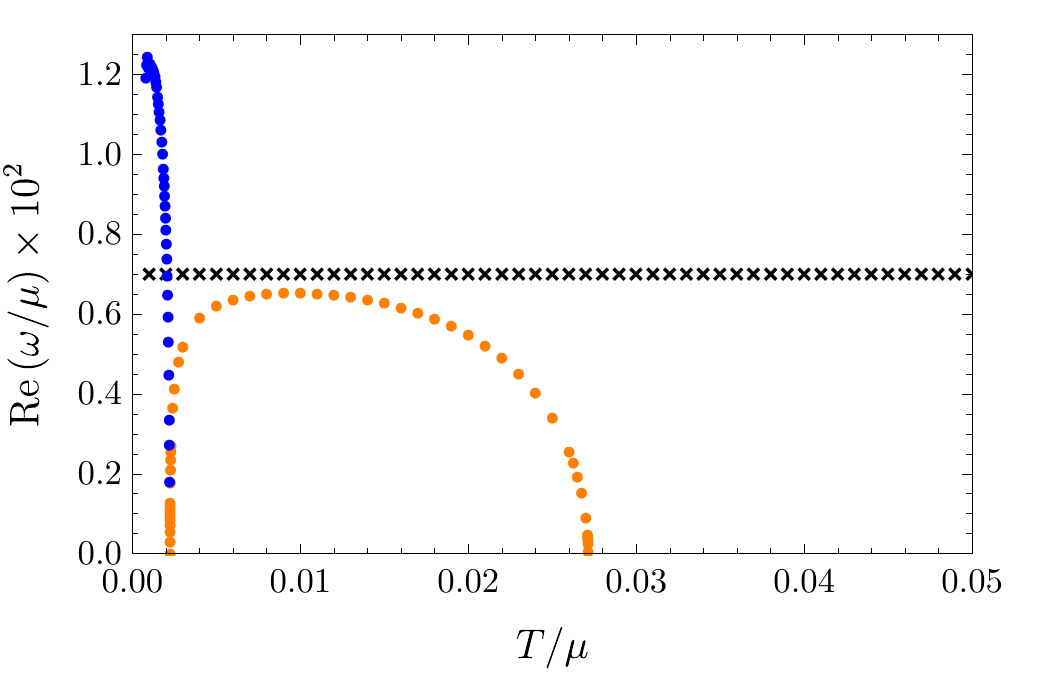}
		\caption{Real part.}
		\label{realimtau10e3a}
	\end{subfigure}
	\begin{subfigure}{0.5\textwidth}
		\includegraphics[width=\linewidth]{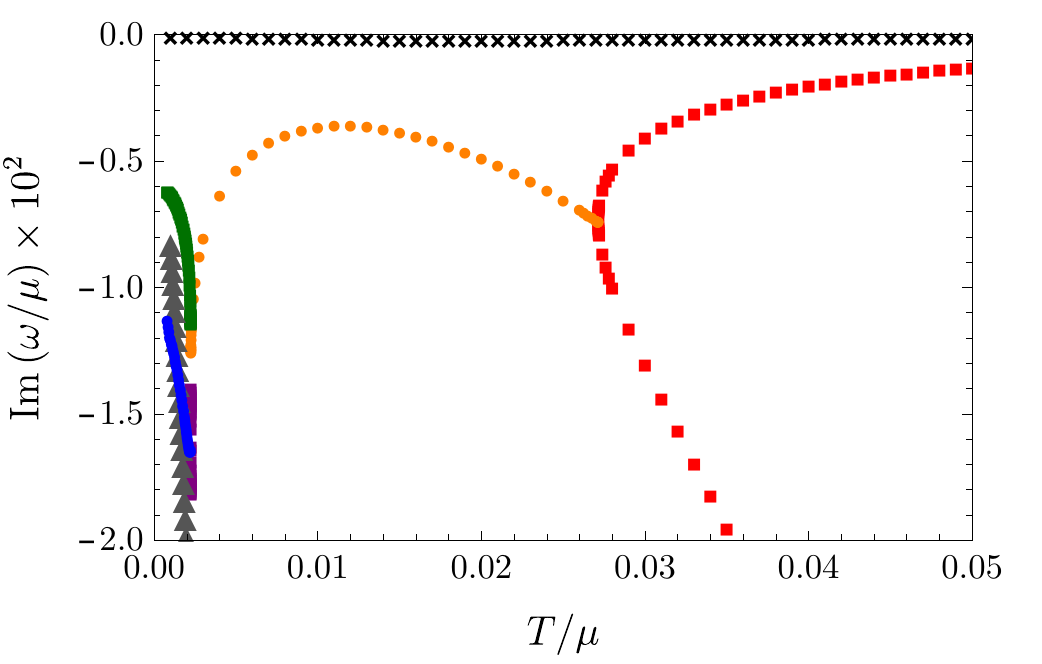}
		\caption{Imaginary part.}
		\label{realimtau10e3b}
		\end{subfigure}
	\caption[Poles in Green's functions of \(J^\m\) and \(T^{\m\n}\) as a function of temperature, at intermediate back-reaction.]{Frequencies of poles of $G_{JJ}$ and $G_{TT}$, with $\tau = 10^{-3}$, $\talpha=1$, $k/\mu = 0.01$, for temperatures in the range \(1.25 \times 10^{-3} \leq T/\m \leq 0.05\). \textbf{(a,\,b,\,c):} The poles in the complex frequency plane at successively larger values of \(T/\m\), with arrows indicating the motion with increasing temperature. Sound poles, indicated by black crosses, exist for all temperatures studied. The motion of the poles and formation of the diffusion pole is significantly more complicated than for smaller \(\t\). \textbf{(d,\,e):} The real and imaginary parts of the frequency as a function of temperature. The colours and shapes of the plot markers are the same as in the plots of the complex plane.
	}
	\label{plane_tau_0p001} 
\end{figure}
Figure~\ref{plane_tau_0p001} shows our numerical results for the poles with larger back-reaction, $\t = 10^{-3}$, still with $\talpha=1$ and $k/\mu = 10^{-2}$, for temperatures in the range $1.25 \times 10^{-3} \leq T/\mu \leq 0.05$. Figures~\ref{plane_tau_0p001_low_T},~\ref{plane_tau_0p001_mid_T}, and~\ref{plane_tau_0p001_high_T} show how the poles move in the complex plane as the temperature is changed, with arrows indicating the pole movement as $T/\mu$ increases. Figures~\ref{realimtau10e3a} and~\ref{realimtau10e3b} show the real and imaginary parts of the pole frequencies as a function of temperature.

At the lowest temperature that we access, \(T/\m = 1.25 \times 10^{-3}\), the closest poles to the real axis are HZS poles. Deeper into the complex plane, we observe two purely imaginary poles (green squares and grey diamonds), and a pair of poles with finite real part (blue dots in figure~\ref{plane_tau_0p001_low_T}). The purely imaginary poles are presumably remnants of a branch cut at \(T=0\).

As the temperature is increased, the sound poles barely move, while the purely imaginary poles move to more negative imaginary part. The blue dots move towards the imaginary axis, eventually colliding at \(T/\m \approx 2.23 \times 10^{-3}\). Crucially, the collision occurs below the green squares. The collision forms two new purely imaginary poles (blue dots in figure~\ref{plane_tau_0p001_mid_T}). One of these new poles moves deeper into the complex plane with increasing temperature. The other moves towards the real axis, eventually colliding with the pole denoted by green squares, at \(T/\m \approx 2.24 \times 10^{-3}\).

This second collision creates two new poles, with non-zero real part (orange dots in figures~\ref{plane_tau_0p001_mid_T} and~\ref{plane_tau_0p001_high_T}). As the temperature is increased further, the new poles move away from the imaginary axis and towards the real axis, appearing to approach the sound poles. They reach a closest approach to the sound poles at \(T/\m \approx 0.01\), before turning around and approaching the imaginary axis again. The orange dots eventually collide on the imaginary axis. This forms two purely imaginary poles (red squares in figure~\ref{plane_tau_0p001_high_T}), one of which becomes charge diffusion at high temperature. The probe limit definition of the hydrodynamic crossover is therefore still viable, despite the more complicated pole motion, with the crossover temperature given by the temperature of this final pole collision,~\(T/\m \approx 0.027\).

The key difference between $\tau=10^{-3}$ and \(\t = 10^{-4}\) is that for \(\t=10^{-3}\), the two low-temperature propagating poles (blue dots) collide on the imaginary axis \textit{below} a purely imaginary pole. Thus, neither of the purely imaginary poles created by the collision can become charge diffusion at high temperatures. Instead, the creation of the charge diffusion pole at high temperatures occurs via the complicated sequence of pole motions described above.  Clearly there is a critical value of \(\t\) between \(10^{-4}\) and \(10^{-3}\), separating the regimes where the pole collision occurs above or below any purely imaginary modes. We find that this critical value is $\tau \approx 9 \times 10^{-4}$.

\begin{figure}
	\begin{center}
		\(\t = 10^{-2}\), \(\talpha = 1\), \(k/\m = 10^{-2}\)
	\end{center}
	\begin{subfigure}{0.5\textwidth}
		\includegraphics[width=\textwidth]{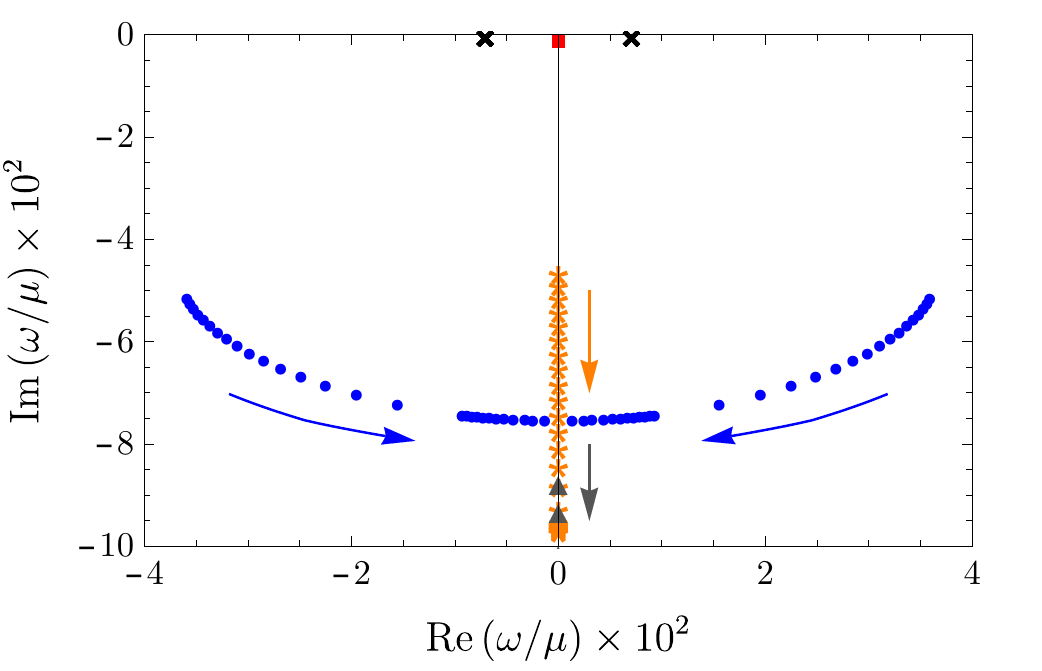}
		\caption{$T/\mu = 5\times 10^{-3}$ to $8.3\times 10^{-3}$.}
		\label{complex_plane_tau0p01_a1_q0p01_t0p005_to_0p0083}
	\end{subfigure}
	\begin{subfigure}{0.5\textwidth}
		\includegraphics[width=\textwidth]{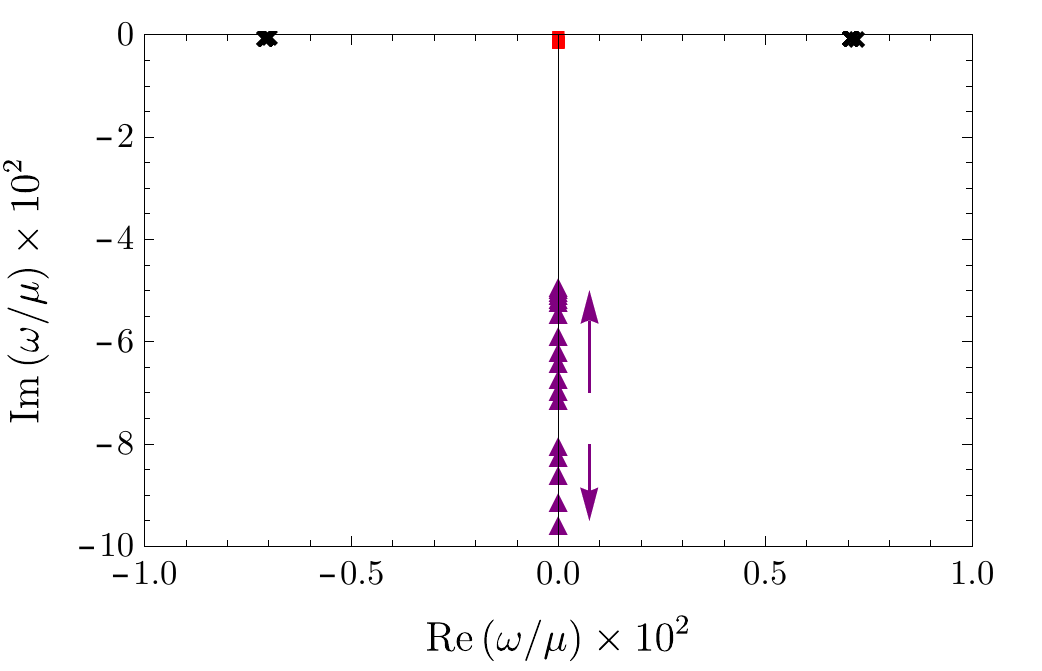}
		\caption{$T/\mu =  8.3\times 10^{-3}$ to $10^{-2}$.}
		\label{complex_plane_tau0p01_a1_q0p01_t0p0083_to_0p01}
	\end{subfigure}
	\begin{subfigure}{0.5\textwidth}
		\includegraphics[width=\textwidth]{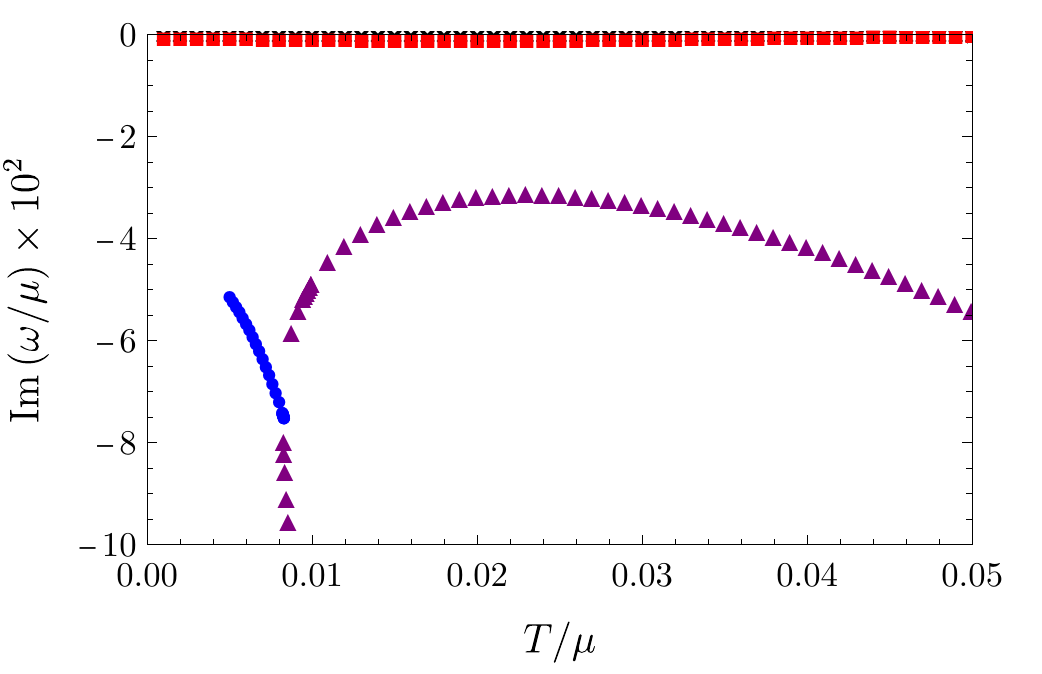}
		\caption{Imaginary part.}
		\label{imaginary_tau_0p01}
	\end{subfigure}
	\begin{subfigure}{0.5\textwidth}
		\includegraphics[width=\textwidth]{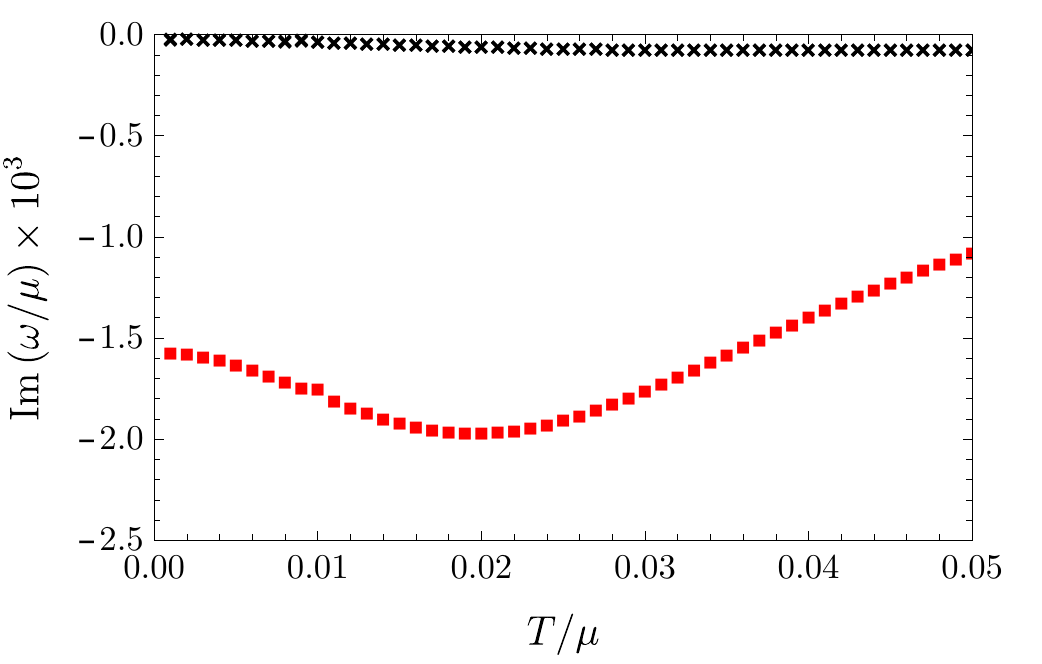}
		\caption{Imaginary part, detail.}
		\label{imaginary_tau0p01_a1_q0p01}
	\end{subfigure}
	\caption[Poles in Green's functions of \(J^\m\) and \(T^{\m\n}\) as a function of temperature, at large back-reaction.]{The frequencies of poles of $G_{JJ}$ and $G_{TT}$ for $\tau=10^{-2}$, $\talpha=1$, $k/\mu=10^{-2}$, and temperatures in the range \(5 \times 10^{-3} \leq T/\m \leq 5 \times 10^{-2}\). \textbf{(a,\,b):} The motion of the poles in the complex frequency plane with changing temperature. Arrows indicate the movement of poles as $T/\mu$ increases. The closest poles to the real axis are always sound poles (the black crosses) and a purely imaginary pole (red squares). Deeper into the complex plane, at low temperatures we observe further purely imaginary poles, and a pair of poles with non-zero real part (the blue dots). As the temperature is raised, the blue dots move towards the imaginary axis. They eventually collide, forming two purely imaginary poles (the purple diamonds). \textbf{(c):} The imaginary parts of the poles as a function of temperature. The orange crosses and grey diamonds have been omitted, for clarity of presentation. \textbf{(d):} A close-up of (c), focusing on the poles closest to the real axis.}
	\label{plane_tau_0p01}
\end{figure}
Figure~\ref{complex_plane_tau0p01_a1_q0p01_t0p005_to_0p0083} shows our numerical results for the pole positions for yet larger back-reaction, $\tau = 10^{-2}$, again with $\talpha=1$ and $k/\mu=10^{-2}$, for temperatures in the range $5 \times 10^{-3}\leq T/\mu \leq 8.3\times 10^{-3}$. At $T/\mu = 5 \times 10^{-3}$, we find that the closest poles to the real axis are HZS poles (black crosses in the figure), with \(\Im \w/\m \approx -2 \times 10^{-5}\). There is also a purely imaginary pole, with \(\Im \w/\m \approx -2 \times 10^{-3}\) (red squares). Significantly deeper into the complex plane, we observe a pair of poles with non-zero real part (blue dots) and two purely imaginary poles (orange stars and grey diamonds) within the range of frequencies that we checked.

As the temperature is increased, the orange stars and grey diamonds move deeper into the complex plane, while the blue dots move towards the imaginary axis. The black crosses and red squares barely move on the scale of figure~\ref{complex_plane_tau0p01_a1_q0p01_t0p005_to_0p0083}. The blue dots eventually collide on the imaginary axis at \(T/\m \approx 8.3 \times 10^{-3}\), below the red squares but above any other imaginary poles.

The collision produces two purely imaginary poles (purple diamonds in figure~\ref{complex_plane_tau0p01_a1_q0p01_t0p0083_to_0p01}), one of which moves deeper into the complex plane as the temperature is increased, while the other moves up towards the real axis. However, as shown in figure~\ref{imaginary_tau_0p01}, the latter pole never collides with the red squares. Instead, it reaches a closest approach, and then begins to to more negative imaginary part. Thus, the red squares exist as the purely imaginary poles closest to the real axis for the full range of temperatures that we study, and at large temperatures become charge diffusion. Figure~\ref{imaginary_tau0p01_a1_q0p01} shows how the imaginary parts of the red squares and black crosses change with temperature.

The poles closest to the real axis are therefore qualitatively similar to those in \ads[4]-Reissner-Nordstr\"om for all temperatures; in particular the charge diffusion pole remains purely imaginary down to low temperature. We find that the critical value of \(\t\), above which the charge diffusion pole is purely imaginary for all temperatures, is \(\t \approx 3.2 \times 10^{-3}\). We have sampled values of \(\t\) up to \(\t = 10^{-1}\), and found that the pole evolution remains qualitatively similar. For $\tau > 3.2 \times 10^{-3}$ we cannot use the probe limit definition of the crossover, since at no point do poles collide on the imaginary axis to produce the hydrodynamic charge diffusion pole.

\vspace{-0.6cm}

\subsubsection{Larger Momentum}

\vspace{-0.2cm}

\begin{figure}
	\begin{center}
		\(\t = 10^{-4}\), \(\talpha = 1\)
	\end{center}
	\begin{subfigure}{0.5\textwidth}
		\includegraphics[width=\textwidth]{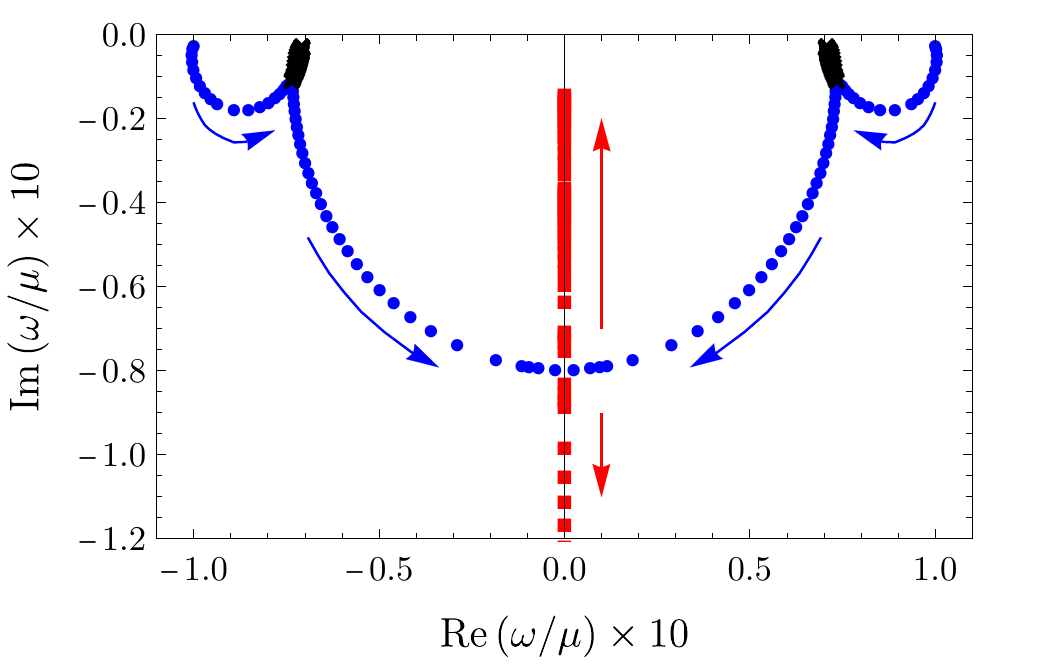}
		\caption{\(k/\m = 0.1\), \(10^{-3} \leq T/\m \leq 0.2\).}
		\label{fig:complex_plane_tau10m4_q0p1}
	\end{subfigure}
	\begin{subfigure}{0.5\textwidth}
		\includegraphics[width=\textwidth]{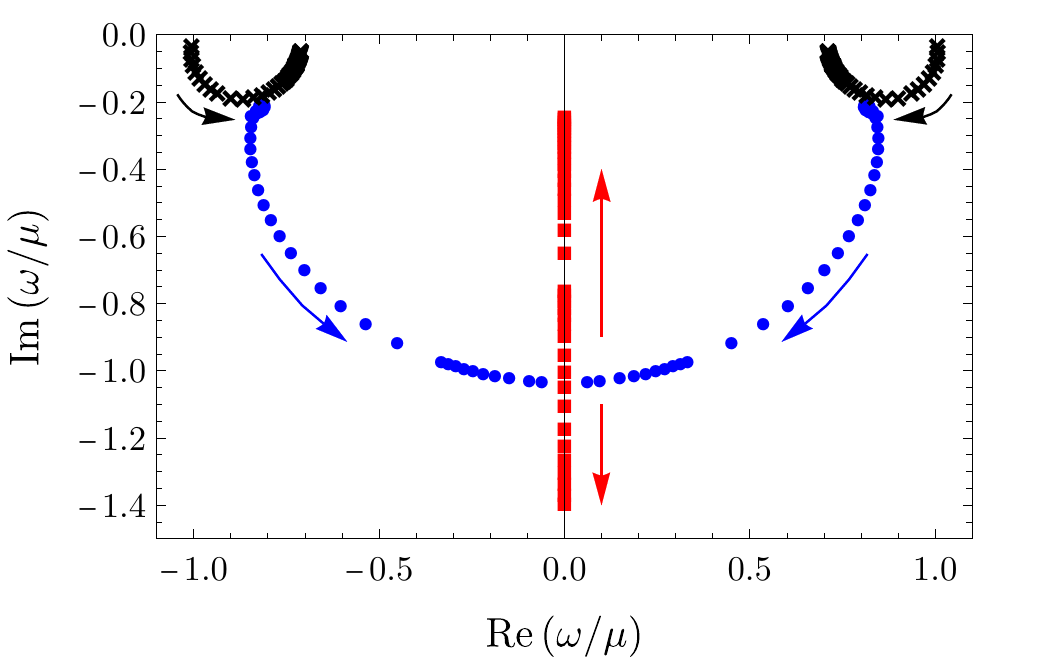}
		\caption{\(k/\m = 1\), \(0.02 \leq T/\m \leq 1\).}
	\end{subfigure}
	\begin{subfigure}{0.5\textwidth}
		\includegraphics[width=\textwidth]{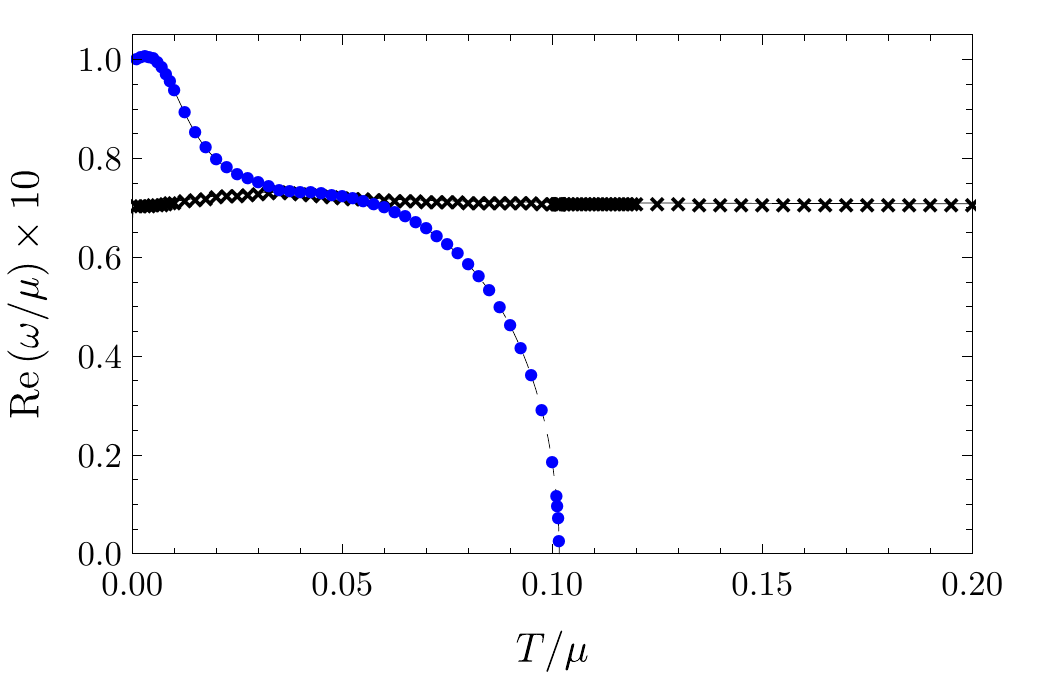}
		\caption{\(k/\m = 0.1\), real part.}
		\label{fig:tau0p0001_a1_k0p1_real}
	\end{subfigure}
	\begin{subfigure}{0.5\textwidth}
		\includegraphics[width=\textwidth]{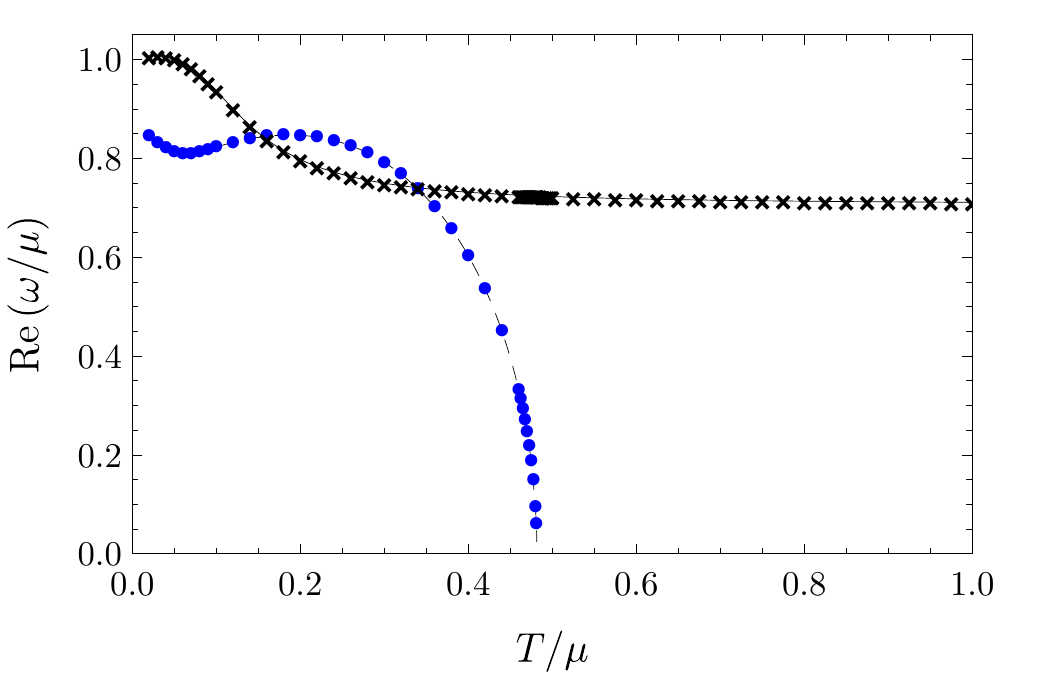}
		\caption{\(k/\m = 1\), real part.}
		\label{fig:tau0p0001_a1_k1_real}
	\end{subfigure}
	\begin{subfigure}{0.5\textwidth}
		\includegraphics[width=\textwidth]{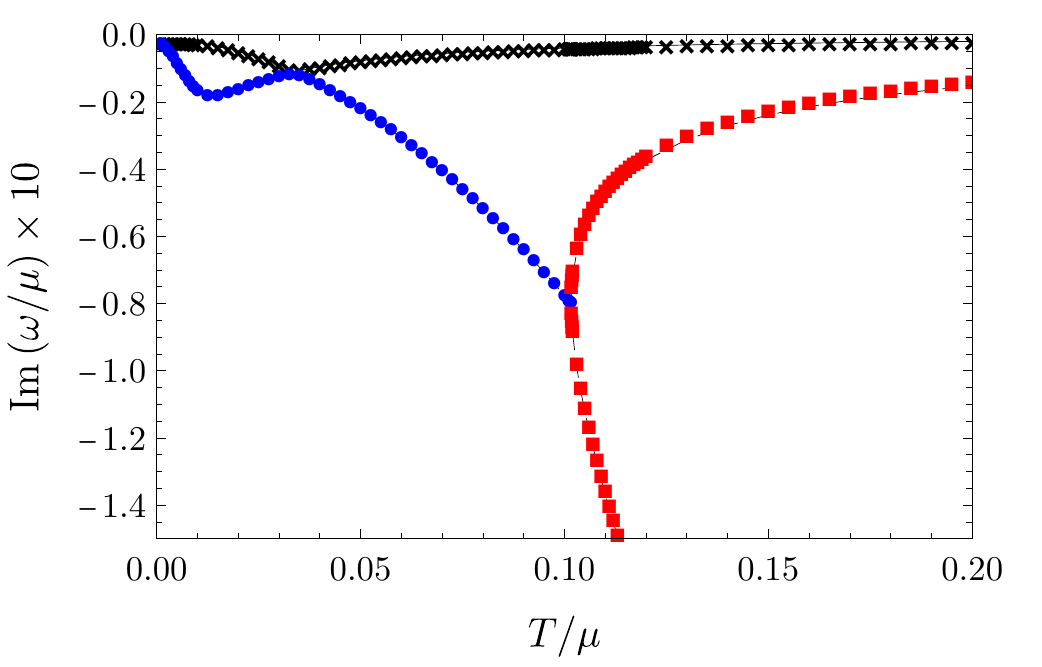}
		\caption{\(k/\m = 0.1\), imaginary part.}
	\end{subfigure}
	\begin{subfigure}{0.5\textwidth}
		\includegraphics[width=\textwidth]{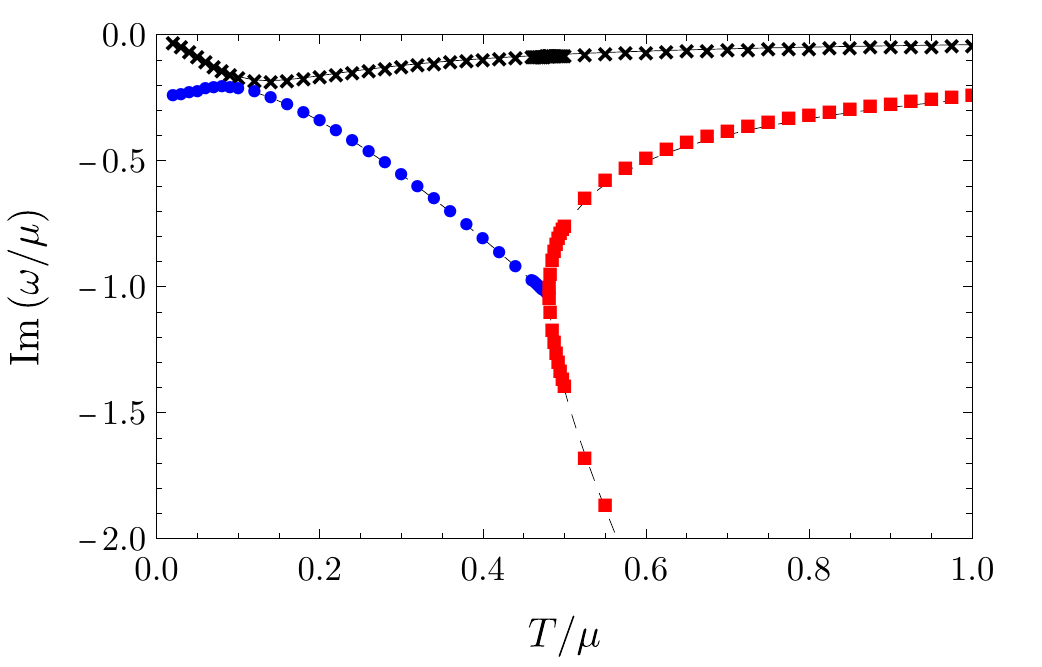}
		\caption{\(k/\m = 1\), imaginary part.}
		\label{fig:tau0p0001_a1_k1_imaginary}
	\end{subfigure}
	\caption[Poles in Green's functions of \(J^\m\) and \(T^{\m\n}\), at small back-reaction and large values of momentum.]{
		Frequencies of the poles in \(G_{JJ}\) and \(G_{TT}\), as a function of temperature, for \(\t = 10^{-4}\), \(\talpha=1\), and \(k/\m = 0.1\) (left column) and \(k/\m = 1\) (right column). The first row shows the positions of the poles in the complex frequency plane, with arrows indicating the motion of the poles with increasing temperatures. The second and third rows show the real and imaginary parts of the poles as a function of temperature, respectively. The solid and dashed black lines show the probe limit results for the poles of \(G_{TT}\) and \(G_{JJ}\) respectively, at the same values of \(k/\m\) and \(\talpha\).
	}
	\label{fig:poles_small_tau_large_momentum}
\end{figure}
We now study the effect on the poles of \(G_{JJ}\) and \(G_{TT}\) of increasing the momentum~\(k/\m\).

Figure~\ref{fig:poles_small_tau_large_momentum} shows the behaviour of the poles with increasing temperature, for \(\t=10^{-4}\), \(\talpha = 1\), and \(k/\m = 0.1\) and \(k/\m = 1\). For \(k/\m = 0.1\), the motion of the poles with temperature is qualitatively similar to that for \(\t = 10^{-4}\) and \(k/\m = 10^{-2}\), plotted in figure~\ref{plane_10Em4}. At low temperatures, there are two pairs of poles with non-zero real part: a pair with \(\Re \w \approx \pm k/\sqrt{2}\) (the black crosses in the figure), and a pair with \(\Re \w \approx \pm k\) (the blue dots).\footnote{The colour coding is the same as in \ref{fig:dispersionsmalltau}; that is, the black crosses have sound-like dispersion at small values of \(k/\m\).} As the temperature is increased, the blue dots move toward the imaginary axis, eventually colliding and forming two purely imaginary poles. The main qualitative difference to the \(k/\m=10^{-2}\) is that the blue dots and black crosses are much closer to each other at their point of closest approach.

For \(k/\m = 1\), the motion of the poles is qualitatively similar to that for \(k/\m = 10^{-2}\) in the probe limit, \(\t = 0\), plotted in figure~\ref{plane_probe}. At low temperature, there is a pair of poles with \(\Re\w \approx \pm k\) (the black crosses in the figure), and another pair of poles with smaller \(|\Re\w|\) (the blue dots). As the temperature is increased, the black crosses move to \(\Re \w \approx k/\sqrt{2}\), becoming the sound poles at large \(T/\m\), while the blue dots collide on the imaginary axis to form the diffusion pole.

\begin{figure}
	\begin{center}
		\(\t = 10^{-2}\), \(\talpha = 1\)
	\end{center}
	\begin{subfigure}{0.5\textwidth}
		\includegraphics[width=\textwidth]{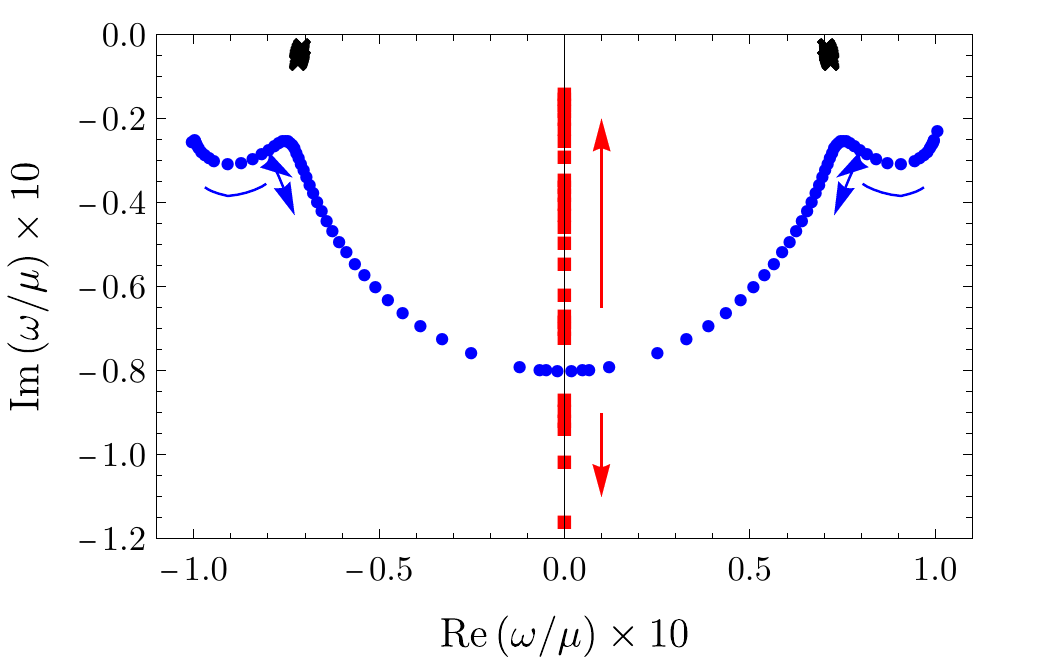}
		\caption{\(k/\m = 0.1\), \(10^{-3} \leq T/\m \leq 0.2\).}
	\end{subfigure}
	\begin{subfigure}{0.5\textwidth}
		\includegraphics[width=\textwidth]{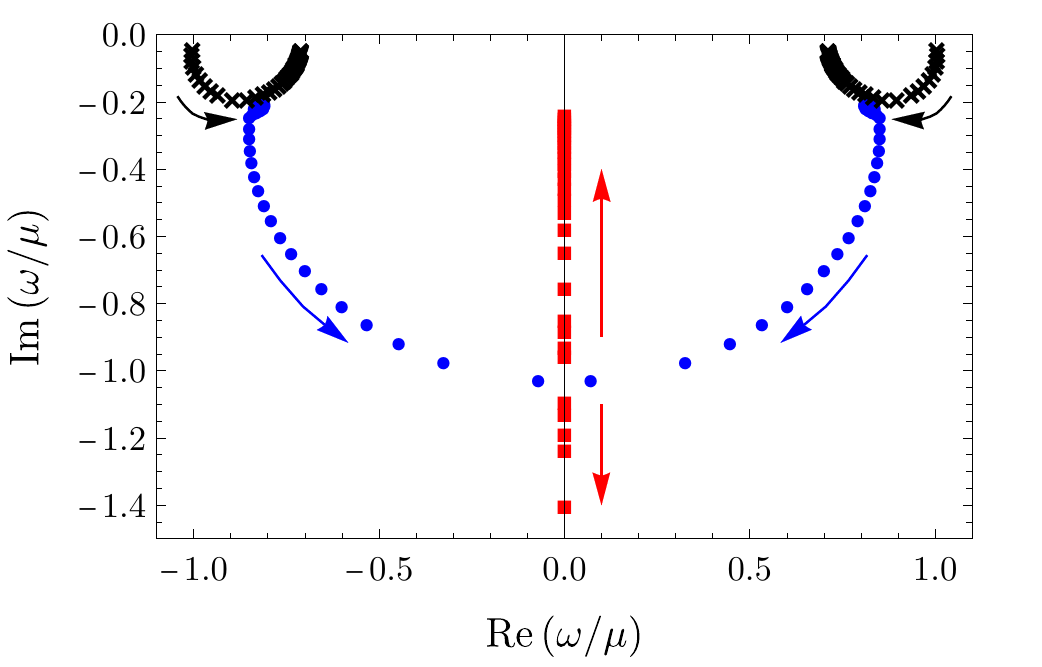}
		\caption{\(k/\m = 1\), \(0.02 \leq T/\m \leq 1\).}
	\end{subfigure}
	\begin{subfigure}{0.5\textwidth}
		\includegraphics[width=\textwidth]{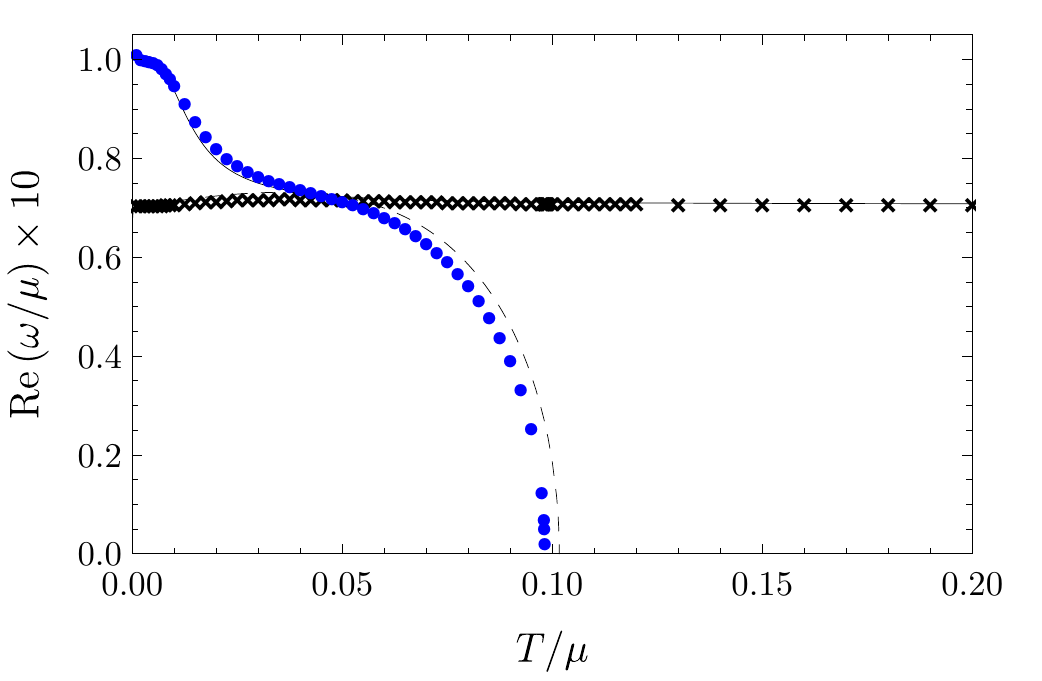}
		\caption{\(k/\m = 0.1\), real part.}
	\end{subfigure}
	\begin{subfigure}{0.5\textwidth}
		\includegraphics[width=\textwidth]{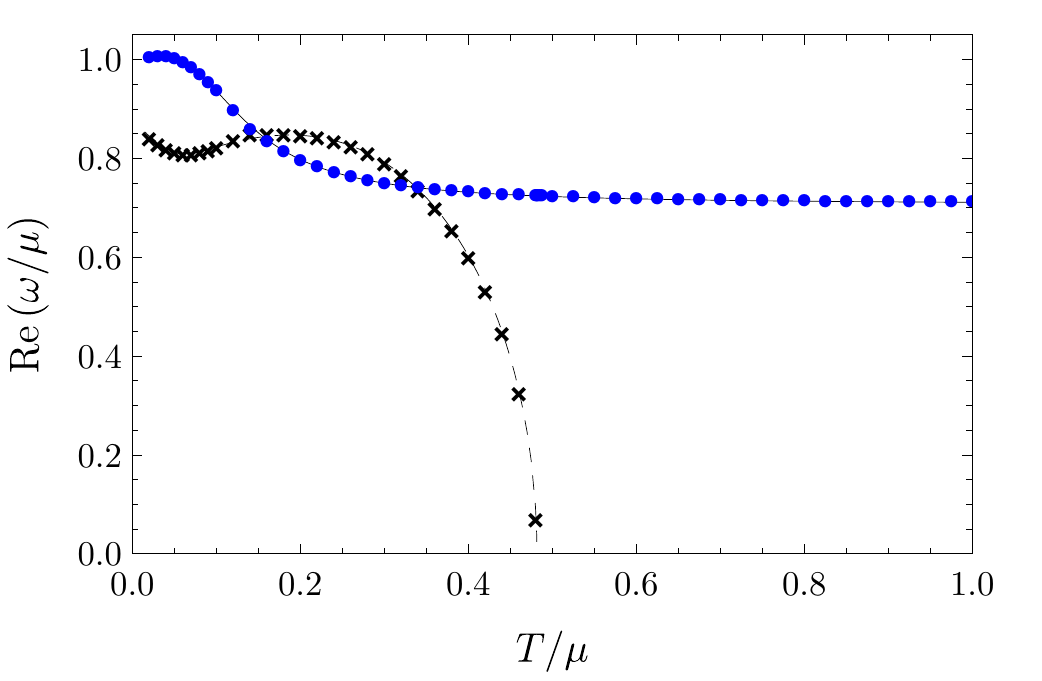}
		\caption{\(k/\m = 1\), real part.}
		\label{fig:tau0p01_a1_k1_real}
	\end{subfigure}
	\begin{subfigure}{0.5\textwidth}
		\includegraphics[width=\textwidth]{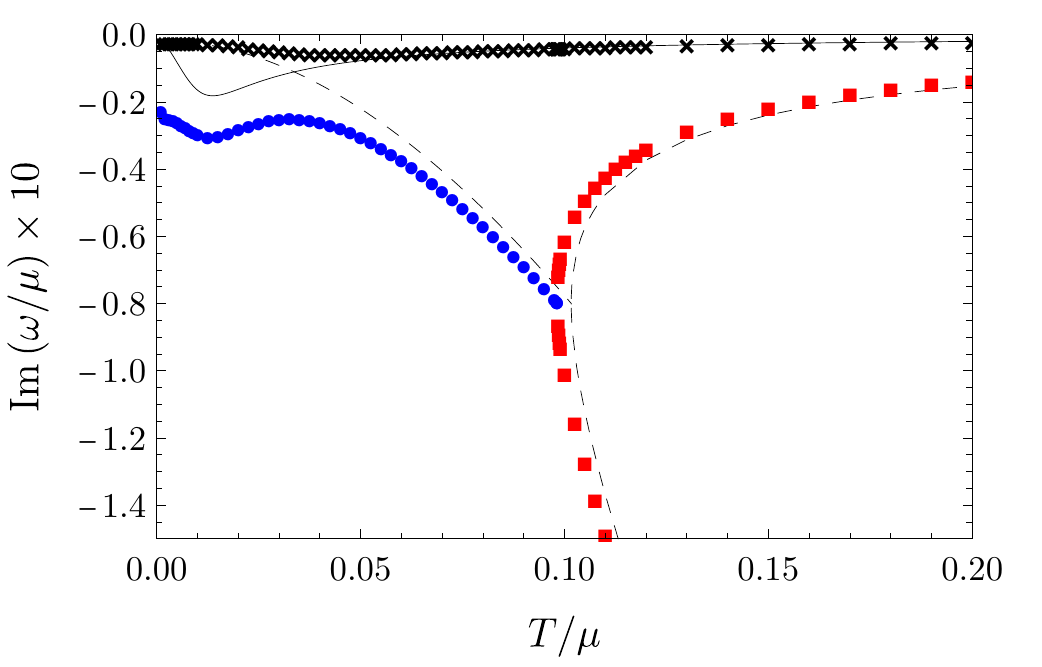}
		\caption{\(k/\m = 0.1\), imaginary part.}
	\end{subfigure}
	\begin{subfigure}{0.5\textwidth}
		\includegraphics[width=\textwidth]{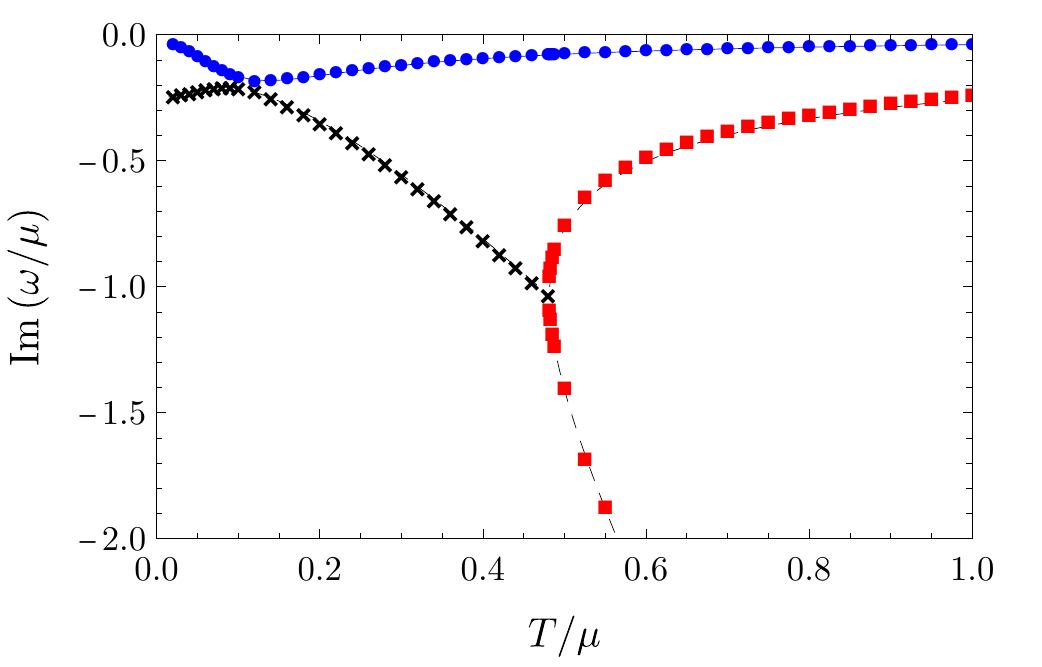}
		\caption{\(k/\m = 1\), imaginary part.}
		\label{fig:tau0p01_a1_k1_imaginary}
	\end{subfigure}
	\caption[Poles in Green's functions of \(J^\m\) and \(T^{\m\n}\), at large back-reaction and large values of momentum.]{
		Frequencies of the poles in \(G_{JJ}\) and \(G_{TT}\), as a function of temperature, for \(\t = 10^{-2}\), \(\talpha=1\), and \(k/\m = 0.1\) (left column) and \(k/\m = 1\) (right column). The first row shows the positions of the poles in the complex frequency plane, with arrows indicating the motion of the poles with increasing temperatures. The second and third rows show the real and imaginary parts of the poles as a function of temperature, respectively. The solid and dashed black lines show the probe limit results for the poles of \(G_{TT}\) and \(G_{JJ}\) respectively, at the same values of \(k/\m\) and \(\talpha\).
	}
	\label{fig:poles_large_tau_large_momentum}
\end{figure}
Figure~\ref{fig:poles_large_tau_large_momentum} shows the frequencies of the poles for \(\t = 10^{-2}\), \(\talpha=1\), and \(k/\m = 0.1\) and \(k/\m = 1\). For both values of \(k/\m\), the behaviour of the poles is qualitatively similar to that for \(\t=10^{-4}\) and the same \(k/\m\). In particular, the purely imaginary pole which becomes hydrodynamic charge diffusion at high temperatures is created by the collision of two propagating poles on the imaginary axis. This is in stark contrast to \(\t=10^{-2}\) and \(k/\m = 10^{-2}\), where no such collision occurs.

For both values of \(\t\), with increasing \(k/\m\) the frequencies of the poles as functions of temperature approach the frequencies found at the same momentum in the probe limit, indicated by the solid and dashed back lines in figures~\ref{fig:poles_small_tau_large_momentum} and~\ref{fig:poles_large_tau_large_momentum}. Evidently, the poles of the Green's functions are less sensitive to the back-reaction when \(k/\m\) is large.

Since the gravity theory is invariant under the combined rescaling $\talpha \to \lambda \talpha$ and $F_{MN} \to \lambda^{-1} F_{MN}$ (and therefore \(\m \to \l^{-1} \m\)), increasing \(k/\m\) for fixed \(\talpha\) is equivalent to decreasing \(\talpha\) for fixed \(k/\m\). Hence, the results of this section also imply that at fixed \(\t\) and \(k/\m\), the poles approach those of the probe limit as \(\talpha\) is decreased. This is intuitive, since for small values of \(\talpha\) we expect to obtain a good approximation by expanding the action~\eqref{eq:hzs_model} up to quadratic order in \(\talpha F\). In this truncated action, the back-reaction parameter \(\t\) appears only in the combination \(\t\talpha^2\). It is therefore unsurprising that one approaches the probe limit as \(\talpha\) is decreased with \(\t\) held fixed.

\subsection{Spectral Functions}
\label{sec:spectral_functions}

In this section we present our numerical results for the charge and energy spectral functions, $\rho_{qq}$ and $\rho_{\ve\ve}$, defined in~\eqref{eq:hzs_spectral_functions}. In both probe brane and Einstein-Maxwell models, at low temperatures \(\r_{qq}\) exhibits a peak at \(\w \approx k/\sqrt{2}\), due to the HZS mode~\cite{Karch:2008fa,Davison:2011ek,Edalati:2010pn,Davison:2011uk}. We find that the same is true in our model, for all levels of back-reaction. At high temperatures, the equations of hydrodynamics imply that \(\r_{qq}\) exhibits a peak near \(\w = 0\). We will examine in detail how \(\r_{qq}\) interpolates between these two regimes.

We will compare our numerical results to an approximation of the Green's functions as a sum over poles,
\begin{equation}
\label{eq:mero}
G_{ab}(\omega,k) \approx \sum_{n}\frac{{\cal R}_{ab}^{(n)}(k)}{\omega-\omega_*^{(n)}(k)},
\end{equation}
where $\omega_*^{(n)}(k)$ are our numerical results for the highest poles, specifically the sound poles and the next highest pole or pair of poles, and ${\cal R}_{ab}^{(n)}(k)$ is a matrix of pole residues, which are generically complex-valued. The method used to compute the spectral functions and the matrix of residues \(\mathcal{R}_{ab}^{(n)}\) is explained in appendix~\ref{app:zero_sound_numerics}.

In principle, the Green's functions need not take such a simple form~\eqref{eq:mero}, and one should use a Mittag-Leffler expansion, for example including terms analytic in frequency~\cite{Solana:2018pbk}. However, a sum over poles provides a good approximation to the spectral functions of extremal \ads[4]-Reissner-Nordstr\"om~\cite{Edalati:2010pn}. We find that the approximation~\eqref{eq:mero} works well for many, but not all, values of temperature, back-reaction, and momentum that we study, indicating that the spectral functions are often dominated by their poles.

\begin{figure}
	\begin{center}
		\(\t = 10^{-5}\), \(\talpha = 1\), \(k/\m = 0.01\)
	\end{center}
	\begin{subfigure}{0.5\textwidth}
		\includegraphics[width=\textwidth]{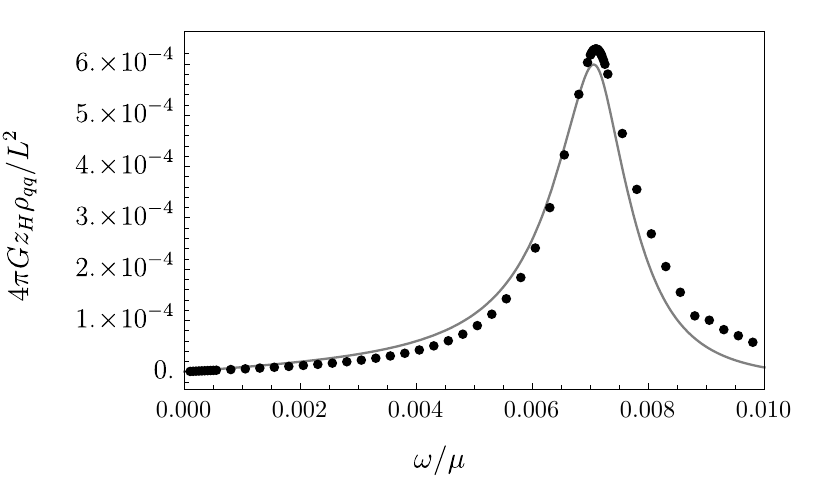}
		\caption{Charge density, \(T/\m = 0.01\).}
	\end{subfigure}
	\begin{subfigure}{0.5\textwidth}
		\includegraphics[width=\textwidth]{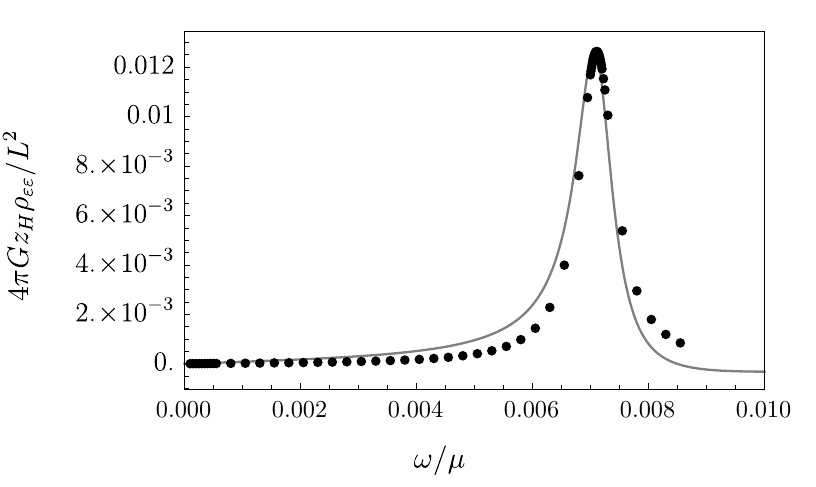}
		\caption{Energy density, \(T/\m = 0.01\).}
	\end{subfigure}
	\begin{subfigure}{0.5\textwidth}
		\includegraphics[width=\textwidth]{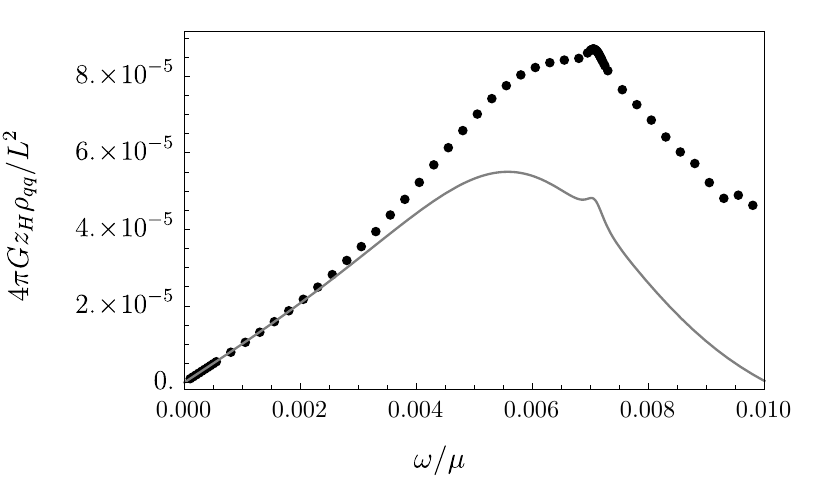}
		\caption{Charge density, \(T/\m = 0.02\).}
	\end{subfigure}
	\begin{subfigure}{0.5\textwidth}
		\includegraphics[width=\textwidth]{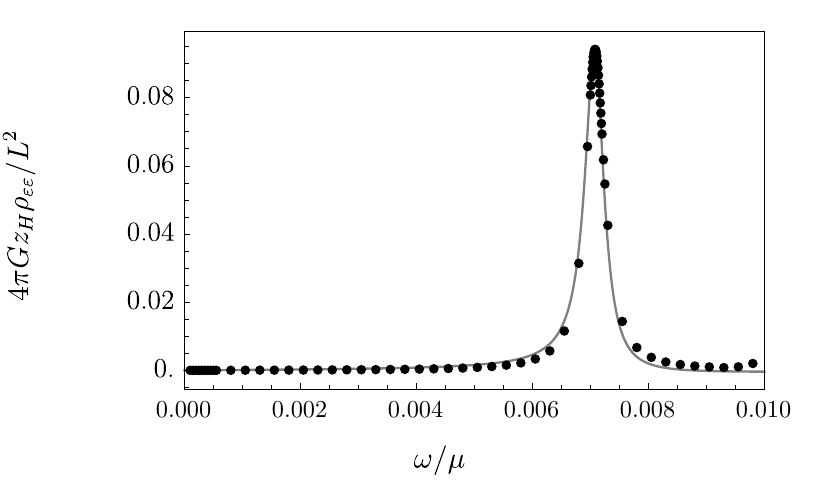}
		\caption{Energy density, \(T/\m = 0.02\).}
	\end{subfigure}
	\begin{subfigure}{0.5\textwidth}
		\includegraphics[width=\textwidth]{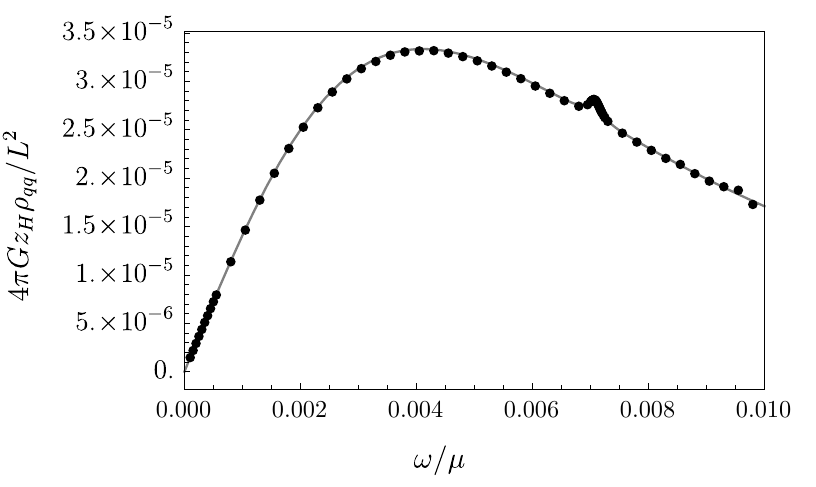}
		\caption{Charge density, \(T/\m = 0.03\).}
	\end{subfigure}
	\begin{subfigure}{0.5\textwidth}
		\includegraphics[width=\textwidth]{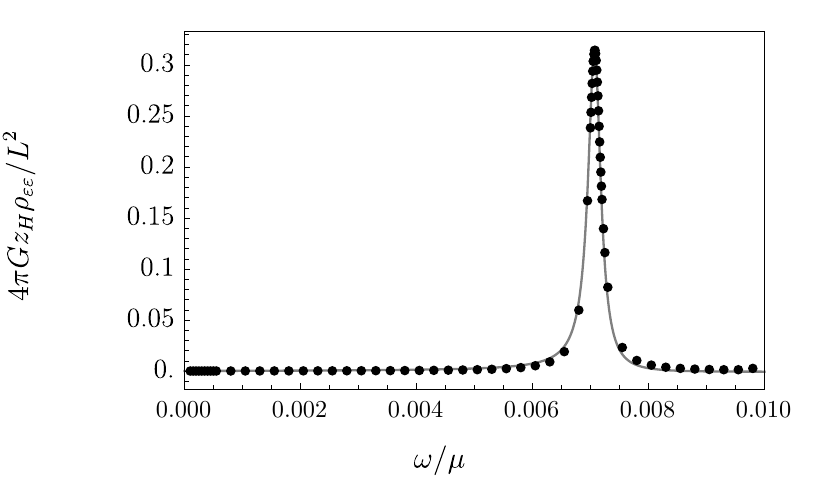}
		\caption{Energy density, \(T/\m = 0.03\).}
	\end{subfigure}
	\caption[Charge and energy density spectral functions for the holographic dual to a spacetime filling brane at very small back-reaction.]{
		Numerical results for the charge density spectral function \(\r_{qq}\) (left column) and energy density spectral function \(\r_{\ve\ve}\) (right column), for \(\t = 10^{-5}\), \(\talpha = 1\), and \(k/\m = 0.01\). The plots show results for \(T/\m = 0.01\) (top row), 0.02 (middle row), and 0.03 (bottom row). The black points are direct numerical results, while the grey curves show the spectral functions obtained using the sum-over-poles approximation~\eqref{eq:mero}. At \(T/\m = 0.01\), both spectral functions exhibit peaks at \(\w \approx k/\sqrt{2}\), due to the sound pole. As the temperature is raised, the energy density spectral function continues to be dominated by the sound pole. The charge density spectral function develops a new, broad peak at smaller values of \(\w/\m\). By \(T/\m = 0.03\) this peak provides the dominant contribution to the spectral function. At large temperatures the new peak becomes the charge diffusion peak.
	}
	\label{tau_0p00001_spectral_functions}
\end{figure}
Figure~\ref{tau_0p00001_spectral_functions} shows our numerical results for $\rho_{qq}$ and $\rho_{\ve\ve}$, for $\tau=10^{-5}$, $\talpha=1$, $k/\mu=0.01$ and $T/\mu=0.01$, $0.02$, and $0.03$. In the figure, the black dots are the result of direct numerical calculation of \(\r_{qq}\) and \(\r_{\ve\ve}\), while the solid grey curves are the spectral functions obtained from the sum-over-poles approximation to the Green's functions in equation~\eqref{eq:mero}.

In both $\rho_{qq}$ and $\r_{\ve\ve}$, at $T/\mu = 0.01$ we find a peak from the sound pole at $\omega \approx k/\sqrt{2}$.\footnote{The positions of the poles for \(\t = 10^{-5}\), \(\talpha=1\), and \(k/\m = 0.01\) are qualitatively similar to the results for \(\t=10^{-4}\) plotted in figure~\ref{planesmalltau}.} As $T/\mu$ increases through the values shown, in $\rho_{qq}$ the sound peak's height decreases, while in $\r_{\ve\ve}$ the height increases, indicating that as $T/\mu$ increases, the sound pole's residue decreases in $G_{J^t J^t}$ but increases in $G_{T^{tt}T^{tt}}$. In addition, as \(T/\m\) increases, the charge density spectral function develops a second, broader peak, due to an additional pole with non-zero real part. As the temperature is increased, this peak becomes taller, eventually providing the dominant contribution to \(\r_{qq}\). The frequency of the peak decreases with increasing \(T/\m\). In terms of the poles in the Green's functions, this transfer of dominance occurs because of the relative sizes of the residues of the poles. The second peak also moves to smaller frequencies with increasing temperature, eventually becoming the charge diffusion peak.

This changeover in poles dominating the charge density spectral function also occurs in \ads-Reissner-Nordstr\"om~\cite{Davison:2011uk}, and contrasts with the probe limit, in which the charge density spectral function only ever exhibits a single peak~\cite{Davison:2011ek}. In principle, one could define the crossover to hydrodynamic behaviour as occurring at the value of \(T/\m\) for which the two peaks in \(\r_{qq}\) have equal height, as in ref.~\cite{Davison:2011uk}. In practice, however, one of the peaks is so broad that it is difficult to identify precisely when this occurs. The definition of the crossover using the spectral functions is therefore unreliable at small back-reaction.

The pole which produces the second peak in \(\r_{qq}\) does not provide a significant contribution to the energy density spectral function, which remains dominated by the sound pole as the temperature is increased. This is because the residue of the sound pole in \(G_{T^{tt} T^{tt}}\) is always much larger than the residue of the second pole.

The sum-over-poles approximation works well for most values of \(T/\m\). However, the approximation  seems to break down for \(\r_{qq}\) around \(T/\m \approx 0.02\). We have not found any other poles which provide a significant contribution to the spectral functions for temperatures in this range, so the failure of the approximation is not due to the truncation to a small number of poles. It appears that for \(T/\m \approx 0.02\) there are significant contributions to \(\r_{qq}\) that are analytic in frequency.

\begin{figure}
	\begin{center}
		\(\t = 10^{-4}\), \(\talpha = 1\), \(k/\m = 0.01\)
	\end{center}
	\begin{subfigure}{0.5\textwidth}
		\includegraphics[width=\textwidth]{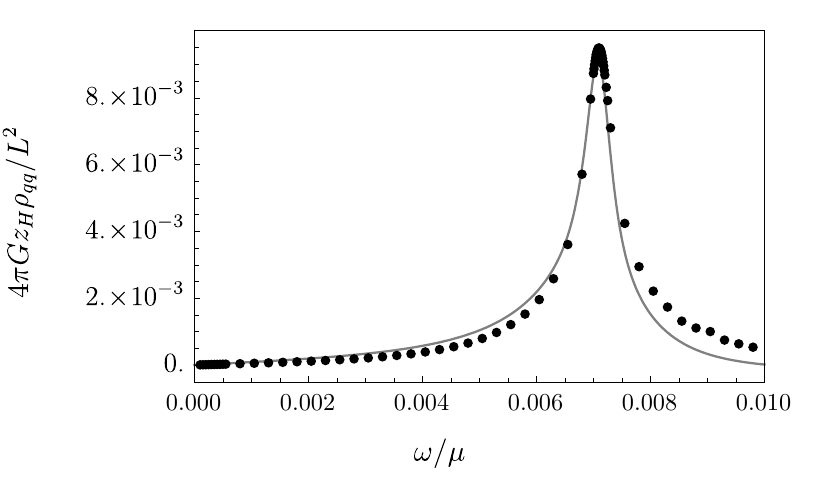}
		\caption{Charge density, \(T/\m = 0.01\).}
	\end{subfigure}
	\begin{subfigure}{0.5\textwidth}
		\includegraphics[width=\textwidth]{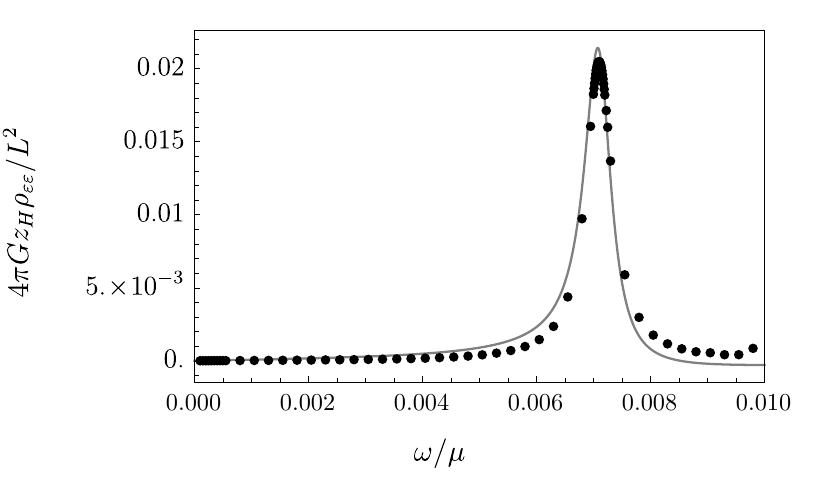}
		\caption{Energy density, \(T/\m = 0.01\).}
	\end{subfigure}
	\begin{subfigure}{0.5\textwidth}
		\includegraphics[width=\textwidth]{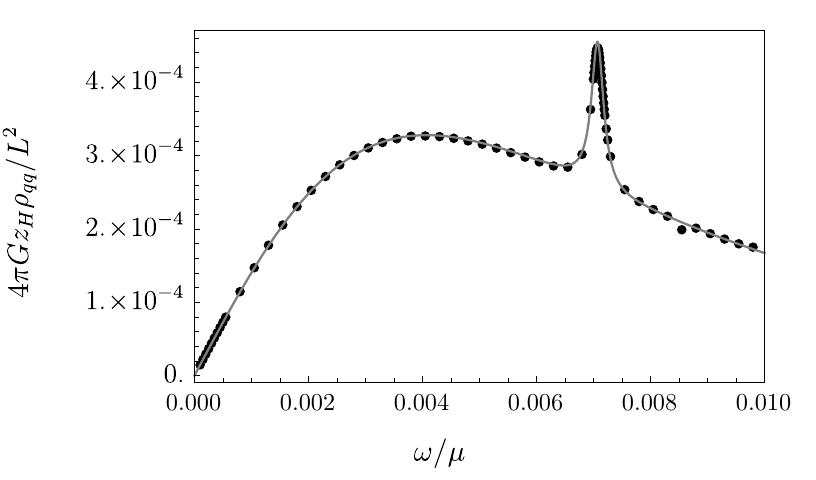}
		\caption{Charge density, \(T/\m = 0.03\).}
	\end{subfigure}
	\begin{subfigure}{0.5\textwidth}
		\includegraphics[width=\textwidth]{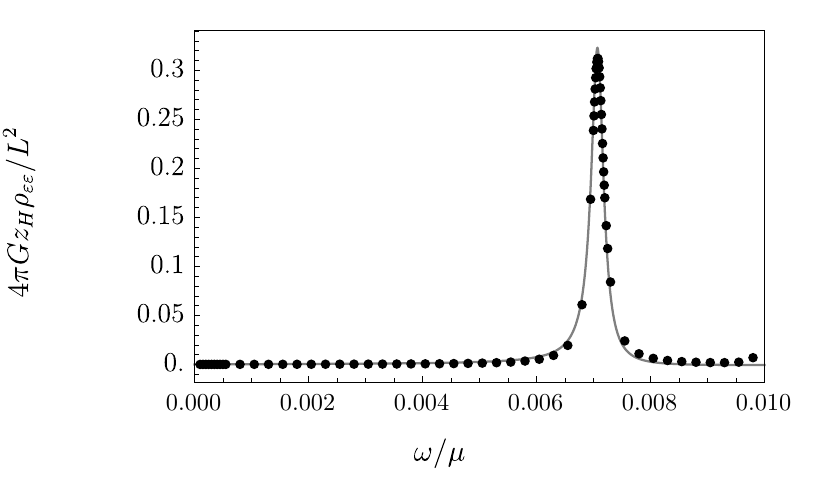}
		\caption{Energy density, \(T/\m = 0.03\).}
	\end{subfigure}
	\begin{subfigure}{0.5\textwidth}
		\includegraphics[width=\textwidth]{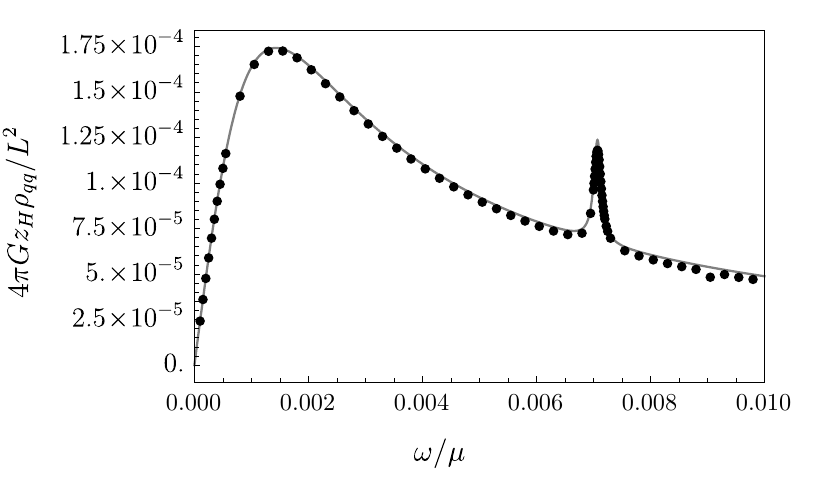}
		\caption{Charge density, \(T/\m = 0.05\).}
	\end{subfigure}
	\begin{subfigure}{0.5\textwidth}
		\includegraphics[width=\textwidth]{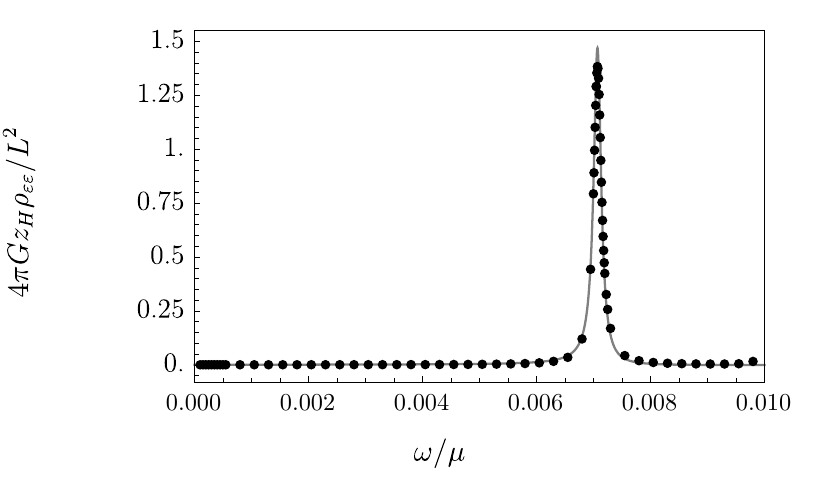}
		\caption{Energy density, \(T/\m = 0.05\).}
	\end{subfigure}
	\caption[The spectral functions at small back-reaction.]{
		Numerical results for the charge density spectral function \(\r_{qq}\) (left column) and energy density spectral function \(\r_{\ve\ve}\) (right column), for \(\t = 10^{-4}\), \(\talpha = 1\), and \(k/\m = 0.01\). The plots show results for \(T/\m = 0.01\) (top row), 0.03 (middle row), and 0.05 (bottom row). The black points are direct numerical results, while the grey curves show the spectral functions obtained using the sum-over-poles approximation~\eqref{eq:mero}. At \(T/\m = 0.01\), both spectral functions exhibit peaks at \(\w \approx k/\sqrt{2}\), due to the sound pole. As the temperature is raised, the energy density spectral function continues to be dominated by the sound pole. The charge density spectral function develops a new, broad peak at smaller values of \(\w/\m\).  The two peaks have equal height for \(T/\m \approx 0.039\). For larger values of \(T/\m\), the peak at smaller \(\w/\m\) provides the dominant contribution to the spectral function. At large temperatures it becomes the charge diffusion peak.
	}
	\label{tau_0p0001_spectral_functions}
\end{figure}

Figure~\ref{tau_0p0001_spectral_functions} shows our numerical results for the spectral functions for $\tau=10^{-4}$, $\talpha=1$, $k/\mu=0.01$, and $T/\mu=0.01$, $0.03$, and $0.05$. The results are qualitatively similar to the \(\t=10^{-5}\) results plotted in figure~\ref{tau_0p00001_spectral_functions}. In particular, the charge density spectral function exhibits a sound peak at \(T/\m = 0.01\), and develops a second peak as the temperature is raised, which eventually dominates at sufficiently large temperature. The separation between these two peaks is greater than for \(\t=10^{-5}\), allowing us to identify the temperature at which they are of equal height as \(T/\mu \approx 0.039\). This is larger than crossover temperature obtained using the probe limit definition --- the temperature of pole collision --- which was \(T/\m \approx 0.029\).

\begin{figure}
	\begin{center}
		\(\t = 10^{-3}\), \(\talpha = 1\), \(k/\m = 0.01\)
	\end{center}
	\begin{subfigure}{0.5\textwidth}
		\includegraphics[width=\textwidth]{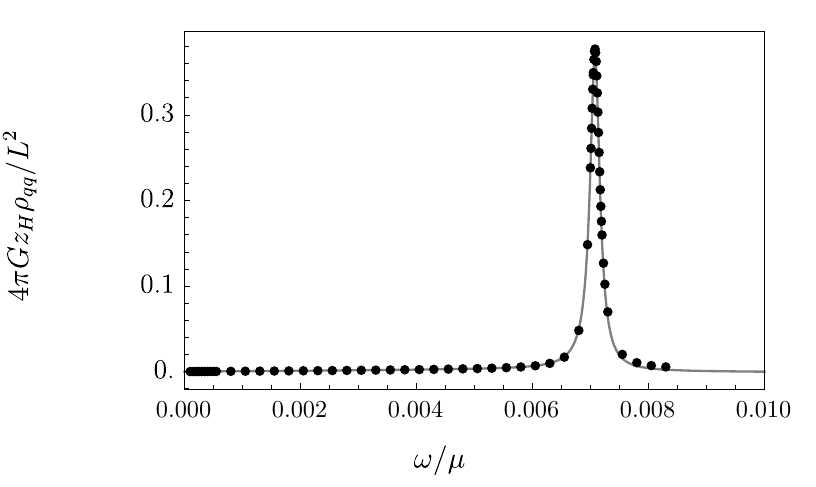}
		\caption{Charge density, \(T/\m = 0.01\).}
	\end{subfigure}
	\begin{subfigure}{0.5\textwidth}
		\includegraphics[width=\textwidth]{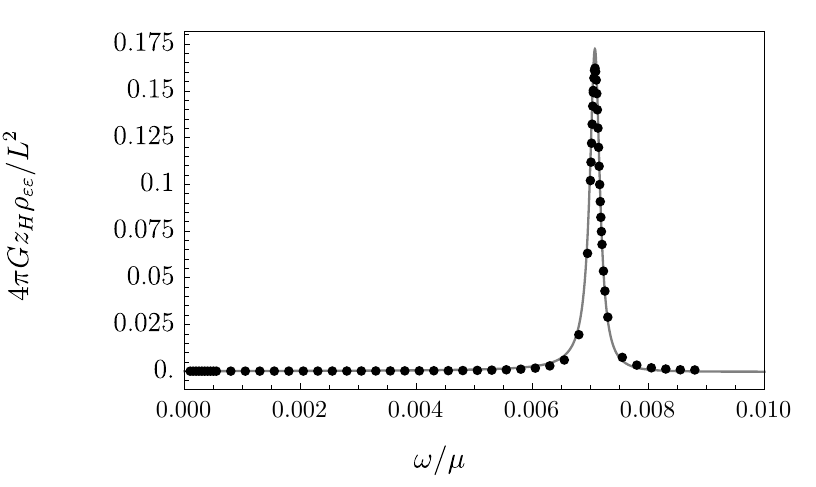}
		\caption{Energy density, \(T/\m = 0.01\).}
	\end{subfigure}
	\begin{subfigure}{0.5\textwidth}
		\includegraphics[width=\textwidth]{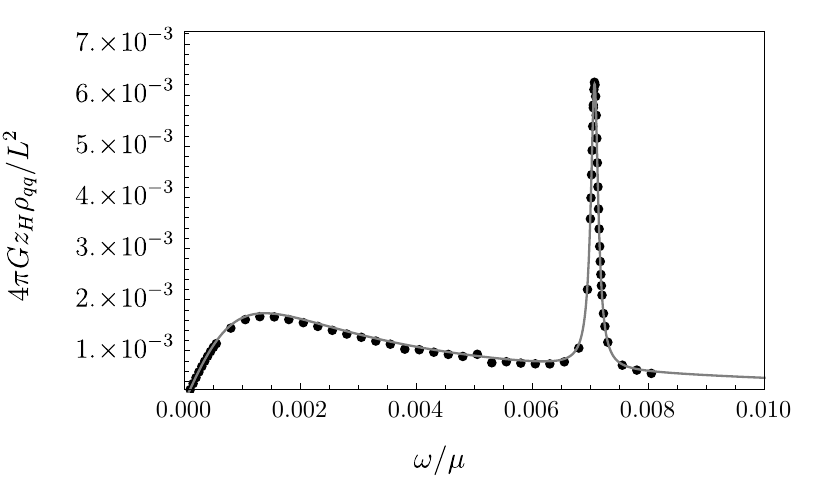}
		\caption{Charge density, \(T/\m = 0.05\).}
	\end{subfigure}
	\begin{subfigure}{0.5\textwidth}
		\includegraphics[width=\textwidth]{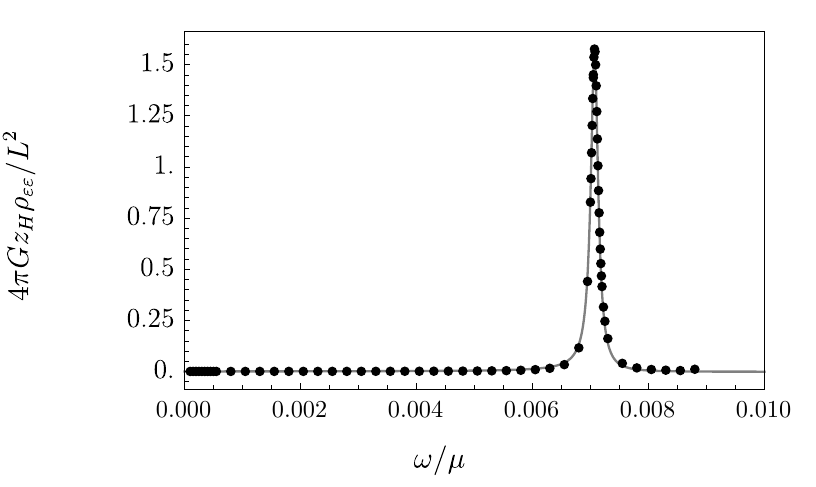}
		\caption{Energy density, \(T/\m = 0.05\).}
	\end{subfigure}
	\begin{subfigure}{0.5\textwidth}
		\includegraphics[width=\textwidth]{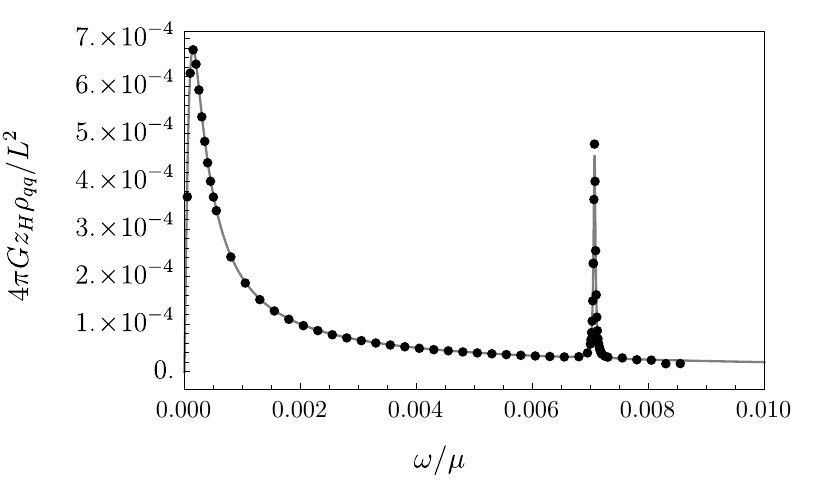}
		\caption{Charge density, \(T/\m = 0.2\).}
	\end{subfigure}
	\begin{subfigure}{0.5\textwidth}
		\includegraphics[width=\textwidth]{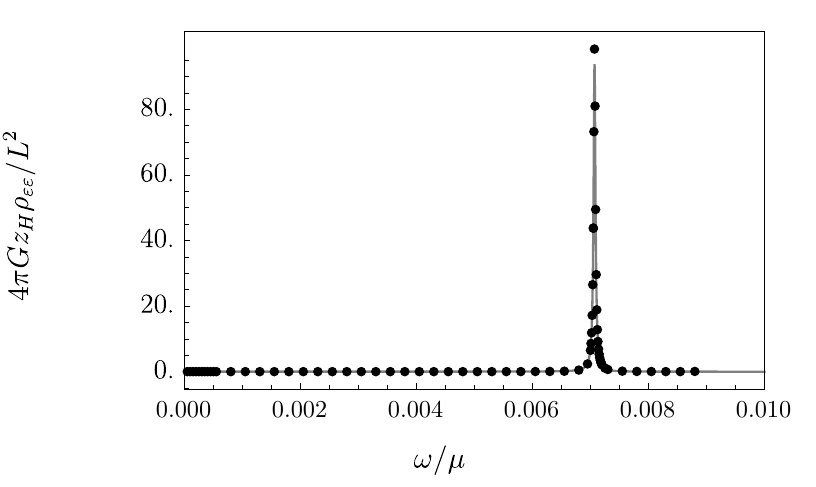}
		\caption{Energy density, \(T/\m = 0.2\).}
	\end{subfigure}
	\caption[The spectral functions at intermediate back-reaction.]{
		Numerical results for the charge density spectral function \(\r_{qq}\) (left column) and energy density spectral function \(\r_{\ve\ve}\) (right column), for \(\t = 10^{-3}\), \(\talpha = 1\), and \(k/\m = 0.01\). The plots show results for \(T/\m = 0.01\) (top row), \(0.05\) (middle row), and \(0.2\) (bottom row). The black points are direct numerical results, while the grey curves show the spectral functions obtained using the sum-over-poles approximation~\eqref{eq:mero}. At \(T/\m = 0.01\), both spectral functions exhibit peaks at \(\w \approx k/\sqrt{2}\), due to the sound pole. As the temperature is raised, the energy density spectral function continues to be dominated by the sound pole. The charge density spectral function develops a new, broad peak at smaller values of \(\w/\m\).  The two peaks have equal height for \(T/\m \approx 0.136\). For larger values of \(T/\m\), the peak at smaller \(\w/\m\) provides the dominant contribution to the spectral function. At large temperatures it becomes the charge diffusion peak.
	}
	\label{tau_0p001_spectral_functions}
\end{figure}

\begin{figure}
	\begin{center}
		\(\t = 10^{-2}\), \(\talpha = 1\), \(k/\m = 0.01\)
	\end{center}
	\begin{subfigure}{0.5\textwidth}
		\includegraphics[width=\textwidth]{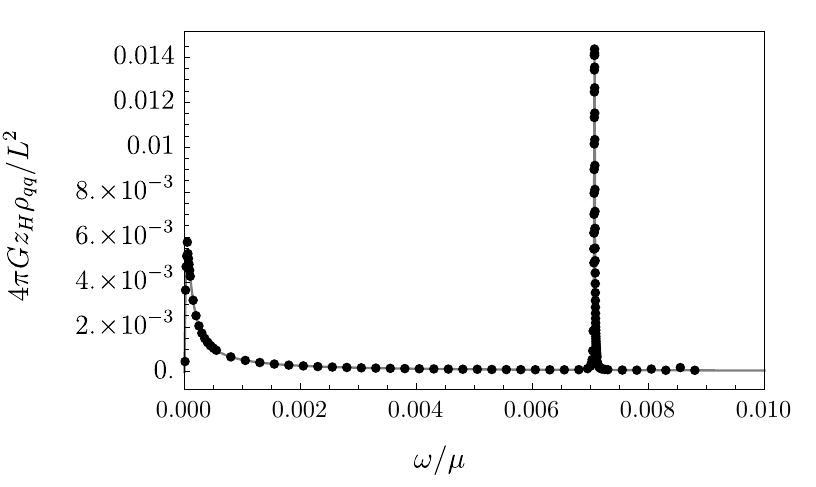}
		\caption{Charge density, \(T/\m = 0.5\).}
	\end{subfigure}
	\begin{subfigure}{0.5\textwidth}
		\includegraphics[width=\textwidth]{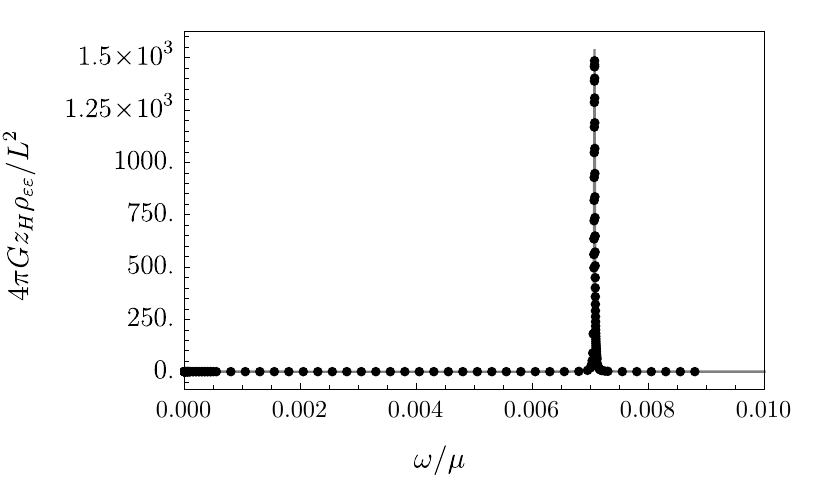}
		\caption{Energy density, \(T/\m = 0.5\).}
	\end{subfigure}
	\begin{subfigure}{0.5\textwidth}
		\includegraphics[width=\textwidth]{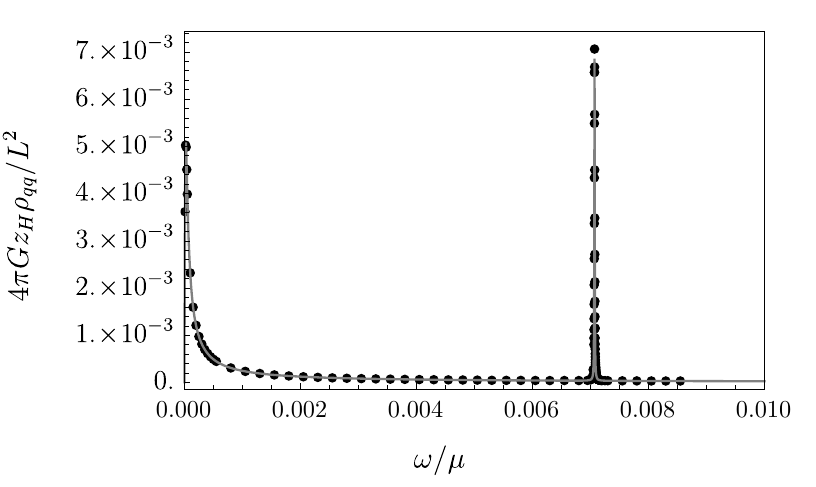}
		\caption{Charge density, \(T/\m = 1\).}
	\end{subfigure}
	\begin{subfigure}{0.5\textwidth}
		\includegraphics[width=\textwidth]{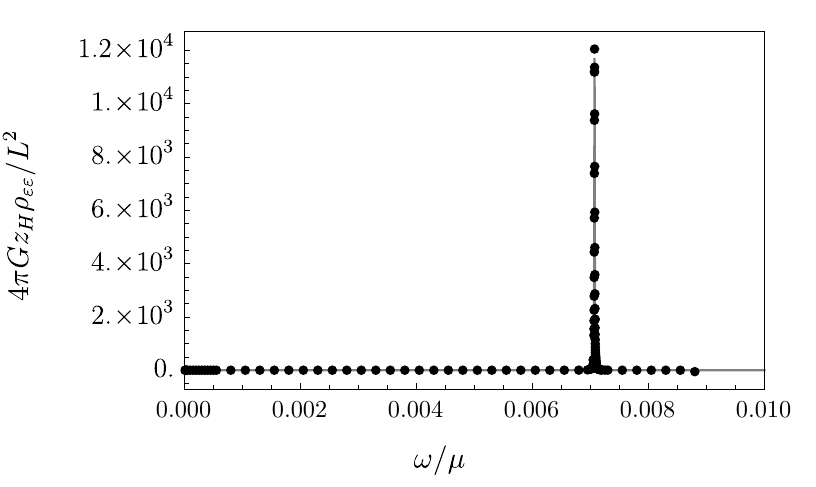}
		\caption{Energy density, \(T/\m = 1\).}
	\end{subfigure}
	\begin{subfigure}{0.5\textwidth}
		\includegraphics[width=\textwidth]{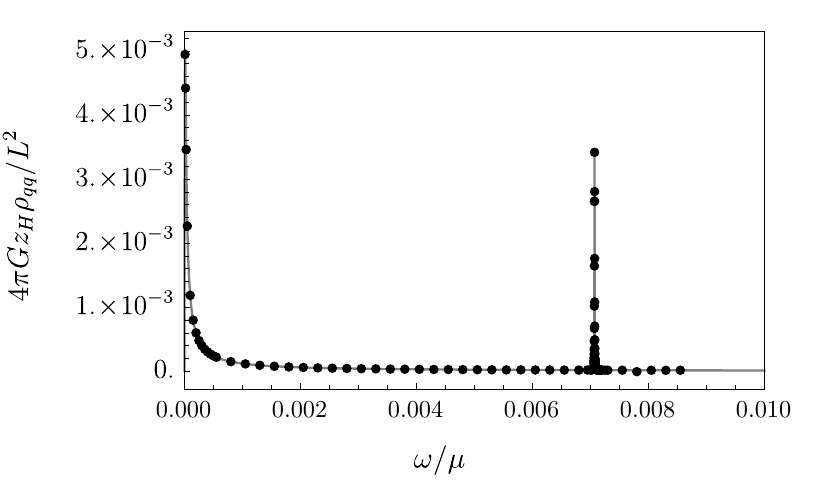}
		\caption{Charge density, \(T/\m = 2\).}
	\end{subfigure}
	\begin{subfigure}{0.5\textwidth}
		\includegraphics[width=\textwidth]{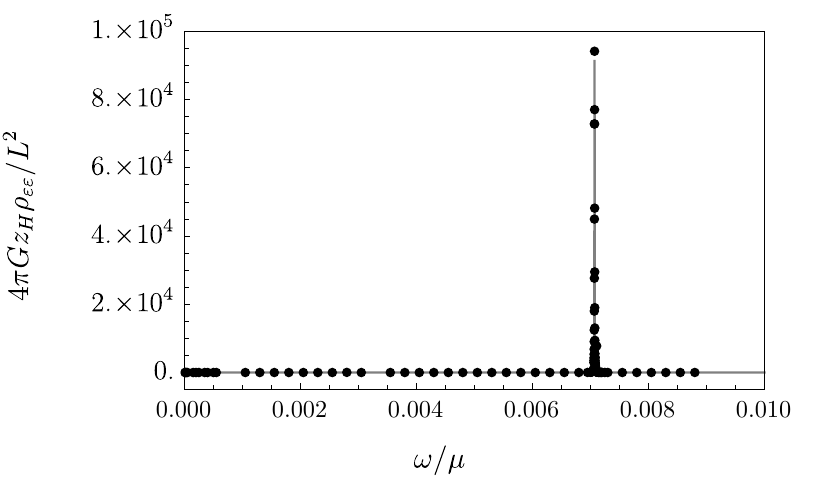}
		\caption{Energy density, \(T/\m = 2\).}
	\end{subfigure}
	\caption[The spectral functions at large back-reaction.]{
		Numerical results for the charge density spectral function \(\r_{qq}\) (left column) and energy density spectral function \(\r_{\ve\ve}\) (right column), for \(\t = 10^{-2}\), and \(\talpha = 1\) and \(k/\m = 0.01\). The plots show results for \(T/\m = 0.5\) (top row), \(1\) (middle row), and \(2\) (bottom row). The black points are direct numerical results, while the grey curves show the spectral functions obtained using the sum-over-poles approximation~\eqref{eq:mero}. At \(T/\m = 0.5\), both spectral functions exhibit peaks at \(\w \approx k/\sqrt{2}\), due to the sound pole. As the temperature is raised, the energy density spectral function continues to be dominated by the sound pole. The charge density spectral function develops a new peak at smaller values of \(\w/\m\).  The two peaks have equal height for \(T/\m \approx 1.45\). For larger values of \(T/\m\), the peak at smaller \(\w/\m\) provides the dominant contribution to the spectral function. At large temperatures it becomes the charge diffusion peak.
	}
	\label{tau_0p01_spectral_functions}
\end{figure}
Figures~\ref{tau_0p001_spectral_functions} and~\ref{tau_0p01_spectral_functions} show the spectral functions for $\tau=10^{-3}$ and \(\t = 10^{-2}\) respectively, with $\talpha=1$ and $k/\mu=0.01$. The results are qualitatively similar to those for \(\t = 10^{-5}\) and \(\t=10^{-4}\). In particular, the charge density spectral function exhibits two distinct peaks for sufficiently large \(T/\m\), while the energy density spectral function exhibits only a single peak. The main qualitative change is that the two peaks in \(\r_{qq}\) become narrower with increasing \(\t\). We find that the peaks have equal height at \(T/\m \approx 0.136\) for \(\t=10^{-3}\), and \(T/\m \approx 1.45\) for \(\t = 10^{-2}\). In contrast, defining the crossover by the collision of poles that produces the charge diffusion pole gave $T/\mu\approx0.027$ for \(\t = 10^{-3}\), while for \(\t = 10^{-2}\) no pole collision occurs.  For both values of \(\t\), the spectral functions are insensitive to the details of the motion of the poles in the complex plane plotted in figures~\ref{plane_tau_0p001} and~\ref{plane_tau_0p01}.

\clearpage

In summary, for \(k/\m = 10^{-2}\), \(\talpha=1\), and all non-zero \(\t\) that we study, the charge density spectral function \(\r_{qq}\) exhibits temperature-dependence similar to in \ads[4]-Reissner-Nordstr\"om; it is dominated by a sound peak at low temperature, and a diffusion peak at high temperature. We may therefore take the crossover to hydrodynamics as occurring when the two peaks have equal height. However, for small \(\t\) we find that the diffusion peak is very broad at low temperatures, which can make precise determination of the crossover temperature difficult.

\begin{figure} 
	\begin{center}
		\(\t = 10^{-4}\), \(\talpha = 1\), \(k/\m = 0.1\)
	\end{center}
	\begin{subfigure}{0.5\textwidth}
		\includegraphics[width=\textwidth]{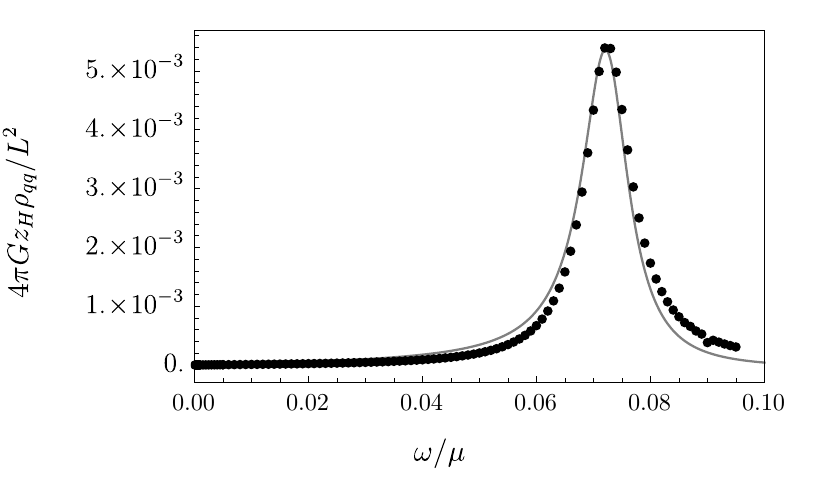}
		\caption{Charge density, \(T/\m = 0.02\).}
	\end{subfigure}
	\begin{subfigure}{0.5\textwidth}
		\includegraphics[width=\textwidth]{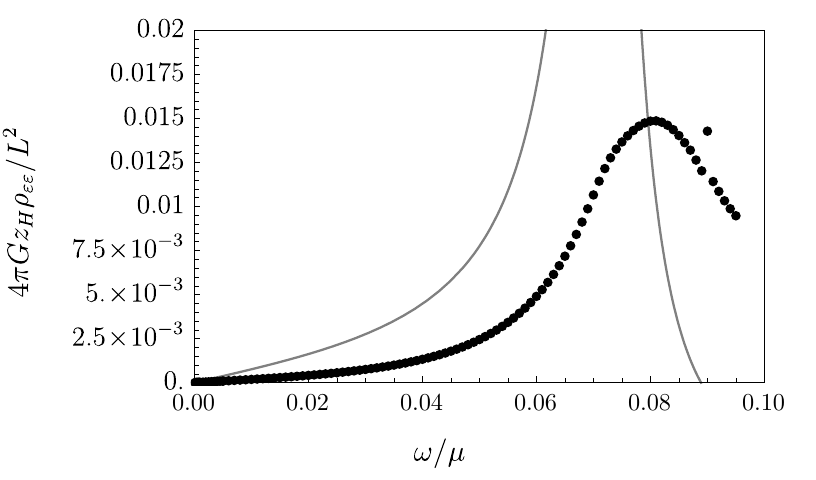}
		\caption{Energy density, \(T/\m = 0.02\).}
		\label{fig:energy_density_tau10m4_q0p1}
	\end{subfigure}
	\begin{subfigure}{0.5\textwidth}
		\includegraphics[width=\textwidth]{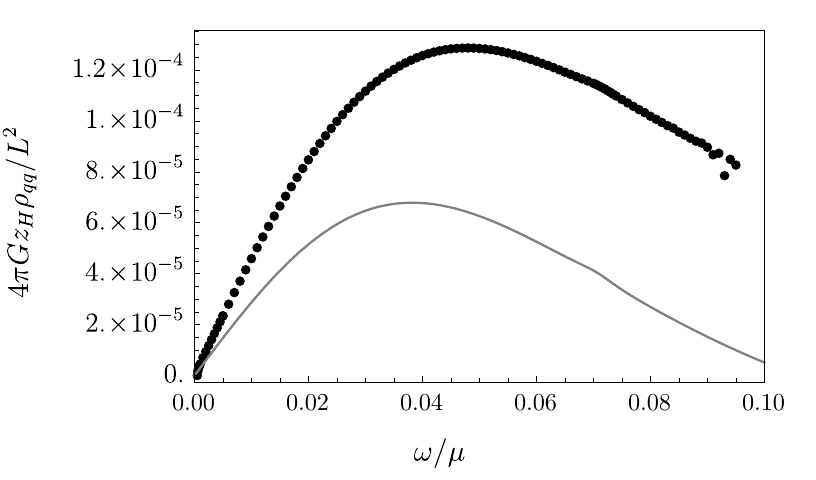}
		\caption{Charge density, \(T/\m = 0.1\).}
	\end{subfigure}
	\begin{subfigure}{0.5\textwidth}
		\includegraphics[width=\textwidth]{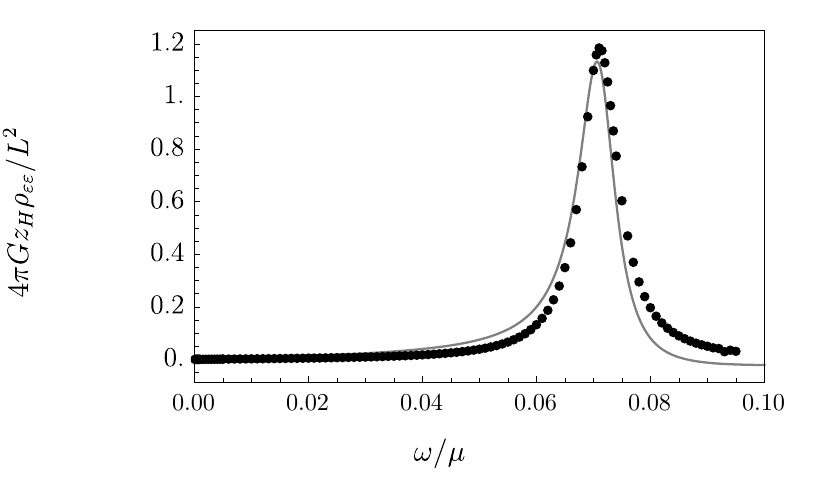}
		\caption{Energy density, \(T/\m = 0.1\).}
	\end{subfigure}
	\begin{subfigure}{0.5\textwidth}
		\includegraphics[width=\textwidth]{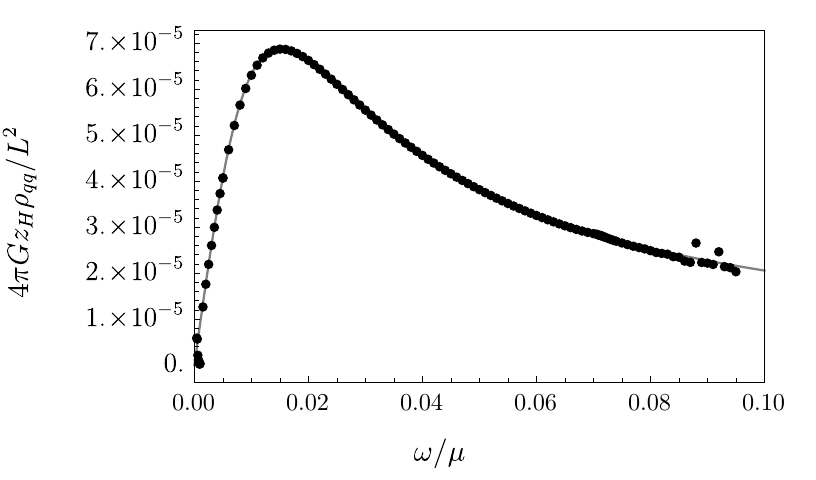}
		\caption{Charge density, \(T/\m = 0.2\).}
	\end{subfigure}
	\begin{subfigure}{0.5\textwidth}
		\includegraphics[width=\textwidth]{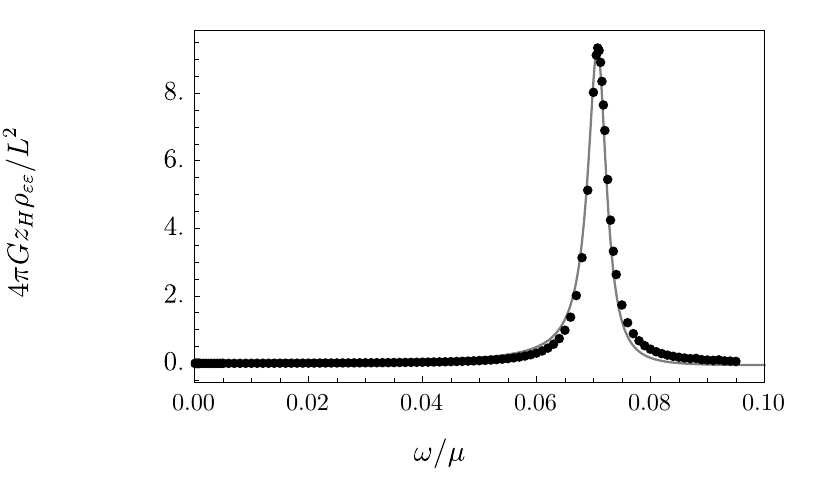}
		\caption{Energy density, \(T/\m = 0.2\).}
	\end{subfigure}
	\caption[The spectral functions at small back-reaction, for larger momentum.]{
		Numerical results for the charge density spectral function \(\r_{qq}\) (left column) and energy density spectral function \(\r_{\ve\ve}\) (right column), for \(\t = 10^{-4}\), and \(\talpha = 1\) and \(k/\m = 0.1\). The plots show results for \(T/\m = 0.02\) (top row), \(0.1\) (middle row), and \(0.2\) (bottom row). The black points are direct numerical results, while the grey curves show the spectral functions obtained using the sum-over-poles approximation~\eqref{eq:mero}. At \(T/\m = 0.02\), the charge density spectral function is dominated by the HZS pole (the black crosses in figure~\ref{fig:complex_plane_tau10m4_q0p1}), with a peak at \(\w/\m \approx k/\sqrt{2}\). As the temperature is raised, the peak moves to smaller frequency, eventually becoming charge diffusion. Unlike previous examples, we do not observe multiple peaks in \(\r_{qq}\) at any temperature, so the \ads-Reissner-Nordstr\"om definition of the crossover may not be used. In this case, the qualitative behaviour of the spectral functions is similar to that in the probe limit~\cite{Davison:2011ek}.
	}
	\label{tau_0p0001_spectral_functions_large_momentum}
\end{figure}
We now give a couple of examples of the behaviour of spectral functions at larger momentum. Figure~\ref{tau_0p0001_spectral_functions_large_momentum} shows the spectral functions for \(\t = 10^{-4}\), \(\talpha = 1\), and \(k/\m = 0.1\), for a range of values of \(T/\m\). In section~\ref{sec:poles} we saw that fixing \(\t\) and increasing \(k/\m\) makes the poles of the Green's functions behave more like in the probe limit. In figure~\ref{tau_0p0001_spectral_functions_large_momentum} we see that the same is true for the spectral functions. Concretely, the charge density spectral function only ever exhibits a single peak over the range of frequencies we plot. At the lowest temperature plotted, \(T/\m = 0.02\), this peak is at \(\w/\m \approx k/\sqrt{2}\). It arises due to the sound pole. As the temperature is increased, the peak moves to smaller values of the frequency, becoming the charge diffusion pole at high temperature.

\begin{figure}
	\begin{center}
		\(\t = 10^{-2}\), \(\talpha = 1\), \(k/\m = 0.1\)
	\end{center}
	\begin{subfigure}{0.5\textwidth}
		\includegraphics[width=\textwidth]{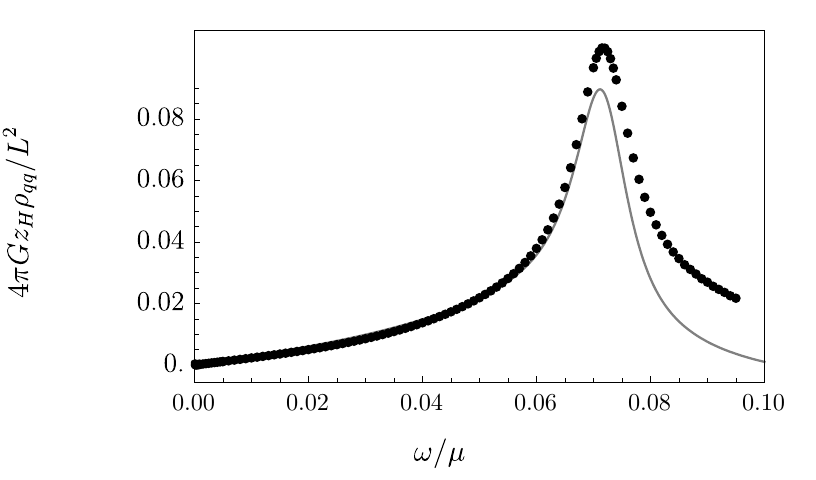}
		\caption{Charge density, \(T/\m = 0.05\).}
	\end{subfigure}
	\begin{subfigure}{0.5\textwidth}
		\includegraphics[width=\textwidth]{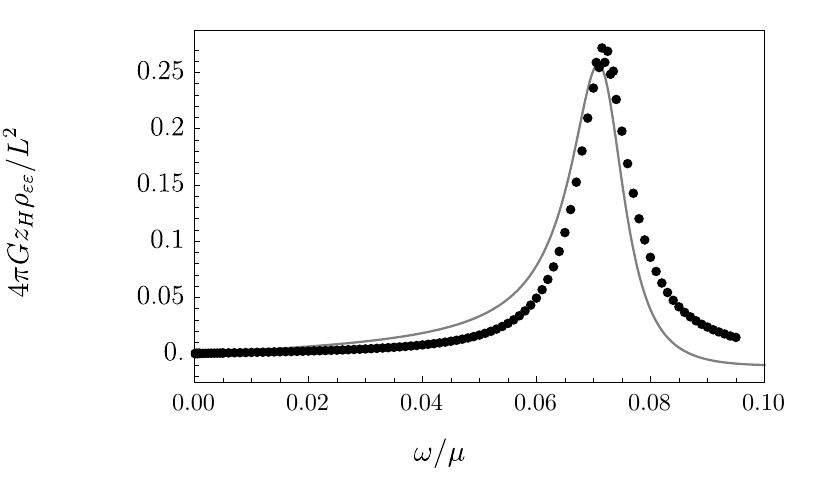}
		\caption{Energy density, \(T/\m = 0.05\).}
	\end{subfigure}
	\begin{subfigure}{0.5\textwidth}
		\includegraphics[width=\textwidth]{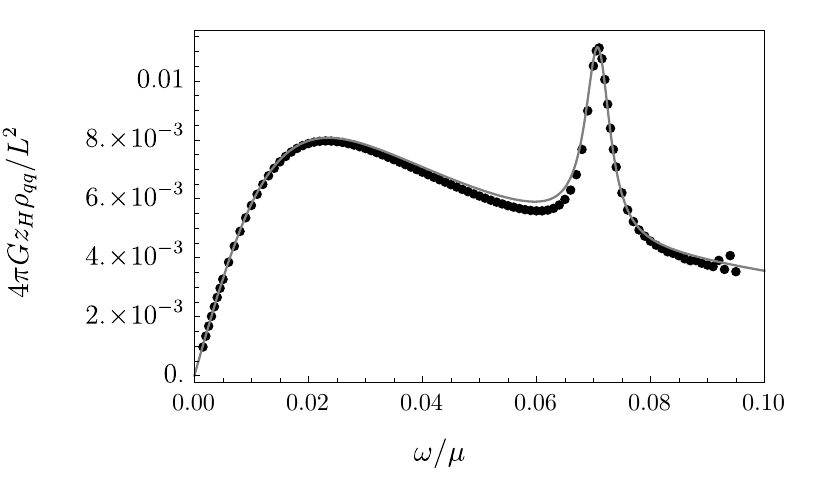}
		\caption{Charge density, \(T/\m = 0.15\).}
	\end{subfigure}
	\begin{subfigure}{0.5\textwidth}
		\includegraphics[width=\textwidth]{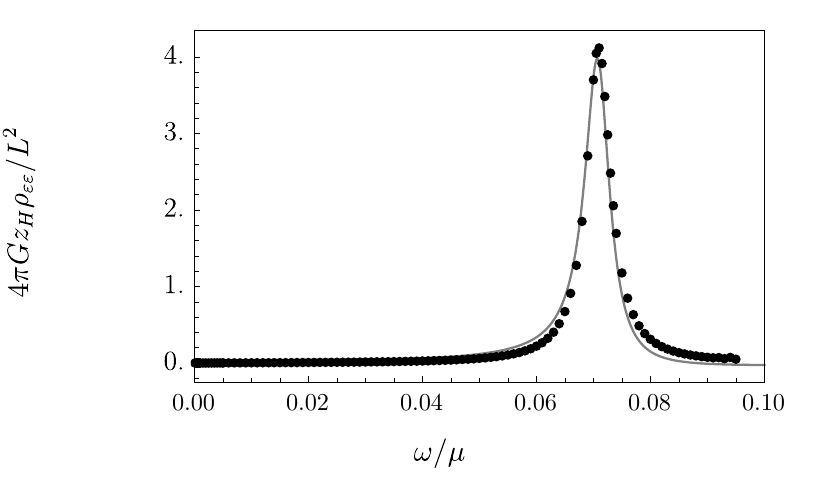}
		\caption{Energy density, \(T/\m = 0.15\).}
	\end{subfigure}
	\begin{subfigure}{0.5\textwidth}
		\includegraphics[width=\textwidth]{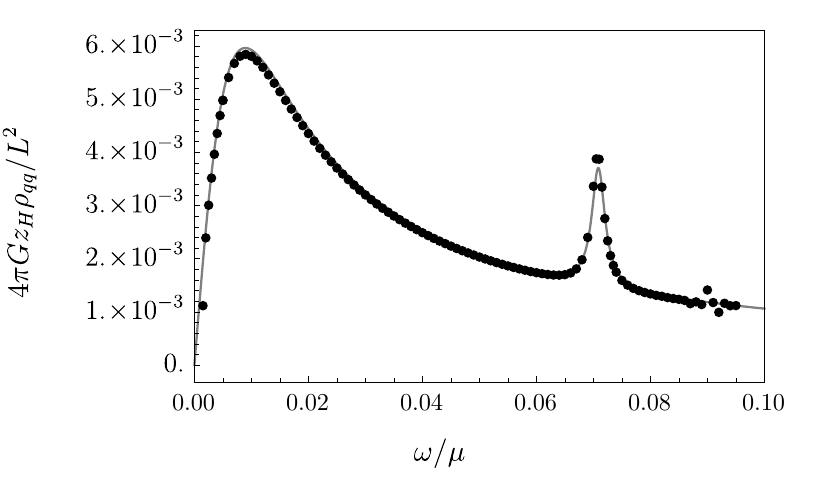}
		\caption{Charge density, \(T/\m = 0.3\).}
	\end{subfigure}
	\begin{subfigure}{0.5\textwidth}
		\includegraphics[width=\textwidth]{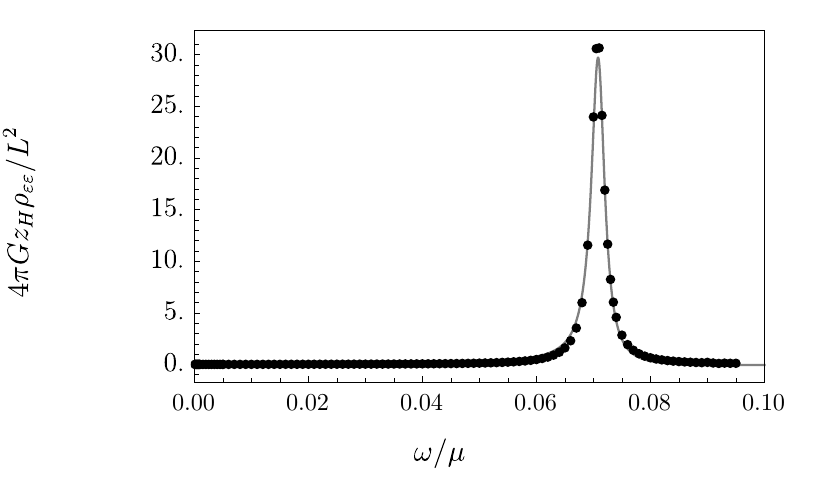}
		\caption{Energy density, \(T/\m = 0.3\).}
	\end{subfigure}
	\caption[The spectral functions at large back-reaction, for larger momentum.]{
		Numerical results for the charge density spectral function \(\r_{qq}\) (left column) and energy density spectral function \(\r_{\ve\ve}\) (right column), for \(\t = 10^{-2}\), and \(\talpha = 1\) and \(k/\m = 0.1\). The plots show results for \(T/\m = 0.05\) (top row), \(0.15\) (middle row), and \(0.3\) (bottom row). The black points are direct numerical results, while the grey curves show the spectral functions obtained using the sum-over-poles approximation~\eqref{eq:mero}. This case is qualitatively similar to the results for \(k/\m = 0.01\), with \(\r_{qq}\) exhibiting two peaks over a range of temperatures. The two peaks have equal height at \(T/\m \approx 0.21\).
	}
	\label{tau_0p01_spectral_functions_large_momentum}
\end{figure}
Figure~\ref{tau_0p01_spectral_functions_large_momentum} shows our results for the spectral functions for \(\t = 10^{-2}\), \(\talpha = 1\), and \(k/\m = 0.1\). This case qualitatively resembles the results for \(k/\m = 10^{-2}\), with the charge density spectral function exhibiting two peaks over a range of temperatures. At low temperature, \(\r_{qq}\) is dominated by a peak at \(\w/\m \approx k/\sqrt{2}\), while at high temperature it is dominated by a peak at small \(\w/\m\). The two peaks have equal height at \(T/\m \approx 0.21\).

Our results show that the definition of the crossover in terms of the spectral functions is viable only for sufficiently large \(\t\) and sufficiently small \(k/\m\). If either the back-reaction is small or the momentum is large, the charge density spectral function behaves as in the probe limit, exhibiting only a single peak at any given temperature.

For many of the parameter values that we have plotted, the sum over poles~\eqref{eq:mero} provides an excellent approximation to the spectral functions. However, for given values of \(\t\), \(\talpha\), and \(k/\m\) there is often a range of temperatures for which the approximation breaks down. In all of the cases where the sum over poles fails to provide a good approximation to a spectral function, we find that there are two poles with similar values of \(\Re \w\) both providing a significant contribution to that spectral function. When this happens, it seems likely that analytic terms in the spectral functions become important due to destructive interference in the contributions of the different poles.

\FloatBarrier

\subsection{Sound Attenuation}
\label{sec:soundatt}

In this section we present our results for the sound attenuation, i.e. the imaginary part of the sound pole (either HZS or hydrodynamic sound), as a function of $\tau$, \(k/\m\) and $T/\mu$. We will also compare the results to the attenuation of Fermi liquid zero sound sketched in figure~\ref{fig:attenuation_cartoon}.

As reviewed in section~\ref{sec:hzs_background}, in probe brane models HZS exhibits a low-temperature regime, where \(\Im \w \propto T^0\) similar to the quantum collisionless regime of zero sound in Fermi liquids, and an intermediate-temperature regime where \(\Im \w \propto T^2\), similar to the thermal collisionless regime. However, due to the probe limit, the HZS poles are poles only of \(G_{JJ}\), not \(G_{TT}\), so HZS cannot crossover to hydrodynamic sound at high temperature. Instead, HZS crosses over to charge diffusion.

On the other hand, HZS in \ads-Reissner-Nordstr\"om \textit{does} crossover to hydrodynamic sound at high temperature. It also obeys \(\Im \w \propto T^0\) at low temperatures, but does not go through a regime with \(\Im \w \propto T^2\) at intermediate temperature. We will demonstrate how turning on a non-zero back-reaction allows us to interpolate between the probe brane and \ads-Reissner-Nordstr\"om behaviours.

\begin{figure}
	\begin{center}
		\(\talpha = 1\), \(k/\m = 0.01\)
		\\
		\includegraphics[width=\textwidth]{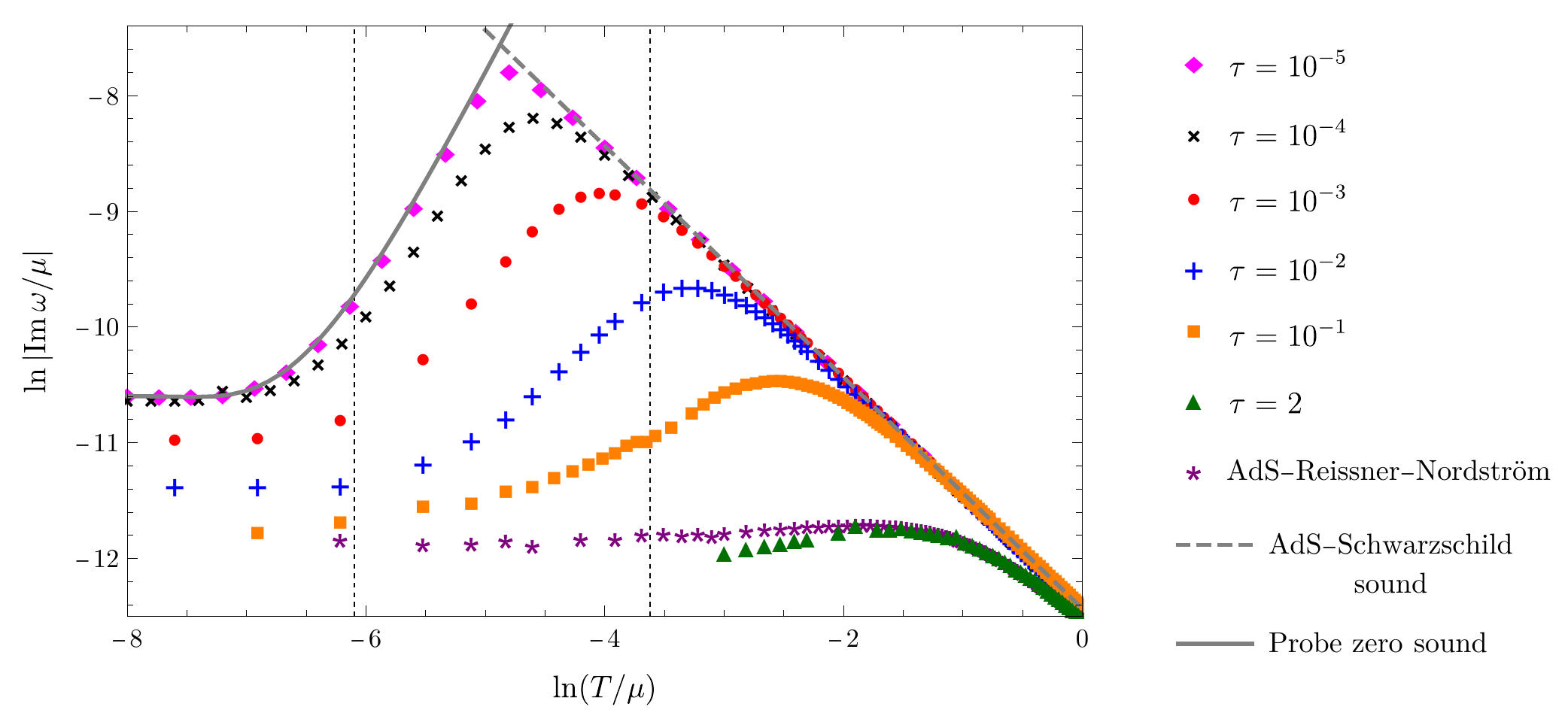}
	\caption[The imaginary part of the sound mode as a function of temperature, for different levels of back-reaction.]{
		Numerical results for the logarithm of the sound attenuation, $\ln | \Im \omega/\mu |$, as a function of $\ln \left(T/\mu\right)$ for $\talpha=1$, $k/\mu=0.01$, and various values of \(\t\), as well as the \ads[4]-Reissner-Nordstr\"om result (the purple stars). The solid grey curve is the attenuation of HZS in the probe limit, while the dashed grey curve is the attenuation of sound modes in \ads[4]-Schwarzschild, i.e. the attenuation of poles in \(G_{TT}\) in the probe limit. The vertical dashed black lines indicate the boundaries between quantum collisionless, thermal collisionless, and hydrodynamic regimes in Landau Fermi liquids, \(\pi T/\m \approx k/\sqrt{2}\,\m\) and $\le(\pi T/\m\ri)^2 \approx k/\sqrt{2}\,\m$ respectively. At sufficiently low temperature, $|\Im \w|\propto T^0$, similar to the quantum collisionless regime of zero sound in Fermi liquids. As \(T/\m\) is increased, the HZS poles eventually enter a regime where \(\ln|\Im \w/\m|\) increases with \(T/\m\), at a rate dependent on \(\t\). At high temperatures, for all non-zero \(\t\) that we study, HZS crosses over to hydrodynamic sound, with \(|\Im \w| \propto T^{-1}\). The value of \(T/\m\) at which the attenuation exhibits a global maximum is the Fermi liquid definition of the hydrodynamic crossover.
	}
	\label{fig:sound_attenuation}
	\end{center}
	\end{figure}
Figure~\ref{fig:sound_attenuation} shows our results for $\ln\left|\Im \omega/\mu\right|$ as a function of $\ln \left(T/\mu\right)$ for $\talpha=1$, $k/\mu=0.01$, and values of \(\t\) between \(\t=10^{-5}\) and \(\t=2\). Also plotted are numerical results for probe brane HZS (the solid grey curve), sound modes in \ads[4]-Schwarzschild (dashed grey)~\cite{Herzog:2003ke,Kovtun:2005ev}, and sound modes in \ads[4]-Reissner-Nordstr\"om (purple stars)~\cite{Davison:2011uk}.

For all of the values of \(\t\) displayed in figure~\ref{fig:sound_attenuation}, at sufficiently low temperature the attenuation of HZS is approximately independent of temperature, similar to the quantum collisionless regime of zero sound in Fermi liquids. Eventually, when the temperature becomes large enough, the attenuation begins to increase, at a rate that depends on \(\t\). For the smallest non-zero value of \(\t\) that we study, \(\t = 10^{-5}\) (the pink diamonds in figure~\ref{fig:sound_attenuation}), the attenuation in this regime is very close to that in the probe limit, with \(|\Im \w| \propto T^2\) similar to the thermal collisionless regime of Fermi liquid zero sound. For larger values of \(\t\) the attenuation increases more slowly with temperature.

For \(\t\neq0\), as the temperature is increased further, HZS smoothly evolves into hydrodynamic sound, with attenuation that decreases with temperature as \(|\Im \w| \propto T^{-1}\). This power of \(T\) contrasts with the \(|\Im \w| \propto T^{-2}\) behaviour in the hydrodynamic regime of a Fermi liquid, and occurs because our holographic model describes a CFT at finite temperature. The attenuation exhibits a global maximum, dividing the regions where the attenuation either increases or decreases with temperature. The temperature at which this maximum occurs is the Landau Fermi liquid definition of the crossover to hydrodynamics.

The vertical dashed black lines in figure~\ref{fig:sound_attenuation} represent the boundaries between quantum collisionless, thermal collisionless, and hydrodynamic regimes in Fermi liquids, \(\pi T/\mu = k/\sqrt{2}\,\mu\) and \( (\pi T/\mu)^2 = k/\sqrt{2}\,\mu\). The sound attenuation in Fermi liquids exhibits a maximum at the latter boundary, as shown in figure~\ref{fig:attenuation_cartoon}. The results in figure~\ref{fig:sound_attenuation} show that changes in behaviour of \(\Im \w\) do not occur at the same values of \(T/\m\) in our holographic model as in Fermi liquids, despite the qualitative similarity in the functional form of \(\Im \w/\m\).

\begin{figure}
	\begin{center}
		\(\t = 10^{-4}\), \(\talpha = 1\)
		\\
		\includegraphics{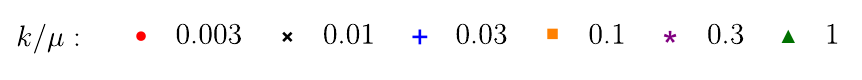}
		\vspace{-2em}
	\end{center}
	\begin{subfigure}{0.5\textwidth}
		\includegraphics[width=\textwidth]{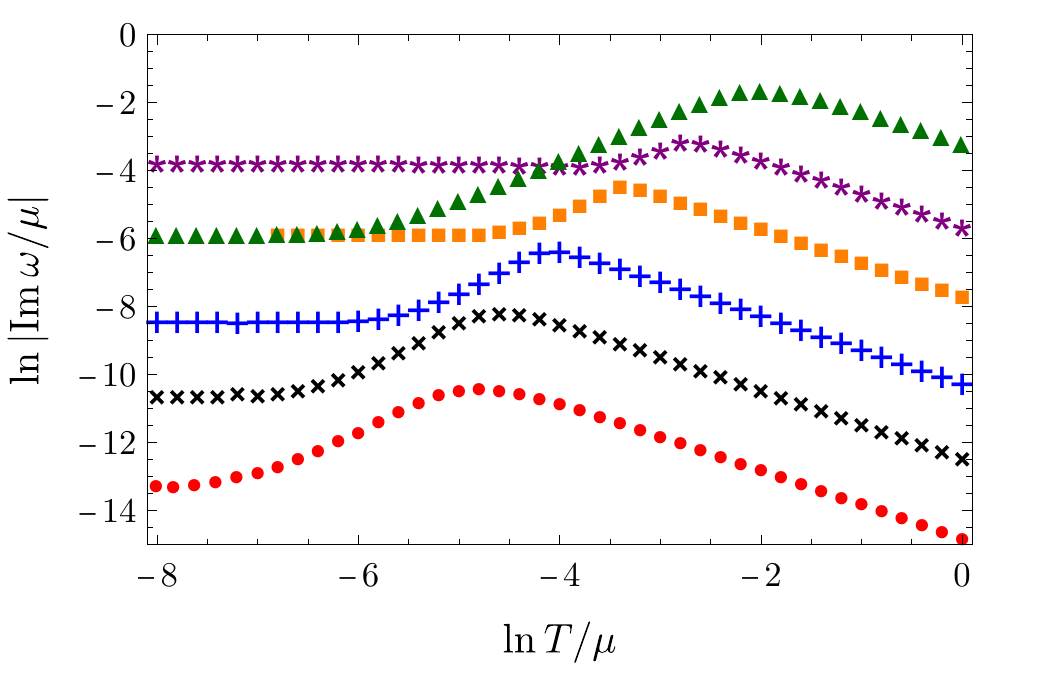}
	\end{subfigure}
	\begin{subfigure}{0.5\textwidth}
		\includegraphics[width=\textwidth]{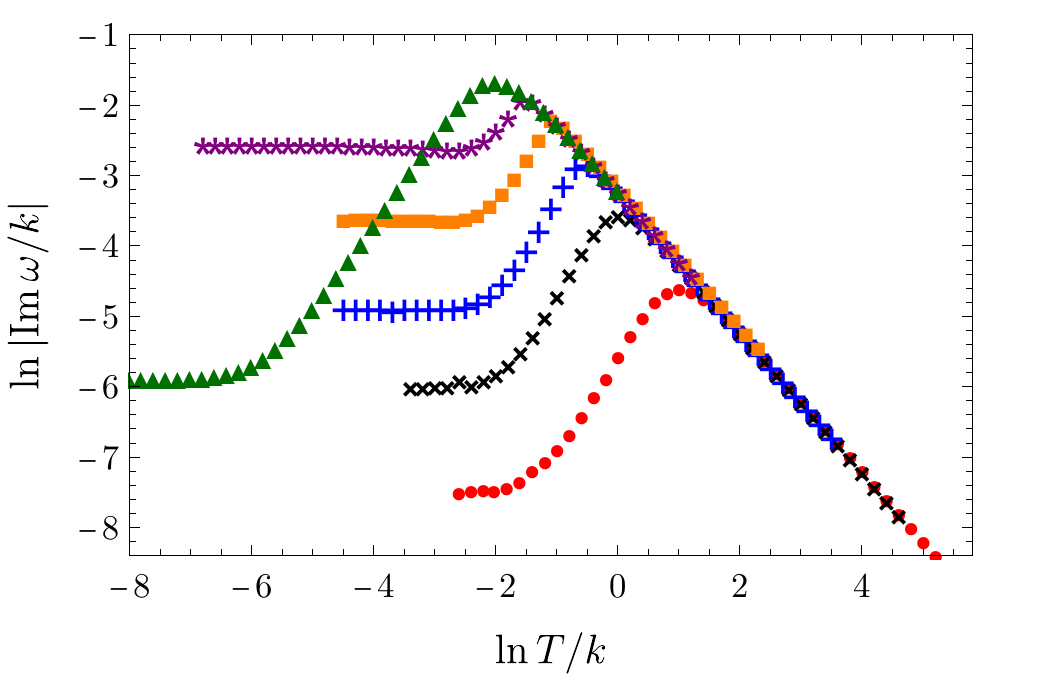}
	\end{subfigure}
	\caption[The imaginary part of the sound mode for different momenta.]{
		The imaginary part of the sound mode as a function of temperature for \(\t = 10^{-4}\), \(\talpha = 1\), and a range of values of \(k/\m\). The plot on the right shows the same data as the plot on the left, but with the frequency and temperature measured in units of \(k\) rather than \(\m\). For all \(k/\m\) that we have checked, the attenuation behaves similarly with temperature. At low temperature, \(|\Im \w| \propto T^0\). As the temperature is increased, \(|\Im \w|\) begins to grow, eventually reaching a maximum, then decreasing as \(\Im \w \propto T^{-1}\).
	}
	\label{fig:sound_attenuation_changing_k}
\end{figure}
We now investigate the effect on the sound attenuation of changing the momentum. Figure~\ref{fig:sound_attenuation} shows the imaginary part of the sound poles as a function of temperature, for \(\t = 10^{-4}\), \(\talpha = 1\), and a range of values of \(k/\m\). For all of the values of \(k/\m\) that we plot, the functional form of \(\Im\w/\m\) is similar to that for \(k/\m = 0.01\). There is a low-temperature regime where \(|\Im \w| \propto T^0\), an intermediate-temperature regime where \(\Im \w\) grows with \(T\), and a high-temperature regime where \(\Im \w \propto T^{-1}\). There is therefore always a maximum in \(\Im \w/\m\) as a function of \(T/\m\), so the Fermi liquid definition of the crossover to hydrodynamics may always be used.

In figure~\ref{fig:sound_attenuation_changing_k}, the results for \(k/\m = 1\) conflict with a pattern of \(\lim_{T\to0} |\Im \w/\m|\) increasing with \(k/\m\). For \(k/\m = 1\), the sound pole smoothly evolves into a pole with \(\Re \w \approx k\) as the temperature is decreased (the black crosses in figure~\ref{fig:tau0p0001_a1_k1_real}) while the diffusion pole evolves into a pole with \(\Re \w < k\) (blue dots in the figure). For smaller values of \(k/\m\), precisely the opposite is true; see figures~\ref{realimsmalltaua} and~\ref{fig:tau0p0001_a1_k0p1_real} for \(k/\m = 0.01\) and \(k/\m = 0.1\) respectively. It is this exchange in the evolution of the hydrodynamic poles to low temperatures which causes the attenuation at \(k/\m = 1\) to break with the pattern in figure~\ref{fig:sound_attenuation_changing_k}.

It was pointed out in~\cite{Davison:2013bxa}, using the numerical results of~\cite{Davison:2011uk}, that in \ads[4]-Reissner-Nordstr\"om the hydrodynamic prediction for sound attenuation holds even at very low temperatures, in the following sense~\cite{Davison:2013bxa,Davison:2013uha}. Hydrodynamics predicts that sound waves in the holographic dual to \ads[4]-Reissner-Nordstr\"om have attenuation
\begin{equation} \label{eq:ads-rn_sound_attenuation}
	\G_s = \frac{\h}{3\ve}.
\end{equation}
This was obtained by setting \(d=3\), \(p=\ve/2\), and \(\z=0\) in~\eqref{eq:hydro_sound_attenuation}.\footnote{
	The holographic dual to \ads-Reissner-Nordstr\"om is a conformal field theory at finite temperature and chemical potential. Conformal field theories have \(p = \ve/(d-1)\) and vanishing bulk viscosity, \(\z\) = 0, due to tracelessness of the stress tensor~\cite{Kovtun:2012rj}.
}
For rotationally invariant holographic models in which the gravitational theory is Einstein gravity coupled to matter, the shear viscosity takes the universal value \(\h = s/4\pi\)~\cite{Policastro:2001yc,Buchel:2003tz,Kovtun:2004de}. We may therefore rewrite the sound attenuation~\eqref{eq:ads-rn_sound_attenuation} entirely in terms of thermodynamic quantities,
\begin{equation} \label{eq:ads-rn_sound_attenuation_entropy}
	\G_s = \frac{s}{12\pi\ve}.
\end{equation}
Ref.~\cite{Davison:2013bxa} observed that the dispersion relation~\eqref{eq:sound_dispersion}, with attenuation given by~\eqref{eq:ads-rn_sound_attenuation_entropy}, agrees with the numerical results for the sound pole frequencies in \ads[4]-Reissner-Nordstr\"om even at low temperature, provided \(k \ll \m\).

\begin{figure}
	\includegraphics[width=\textwidth]{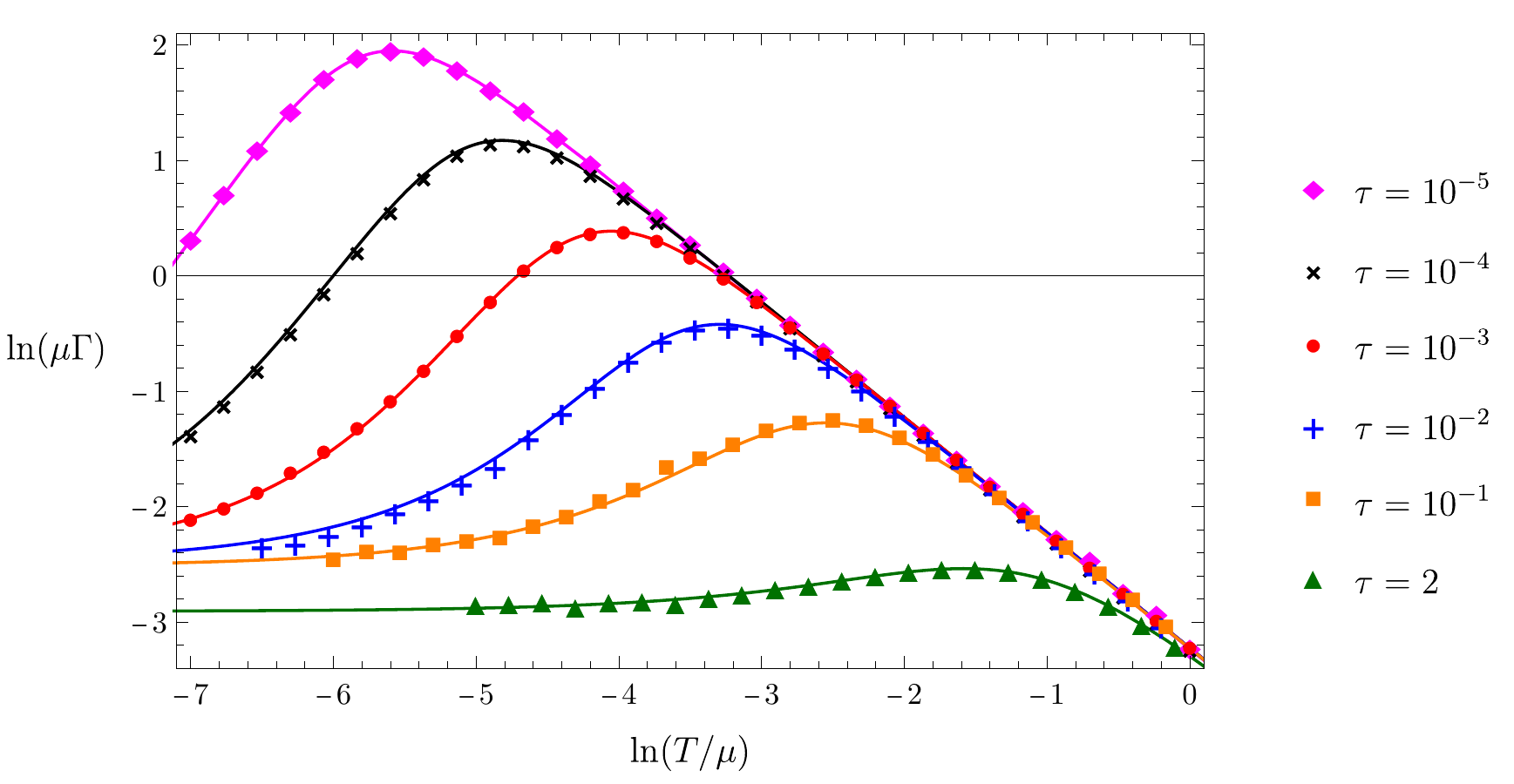}
	\caption[Comparison of the sound attenuation to the hydrodynamical prediction.]{Numerical results for $\ln\left(\m\G\right)$ versus $\ln \left(T/\mu\right)$, for $\talpha=1$. We obtain $\Gamma$ by numerically fitting $-\G k^2 + \Delta k^4$ to the sound pole's  $\textrm{Im}\left(\omega\right)$ over the range of \(k/\m\) listed in table~\ref{tab:momenta}. The curves show the hydrodynamic prediction~\eqref{eq:ads-rn_sound_attenuation_entropy} for the sound attenuation \(\G_s\), which agrees well with our numerical results.}
	\label{hydro_prediction}
\end{figure}
%
We find that the same is true in our model, as shown in figure~\ref{hydro_prediction}. The points show our numerical results for the sound attenuation \(\G\) as a function of temperature for a range of \(\t\). Each point is obtained by fitting a functional form \(-\G k^2 + \Delta k^4\) to the imaginary part of \(\w\). The ranges of values of \(k/\m\) used for the fits are listed in table~\ref{tab:momenta}. The solid lines are the hydrodynamic predictions \(\G_s\) for each value of \(\t\), given by~\eqref{eq:ads-rn_sound_attenuation_entropy} with entropy and energy densities given by~\eqref{entropy} and~\eqref{energy}, respectively. The numerical results and the hydrodynamic prediction agree well over the range of \(T/\m\) plotted, even when \(k/T > 1\).

\begin{table}
	\begin{center}
	\begin{tabular}{c | c | c}
		\(\t\) & Range of \(k/\m\) & Smallest value of \(T/\m\)
		\\ \hline \rule{0pt}{1\normalbaselineskip}
		\(10^{-5}\), \(10^{-4}\), \(10^{-3}\) & \(10^{-3}\) to \(5 \times 10^{-3}\) & \(9.1 \times 10^{-4}\)
		\\
		\(10^{-2}\) & \(2.5 \times 10^{-3}\) to \(10^{-2}\) & \(1.5 \times 10^{-3}\)
		\\
		\(10^{-1}\) & \(5 \times 10^{-3}\) to \(2 \times 10^{-2}\) & \(2.5 \times 10^{-3}\)
		\\
		\(2\) & \(10^{-2}\) to \(5 \times 10^{-2}\) & \(6.7 \times 10^{-3}\)
	\end{tabular}
	\end{center}
	\caption[The range of momenta used to fit the attenuation of holographic zero sound.]{
		The middle column of this table lists the range of momenta used to fit \(\G\) for the results presented in figure~\ref{hydro_prediction}, for each values of \(\t\). Four or five values of the momentum in each range were used for each fit. The column on the right is the smallest value of \(T/\m\) for which the fit was performed. The small \(T/\m\) points in figure~\ref{hydro_prediction} arise from fits to \(|\Im \w/\m|\) at \(k/T > 1\).
	}
	\label{tab:momenta}
\end{table}

Ref.~\cite{Davison:2013bxa} also found that the diffusion mode in the shear channel of \ads[4]-Reissner-Nordstr\"om is also well described by hydrodynamics even at low temperatures. The same was found to be true in our Einstein-DBI model in ref.~\cite{Gushterov:2018nht}. It therefore appears that hydrodynamics provides a good effective description of both the sound and shear diffusion poles of this model, even for momenta large compared to the temperature, provided \(k \ll \m\). Similar behaviour has been observed in a holographic model of a Fermi liquid~\cite{Davison:2013uha}.

\section{Discussion}
\label{summary}

We have computed the poles of the sound-channel retarded Green's functions of \(J^\m\) and \(T^{\m\n}\) in the holographic dual to the Einstein-DBI charged black brane. For all levels of back-reaction, we observe a  mode with sound-like dispersion at low temperatures. The way in which the hydrodynamic charge diffusion pole emerges at high temperature is very different at different levels of back-reaction.

We have studied three different definitions of the temperature of crossover to hydrodynamic behaviour: the temperature at which two poles collide on the imaginary axis to produce the charge diffusion pole (the probe limit definition), the temperature at which the charge density spectral function becomes dominated by charge diffusion (the \ads-Reissner-Nordstr\"om definition), and the temperature of maximum sound attenuation (the Fermi liquid definition).

\begin{figure}
	\begin{center}
		\(\talpha = 1\), \(k/\m = 0.01\)
		\\
		\includegraphics{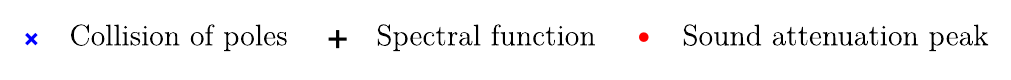}
		\vspace{-2em}
	\end{center}
		\begin{subfigure}{0.5\textwidth}
			\includegraphics[width=\textwidth]{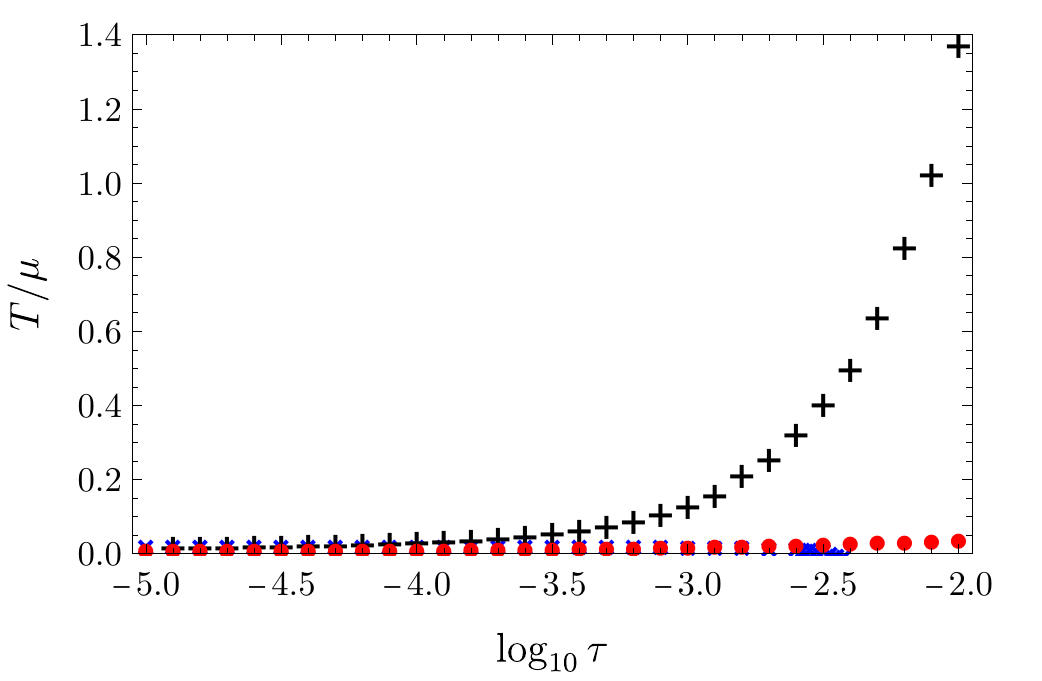}
			\caption{Comparison of crossover definitions.}
			\label{fig:compare} 
		\end{subfigure}
		\begin{subfigure}{0.5\textwidth}
			\includegraphics[width=\textwidth]{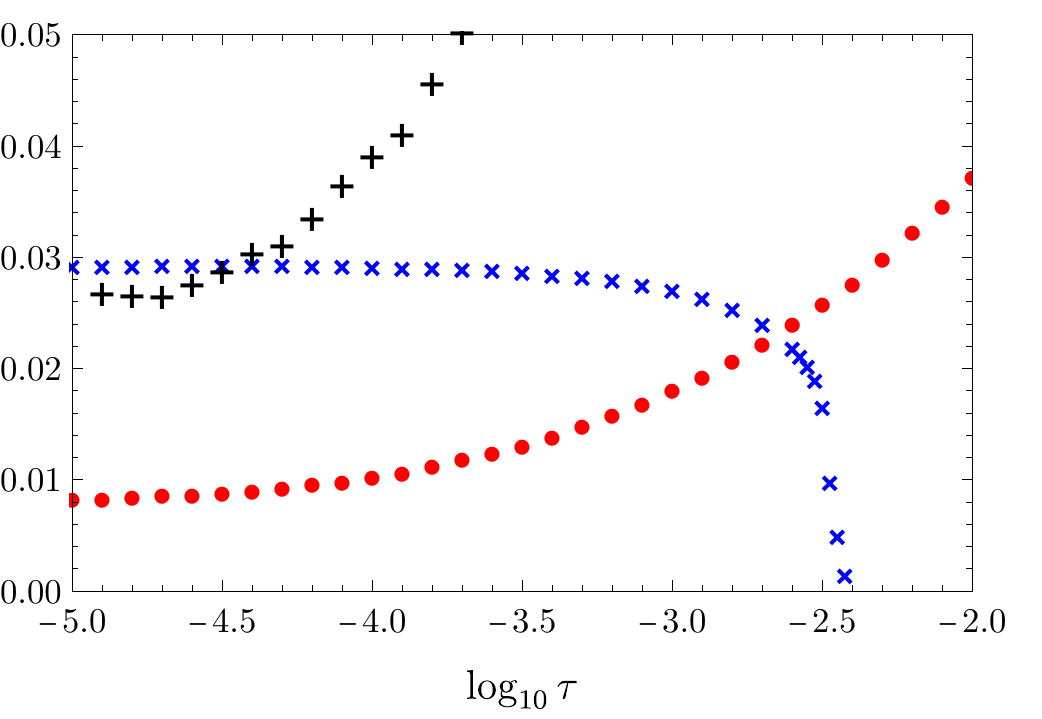}
			\caption{Close-up of (a)}
			\label{fig:comparecu} 
		\end{subfigure}
		\caption[Comparison of the different definitions of the crossover to hydrodynamics in holographic zero sound.]{Our numerical results for the value of $T/\mu$ at the crossover to hydrodynamics as a function of \(\t\) for $\talpha=1$ and \(k/\m = 0.01\), using the three different definitions: the value of \(T/\m\) at which poles collide on the imaginary axis to produce charge diffusion pole (blue crosses), the value of \(T/\m\) at which the sound and diffusion peaks in the charge density spectral function have equal height (black plus signs), and the value of \(T/\m\) at which the sound attenuation is greatest (red dots).}
		\label{fig:comparedefs}
\end{figure}
For given values of \(\t\), \(\talpha\), and \(k/\m\), each of these definitions gives different values of the crossover temperature. For example, figure~\ref{fig:comparedefs} shows our numerical results for the three different crossover temperatures as a function of \(\t\), for \(\talpha = 1\) and \(k/\m = 0.01\).

Not all of the definitions of the crossover temperature may be viable for given values of the parameters. We found that the collision of poles only occurs for sufficiently small \(\t\). In contrast, it is numerically difficult to distinguish the separate peaks in the spectral function at small \(\t\), hence the apparent numerical noise in the spectral function results in figure~\ref{fig:comparecu} for \(\log_{10}\t \lesssim -4.5\). The maximum in the sound attenuation appears to exist for all \(\t \neq 0\).

In section~\ref{sec:soundatt}, we found that the hydrodynamic result for the sound attenuation works well even at very small values of \(T/\m\), provided \(k/\m\) is also sufficiently small (although \(k\) may be large compared to \(T\)). Why, then, does it make sense to define a crossover to hydrodynamic behaviour? The various definitions of the crossover are defined at a given value of \(k/\m\), so they should be interpreted as guidelines as to the minimum temperature at which hydrodynamics applies at that specific value of the momentum. For example, at \(\t = 10^{-4}\), \(\talpha = 1\), \(k/\m = 0.01\), and \(T/\m = 0.01\) the Green's functions are clearly not well described by hydrodynamics since there is no charge diffusion pole, see figure~\ref{planesmalltau}, and the charge density spectral function is dominated by the sound pole, figure~\ref{tau_0p0001_spectral_functions}. On the other hand, at \(T/\m = 0.05\) there is a purely imaginary diffusion pole which dominates the charge density spectral function.

In the probe limit, \(\t = 0\), the poles in \(G_{TT}\) and \(G_{JJ}\) decouple. The effective theory governing the poles in \(G_{TT}\) closest to the real axis is uncharged hydrodynamics~\cite{Kovtun:2005ev}, while at low temperature the effective theory governing the poles in \(G_{JJ}\) appears to be the quasihydrodynamics of a weakly conserved current~\cite{Chen:2017dsy,Grozdanov:2018fic}. At high temperature, the poles of both \(G_{TT}\) and \(G_{JJ}\) are well described by hydrodynamics with a \(\U(1)\) current.

In our Einstein-DBI model, for \(\t \neq 0\) we find that hydrodynamics works well even at low temperature, provided \(k/\m\) is sufficiently small, similar to \ads-Reissner-Nordstr\"om. If \(k/\m\) is increased, the spectral functions and their poles tend to those of the probe limit, so the appropriate effective description at low temperature appears to become quasihydrodynamics. The range of \(k/\m\) for which hydrodynamics provides a good effective description appears to increase with increasing \(\t\). Recalling that \(\t\) should measure the fraction of degrees of freedom charged under the \(\U(1)\) global symmetry, it appears that the effective description of low temperature sound modes at a given momentum is controlled by the number of charged degrees of freedom, at least in the class of models considered in this chapter.

Many holographic models of non-Fermi liquids support low temperature sound modes. The most important question to address is whether any real non-Fermi liquids, such as the cuprates and graphene similarly support low temperature sound modes.

As a step towards answering this question, low-temperature sound modes in other holographic models of non-Fermi liquids should be analysed. If every such model supports low temperature sound modes, this would suggest that HZS is a universal feature of holographic compressible quantum matter. Alternatively, if HZS is not universal, one could look for criteria that distinguish models with and without low temperature sound, and test whether the same criteria hold outside of holography.

For example, it has been observed that HZS appears in models where the charge density has non-zero spectral weight at zero temperature and non-zero momentum~\cite{Hartnoll:2016apf}.\footnote{The spectral weight \(\r(k)\) of an operator \(\cO\) is defined as
\(
	\r(k) \equiv \lim_{\w\to0} \frac{\Im G^\mathrm{R}_{\cO\cO}(\w,k)}{\w},
\)
where \(G^\mathrm{R}_{\cO\cO}(\w,k)\) is the momentum-space retarded Green's function of the operator \(\cO\).
} This is particularly striking in probe brane models, where the spectral weight decreases exponentially with increasing momentum, becoming extremely small at momentum large compared to a characteristic scale set by the charge density. In ref.~\cite{Hartnoll:2016apf} this was given a possible interpretation as a Fermi surface smeared out by strong interactions.

If HZS turns out to be universal in holographic models, this would hint that low temperature sound modes may exist in compressible quantum matter beyond the traditional systems of Fermi liquids, solids, and superfluids. A comprehensive understanding of HZS may provide useful guidance in the study of low-temperature sound in real-world systems.

\chapter{Wilson surfaces and RG flows}
\label{chap:probe_m5}

In this chapter we will use gauge/gravity duality to compute entanglement entropy for surface defects in the QFT describing the low energy excitations of a stack of M5-branes. We first provide some background on the M5-brane QFT, the surface defects we study, and their holographic duals in sections~\ref{sec:wilson_surfaces_intro},~\ref{sec:defect_ee}, and~\ref{sec:actions}. Then, in each of sections~\ref{sec:m2},~\ref{sec:antisymmetric}, and~\ref{sec:symmetric}, we review probe brane solutions holographically dual to different surface defects in the M5-brane QFT, before presenting our results for entanglement entropy. The solutions studied in sections~\ref{sec:antisymmetric} and~\ref{sec:symmetric} are holographically dual to defect RG flows, and in these cases we test the monotonicity of the on-shell action in the entanglement wedge proposed in ref.~\cite{Kumar:2017vjv}.

\section{M-theory and \(\cN = (2,0)\) theory}
\label{sec:wilson_surfaces_intro}

As described in chapter~\ref{chap:ads_cft}, the low energy excitations of a stack of coincident D-branes are described by supersymmetric Yang-Mills theory \cite{Witten:1995im}. Similarly, in M-theory the low energy excitations of a stack of M2-branes are described by Aharony-Bergman-Jafferis-Maldacena (ABJM) theory, maximally supersymmetric Chern-Simons theory coupled to matter~\cite{Bagger:2006sk,Gustavsson:2007vu,Bagger:2007jr,Bagger:2007vi,Aharony:2008ug}.

The theory describing the low energy excitations of a stack of \(N_5 > 1\) M5-branes is not known. However, some information about the theory may be deduced from supergravity. The theory is six-dimensional, since M5-branes are six-dimensional, and from the supersymmetries preserved by the supergravity solution~\eqref{eq:m_brane_solution} one finds that the M5-brane theory possesses \(\cN=(2,0)\) supersymmetry.\footnote{The notation ``\((2,0)\)'' means there are two left- and no right-handed supersymmetry generators.}

The massless fields of the M5-brane theory are the Goldstone modes of the symmetries of 11D SUGRA spontaneously broken by the M5-brane solution~\eqref{eq:m_brane_solution}~\cite{Adawi:1998ta}. The field content for a single M5-brane is:
\begin{itemize}
    \item Five real scalar fields, from the partial breaking of translational symmetry in the five directions normal to the brane.
    \item A two-form gauge field, with self-dual field strength, from the breaking of large gauge transformations of the bulk three-form gauge field \(\C3\).
    \item Four symplectic Majorana-Weyl spinors, arising from the broken supersymmetry.
\end{itemize}
These fields form the tensor multiplet of \(\cN=(2,0)\) supersymmetry. For \(N_5\) coincident M5-branes, these fields are expected to become valued in the adjoint representation of an \(\mathfrak{su}(N_5)\) gauge algebra~\cite{Strominger:1995ac}.

The \(\cN=(2,0)\) theory is clearly not an ordinary gauge theory, since the gauge field is a two-form. To date, no action for the non-abelian (\(N_5 > 1\)) theory has been found.\footnote{We will discuss an action for the abelian theory in section~\ref{sec:actions}.} Indeed, various arguments have been made that no such action exists, i.e. that the non-abelian \(\cN=(2,0)\) theory is non-Lagrangian, see for example ref.~\cite{Lambert:2019khh} and references therein. Gauge/gravity duality provides a powerful tool to study the M5-brane theory. When \(N_5 \gg 1\), the world volume theory is expected to be holographically dual to 11D SUGRA on \(\ads[7] \times \sph[4]\) \cite{Maldacena:1997re}; in particular the theory is expected to be a superconformal field theory (SCFT).

Beyond the important role that the \(\mathcal{N}=(2,0)\) theory plays in M-theory, it is notable among quantum field theories as it is the maximally supersymmetric theory in six-dimensions, which is the largest number of dimensions in which superconformal symmetry is possible~\cite{Nahm:1977tg}. Study of compactifications of this theory and other 6D SCFTs has revealed intriguing properties of lower dimensional quantum field theories, such as the conjectured relation between four-dimensional \(\mathcal{N}=2\) gauge theories and Liouville or Toda theories in two dimensions \cite{Alday:2009aq,Wyllard:2009hg}.

Various supergravity calculations, holographic and otherwise, indicate that the number of massless degrees of freedom in \(\mathfrak{su}(N_5)\) \(\cN=(2,0)\) theory scales as \(N_5^3\) at large \(N_5\)~\cite{Klebanov:1996un,Freed:1998tg,Henningson:1998gx,Henningson:1998ey,Harvey:1998bx}. For example, a holographic calculation shows that the thermodynamic entropy density \(s\) of the \(\cN = (2,0)\) theory at finite temperature \(T\) is \(s = 2^7 3^{-6} \pi^3 N_5^3 T^5\) for \(N_5 \gg 1\). This contrasts with \(N \gg 1\) D-branes, described by supersymmetric Yang-Mills theory, for which the number of degrees of freedom scales as \(N^2\) as discussed in section~\ref{sec:duality_statement}.

The observables of \(\cN = (2,0)\) theory are believed to be Wilson surfaces~\cite{Ganor:1996nf}, the holonomy of the two-form gauge field \(A\) on a two-dimensional surface \(\Sigma\). In the abelian theory, a term involving the scalar fields of the tensor multiplet may be added to create the half-BPS Wilson surface, similar to the half-BPS Wilson loop~\eqref{eq:bps_wilson_loop}. See ref.~\cite{Bullimore:2014upa} for an explicit construction. It is believed that the non-Abelian Wilson surface may be similarly deformed into a half-BPS Wilson surface, although there is no explicit formula for the operator since the non-Abelian \(\cN=(2,0)\) theory has not been formulated. There is a natural candidate for the holographic dual to the non-Abelian half-BPS Wilson suface: supersymmetric configurations of M2-branes that intersect the boundary of \ads[7] along the surface \(\Sigma\)~\cite{Maldacena:1998im,Lunin:2007ab,Chen:2007ir}. This is similar to the duality between strings reaching the boundary of \ads[5] and Wilson loops in \(\cN = 4\) SYM, discussed in section~\ref{sec:wilson_loops}.

\section{Defect entanglement entropy}
\label{sec:defect_ee}

We will use gauge/gravity duality to compute entanglement entropy in \(\cN=(2,0)\) theory in the presence of two-dimensional defects. A subset of the defects we study are the half-BPS Wilson surface operators~\cite{Maldacena:1998im,Lunin:2007ab,Chen:2007ir,Mori:2014tca}, and in these cases our results reproduce the probe limit of the calculations in refs.~\cite{Gentle:2015jma, Estes:2018tnu}. We will also study holographic solutions dual to defect renormalisation group (RG) flows, in which Wilson surfaces appear as fixed points.

\begin{figure}
	\begin{center}
	\includegraphics{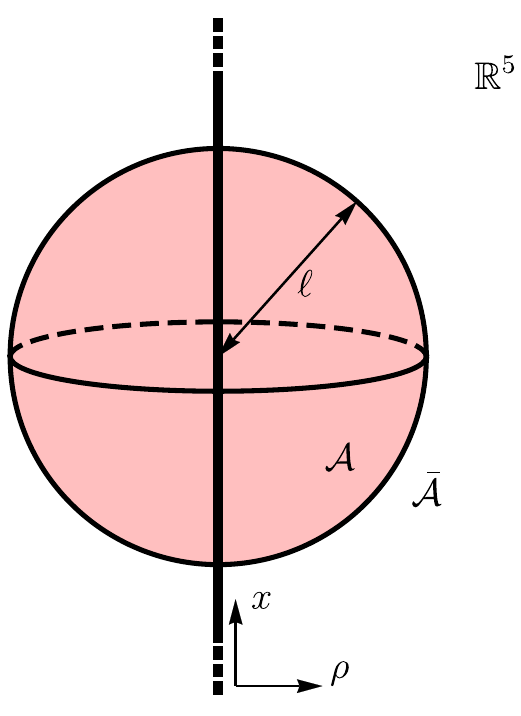}
	\caption[The entangling region we use to study defects in \(\cN = (2,0)\) theory.]{In this chapter we use a spherical entangling region, of radius \(\ell\) (shaded red in the diagram), centered on a planar (1+1)-dimensional defect (the thick, vertical line, with dots indicating that the defect extends to infinity). The intersection of the entangling region with the defect is an interval of length \(2\ell\). For conformal defects, the contribution of the defect to the entanglement entropy takes the form~\eqref{eq:defect_ee} of the entanglement entropy for a single interval in a 2D CFT. The coordinates \(x\) and \(\r\) are defined in section~\ref{sec:actions}.}
	\label{fig:entangling_region}
	\end{center}
\end{figure}

For simplicity, we will restrict to planar defects, and choose the entangling region to be a sphere of radius \(\ell\) centered at a point on the defect, as illustrated in figure~\ref{fig:entangling_region}. In a CFT in six dimensions, the entanglement entropy \(\see\) of a spherical subregion takes the form~\eqref{eq:pure_ads_sphere_ee}~\cite{Ryu:2006bv,Ryu:2006ef}
\begin{equation} \label{eq:entanglement_general_form}
	\see^\mathrm{sphere} = \frac{2 \pi^2 L^5}{3 \gn} \le[ p_4 \frac{\ell^4}{\e^4} + p_2 \frac{\ell^2}{\e^2}  + p_0  + p_L \ln\le( \frac{2\ell}{\e} \ri) \ri] + \dots,
\end{equation}
where \(\e\) is an ultraviolet cutoff. The coefficients \(p_0\), \(p_2\), and \(p_4\) are scheme dependent --- they are not invariant under multiplicative changes in the cutoff --- while the coefficient \(p_L\) of the logarithm is scheme independent. For the vacuum of the \(\mathcal{N} = (2,0)\) theory, holographically dual to 11D SUGRA on \(\ads[7] \times \sph[4]\), it is given by~\cite{Ryu:2006bv}
\begin{equation} \label{eq:m5_entanglement_coefficient}
	\frac{2 \pi^2 L^5}{3 \gn}p_L = \frac{4}{3} N_5^3.
\end{equation}

The presence of a two-dimensional planar defect will modify the coefficient of the logarithmic term in equation~\eqref{eq:entanglement_general_form}. Defining \(\see^{(0)}\) as the entanglement entropy without the defect, i.e. \eqref{eq:entanglement_general_form} with in particular the coefficient of the logarithm given by \eqref{eq:m5_entanglement_coefficient}, we define the contribution to entanglement entropy from the defect as\footnote{We assume that the same regularisation prescription is used for the entanglement entropy in the defect CFT as in the theory without the defect.}
\begin{equation} \label{eq:defect_ee}
	\see^{(1)} \equiv \see - \see^{(0)} = \frac{c}{3} \ln \le( \frac{2\ell}{\e} \ri) + \mathcal{O}(\e^0),
\end{equation}
with a coefficient \(c\) to be determined.

The entanglement entropy of an interval of length \(\propto \ell\) in a two-dimensional CFT takes the same form as~\eqref{eq:defect_ee}, with \(c\) the central charge \(c\) of the CFT~\cite{Calabrese:2004eu}. The central charge of a 2D CFT measures the number of degrees of freedom in the theory. One of the questions we will seek to address is whether \(c\) similarly measures degrees of freedom on two-dimensional defects.

For defect RG flows, we may naturally define an \(\ell\)-dependent function
\begin{equation} \label{eq:b_function}
    C(\ell) \equiv 3 \ell \frac{ \diff \see^{(1)} }{ \diff \ell }.
\end{equation}
In a CFT, \(C(\ell)=c\) is a constant.\footnote{Note that in a CFT the \(\mathcal{O}(\e^0)\) term in \eqref{eq:defect_ee} must be independent of \(\ell\), since the entanglement entropy is dimensionless and there are no other scales which may be combined with \(\ell\) to yield a dimensionless quantity.} Along an RG flow, \(C(\ell)\) interpolates between the values of \(c\) at the fixed points,
\begin{equation}
    \lim_{\ell\to0} C(\ell) = c_\mathrm{UV},
    \quad
    \lim_{\ell\to\infty} C(\ell) =  c_{\mathrm{IR}},
\end{equation}
where \(c_\mathrm{UV}\) and \(c_\mathrm{IR}\) are the coefficients at the UV and IR fixed points, respectively.

In two dimensions, the logarithmic derivative of the single-interval entanglement entropy with respect to the length of the interval satisfies a strong monotonicity theorem \cite{Casini:2006es}, so it decreases monotonically along any RG flow. On the other hand, we find that for the flows we study \(C(\ell)\) is not monotonic, and may be larger in the IR than in the UV. This provides an obstruction to interpreting \(c\) as a measure of degrees of freedom. This is consistent with the observations of ref.~\cite{Kobayashi:2018lil}, who found similar behaviour for other defect RG flows.

Another candidate measure of defect degrees of freedom is the type A anomaly coefficient, \(b\), of of the defect contribution to the Weyl anomaly~\eqref{eq:defect_weyl_anomaly}. As discussed in section~\ref{sec:defects_background}, in ref.~\cite{Jensen:2015swa} it was shown that \(b\) obeys a weak monotonicity theorem, \(b_\mathrm{UV} \geq b_\mathrm{IR}\), for defect RG flows triggered by a source for a relevant operator. We find that \(b\) obeys this inequality for the holographic flows that we construct in sections~\ref{sec:antisymmetric} and~\ref{sec:symmetric}, despite the fact that the theorem of ref.~\cite{Jensen:2015swa} does not necessarily hold since the flows we study are triggered by a VEV rather than a source. We also find that another Weyl anomaly coefficient, \(d_2\), decreases along these flows. We note that in ref.~\cite{Jensen:2018rxu} it was found that \(c\) is related to \(b\) and \(d_2\) by
\begin{equation} \label{eq:defect_entropy_weyl_relation}
	c = b - \frac{3}{5} d_2,
\end{equation}
for a two-dimensional defect in a six-dimensional CFT.

\section{Probe branes in AdS\(_7  \times \) S\(^4\)}
\label{sec:actions}

The equations of motion of \(D=11\) SUGRA, which follow from the action~\eqref{eq:11d_sugra_action}, admit the solution~\eqref{eq:m_brane_solution}, which for \(p = 5\) corresponds to a flat stack of \(N_5 = L^3/\pi\lp^3\) M5-branes. Taking the near horizon limit, \(r \ll L\), the metric becomes that of \(\ads[7]\times\sph[4]\). It will be convenient to define a new radial coordinate, \(z\), by \(r = 4 L^3 / z^2\), in terms of which the near-horizon solution becomes
\begin{subequations} \label{eq:m5_brane_near_horizon}
	\begin{align}
		\diff s^2 &= \frac{4L^2}{z^2} \le(\h_{\m\n} \diff x^\m \diff x^\n + \diff z^2\ri) + L^2 \diff s_{\sph[4]}^2 ,
		\\
		\CF4 &= - 3 L^3 \diff s_{\sph[4]},
	\end{align}
\end{subequations}
where \(\h_{\m\n}\) is the six-dimensional Minkowski metric and \(\CF4 \equiv \diff \C3\). In our notation, the \sph[4] factor has radius \(L\), while the \ads[7] factor has radius \(2L\). For \(L/\lp \gg 1\), 11D SUGRA on this background is conjectured to be holographically dual to the \(\cN = (2,0)\) theory with \(\mathfrak{su}(N_5)\) gauge algebra, with \(N_5\gg1\).

We will study \((1+1)\)-dimensional planar defects, which we will take to span the \((x^0,x^1)\) plane. We will work in cylindrical coordinates for the spatial directions, with axis oriented along the defect. Let us define \(t \equiv x^0\), \(x \equiv x^1\), and \(\r^2 \equiv \sum_{i=2}^5 (x^i)^2\). After these coordinate transformations, the near-horizon solution~\eqref{eq:m5_brane_near_horizon} becomes
\begin{subequations}\label{eq:ads_solution}
\begin{align}
    \diff s^2 &= \frac{4L^2}{z^2} \le(-\diff t^2 + \diff x^2 + \diff \r^2 + \r^2  \diff s_{\sph[3]}^2 + \diff z^2\ri) + L^2 \diff s_{\sph[4]}^2 ,
    \label{eq:ads_7_metric}
	\\
	\CF4 &= - 3 L^3 \diff s_{\sph[4]},
	\label{eq:ads_7_field_strength}
\end{align}
\end{subequations}
with the boundary of \ads\ at \(z=0\). The metric factor \(\diff s_{\sph[3]}^2\) is the metric on a unit, round \sph[3] parameterised by angles \(\f_{1,2,3}\). We parameterise the \sph[4] by a polar angle \(\q\) and three azimuthal angles \(\chi_{1,2,3}\). We will choose a gauge in which
\begin{subequations}
\begin{align}
	\C3 &= L^3 \le(3 \cos \q - \cos^3 \q - 2\ri) \sin^2\c_1 \sin\c_2 \, \diff \c_1 \wedge \diff \c_2 \wedge \diff \c_3,
	\label{eq:c3}
	\\
	\C6 &= \le( \frac{2 L}{z} \ri)^6 \r^3 \sin^2 \f_1 \sin \f_2 \, \diff t \wedge \diff x \wedge \diff \r \wedge \diff \f_1 \wedge \diff \f_2 \wedge \diff \f_3.
	\label{eq:flat_slicing_c6}
\end{align}
\end{subequations}
In particular, \(\C3\) vanishes at the north pole of the \sph[4], \(\q = 0\), so as to match the conventions used for the flux quantization condition of ref.~\cite{Camino:2001at}.

\subsection{M-brane actions}

In the presence of an M2- or M5-brane, the action for 11D SUGRA becomes
\begin{equation}
	S = S_\mathrm{bulk} + S_\mathrm{brane},
\end{equation}
where \(S_\mathrm{bulk}\) is the bulk action~\eqref{eq:11d_sugra_action} for the eleven-dimensional supergravity fields, and \(S_\mathrm{brane}  = S_\mathrm{M2}\) or \(S_\mathrm{M5}\) is a contribution localized to the brane. We will always work in the probe limit described in section~\ref{sec:probe_branes}, in which it is a good approximation to neglect the back-reaction of the brane on the metric and gauge field, which we may therefore take to be the \(\ads[7] \times \sph[4]\) solution \eqref{eq:ads_solution}. The brane action \(S_\mathrm{brane}\) is then an action for the world volume fields of the brane, decoupled from the bulk supergravity fields.

For a single M2-brane, the bosonic world volume fields are eight real scalars, which determine the embedding of the brane. They are described by the action \cite{Bergshoeff:1987cm}
\begin{equation} \label{eq:m2_action}
	S_\mathrm{M2} = - \ttwo \int_\Sigma \diff^3 \xi \sqrt{-\det g} + \ttwo \int_\Sigma P[\C3],
\end{equation}
where \(P\) denotes the pullback onto the brane of a bulk supergravity field, \(g \equiv P[G]\), and \(\xi\) are coordinates on the brane world volume \(\Sigma\). The tension \(\ttwo\) is related to the Planck length by \(\ttwo = 1/4\pi^2\lp^3\).

For a single M5-brane, the bosonic fields are five real scalar fields and an abelian two-form gauge field \(A\), with self-dual field strength \(F_3 \equiv \diff A\). Various formulations of M5-brane dynamics exist, which impose the self-duality constraint in different ways ~\cite{Perry:1996mk,Pasti:1997gx,Bandos:1997ui,Aganagic:1997zq,Townsend:1995af,Schwarz:1997mc, Ko:2013dka}. The different formulations are believed to be equivalent, at least classically \cite{Bandos:1997gm,Ko:2013dka}. We will use the approach of  Pasti, Sorokin and Tonin (PST) \cite{Pasti:1997gx,Bandos:1997ui,Aganagic:1997zq}, which we find to be the simplest for our purposes. In this approach, the self-duality constraint is imposed by an additional local symmetry due to the presence of an auxiliary scalar field \(a\). The bosonic part of the PST action for a single M5-brane is \cite{Pasti:1997gx}
\begin{align} \label{eq:pst_action}
	S_\mathrm{M5} &= - \tfive \int_\Sigma \diff^6 \xi \le[
		\sqrt{-\det\le(g + i \tilde H\ri)} + \frac{\sqrt{-\det g}}{4 (\p a)^2} \p_m a H^{*mnl} H_{mnp} \p^p a
	\ri]
	\nonumber \\ & \hspace{4cm}
	+ \tfive \int_\Sigma \le(P[C_6] + \frac{1}{2} F_3 \wedge P[\C3] \ri).
\end{align}
where \(H \equiv F_3 + P[\C3]\), \(H^{*mnl} \equiv \frac{1}{6\sqrt{-g}}\e^{mnlpqr} H_{pqr}\), and \(\tilde H_{mn} \equiv H^{*}_{mn}{}^l \p_l a /\sqrt{(\p a)^2}\). The tension is given in terms of the Planck length by \(\tfive = 1/(2\pi)^5 \lp^6\).

We seek brane embeddings that span the defect (the (\(t,x\)) plane) at the boundary. We will study only a single M2-brane embedding, with \ads[3] world volume. For the M5-brane embeddings, near the boundary the geometry of the brane's world volume will be \(\ads[3] \times \sph[3]\), where the \sph[3] is either the \sph[3] inside \ads[7] parameterised by the \(\f_i\), or is internal to the \sph[4], parameterised by the \(\chi_i\). This \sph[3] is supported by flux of the world volume gauge field \(A\), sourced by M2-brane charge dissolved within the M5-brane. The total number of dissolved M2-branes \(N_2\) is given by the flux quantization condition~\cite{Camino:2001at}
\begin{equation} \label{eq:flux_quantization}
	N_2 = \frac{T_\mathrm{M2}}{2\pi} \int_{\sph[3]} F_3.
\end{equation}

A subset of the M5-brane embeddings we consider are believed to be dual to half-BPS Wilson surface operators, in representations described by Young tableaux with number of boxes \(N_2\) of the order of the rank (\(\sim N_5\)) of the gauge algebra or smaller \cite{Lunin:2007ab,Chen:2007ir,Mori:2014tca}. This is analogous to the holographic description of Wilson lines in \(\mathcal{N}=4\) super Yang-Mills theory (SYM) by D-branes \cite{Drukker:1999zq,Drukker:2005kx,Gomis:2006sb,Gomis:2006im}. The probe approximation should hold provided the condition~\eqref{eq:probe_limit_general} is satisfied, which for \(N_2\) M2-branes implies \(N_2 \ll N_5^2\). We will always assume that this is the case. The holographic entanglement entropy for Wilson surfaces of arbitrary shape, but with \(N_2 \gg 1\), is calculated in refs.~\cite{Gentle:2015jma,Estes:2018tnu}.

To holographically describe Wilson lines in SYM, one must add boundary terms to the D-brane action which implement a Legendre transformation with respect to the brane's position and gauge field \cite{Drukker:1999zq,Drukker:2005kx}. The former is needed because a string describing a Wilson line obeys complementary boundary conditions to a string ending on a D-brane. The latter fixes the total amount of fundamental string charge dissolved in the brane, and thus the representation of the Wilson line.

For M2- and M5-branes, we will use an analogous boundary term,
\begin{equation} \label{eq:boundary_term}
    S_\mathrm{bdy} = -\int_{\p\Sigma} \diff^p\s \, r \frac{\d S_\mathrm{brane}}{\d(\p_n r)} = \frac{1}{2} \int_{\p\Sigma} \diff^p\s \, z \frac{\d S_\mathrm{brane}}{\d(\p_n z)},
\end{equation}
where \(\s\) are coordinates on \(\p\Sigma\), the intersection of the brane with the boundary of \ads, and \(p = 2\) or 5 for an M2- or M5-brane, respectively. This implements a Legendre transformation with respect to the position \(r\) of the end of the brane at infinity. There is no need to Legendre transform with respect to the gauge field, as the dissolved M2-brane charge is already fixed by the flux quantization condition \eqref{eq:flux_quantization}.

\subsection{Entanglement entropy and probe branes}

In order to compute the defect contribution to entanglement entropy, we will use the methods of refs.~\cite{Casini:2011kv,Jensen:2013lxa,Lewkowycz:2013nqa,Karch:2014ufa}, reviewed in sections~\ref{sec:casini_huerta_myers} and~\ref{sec:generalised_gravitational_entropy}, which allow the leading order contribution in the probe limit to be obtained without any knowledge of back-reaction.

In terms of the cylindrical coordinate system~\eqref{eq:ads_7_metric}, the map~\eqref{eq:hyperbolic_slicing_diffeomorphism}  hyperbolic space may be written as
\begin{alignat}{2} \label{eq:map_to_hyperbolic}
		t &=  \Omega^{-1} \ell \sqrt{v^2 - 1} \sinh \tau,
		\qquad
		& & z = \Omega^{-1} \ell,
		\nonumber \\
		\r &= \Omega^{-1} \ell v \sinh  u \sin \f_0,
		& & x = \Omega^{-1} \ell v \sinh  u \cos \f_0,
\end{alignat}
where \(\Omega = v \cosh u + \sqrt{v^2 - 1} \cosh \tau\), with other coordinates unchanged. The gauge field strength \(F_4\) is unchanged under this transformation, while the \ads[7] is put into hyperbolic slicing~\eqref{eq:hyperbolic_slicing_metric}. The full \(\ads[7] \times \sph[4]\) metric in these coordinates is
\begin{equation} \label{eq:hyperbolic_slicing}
	\diff s^2 = 4L^2 \le[
		\frac{\diff v^2}{f(v)} - f(v) \diff \tau^2 + v^2 \diff u^2 + v^2 \sinh^2 u \, \diff \f_0^2 + v^2 \sinh^2 u \sin^2 \f_0 \diff s_{\sph[3]}^2
	\ri]
	+ L^2 \diff s_{\sph[4]}^2,
\end{equation}
where \(f(v) = v^2 - 1\). The metric \eqref{eq:hyperbolic_slicing} and gauge field \eqref{eq:ads_7_field_strength} remain a solution to the 11D SUGRA equations of motion with the more general metric function~\eqref{eq:hyperbolic_metric_function} (with \(d=6\)), which changes the inverse temperature \(\b=1/T\) to that given by~\eqref{eq:hyperbolic_horizon_position}.

At leading order in the probe limit, the contribution of the brane to the free energy in the dual CFT in hyperbolic space is
\begin{equation} \label{eq:hyperbolic_free_energy}
	F^{(1)}(\b) = \b^{-1} I^\star_\mathrm{brane}(\b),
\end{equation}
where \(I_\mathrm{brane}^\star(\b)\) is the on-shell action of the brane in Euclidean signature,\footnote{We put the metric \eqref{eq:hyperbolic_slicing} in Euclidean signature by a Wick rotation \(\t \to i \tilde \t\). We will abuse notation slightly by dropping the tilde on the Euclidean time coordinate.} with \(\t \sim \t + \b \). The contribution from the brane to the R\'enyi and entanglement entropies are then given by~\eqref{eq:probe_brane_renyi} and~\eqref{eq:probe_brane_entropy}, respectively.

When the brane embedding breaks conformal symmetry, it is no longer possible to perform the conformal transformation to hyperbolic space. However, the bulk coordinate change to hyperbolic slicing, without changing the defining function, remains a convenient way of computing generalised gravitational entropy~\eqref{eq:probe_generalised_gravitational_entropy_beta}~\cite{Lewkowycz:2013nqa,Karch:2014ufa}.

A subtlety that affects the results of section~\ref{sec:symmetric} comes from the six-form gauge field. In hyperbolic slicing, its field strength is given at all temperatures by
\begin{equation}
	F_7 = 6 (2L)^6 v^5 \sinh^4 u \sin^3 \f_0 \sin^2 \f_1 \sin \f_2 \, \diff \t \wedge \diff v \wedge \diff u \wedge  \diff \f_0 \wedge \diff \f_1 \wedge \diff \f_2 \wedge \diff \f_3. 
\end{equation}
We consider M5-brane embeddings with boundaries. The on-shell action of such an M5-brane may change by boundary terms under gauge transformations of \(C_6\)~\cite{Drukker:2005kx}. The consequence for us is that the entanglement entropy for solutions presented in section~\ref{sec:symmetric} will depend on the choice of gauge for \(C_6\) in the hyperbolic slicing,\footnote{It is plausible that there exists some boundary term which cancels the gauge dependence, but the form of this boundary term is not known to us.} so we must be careful to choose the appropriate gauge. The same phenomenon occurs in the computation of entanglement entropy for defects dual to D3-branes in type IIB SUGRA~\cite{Kumar:2017vjv}.

We will choose a gauge which is quite natural given the manifest symmetries of the hyperbolic slicing,\footnote{The authors of~\cite{Kumar:2017vjv} chose a gauge which was natural in the Rindler slicing of \(\ads\). We have checked that doing so in our case does not change our results for the entanglement entropy.}
\begin{equation} \label{eq:hyperbolic_c6_gauge}
	C_6 = (2L)^6 (v_H^6 - v^6) \sinh^4 u \sin^3 \f_0 \sin^2 \f_1 \sin \f_2 \, \diff \t \wedge \diff u \wedge  \diff \f_0 \wedge \diff \f_1 \wedge \diff \f_2 \wedge \diff \f_3. 
\end{equation}
Note that this gauge is not the result of performing the coordinate transformation \eqref{eq:map_to_hyperbolic} on the flat slicing gauge potential \eqref{eq:flat_slicing_c6}. In section~\ref{sec:symmetric}, we confirm that with this gauge choice we obtain the same result for the entanglement entropy of a symmetric representation Wilson surface as that computed in ref.~\cite{Estes:2018tnu, Gentle:2015jma} using the Ryu-Takayanagi prescription in the fully back-reacted geometry. The latter calculation is independent of the choice of gauge for \(C_6\), so this agreement supports~\eqref{eq:hyperbolic_c6_gauge} as the correct gauge.

\section{Single M2-brane}
\label{sec:m2}

In this section, we compute the entanglement entropy contribution from a single M2-brane, believed to be holographically dual to a Wilson surface operator in the fundamental representation. We begin by reviewing the embedding of the M2-brane in flat slicing \cite{Lunin:2007ab}. 

The M2-brane is described by the action \eqref{eq:m2_action}. We choose static gauge, parameterising the brane by \(\xi = (t,x,z)\), and take as an ansatz
\(
	\r = \r(z),
\)
with boundary condition \(\lim_{z\to0}\r = 0\). The pullback of \(\C3\) \eqref{eq:c3} onto the brane vanishes with this ansatz, and the action reduces to
\begin{equation} \label{eq:m2_action_ansatz}
	S_\mathrm{M2} = - T_\mathrm{M2} \int \diff t \diff x \diff z \frac{8 L^3}{z^3} \sqrt{1 + \r'(z)^2}.
\end{equation}
This is minimised for constant \(\r\), so the solution obeying the boundary condition \(\r = 0\) at \(z=0\) is
\begin{equation}
	\r(z) = 0.
\end{equation}

Substituting this solution into~\eqref{eq:m2_action_ansatz}, we find the on-shell action
\begin{equation}
	S^\star_\mathrm{M2} = -8 T_\mathrm{M2} L^3 \int \diff t \diff x \diff z \frac{1}{z^3} = - \frac{N_5}{\pi \e^2} \int \diff t \diff x,
\end{equation}
where on the right hand side we have performed the integral over \(z\). Since this integral is UV divergent, we have implemented a cutoff at small \(z = \e\), the same cutoff as used in \eqref{eq:entanglement_general_form}. The boundary term \eqref{eq:boundary_term} turns out to precisely cancel the bulk contribution to the action, so the full on-shell action vanishes,
\begin{equation}
    S^\star_\mathrm{M2} + S^\star_\mathrm{bdy} = 0.
\end{equation}

Mapping to hyperbolic slicing using \eqref{eq:map_to_hyperbolic}, the solution spans \((\t,v,u)\) and obeys \(\sin\f_0 = 0\). It is straightforward to check that this remains a solution for all temperatures in the hyperbolic slicing. Substituting this solution into the Euclidean action for the M2-brane, we obtain the on-shell action as a function of temperature
\begin{equation} \label{eq:m2_hyperbolic_on_shell_action}
	I_\mathrm{M2}^\star = \frac{4 N_5}{\pi} \int_0^\b \diff \tau \int_{v_H}^\Lambda \diff v \int_0^{u_c} \diff u \, v
	= \frac{2 N_5}{\pi} \b \le(\Lambda^2 - v_H^2\ri) u_c.
\end{equation}
We have imposed upper limits \(\Lambda\) and \(u_c\) to regulate the integrals over \(v\) and \(u\). The term which diverges as \(\Lambda \to \infty\) is removed by the boundary term \eqref{eq:boundary_term},
\begin{equation} \label{eq:m2_hyperbolic_on_shell_action_boundary}
	I_\mathrm{bdy}^\star = - \frac{2 N_5}{\pi} \b \Lambda^2 u_c.
\end{equation}
The divergence arising from the limit \(u_c \to \infty\) is not removed by this boundary term. In fact, this divergence is physical, and leads to the expected logarithmic divergence in the entanglement entropy. In terms of the cutoff at small \(z=\e\), the large \(u\) cutoff is given by~\cite{Jensen:2013lxa}
\begin{equation} \label{eq:cutoff_identification}
	u_c = \ln\le( \frac{2\ell}{\e} \ri) + \mathcal{O}\le( \frac{\e^2}{\ell^2} \ri).	 
\end{equation}

Making use of equation \eqref{eq:hyperbolic_horizon_position} for the position of the horizon, we obtain the contribution of the brane to the free energy as a function of inverse temperature,
\begin{equation} \label{eq:m2_hyperbolic_free_energy}
	F^{(1)}(\b) = \b^{-1}(I_\mathrm{M2}^\star + I_\mathrm{bdy}^\star)= - \frac{2 N_5}{9 \pi \b^2}\le(
	\pi + \sqrt{\pi^2 + 6 \b^2}
	\ri)^2 u_c.
\end{equation}
Substituting this into \eqref{eq:probe_brane_renyi}, and identifying the large \(u\) and small \(z\) cutoffs using~\eqref{eq:cutoff_identification}, we find the contribution of the brane to the R\'enyi entropies to be
\begin{equation} \label{eq:m2_renyi}
	S_q^{(1)} = \frac{2}{9} N_5 \frac{1 - 6 q^2 + \sqrt{1 + 24 q^2} }{
		q(1-q)
	}\ln\le(
		\frac{2 \ell}{\e}
	\ri) + \mathcal{O}\le( \frac{\e^2}{\ell^2} \ri).
\end{equation}
This calculation matches the result in equation (3.33) of \cite{Jensen:2013lxa},\footnote{To obtain our result, set \(d=6\) and \(n=4\) in their formula.} which applies to probe branes in \(\ads\) described by the same bulk action but with different boundary terms. The two calculations agree because the boundary terms are equal when the equations of motion are satisfied.

The entanglement entropy is obtained from the limit \(q \to 1\) of the R\'enyi entropies~\eqref{eq:m2_renyi},
\begin{equation} \label{eq:m2_ee}
	\see^{(1)} = \frac{8}{5} N_5 \ln \le(\frac{2 \ell}{\e} \ri) + \mathcal{O}\le( \frac{\e^2}{\ell^2} \ri).
\end{equation}
Other physically interesting limits are \(q \to 0\) and \(q \to \infty\),
\begin{subequations}
\begin{align}
	S_{q\to 0}^{(1)} &= \frac{4}{9q} N_5 \ln \le( \frac{2 \ell}{\e} \ri) + \mathcal{O}(q^0),
	\\
	\lim_{q\to\infty} S_{q}^{(1)} &= \frac{4}{3} N_5 \ln \le( \frac{2\ell}{\e} \ri),
\end{align}
\end{subequations}
where we neglect terms which vanish as \(\e\to0\).

From the coefficient of the logarithm in the entanglement entropy~\eqref{eq:m2_ee} we obtain
\begin{equation} \label{eq:m2_central_charge}
	c = \frac{24}{5} N_5,
\end{equation}
reproducing the result of \cite{Gentle:2015jma,Estes:2018tnu} in the fundamental representation. This central charge suggests that the number of massless degrees of freedom of a fundamental representation Wilson surface scales as \(N_5\), as opposed to the \(N_5^3\) scaling of the degrees of freedom in the bulk \(\mathcal{N}=(2,0)\) theory. This is the same scaling found from the chiral R-symmetry anomaly for a single M2-brane stretched between parallel M5-branes~\cite{Berman:2004ew}.

As a check of our results, we compare the expression~\eqref{eq:m2_central_charge} for \(c\) to that obtained from the Weyl anomaly~\eqref{eq:defect_entropy_weyl_relation}. The holographic Weyl anomaly for a fundamental representation Wilson surface was calculated in ref.~\cite{Graham:1999pm}. In our notation, their result~is
\begin{equation} \label{eq:graham_witten_anomaly}
	b = d_1 = d_2 = 12 N_5.
\end{equation}
Substituting \(b\) and \(d_2\) into the identity~\eqref{eq:defect_entropy_weyl_relation}, we find
\begin{equation} 
	c = b - \frac{3}{5} d_2 = \frac{24}{5}N_5,
\end{equation}
reproducing~\eqref{eq:m2_central_charge}.

\section{M5-branes wrapping \sph[3] \(\subset\) \sph[4]}
\label{sec:antisymmetric}

In this section we will seek solutions wrapping an \sph[3] internal to the \sph[4] factor of the background geometry. When the world volume of the brane takes the form \(\ads[3] \times \sph[3]\), such a solution is expected to be dual to a Wilson surface in an antisymmetric representation~\cite{Lunin:2007ab}, corresponding to a Young tableau consisting of a single column. The number of boxes \(N_2\) in the tableau is equal to the amount of M2-brane charge dissolved in the M5-brane, determined from the flux quantization condition~\eqref{eq:flux_quantization}.

\subsection{The solution in flat slicing}

Let us parameterise the brane by \(\xi = (t,x,z,\c_1,\c_2,\c_3)\), gauge fix the auxiliary scalar field to \(a = z\), and employ an ansatz
\begin{equation}
    \q = \q(z),
    \quad
    F_3 = \frac{4 L^3 N_2}{N_5} \sin^2 \c_1 \sin \c_2 \, \diff \c_1 \wedge \diff \c_2 \wedge \diff \c_3,
\end{equation}
with \(\r = 0\). One can verify that this ansatz satisfies the equations of motion for the gauge field. Substituting the ansatz and integrating over \(\c_{1,2,3}\), the PST action~\eqref{eq:pst_action} becomes
\begin{equation} \label{eq:flow_action}
	S_\mathrm{M5} = - \frac{N_5^2}{4\pi} \int \diff t \diff x \diff z \frac{1}{z^3} \sqrt{
		\le(D(\q)^2 + \sin^6 \q\ri) \le(4 + z^2 \q'^2\ri)
	},
\end{equation}
where
\begin{equation}
	D(\q) = 3 \cos\q - \cos^3 \q - 2 + \frac{4 N_2}{N_5}.
\end{equation}

The Euler-Lagrange equation for \(\q\) is
\begin{equation}
	\p_z \le(
		\frac{\q'}{z} \sqrt{
			\frac{ D(\q)^2 + \sin^6 \q}{4 + z^2 \q'^2}
		}
	\ri)
	+ \frac{3 \sin^3 \q}{z^3} \le( D(\q) - \cos \q \sin^2\q \ri) \sqrt{
		\frac{4 + z^2 \q'^2}{D(\q)^2 + \sin^6 \q}
	} = 0.
\end{equation}
This is satisfied by any solution to the first order BPS condition \cite{Camino:2001at,Gomis:1999xs}
\begin{equation} \label{eq:flow_bps}
	\q' = - \frac{2}{z} \frac{
			\p_{\q} \le( D(\q) \cos \q + \sin^4 \q \ri)
		}{
			D(\q) \cos \q + \sin^4 \q
		},
\end{equation}
which ensures that the brane embedding preserves one quarter of the supersymmetries of the background solution.

The BPS condition \eqref{eq:flow_bps} possesses two classes of solutions with constant \(\q\). One class is the antisymmetric Wilson surface~\cite{Lunin:2007ab,Chen:2007ir,Mori:2014tca}, corresponding to a representation of \(\mathfrak{su}(N_5)\) with a Young tableau consisting of \(N_2\) boxes. For these solutions,
\begin{equation} \label{eq:antisymmetric_wilson_surface_angle}
	\cos\q = 1 - 2N_2/N_5.
\end{equation}
The other class of solution sits at either the north or south pole of the \sph[4], \(\q = 0\) or \(\pi\), with arbitrary \(N_2\). The wrapped \sph[3] therefore collapses to zero size and this solution corresponds to a bundle of \(N_5\) M2-branes~\cite{Gomis:1999xs}.

\begin{figure}
\begin{subfigure}{0.5\textwidth}
	\begin{center}
	\includegraphics[height=4cm]{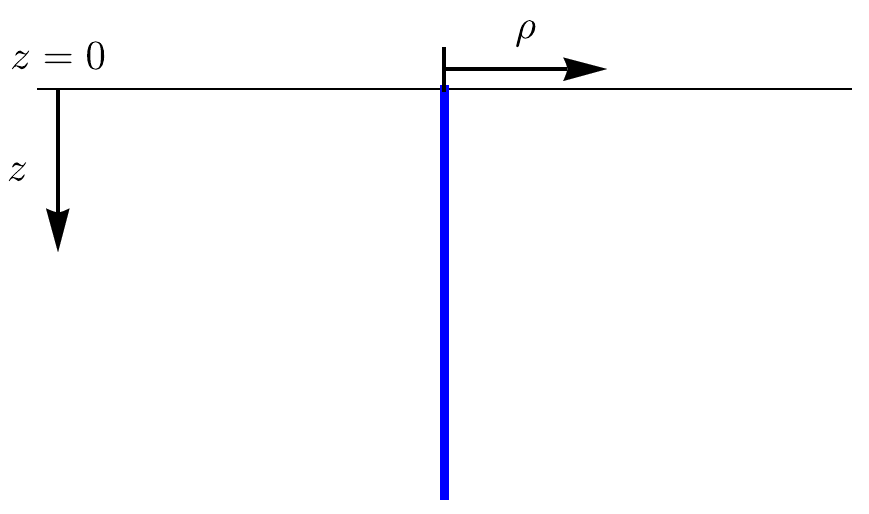}
	\caption{\ads[7]}
	\end{center}
\end{subfigure}
\begin{subfigure}{0.5\textwidth}
	\begin{center}
	\includegraphics[height=4cm]{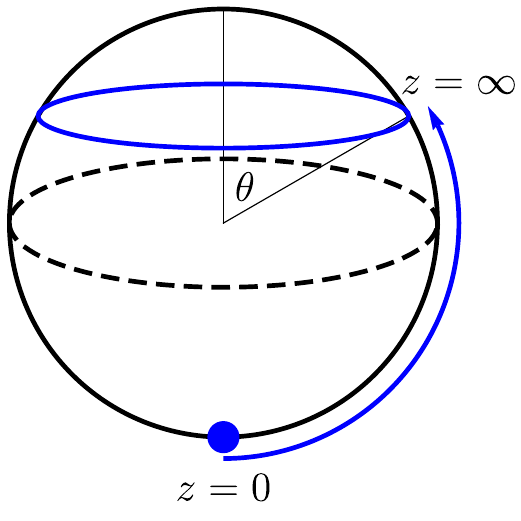}
	\caption{\sph[4]}
	\end{center}
\end{subfigure}
\caption[The brane embedding corresponding to the antisymmetric flow solution.]{Cartoon of the antisymmetric flow M5-brane embedding in \(\ads[7] \times \sph[4]\). \textbf{(a):}~In the \ads[7] factor of the geometry, the brane (the thick blue line) spans the directions \((t,x,z)\) (the \(t\) and \(x\) directions are suppressed in this figure) and occupies \(\r=0\). The thin, horizontal, black line in this figure is the boundary of \ads[7]. \textbf{(b):}~In the UV (\(z\to0\)) the M5-brane collapses to the south pole of the \sph[4].  For non-zero \(z\), the M5-brane wraps an \sph[3] at a polar angle \(\q(z)\) which decreases with increasing \(z\). In the IR (\(z \to \infty\)) this angle tends to a value~\eqref{eq:antisymmetric_wilson_surface_angle}, determined by the dissolved M2-brane charge. The UV and IR solutions correspond to a bundle of M2-branes and the antisymmetric Wilson surface, respectively.}
\label{fig:antisymmetric_flow_cartoon}
\end{figure}

To obtain solutions where \(\q\) depends non-trivially on \(z\), we integrate the BPS condition~\eqref{eq:flow_bps} to obtain
\begin{equation} \label{eq:antisymmetric_flow_solution}
	\cos \q(z) = m^4 z^4 - \sqrt{
		\le(1 - m^4 z^4 \ri)^2 + \frac{4 N_2}{N_5} m^4 z^4
	},
\end{equation}
where \(m\) is an integration constant. This solution tends to each of the constant \(\q\) solutions in opposite limits. In the UV, \(m z \to 0\), \(\q  \to \pi\) and the solution collapses to the bundle of M2-branes. In the IR, \(m z \gg 1\), the solution becomes \(\cos\q \approx 1 - 2N_2/N_5\), the antisymmetric Wilson surface. The solution therefore describes an RG flow from the bundle of M2-branes to the antisymmetric Wilson surface, which we will refer to as the antisymmetric flow.  We sketch this embedding in figure~\ref{fig:antisymmetric_flow_cartoon}. 
Expanding \(\q(z)\) for small \(z\) and using~\eqref{eq:generic_scalar_expansion}, we find that that \(\q\) is dual to an operator with \(\D=2\), and that the flow is driven by a non-zero VEV for this operator.

The antisymmetric flow only exists for \(N_2 < N_5\). One way to see this is to rearrange the solution~\eqref{eq:antisymmetric_flow_solution} into the form
\begin{equation}
	2 m^4 z^4 = \frac{\sin^2\q}{1 - \cos \q - 2N_2/N_5}.
\end{equation}
The left-hand side is manifestly positive, while the right hand side is negative for all \(\q\) unless \(N_2 \leq N_5\). If \(N_2 = N_5\), then the solution reduces to the bundle of M2-branes.

Substituting the BPS condition \eqref{eq:flow_bps} into the action \eqref{eq:flow_action}, the on-shell action may be written in the form
\begin{align} \label{eq:antisymmetric_flow_on_shell_action_bulk}
	S_\mathrm{M5}^\star &= - \frac{N_5^2}{2 \pi} \int \diff t \diff x \diff z \frac{1}{z^3} \frac{D(\q)^2 + \sin^6 \q}{D(\q) \cos \q + \sin^4\q}
	\nonumber \\
	&= \frac{N_5^2}{4\pi} \int \diff t \diff x \diff z \, \p_z \le[
		\frac{1}{z^2} \le(
		D(\q) \cos\q + \sin^4 \q
	\ri)
	\ri]
\end{align}
The boundary term \eqref{eq:boundary_term} evaluates to
\begin{equation} \label{eq:antisymmetric_flow_on_shell_action_boundary}
	S_\mathrm{bdy} = \frac{N_5^2}{4\pi} \int \diff t \diff x \frac{1}{\e^2} \le.\le(
		D(\q) \cos\q + \sin^4 \q
	\ri)\ri|_{z=\e}.
\end{equation}
Noting that the integration over \(z\) in~\eqref{eq:antisymmetric_flow_on_shell_action_bulk} has limits \([\e,\infty)\), and that the contents of the square brackets vanish in the limit \(z \to \infty\), we see that the bulk and boundary contributions cancel. Hence the contribution of the brane to the on-shell action vanishes in flat slicing.

The Weyl anomaly coefficients of the defect dual to the bundle of M2-branes are given by \((N_5 - N_2)\) times those for the fundamental representation Wilson surface~\eqref{eq:graham_witten_anomaly},
\begin{equation} \label{eq:antisymmetric_weyl_anomaly}
	b = d_1 = d_2 = 12 N_5 (N_5 - N_2).
\end{equation}
For the antisymmetric representation Wilson surface, two of the Weyl anomaly coefficients, \(b\) and \(d_2\), were calculated holographically in ref.~\cite{Jensen:2018rxu},
\begin{equation}
	b = d_2 = 12 N_2 (N_5 - N_2).
\end{equation}
Since the antisymmetric flow only exists for \(N_5 > N_2\), we find that both \(b\) and \(d_2\) are larger in the UV than in the IR.

\subsection{Entanglement entropy of the antisymmetric Wilson surface}

In this section we compute the entanglement entropy contribution from the M5-brane embedding with constant \(\q = \cos^{-1} \le(1 - 2N_2/N_5\ri)\), the antisymmetric Wilson surface.

In hyperbolic slicing, the solution at inverse temperature \(\b_0 = 2\pi\) may be obtained by a coordinate transformation from flat space. It spans \((\t,v,u)\) and satisfies \(\sin \f_0 = 0\). It is straightforward to verify that this is still a solution for arbitrary temperatures in the hyperbolic slicing.

Substituting this solution into the PST action, and Wick rotating to Euclidean signature, we find the bulk contribution to the Euclidean on-shell action to be
\begin{equation} \label{eq:antisymmetric_hyperbolic_action}
	I_\mathrm{M5}^\star = \frac{4 N_2 (N_5 - N_2)}{\pi} \int_0^\b \diff \tau \int_{v_H}^\Lambda \diff v \int_0^{u_c} \diff u \, v
	= \frac{2N_2(N_5 - N_2)}{\pi} \b \le(\Lambda^2 - v_H^2\ri) u_c.
\end{equation}
This is \(N_2(N_5-N_2)/N_5\) times the result for the M2-brane~\eqref{eq:m2_hyperbolic_on_shell_action}. The same is true for the boundary term,
\begin{equation} \label{eq:antisymmetric_hyperbolic_action_boundary}
	I_\mathrm{bdy}^\star = - \frac{2 N_2 (N_5-N_2)}{\pi} \b \Lambda^2 u_c,
\end{equation}
and therefore the contribution from the brane to the free energy in hyperbolic slicing is given by
\begin{equation} \label{eq:antisymmetric_hyperbolic_free_energy}
	F^{(1)}(\b) = - \frac{2 N_2 (N_5-N_2)}{9 \pi \b^2} \le(
		\pi + \sqrt{\pi^2 + 6 \b^2}
	\ri)^2 u_c.
\end{equation}

Substituting the free energy into \eqref{eq:probe_brane_renyi} and using \eqref{eq:cutoff_identification} to relate \(u_c\) and \(\e\), we find that the contribution of the antisymmetric Wilson surface to the \(q\)-th R\'enyi entropy is given by
\begin{equation}
    S_q^{(1)} = \frac{2}{9} N_2(N_5-N_2) \frac{1 - 6 q^2 + \sqrt{1 + 24 q^2} }{q (1-q)} \ln \le( \frac{2 \ell}{\e} \ri) + \cO\le( \frac{\e^2}{\ell^2} \ri).
\end{equation}
Taking the limit \(q \to 1\), the entanglement entropy contribution from the Wilson surface is
\begin{equation} \label{eq:antisymmetric_ee}
    \see^{(1)} = \frac{8}{5} N_2 (N_5-N_2) \ln \le( \frac{2 \ell}{\e} \ri) + \cO\le( \frac{\e^2}{\ell^2} \ri).
\end{equation}
In the limits of small and large \(q\), we find respectively
\begin{subequations}
\begin{align}
	S_{q\to 0}^{(1)} &= \frac{4}{9q} N_2(N_5 - N_2) \ln \le( \frac{2 \ell}{\e} \ri) + \mathcal{O}(q^0),
	\\
	\lim_{q\to\infty} S_{q}^{(1)} &= \frac{4}{3} N_2 (N_5 - N_2) \ln \le( \frac{2\ell}{\e} \ri).
\end{align}
\end{subequations}

From the entanglement entropy~\eqref{eq:antisymmetric_ee} we find
\begin{equation} \label{eq:antisymmetric_central_charge}
    c = \frac{24}{5} N_5 (N_2 - N_5).
\end{equation}
This reproduces the central charge obtained for an antisymmetric representation in \cite{Gentle:2015jma,Estes:2018tnu}. It is invariant under the replacement \(N_2 \to (N_5 - N_2)\), corresponding to complex conjugation of the representation of \(\mathfrak{su}(N_5)\), and reduces to \(N_2\) times the central charge~\eqref{eq:m2_central_charge} of a single M2-brane for \(N_2  \ll N_5\). The entanglement entropy~\eqref{eq:antisymmetric_central_charge} and Weyl anomaly coefficients~\eqref{eq:antisymmetric_weyl_anomaly} satisfy the relation~\eqref{eq:defect_entropy_weyl_relation}.

\subsection{Entanglement entropy of the antisymmetric flow solution}
\label{sec:antisymmetric_on_shell}

In hyperbolic slicing, we parameterise the brane by \((\t,v,u,\c_1,\c_2,\c_3)\), and gauge fix the auxiliary scalar field to be given by \(a = v\). The embedding will be specified by the function \(\q = \q(\t,v,u)\). As before, the gauge field strength is determined by the flux quantization condition,
\begin{equation}
    F_3 = \frac{4 L^3 N_2}{N_5} \sin^2\c_1 \sin\c_2 \, \diff \c_1 \wedge  \diff \c_2 \wedge\diff \c_3.
\end{equation}

Substituting this ansatz into the action and integrating out the wrapped \sph[3], we find that the Euclidean action for the M5-brane, with arbitrary \(v_H\) but with \(\t \sim \t + 2\pi\), is
\begin{equation} \label{eq:antisymmetric_flow_hyperbolic_action}
    [I_\mathrm{brane}(\b)]_{2\pi} = \int_0^{2\pi} \diff \tau \int_{v_H}^\infty \diff v \int_0^{u_c} \diff u \, \mathcal{L},
\end{equation}
where
\begin{equation} \label{eq:antisymmetric_flow_hyperbolic_lagrangian}
    \mathcal{L} = \frac{N_5^2}{4\pi} v \sqrt{
        \le( D(\q)^2 + \sin^6 \q \ri)
        \le(
            4 + \frac{1}{f(v)} (\p_\t \q)^2 + f(v) (\p_v \q)^2 + \frac{1}{v^2} (\p_u \q)^2
        \ri)
    }.
\end{equation}
The entanglement entropy is obtained using \eqref{eq:probe_generalised_gravitational_entropy_beta}; we differentiate the off-shell action with respect to \(\b\), set \(\b = 2\pi\), and take \(\q\) on-shell.

The resulting integral for the entanglement entropy must be performed numerically. In appendix \ref{app:antisymmetric_flow} we give some details on how we manipulate the integrals into a form suitable for numerical evaluation. As for similar embeddings of D5-branes in \(\ads[5]\times \sph[5]\)~\cite{Kumar:2017vjv}, it is convenient to perform a coordinate transformation back to flat slicing, where the embedding is much simpler. The result is that the entanglement entropy is given by\footnote{We use \(x^0\) to denote the Euclidean time coordinate in flat slicing.}
\begin{align} \label{eq:antisymmetric_flow_ee}
    \see^{(1)} &= \frac{2 N_5^2}{5} \int_\e^\ell \diff z \frac{\ell}{z \sqrt{\ell^2 - z^2}} \frac{D(\q)^2 + \sin^6\q}{D(\q) \cos\q + \sin^4\q}
    \nonumber \\ &\phantom{=}
    - \frac{8 N_5^2 \ell^4}{5\pi} \int_{z\geq\e} \diff x^0 \diff x \diff z \frac{z\mathbf{N}_1}{\mathbf{D}_1} \frac{
        \le( D(\q) \sin \q - \cos \q \sin^3 \q \ri)^2
        }{
        D(\q) \cos\q + \sin^4 \q
    },
\end{align}
where \(\q\) is given by the solution \eqref{eq:antisymmetric_flow_solution}, and
\begin{subequations} \label{eq:antisymmetric_flow_ee_factors}
    \begin{align}
       \mathbf{N}_1 &=
            \left[\ell^4+2 \ell^2 \left(x_0-x\right) \left(x+x_0\right)+\left(x^2+x_0^2\right)^2\right]^2
            \nonumber \\ &\phantom{=}
            -2 z^4 \left[\ell^4+\ell^2 \left(6
            x_0^2-2 x^2\right)+x^4+6 x^2 x_0^2+5 x_0^4\right]
            \nonumber \\ &\phantom{=}
            -4 x_0^2 z^2 \left[(\ell-x)^2+x_0^2\right] \left[(\ell+x)^2+x_0^2\right]-4
            x_0^2 z^6+z^8,
       \\
       \mathbf{D}_1 &=
            \left[(\ell-x)^2+x_0^2+z^2\right]^2 \left[(\ell+x)^2+x_0^2+z^2\right]^2
            \nonumber \\ &\phantom{=}
            \times \left[\ell^4+2 \ell^2 \left(-x^2+x_0^2-z^2\right)+\left(x^2+x_0^2+z^2\right)^2\right]^2.
    \end{align}
\end{subequations}

The entanglement entropy \eqref{eq:antisymmetric_flow_ee} is logarithmically divergent at small \(\e\),
\begin{equation}
    \see^{(1)} = \frac{8}{5} N_5 (N_5 - N_2) \ln \le( \frac{2\ell}{\e} \ri) + \mathcal{O}(\e^0).
\end{equation}
The divergent term is the entanglement entropy of the UV solution, namely \((N_5-N_2)\) times the entanglement entropy~\eqref{eq:m2_ee} of a single M2-brane. We will obtain a UV-finite quantity by subtracting this contribution, yielding the difference \(\Delta \see^{(1)}\) between the entanglement entropy of the flow solution and the bundle of M2-branes,
\begin{equation}
    \Delta \see^{(1)} = \see^{(1)} - \frac{8}{5} N_5 (N_5 - N_2) \ln \le( \frac{2 \ell}{\e} \ri).
\end{equation}
Differentiating our numerical results for \(\D\see^{(1)}\) using a finite difference method, we obtain \(C(\ell)\) using~\eqref{eq:b_function}.

\begin{figure}[t!]
	\begin{subfigure}{0.5\textwidth}
		\includegraphics[width=\textwidth]{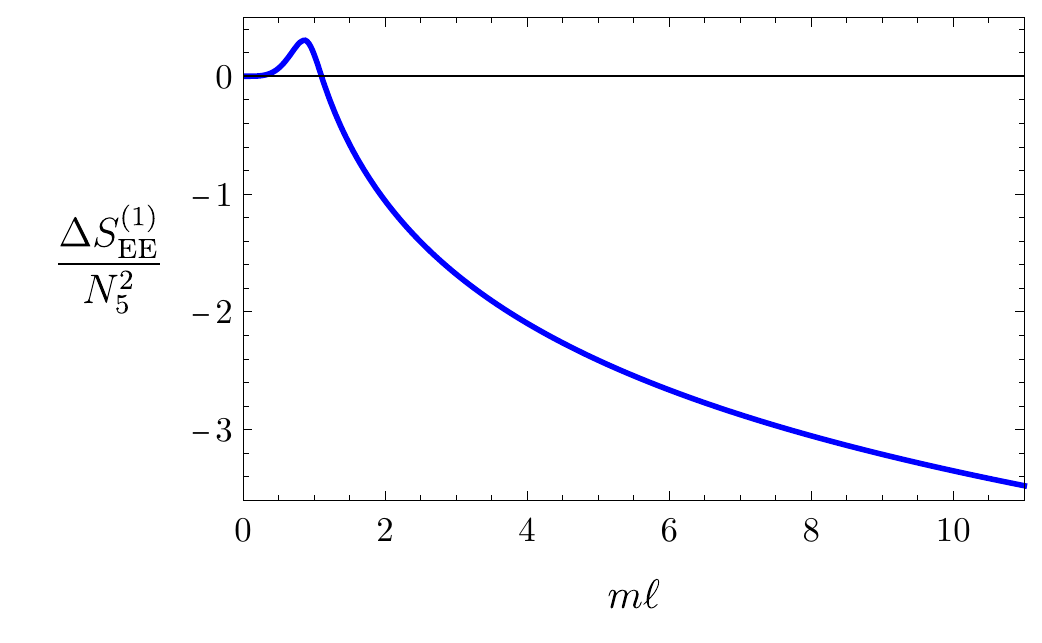}
		\caption{\(\dfrac{N_2}{N_5} = \dfrac{1}{10}\)}
	\end{subfigure}
	\begin{subfigure}{0.5\textwidth}
		\includegraphics[width=\textwidth]{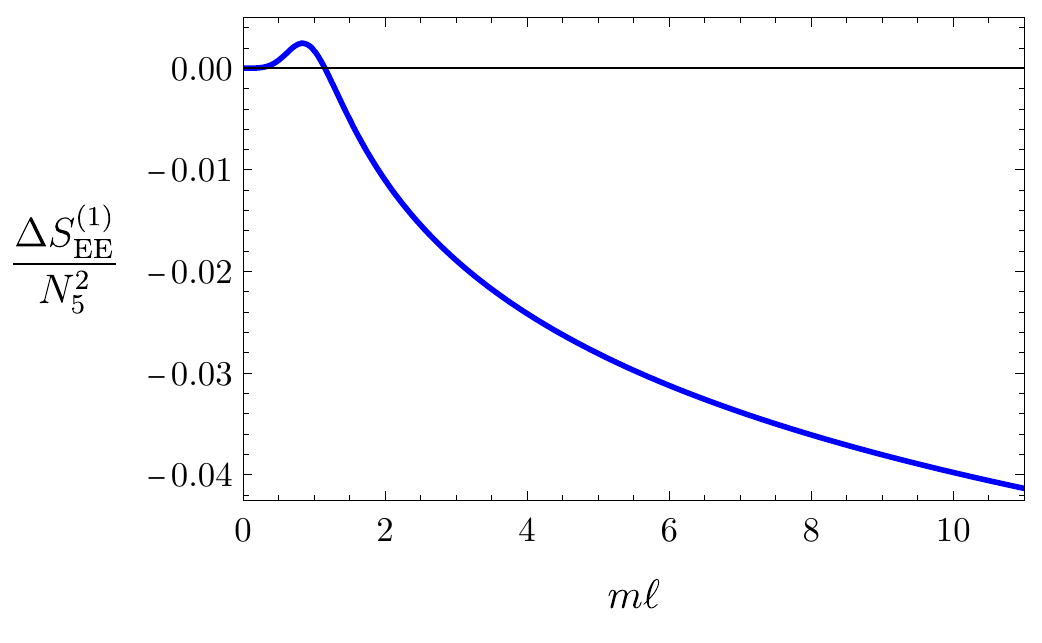}
		\caption{\(\dfrac{N_2}{N_5} = \dfrac{9}{10}\)}
	\end{subfigure}
	\\[1em]
	\begin{subfigure}{0.5\textwidth}
		\includegraphics[width=\textwidth]{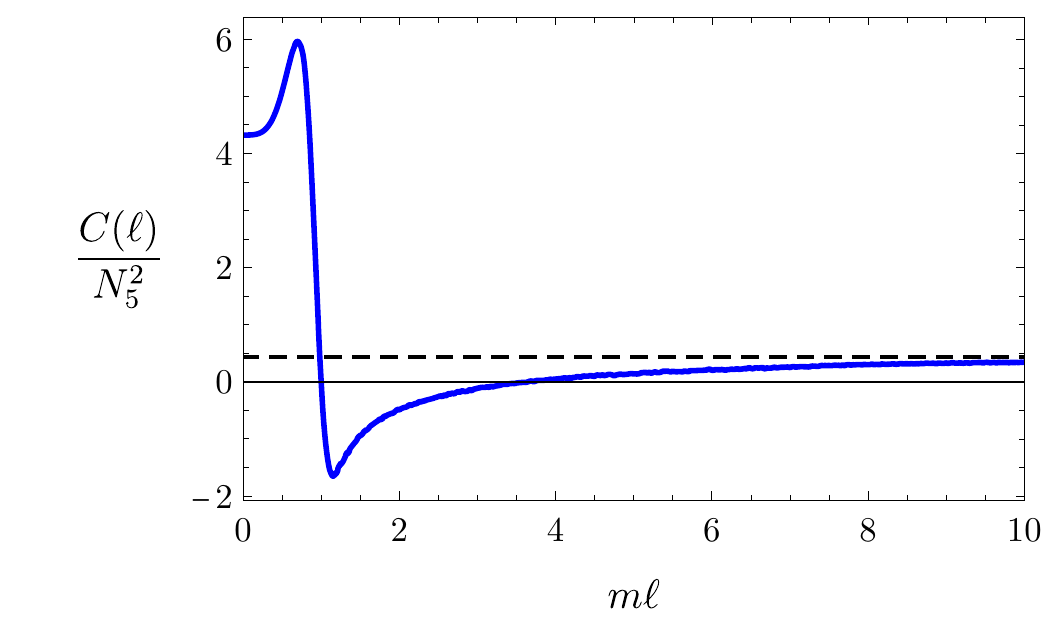}
		\caption{\(\dfrac{N_2}{N_5} = \dfrac{1}{10}\)}
	\end{subfigure}
	\begin{subfigure}{0.5\textwidth}
		\includegraphics[width=\textwidth]{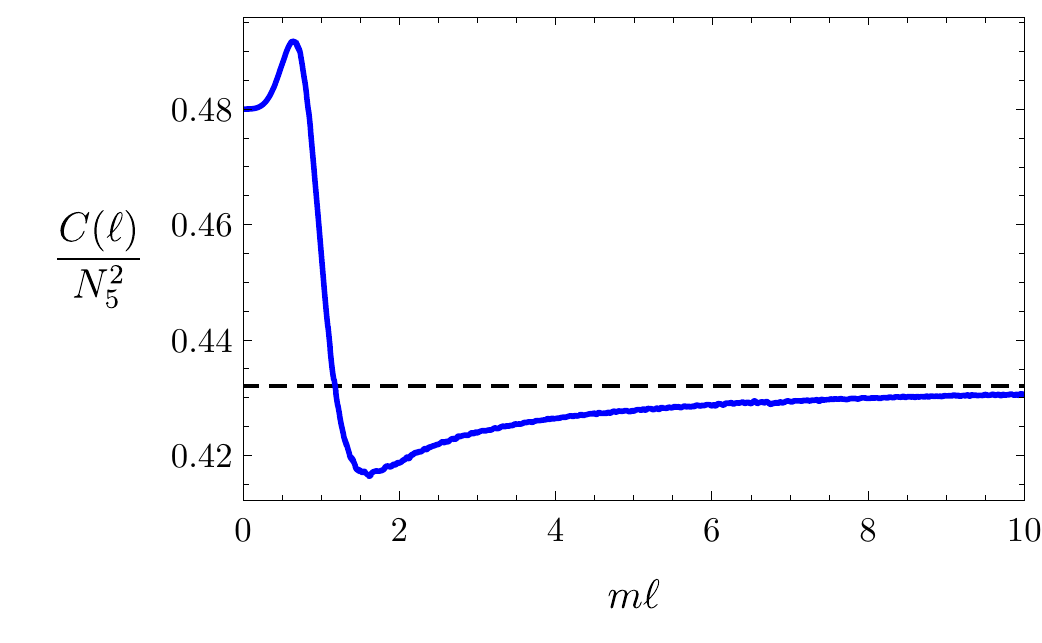}
		\caption{\(\dfrac{N_2}{N_5} = \dfrac{9}{10}\)}
	\end{subfigure}
	\caption[The entanglement entropy of the antisymmetric M5-brane flow.]{Numerical results for the entanglement entropy (top row) and \(C(\ell)\) defined in~\eqref{eq:b_function} (bottom row), for the defect RG flow from a bundle of \(N_2\) M2-branes in the UV to the antisymmetric Wilson surface in the IR. For small values of \(m\ell\), \(C(\ell)\) is given by the \((N_5-N_2)\) times the entanglement coefficient \(c\) for a single M2-brane~\eqref{eq:m2_central_charge}. As \(m\ell \to \infty\), \(C(\ell)\) tends to the value of \(c\) for the antisymmetric Wilson surface~\eqref{eq:antisymmetric_central_charge}, indicated by the horizontal dashed lines in the plots.}
	\label{fig:antisymmetric_flow_ee}
\end{figure}
In figure~\ref{fig:antisymmetric_flow_ee} we plot our numerical results for \(\Delta \see^{(1)}\) and \(C(\ell)\), both as functions of the dimensionless combination \(m\ell\). The difference in  entanglement entropy, \(\Delta \see^{(1)}\), vanishes in the limit \(m\ell \to 0\), by definition. Increasing \(m\ell\) from zero, \(\Delta \see^{(1)}\) at first increases, before reaching a maximum and then decreasing, apparently without bound.

For \(m\ell \to 0\), \(C(\ell)\) tends to the entanglement entropy coefficient \(c\) of the UV solution, namely \((N_5 - N_2)\) times \(c\) for a single M2-brane~\eqref{eq:m2_central_charge}, explicitly \(C(\ell=0) \equiv c_\mathrm{UV} = \frac{24}{5} N_5 (N_5 - N_2)\). Similarly, for \(m\ell \to \infty\), \(C(\ell)\) tends to the central charge of the IR solution --- the antisymmetric Wilson surface. Therefore \(\lim_{m\ell\to\infty} C(\ell) \equiv c_\mathrm{IR} = \frac{24}{5} N_2 (N_5 - N_2)\). We find that \(C(\ell)\) interpolates between these two limits non-monotonically.

Since the antisymmetric flow only exists for \(N_2 \leq N_5\), the central charge is manifestly larger in the UV than in the IR,
\begin{equation} \label{eq:b_inequality}
	c_\mathrm{UV} \geq c_\mathrm{IR}.
\end{equation}
This would appear to support the interpretation of \(c\) as a measure of the massless degrees of freedom on the brane. However, in section~\ref{sec:symmetric} we will see that for M5-brane flow solutions wrapping an \sph[3] internal to \ads[7] instead of \sph[4], the inequality \eqref{eq:b_inequality} does not hold.

\subsection{On-shell action}

It has been argued \cite{Kobayashi:2018lil} that for defect RG flows the free energy on a sphere or in hyperbolic space serves as a better candidate to count degrees of freedom than the entanglement entropy. For the bundle of M2-branes in hyperbolic slicing, the free energy is \((N_5-N_2)\) times that of a single M2-brane, given in~\eqref{eq:m2_hyperbolic_free_energy}, while the free energy of the antisymmetric Wilson surface was computed in \eqref{eq:antisymmetric_hyperbolic_free_energy}. Setting \(\b = \b_0 = 2\pi\), we find
\begin{equation}
	- F^{(1)}(\b_0) = \begin{cases}
		\dfrac{2}{\pi} N_5(N_5-N_2) u_c, \quad &\text{for the bundle of M2-branes},
		\\[1em]
		\dfrac{2}{\pi} N_2(N_5-N_2) u_c, \quad &\text{for the antisymmetric Wilson surface}.
	\end{cases}
\end{equation}
Since these flows only exist for \(N_2 \leq N_5\), we find that \(-F^{(1)}\) is indeed larger in the UV than the IR, consistent with the expectations of ref.~\cite{Kobayashi:2018lil}.

However, since the antisymmetric flow is triggered by a VEV rather than a source, the flow and the bundle of M2-branes describe two different states of the same theory. As pointed out in ref.~\cite{Kobayashi:2018lil}, the difference between the hyperbolic space free energies of the IR and UV solutions is equal to the relative entropy of the two states, and is guaranteed to be positive due to positivity of relative entropy~\cite{Blanco:2013joa}. 

A candidate quantity which we can study along the entire flow is provided by ref.~\cite{Kumar:2017vjv}, in which the contribution of probe D-brane solutions to the on-shell action in the entanglement wedge was observed to decrease monotonically along a defect RG flow. We will now test whether the same is true for the antisymmetric M5-brane flow.

The entanglement wedge on-shell action, which we will denote \(S_\mathcal{W}^\star\), is given by~\eqref{eq:antisymmetric_flow_on_shell_action_bulk} and \eqref{eq:antisymmetric_flow_on_shell_action_boundary} but with the domain of integration replaced by the region
\begin{equation}
	\mathcal{W}'_\e = \{t^2 + x^2 +z^2 \leq \ell^2\} \cap \{z \geq \e\}.
\end{equation}
This is the intersection of the probe brane with the entanglement wedge, with the cutoff region at \(z < \e\) excised. Explicitly
\begin{multline} \label{eq:antisymmetric_flow_entanglement_wedge_action}
	S_\mathcal{W}^\star = \frac{N_5^2}{2\pi} \int_{\mathcal{W}'_\e} \diff x^0 \diff x \diff z \frac{1}{z^3} \frac{D(\q)^2 + \sin^6 \q}{D(\q) \cos\q + \sin^4 \q}
	\\
	- \frac{N_5^2}{4\pi} \int_{\p \mathcal{W}_\e'} \diff t \diff x \frac{1}{\e^2} \le[D(\q) \cos\q + \sin^4\q\ri]_{z=\e},
\end{multline}
where \(\p\mathcal{W}_\e'\) is the region of the boundary of \(\mathcal{W}_\e'\) intersecting the cutoff surface at \(z = \e\). Note that we have assumed that there are no boundary terms arising from the change in the bulk 11D SUGRA action due to the back-reaction of the brane.

The action \eqref{eq:antisymmetric_flow_entanglement_wedge_action} diverges logarithmically as \(\e\to 0\). For the solutions which preserve defect conformal symmetry, we find
\begin{equation} \label{eq:antisymmetric_flow_action_fixed_points}
	S_\mathcal{W}^\star = \begin{cases}
		2 N_5 (N_5 - N_2) \ln \le(\dfrac{\ell}{\e}\ri) + \mathcal{O}(\e^1), \quad &\text{for the bundle of M2-branes},
		\\[1em]
		2N_2(N_5 - N_2)\ln \le(\dfrac{\ell}{\e}\ri) + \mathcal{O}(\e^1), \quad &\text{for the antisymmetric Wilson surface}.
	\end{cases}
\end{equation}
For the flow solution, the coefficient of the logarithmic divergence is the same as for the bundle of M2-branes, but the \(\mathcal{O}(\e^0)\) term will be different.

\begin{figure}[t!]
	\begin{subfigure}{0.5\textwidth}
		\includegraphics[width=\textwidth]{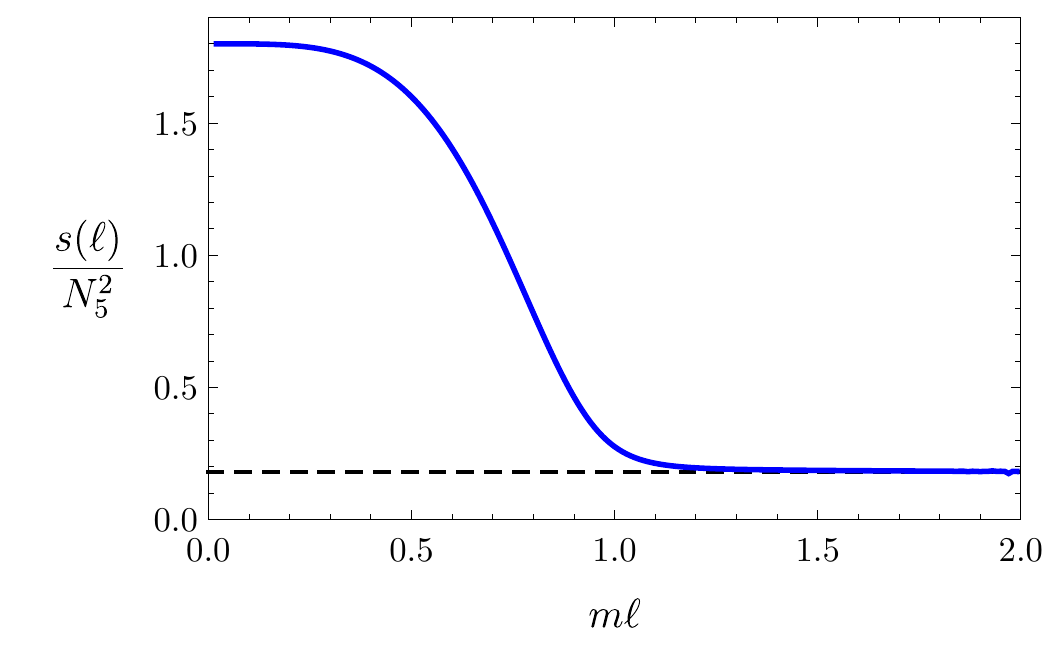}
		\caption{\(\dfrac{N_2}{N_5} = \dfrac{1}{10}\)}
	\end{subfigure}
	\begin{subfigure}{0.5\textwidth}
		\includegraphics[width=\textwidth]{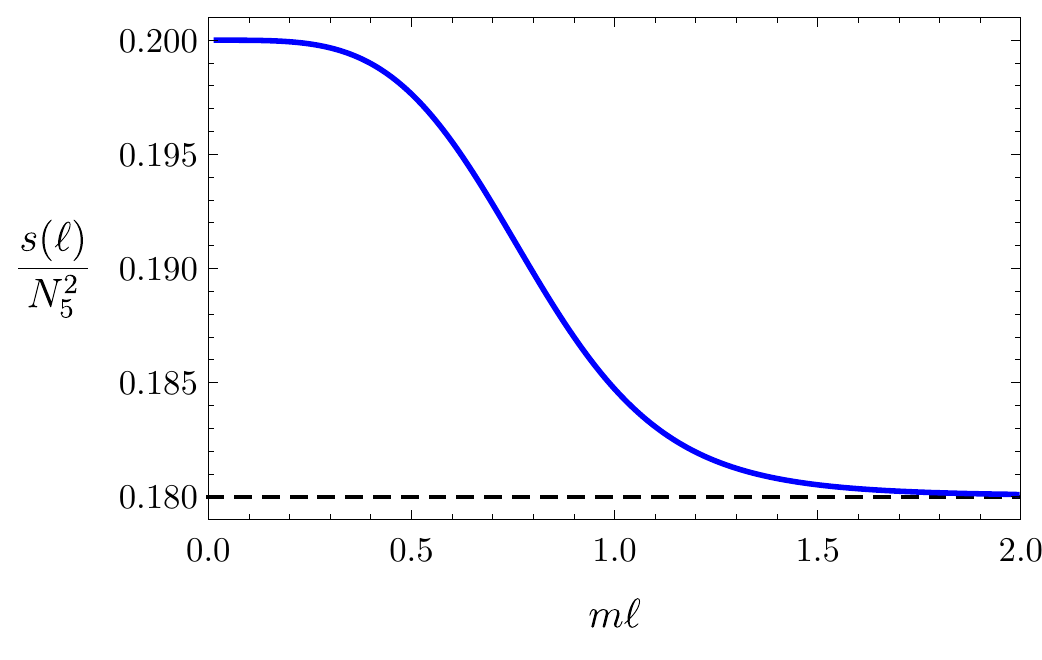}
		\caption{\(\dfrac{N_2}{N_5} = \dfrac{9}{10}\)}
	\end{subfigure}
	\caption[The contribution of the antisymmetric M5-brane flow to the on-shell action in the entanglement wedge.]{Numerical results for the function \(s(\ell)\) for the antisymmetric flow, defined in~\eqref{eq:s_definition} as minus the logarithmic derivative with respect to \(\ell\) of the brane's contribution to the on-shell action in the entanglement wedge of a spherical subregion of radius \(\ell\). For small \(m\ell\) the derivative tends to the coefficient of the logarithmic divergence of the on-shell action for the UV solution, the bundle of M2-branes. For large \(m\ell\) the derivative tends to the coefficient for the IR solution, the antisymmetric Wilson surface (indicated by the horizontal dashed lines). In between these limits, \(s(\ell)\) decreases monotonically.}
	\label{fig:antisymmetric_flow_action}
\end{figure}
In general, we must evaluate~\eqref{eq:antisymmetric_flow_entanglement_wedge_action} numerically. To obtain a UV finite quantity we take a logarithmic derivative with respect to the sphere radius \(\ell\), defining a function
\begin{equation} \label{eq:s_definition}
	s(\ell) \equiv \ell\frac{\diff S_\mathcal{W}^\star}{\diff \ell}.
\end{equation}
Figure~\ref{fig:antisymmetric_flow_action} shows our results, for two sample values of \(N_2/N_5\). 

In the limits \(\ell \to 0\) or \(\infty\), \(s(\ell)\) tends to the values at the UV or IR fixed points, respectively, given by the coefficients of the logarithms in~\eqref{eq:antisymmetric_flow_action_fixed_points}. Since the flows only exist for \(N_2 \leq N_5\), this implies that \(s\) is smaller in the IR than in the UV. For all values of \(N_2/N_5\) that we have checked, \(s(\ell)\) appears to decrease monotonically along the flow.

\section{M5-branes wrapping S\(^3\) \(\subset\) AdS\(_7\)}
\label{sec:symmetric}

In this section we will seek solutions wrapping an \sph[3] internal to the \ads[7] factor of the background geometry. This includes the symmetric representation Wilson surface~\cite{Lunin:2007ab}, corresponding to a Young tableau consisting of a single row of \(N_2\) boxes, with \(N_2\) determined from~\eqref{eq:flux_quantization}. We will also study flows from the symmetric representation Wilson surface to a bundle of M2-branes.

\subsection{The solution in flat slicing}

We begin by working in a supergravity background of the form,
\begin{align}
	\diff s^2 &= h(r)^{-1/3} \h_{\m\n} \diff x^\m \diff x^\n + h(r)^{2/3} \le( \diff r^2 + r^2 \diff s_{\sph[4]}^2	\ri)
	\nonumber \\
	\CF4 &= r^4 h'(r) \diff s_{\sph[4]},
\end{align}
where \(\h_{\m\n}\) is the six-dimensional Minkowski metric, and for now we leave the function \(h(r)\) arbitrary. If \(h(r) = 1 + L^3/r^3 \), the background is the M5-brane solution~\eqref{eq:m_brane_solution}. If \(h(r) = L^3/r^3\) the background is \(\ads[7] \times \sph[4]\). We employ static gauge on the probe M5-brane, \(\xi = (x^0, x^1, x^\a)\), where \(\a\) runs from 2 to 5. With the following ansatz for the world volume fields
\begin{equation}
	r = r(x^\a),
	\quad
	F_{3,\a\b\g} = \e_{\a\b\g\d} \d^{\d\e} \h_\e(x^\a),
	\quad
	a = a(x^1),
\end{equation}
we find that the Euler-Lagrange equations for the world volume fields are satisfied if \(\h_\a = \p_\a r\) and
\begin{equation}
	\d^{\a\b} \p_\a \p_\b r(x^\a) = 0.
\end{equation}
This is just the four-dimensional flat-space Laplace equation, so we find an infinite family of solutions
\begin{equation} \label{eq:multi_center_spike}
	r = r_0 + \frac{2 L^3}{N_5} \sum^{n}_{a=1} \frac{N^{(a)}}{\d^{\a\b} \le(x_\a - y^{(a)}_\a\ri)\le(x_\b - y^{(a)}_\b\ri)}
\end{equation}
With constants \(r_0\), \(N^{(a)}\) and \(y^{(a)}_\a\) determined by the boundary conditions.

Such solutions are well known in flat space, they describe an M5-brane at \(r=r_0\), with \(n\) infinite tension self-dual strings with \(N_a\) units of charge at positions \(y^{(a)}\) \cite{Howe:1997ue}. The solution \eqref{eq:multi_center_spike}, derived in ref.~\cite{Schwarz:2014rxa}, is the generalisation for a probe M5-brane embedded in the geometry produced by a stack of parallel M5-branes.

Let us take take the \(\ads[7] \times \sph[4]\) background, so that \(h(r) = L^3/r^3\), and consider the case \(n = 1\), \(N^{(1)} = N\), and \(y^{(1)} = 0\). The solution \eqref{eq:multi_center_spike} reduces to
\begin{equation}
	r = r_0 + \frac{2 L^3 N_2}{N_5 \r^2},
\end{equation}
where \(\r = \sqrt{\d_{\a\b} x^\a x^\b}\). Substituting \(r = 4 L^3/z^2\) and solving for \(\r\), we obtain the embedding in the \(\ads[7] \times \sph[4]\) metric~\eqref{eq:ads_solution},
\begin{equation} \label{eq:spike_solution}
	\r(z) = \sqrt{ \frac{N_2}{2 N_5} }\frac{z}{\sqrt{1 + \mt z^2}},
\end{equation}
where \(\mt = - r_0/4L^3\). The corresponding field strength is given by
\begin{equation} \label{eq:spike_field_strength}
	F_3 = \frac{4L^3 N_2}{N_5} \sin^2 \f_1 \sin \f_2 \, \diff \f_1 \wedge \diff \f_2 \wedge \diff \f_3.
\end{equation}

Substituting the solution~\eqref{eq:spike_solution} into the full bulk action~\eqref{eq:spike_action_ansatz} for the brane, we find that the on-shell PST action in flat slicing is
\begin{equation}
	S_\mathrm{M5}^\star = - \frac{N_5 N_2}{\pi \e^2} \int \diff t \diff x.
\end{equation}
As for the antisymmetric flow solution, this is completely cancelled by the boundary term~\eqref{eq:boundary_term}, so the renormalised on-shell action in flat slicing vanishes.

\begin{figure}
\begin{subfigure}{0.325\textwidth}
	\includegraphics[width=\textwidth]{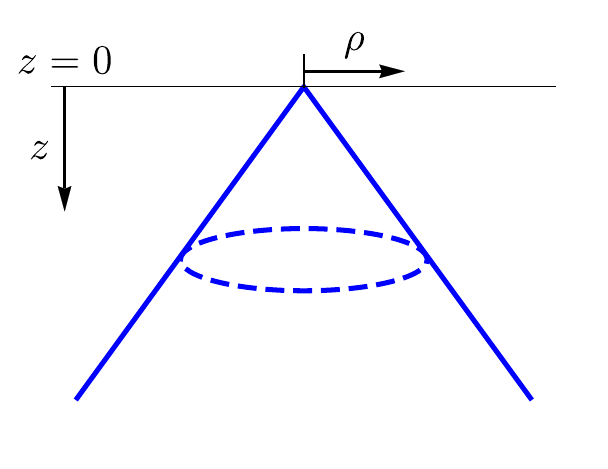}
	\caption{\(\mt = 0\): Wilson surface}
\end{subfigure}
\begin{subfigure}{0.325\textwidth}
	\includegraphics[width=\textwidth]{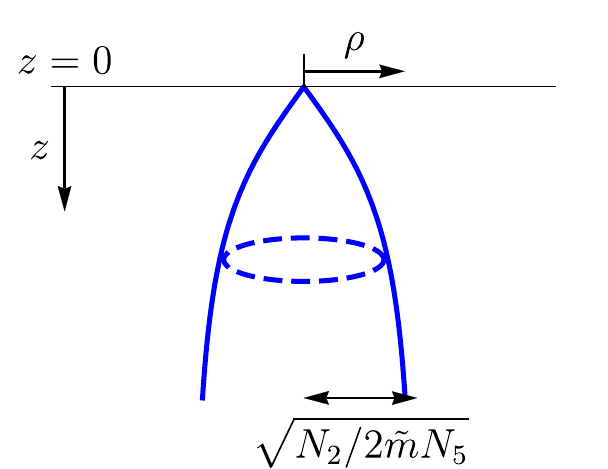}
	\caption{\(\mt > 0\): Symmetric flow}
\end{subfigure}
\begin{subfigure}{0.325\textwidth}
	\includegraphics[width=\textwidth]{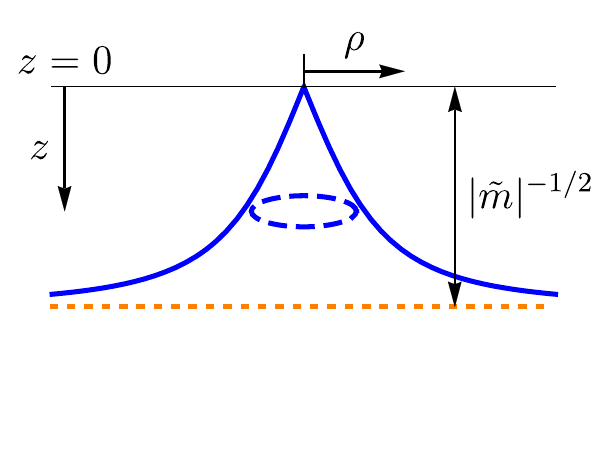}
	\caption{\(\mt < 0\): Funnel}
\end{subfigure}
\caption[Cartoons of different M5-brane embeddings in $\mathrm{AdS}_7$.]{Cartoons of the different M5-brane embeddings wrapping an \sph[3] internal to \ads[7]. In each case the horizontal black line denotes the boundary of \ads[7] at \(z = 0\), and the dashed blue line denotes the \sph[3] wrapped by the brane. \textbf{(a):} For \(\mt = 0\) the brane is dual to a symmetric representation Wilson surface. \textbf{(b):} For \(\mt > 0\) the brane describes a defect RG flow from a symmetric representation Wilson surface to a bundle of M2-branes. In terms of the coordinate \(\r\), as \(z \to \infty\) the radius of the wrapped \sph[3] tends to a finite value \(\r = \sqrt{N_2/2 \mt N_5 }\), so the proper radius vanishes in this limit. \textbf{(c):} For \(\mt < 0\) the solution is a funnel, created by M2-branes ending on a Coulomb branch M5-brane at \(z = |\mt|^{-1/2}\).}
\label{fig:symmetric_cartoon}
\end{figure}

The interpretation of the solution depends on the sign of \(\mt\), as sketched in figure~\ref{fig:symmetric_cartoon}. When \(\mt = 0\), the induced metric on the M5-brane world volume is that of \(\ads[3] \times \sph[3]\)
\begin{equation}
	\diff s_\mathrm{M5}^2 = \frac{4L^2}{z^2} \le[
		-\diff t^2 + \diff x^2 + \le(1 + \frac{N_2}{2N_5}\ri) \diff z^2
	\ri]
	+ \frac{2 L^2 N_2}{N_5} \diff s_{\sph[3]}^2,
\end{equation}
where the radius of the \ads[3] is \(2L \sqrt{1 + N_2/2N_5}\) and the radius of the \sph[3] is \(L\sqrt{2 N_2/N_5}\). The presence of the \ads[3] indicates that the defect preserves the global subgroup of two-dimensional conformal symmetry. Indeed, the solution with \(\mt=0\) is expected to be holographically dual to a Wilson surface in a symmetric representation \cite{Lunin:2007ab,Chen:2007ir,Mori:2014tca}.\footnote{See also ref.~\cite{Chen:2007tt} for a similar M5-brane embedding in \(\ads[4] \times \sph[7]\)}.

When \(\mt \neq 0\) the M5-brane world volume no longer includes an \ads[3] factor, so the defect conformal symmetry is broken. Near the boundary, where \(|\mt| z^2 \ll 1\), the solution approaches the Wilson surface solution. For \(\mt < 0\), \(\r(z)\) becomes infinite at a finite value \(z=|\mt|^{-1/2}\). We interpret this solution as a Coulomb branch brane at \(z=|\mt|^{-1/2}\), probed by an infinite tension self-dual string. We will refer to this as the M5-brane funnel.

For \(\mt > 0\), \(\r(z)\) remains finite for all \(z\). In the infrared, \(\mt z^2 \gg 1\) the world volume again has an \ads[3] factor but with radius \(2L\), indicating that the solution with positive \(\mt\) is dual to a defect RG flow. At large \(z\) the induced metric takes the form
\begin{equation}
	\diff s_\mathrm{M5}^2 = \frac{4 L^2}{z^2} \le(-\diff t^2 + \diff x^2 + \diff z^2 \ri) + \frac{2 L^2 N_2}{N_5 \mt z^2} \diff s_{\sph[3]}^2 + \ldots\;,
\end{equation}
where the dots indicate corrections of higher order in \(\mt z^2\). The proper radius of the \sph[3] shrinks to zero as \(z\to\infty\), and a natural guess is that the infrared is a bundle of M2-branes. Similar D3-brane solutions in \(\ads[5] \times \sph[5]\), flowing from a symmetric representation Wilson surface to a bundle of strings, were studied in \cite{Kumar:2016jxy,Kumar:2017vjv}. Expanding the solution \eqref{eq:spike_solution} for small \(z\), we find that as in the case of the flow involving the antisymmetric representation, the flow is triggered by the VEV of an operator  with conformal dimension \(\Delta = 2\).

To support the intuition that the infrared is a bundle of non-interacting M2-branes, let us carry out a calculation in the style of section 2.4 of \cite{Kobayashi:2006sb}. Substituting the field strength~\eqref{eq:spike_field_strength} into the M5-brane action, along with the ansatz \(\r = \r(z)\), we obtain
\begin{equation} \label{eq:spike_action_ansatz}
	S_\mathrm{M5} = - \frac{4 N_5^2}{\pi} \int \diff t \diff x \diff z \frac{1}{z^6} \le[
		\sqrt{
			\le(\frac{N_2^2}{4 N_5^2}z^6  + \r^6\ri)
			\le(1 + \r'^2\ri)
			} - \r^3 \r'
	\ri].
\end{equation}
As \(z \to \infty\), \(\r\) remains finite and \(\r' \to 0\) when evaluated on the solution \eqref{eq:spike_solution}. To leading order at large \(z\), we may therefore neglect the \(\r^3 \r'\) term compared to the square root, and the \(\r^6\) term inside the square root. Thus for large \(z\)
\begin{equation}
	S_\mathrm{M5} \approx - \frac{2 N_5 N_2}{\pi} \int \diff t \diff x \diff z \frac{1}{z^3} \sqrt{1 + \r'^2}.
\end{equation}
This is \(N_2\) times the action~\eqref{eq:m2_action_ansatz} for a single M2-brane, as expected. Note that in the calculation of ref.~\cite{Kobayashi:2006sb} it was necessary to Legendre transform the action with respect to the gauge field to fix the total amount of fundamental string charge dissolved in the brane. In our case there is no need to Legendre transform, since the M2-brane charge is already fixed by the flux quantization condition~\eqref{eq:flux_quantization}.

The Weyl anomaly coefficients \(b\) and \(d_2\) for a symmetric representation Wilson surface are~\cite{Jensen:2018rxu}
\begin{equation} \label{eq:symmetric_rep_weyl_anomaly}
	b = 12 N_2 \le( N_5 + \frac{N_2}{4} \ri),
	\quad
	d_2 = 12 N_2 \le( N_5 + \frac{N_2}{2} \ri),
\end{equation}
while the Weyl anomaly coefficients for the bundle of M2-branes are given by \(N_2\) times those of a fundamental representation Wilson surface~\eqref{eq:graham_witten_anomaly},
\begin{equation}
	b = d_1 = d_2 = 12 N_2 N_5.
\end{equation}
Both \(b\) and \(d_2\) are larger in the UV than in the IR for the symmetric flow.

\subsection{Entanglement entropy of the symmetric representation Wilson surface}

In the hyperbolic slicing of \ads[7], the solution \eqref{eq:spike_solution} with \(\mt = 0\) becomes
\begin{equation} \label{eq:symmetric_wilson_surface_hyperbolic}
	\Phi \equiv \sin \f_0 = \frac{\kappa}{v \sinh u},
\end{equation}
where \(\kappa^2 \equiv N_2/2N_5\). We have not been able to analytically find the generalisation of this solution for arbitrary temperatures in the hyperbolic slicing. This means we cannot compute the contribution of the Wilson surface to the R\'enyi entropies, but we may still obtain the entanglement entropies by differentiating the off-shell action with respect to the inverse temperature and using \eqref{eq:probe_generalised_gravitational_entropy_beta}.

To write the off-shell action, we parameterise the brane by \(\xi = (\t,v,u,\f_1,\f_2,\f_3)\) and take as an ansatz \(\Phi = \Phi(\t,v,u)\). Substituting this into the PST action and integrating over the \sph[3] parameterised by \((\f_1,\f_2,\f_3)\), we obtain
\begin{equation}
	I_\mathrm{M5} = \int_0^{2\pi} \diff \t \int_{v_H}^\infty \diff v \int_{u_\mathrm{min}}^{u_c} \diff u \, \mathcal{L},
\end{equation}
where
\begin{align} \label{eq:symmetric_off_shell_lagrangian}
	\mathcal{L} &= \frac{8 N_5^2}{\pi} (1 - \Phi^2)^{-1/2} \biggl[v  \le(
		\kappa^4 + \Phi^6 v^6 \sinh^6 u
	\ri)^{1/2} \biggl(
		1 - \Phi^2 + v^2 \sinh^2 u \, \hat{g}^{ab} \p_a \Phi \p_b \Phi
	\biggr)^{1/2}
	\nonumber \\ &\phantom{= \frac{8M^2}{\pi} (1 - \Phi^2)^{-1/2} \biggl[ }
	+ (v^6 - v_H^6) \sinh^4 u \Phi^3 \p_v \Phi\biggr].
\end{align}
The metric \(\hat{g}\) is defined such that
\begin{equation}
	\hat{g}^{ab} \p_a \Phi \p_b \Phi = \frac{1}{f(v)} (\p_\t \Phi)^2 + f(v) (\p_v \Phi)^2
	+ \frac{1}{v^2} (\p_u \Phi)^2.
\end{equation}
When \(v_H=1\) this is the metric of unit-radius \ads[3].

The lower limit \(u_\mathrm{min}\) on the integration over \(u\) is a function of \(v\), determined by the requirement that \(\sin^2\f_0 \leq 1\). From the solution~\eqref{eq:symmetric_wilson_surface_hyperbolic}, we find
\begin{equation}
	\sinh u_\mathrm{min}(v) = \frac{\kappa}{v}.
\end{equation}
Differentiating the off-shell action with respect to \(\b\), taking the limit \(\b \to 2\pi\), and substituting the solution \eqref{eq:symmetric_wilson_surface_hyperbolic}, we find that the contribution to the entanglement entropy from the symmetric representation Wilson surface is given by the integral
\begin{align}
	\see^{(1)} &= \frac{4}{5} N_2 (N_2 + 2N_5) \int_{u_\mathrm{min}(1)}^{u_c} \diff u \frac{\sinh u}{\sqrt{\sinh^2 u - \kappa^2}}
	\nonumber \\ & \phantom{=}
	+ 4 N_2^2 \int_1^\infty \diff v \int_{u_\mathrm{min}(v)}^{u_c} \diff u \frac{\sinh u}{\sqrt{v^2 \sinh^2 u - \kappa^2}}.
\end{align}

Performing the integrals, and identifying the cutoff \(u_c\) with the small \(z\) cutoff \(\e\) using~\eqref{eq:cutoff_identification}, we find
\begin{equation}
	\see^{(1)} = \frac{8}{5} N_2 \le( N_5 - \frac{N_2}{8} \ri) \ln \le(
		\frac{2 \ell}{\e}
	\ri) + \mathcal{O}(\e^0),
\end{equation}
reproducing the result of~\cite{Gentle:2015jma,Estes:2018tnu} for the symmetric representation. From this we obtain \(c\) for a Wilson surface with representation determined by a Young tableau consisting of a single row of \(N_2\) boxes:
\begin{equation} \label{eq:symmetric_central_charge}
	c = \frac{24}{5} N_2 \le(N_5 - \frac{N_2}{8}\ri).
\end{equation}
This matches the appropriate limit of the results of~\cite{Gentle:2015jma,Estes:2018tnu}. This is true even when in the limit \(N_2 \gg N_5\), in which the probe limit is unreliable. The only requirement is that the Young tableau is a single row. The central charge vanishes at a critical value \(N_2 = 8 N_5\), and is negative for larger \(N_2\). We have not observed anything else special about this particular value of \(N_2\). Using the Weyl anomaly coefficients~\eqref{eq:symmetric_rep_weyl_anomaly}, one finds that the relation~\eqref{eq:defect_entropy_weyl_relation} is satisfied.

\subsection{Entanglement entropy of the non-conformal solutions}

We now compute the entanglement entropy contribution from the solutions with \(\mt \neq 0\). We leave the details of the calculation to appendix \ref{app:spike_entanglement}. The final result is that the entanglement entropy is given by the integral
\begin{align} \label{eq:spike_entanglement_integral}
	\see^{(1)} &= 
	\frac{8 N_5 N_2}{5} \int_{\e}^{z_*} \diff z \frac{
		\sqrt{1 + \mt z^2}
		}{
			z \sqrt{1 - (1 + \kappa^2 - \mt \ell^2) z^2 - \mt z^4 / \ell^2}
		} \le[
			1 + \frac{\kappa^2}{(1 + \mt z^2)^3}
		\ri]
	\nonumber \\ &\phantom{=}
	- \frac{8 N_5^2}{5 \pi} \int_{z\geq\e} \diff x^0 \diff x \diff z \frac{8 \ell^6 \r^6 \mathbf{N}_2 }{
		\kappa^2 z^3 \le[
			\le(\ell^2 - x_0^2 - x^2 - \r^2 - z^2 \ri)^2 + 4 \ell^2 \le(x_0^2 + z^2\ri)
		\ri]^3
	},
\end{align}
where \(\r\) is the solution \eqref{eq:spike_solution} and \(z_*\) is the value of the radial coordinate \(z\) at the intersection between the brane and the RT surface, given explicitly by
\begin{equation} \label{eq:zmax}
	z_*^2 = \frac{1}{2|\mt|} \le[
		\sqrt{\le(1 + \frac{N_2}{2N_5} - \mt \ell^2\ri)^2 + 4 \mt \ell^2} - \le(1 + \frac{N_2}{2N_5} - \mt \ell^2\ri)
	\ri].
\end{equation}
The factor \(\mathbf{N}_2\) appearing in the last integral in~\eqref{eq:spike_entanglement_integral} is given by
\begin{align} \label{eq:spike_entanglement_coeff}
	\mathbf{N}_2 &= 
	\frac{
		4 x_0^2 z^2 (\r-\kappa  z)^2 (\r+\kappa  z)^2 \left[\left(\r^2-\ell^2+x^2+x_0^2+z^2\right)^2+4 \ell^2 \left(x_0^2+z^2\right)\right]
	}{
		\left[2 x_0^2 \left(\r^2+\ell^2+x^2+z^2\right)+\left(\r^2-\ell^2+x^2+z^2\right)^2+x_0^4\right]^2
	}
	\nonumber \\ &\phantom{=}
	+\frac{
		6 \r^2 \left[2 \kappa ^2 z^4 \left(\r^2-\ell^2+x^2+x_0^2+z^2\right)+\r^2 \left(\r^2-\ell^2+x^2+x_0^2\right)^2+4 \r^2 \ell^2 x_0^2-\r^2 z^4\right]
	}{
		\left(\r^2-\ell^2+x^2+x_0^2+z^2\right)^2+4 \ell^2 x_0^2
	}
	\nonumber \\ &\phantom{=}
	-\frac{\left[2 \kappa ^2 z^4 \left(\r^2-\ell^2+x^2+x_0^2+z^2\right)+\r^2 \left(\r^2-\ell^2+x^2+x_0^2\right)^2+4 \r^2 \ell^2 x_0^2-\r^2 z^4\right]^2}{\left[\left(\r^2-\ell^2+x^2+x_0^2+z^2\right)^2+4 \ell^2 x_0^2\right]^2}.
\end{align}

\begin{figure}
\begin{subfigure}{0.5\textwidth}
    \includegraphics[width=\textwidth]{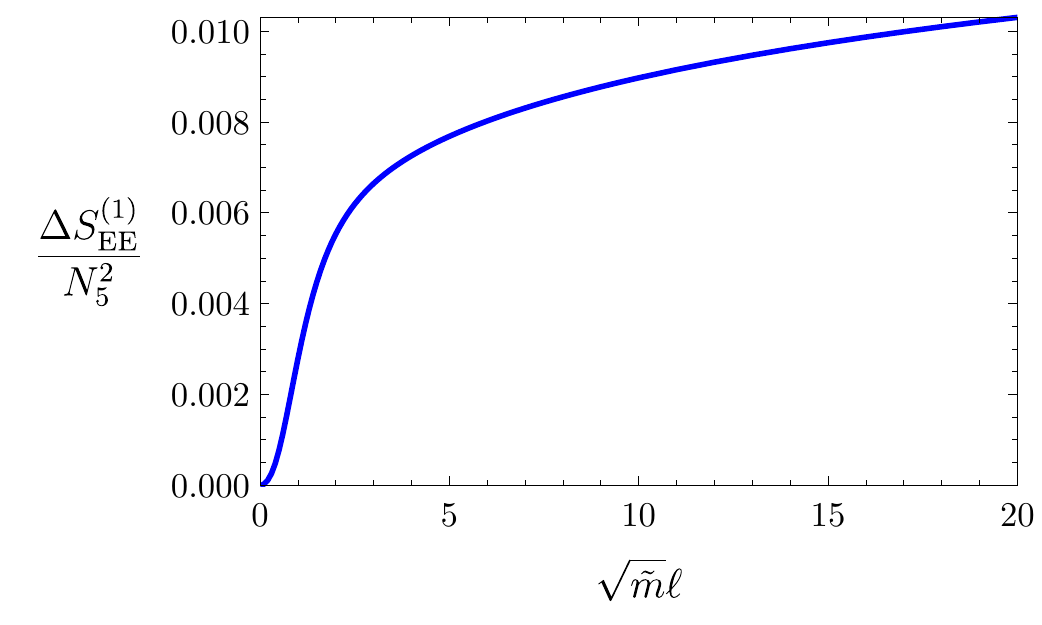}
    \caption{\(\mt > 0,~\dfrac{N_2}{N_5} = \dfrac{1}{10}\)}
\end{subfigure}
\begin{subfigure}{0.5\textwidth}
    \includegraphics[width=\textwidth]{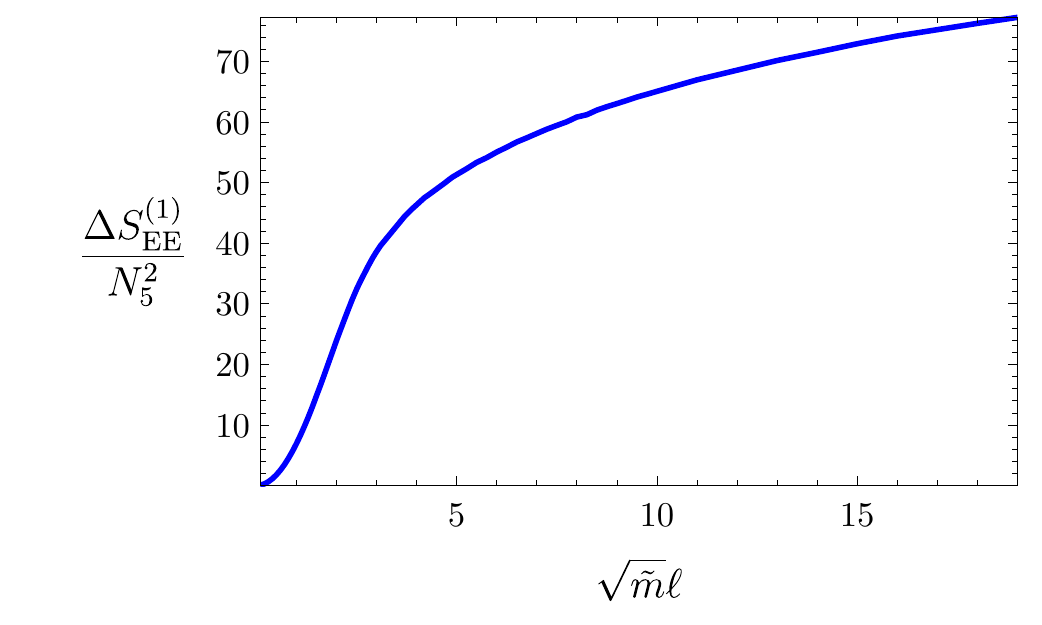}
    \caption{\(\mt > 0,~\dfrac{N_2}{N_5} = 10\)}
\end{subfigure}
\\[1em]
\begin{subfigure}{0.5\textwidth}
    \includegraphics[width=\textwidth]{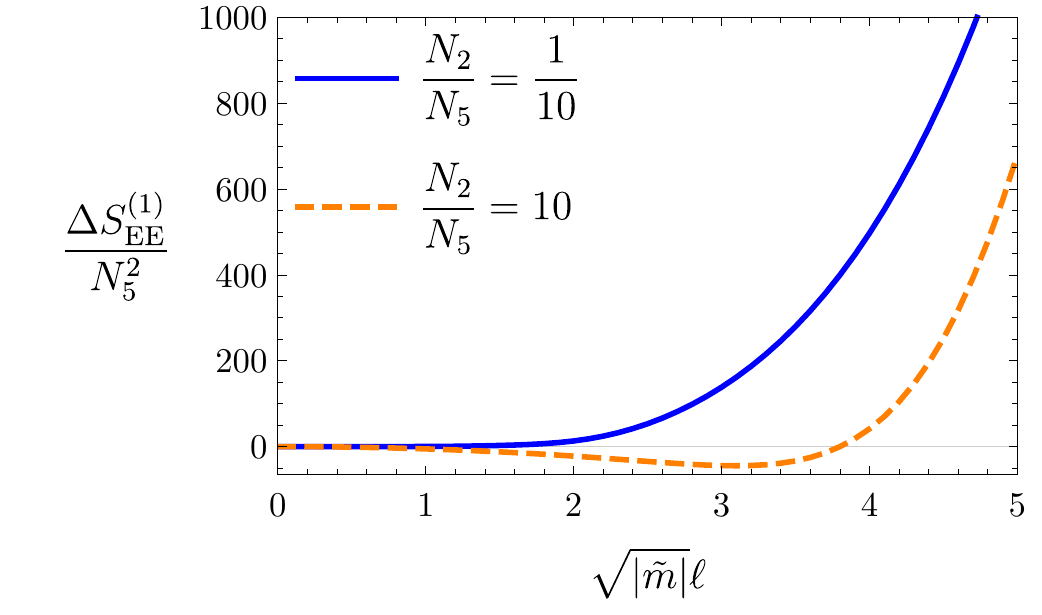}
    \caption{\(\mt < 0\)}
\end{subfigure}
\begin{subfigure}{0.5\textwidth}
    \includegraphics[width=\textwidth]{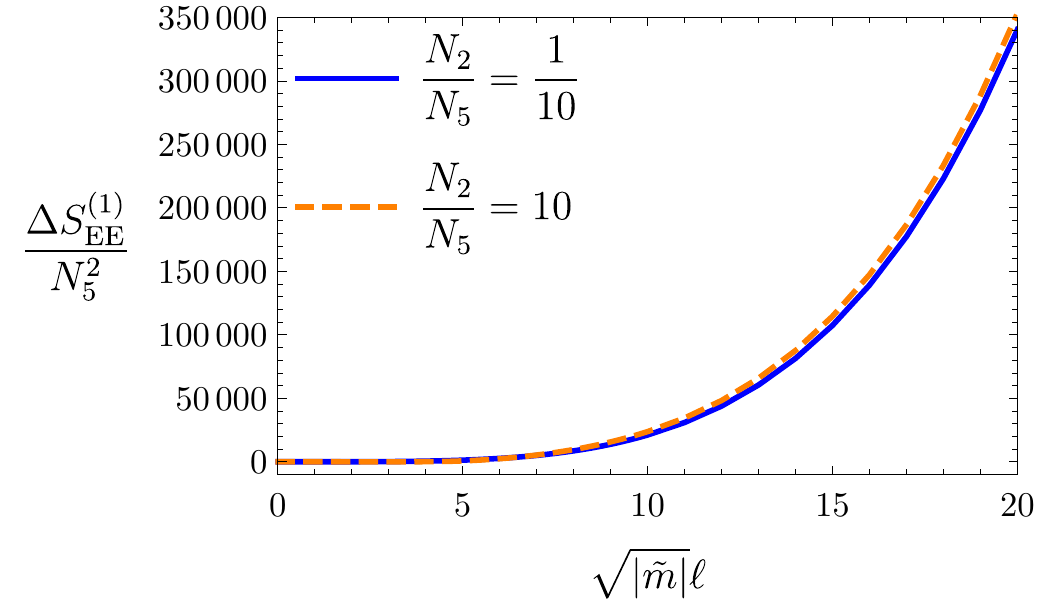}
    \caption{\(\mt < 0\), larger \(\ell\)}
\end{subfigure}
\caption[Entanglement entropy of defects dual to M5-branes in $\mathrm{AdS}_7$.]{
    	The defect contribution to entanglement entropy as a function of the radius \(\ell\) of the entangling region, for the M5-brane solutions which wrap an \sph[3] internal to \ads[7]. The top row shows the entanglement entropy for the symmetric flow, while the bottom row shows the entanglement entropy for the M5-brane funnel. To obtain a UV finite quantity we have calculated the difference between the entanglement entropy of the full solution and that of the symmetric representation Wilson surface. For either sign of \(\mt\) the entanglement entropy appears to grow without bound at large \(\sqrt{|\mt|} \ell\), although it grows much more rapidly for \(\mt < 0\).
	}    
	\label{fig:spike_entanglement}
\end{figure}

Taking the limit \(\sqrt{|\mt|}\ell \to 0\), the formula \eqref{eq:spike_entanglement_integral} for the entanglement entropy reduces to the entanglement entropy of a symmetric representation Wilson surface. For non-vanishing \(\mt\), evaluating the integral \eqref{eq:spike_entanglement_integral} requires numerics. As for the antisymmetric flow solutions, we obtain a finite quantity by subtracting the UV contribution, obtaining the excess due to the flow
\begin{equation}
	\Delta \see^{(1)} = \see^{(1)} - \le.\see^{(1)}\ri|_\mathrm{symmetric}.
\end{equation}
Numerical results for \(\Delta \see^{(1)}\), for both signs of \(\mt\) and sample values of \(N_2/N_5\), are plotted in figure~\ref{fig:spike_entanglement}.

\begin{figure}
    \begin{subfigure}{0.5\textwidth}
        \includegraphics[width=\textwidth]{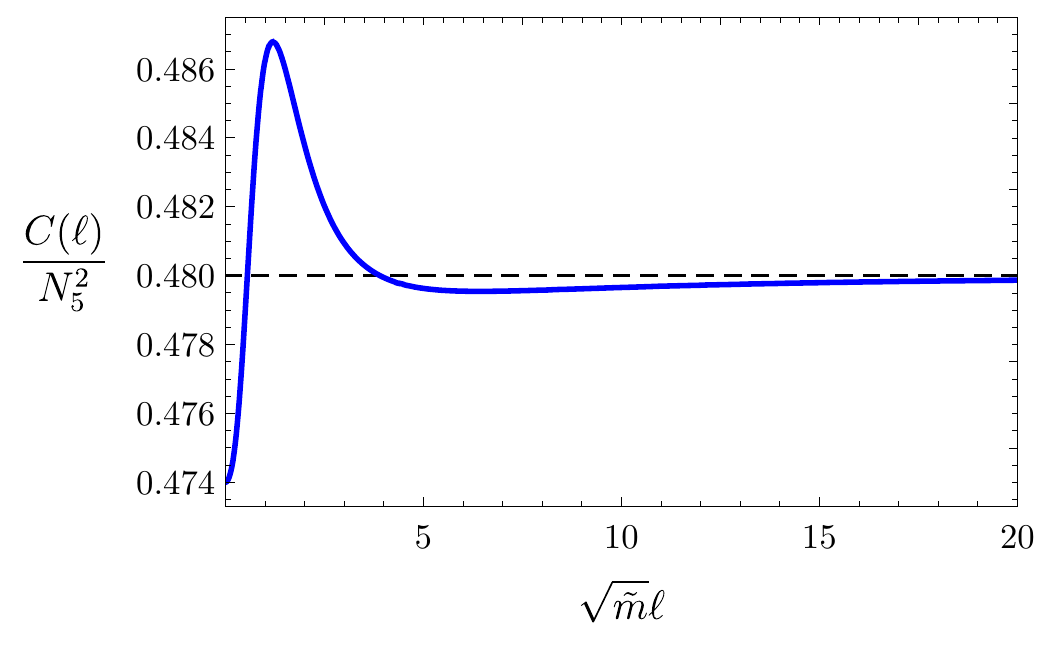}
        \caption{\(\mt > 0,~\dfrac{N_2}{N_5} = \dfrac{1}{10}\)}
    \end{subfigure}
    \begin{subfigure}{0.5\textwidth}
        \includegraphics[width=\textwidth]{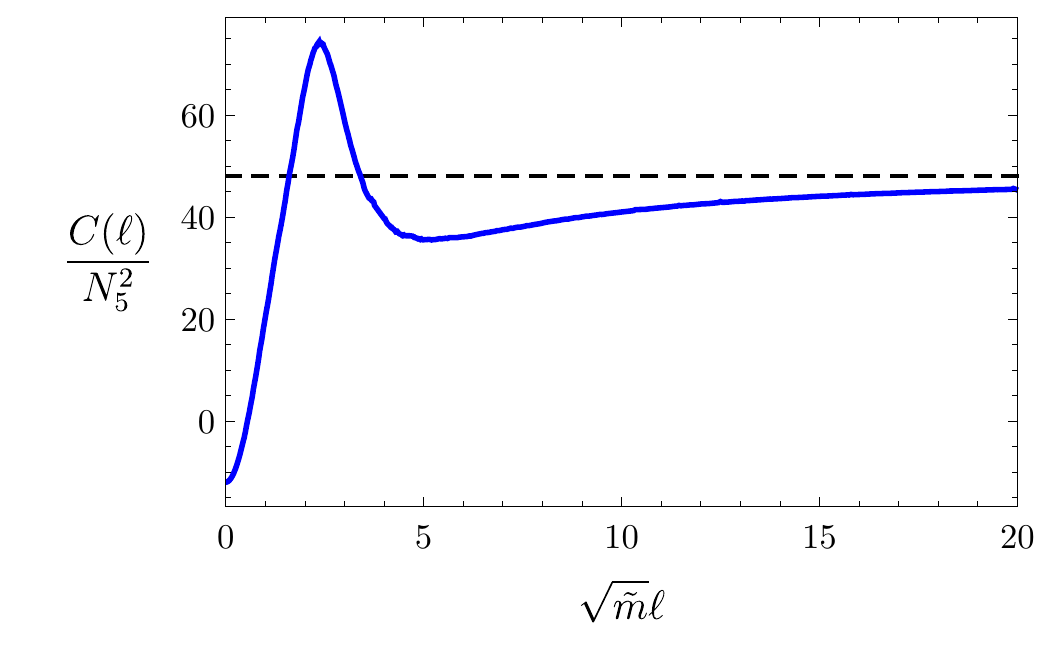}
        \caption{\(\mt > 0,~\dfrac{N_2}{N_5} = 10\)}
	\end{subfigure}
	\\[1em]
    \begin{subfigure}{0.5\textwidth}
        \includegraphics[width=\textwidth]{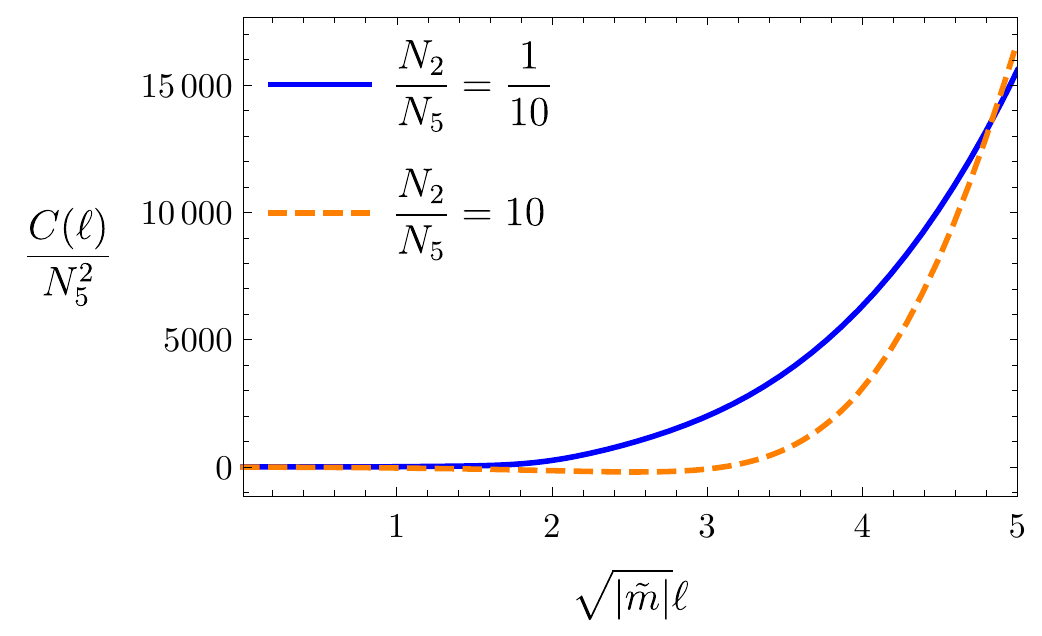}
        \caption{\(\mt < 0\)}
    \end{subfigure}
    \begin{subfigure}{0.5\textwidth}
        \includegraphics[width=\textwidth]{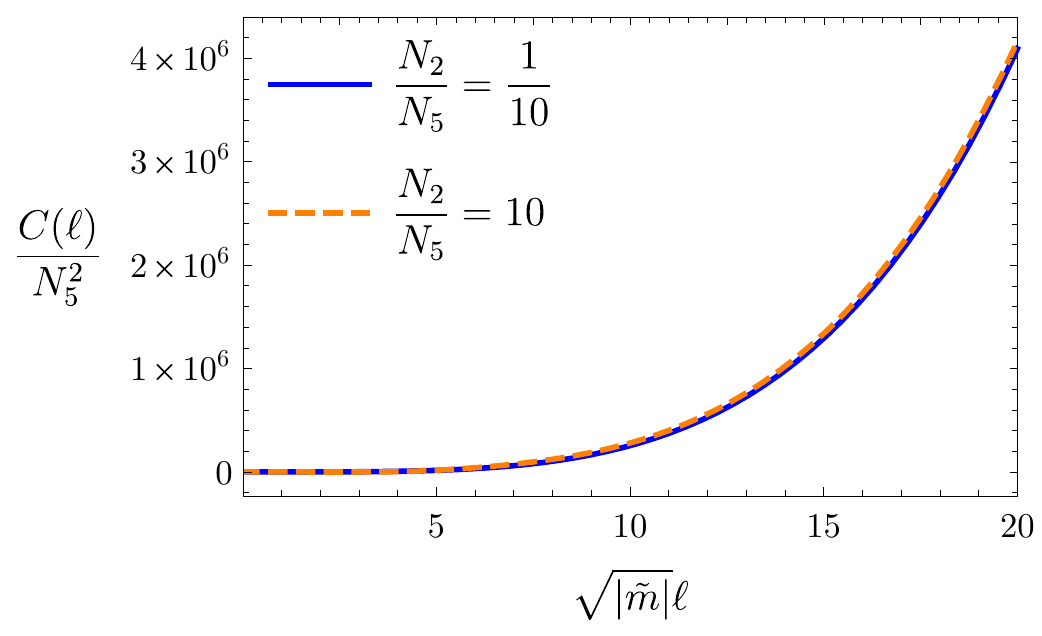}
        \caption{\(\mt < 0\), larger \(\ell\)}
    \end{subfigure}
    \caption[The coefficient of the logarithmic term in entanglement entropy along RG-flows dual to M5-branes in $\mathrm{AdS}_7$.]{
		The function \(C(\ell)\) for solutions wrapping \(\sph[3] \subset \ads[7]\). The top row shows \(C(\ell)\) for the symmetric flow, while the bottom row shows \(C(\ell)\) for the M5-brane funnel. For both signs of \(\mt\), \(C(\ell)\) tends to the UV central charge \eqref{eq:symmetric_central_charge} as \(\sqrt{|\mt|} \ell \to 0\). For \(\mt>0\), dual to a defect RG flow, as \(\sqrt{\mt}\ell \to \infty\) we find that \(C(\ell)\) tends toward the IR value \eqref{eq:symmetric_flow_ir_central_charge}, as indicated by the horizontal dashed line. In between, \(C(\ell)\) is not monotonic. For \(\mt < 0 \) we find that \(C(\ell)\) appears to increase without bound.
	}
	\label{fig:spike_b_function}
\end{figure}
From the entanglement entropy, we compute \(C(\ell)\) as defined in \eqref{eq:b_function}. The numerical results are shown in figure \ref{fig:spike_b_function}. In the limit \(\sqrt{|\mt|}\ell \to 0\), \(C(\ell)\) is given by \(c\) for the symmetric representation Wilson surface~\eqref{eq:symmetric_central_charge}. When \(\mt>0\), the solution flows to a bundle of \(N_2\) M2-branes in the IR. In the limit \(\sqrt{\mt} \ell \to 1\), \(C(\ell)\) approaches the expected infrared value, namely \(N_2\) times \(c\) for a fundamental representation Wilson surface,
\begin{equation} \label{eq:symmetric_flow_ir_central_charge}
	c_\mathrm{IR} = \frac{24}{5} N_5 N_2.
\end{equation}
This is greater than the value in the UV \eqref{eq:symmetric_central_charge}. When \(\mt < 0\), \(C(\ell)\) appears to increase without bound for large \(\mt \ell\). 

\subsection{On-shell action}

We now repeat the analysis of section~\ref{sec:antisymmetric_on_shell} for the solutions wrapping an \sph[3] internal to \ads[7]. In the UV, these solutions tend to the symmetric representation Wilson surface for which the on-shell action in hyperbolic slicing is determined by substituting the solution~\eqref{eq:symmetric_wilson_surface_hyperbolic} into the action \eqref{eq:symmetric_off_shell_lagrangian}. Explicitly, we find
\begin{equation}
	- F^{(1)} = \frac{2}{\pi} N_2 \le( N_5 + \frac{N_2}{4} \ri) u_c + \mathcal{O}(u_c^0).
\end{equation}
The infrared of the symmetric flow is a bundle of \(N_2\) M2-branes, with free energy given by \(N_2\) times the free energy~\eqref{eq:m2_hyperbolic_free_energy} of a single M2-brane. Thus \(- F^{(1)}\) is larger in the UV than in the IR, as guaranteed by positivity of relative entropy.

We now turn to the evaluation of the contribution of the probe brane to the Lorentzian signature on-shell action inside the entanglement wedge. This is given by the integral
\begin{equation}
	S_\mathcal{W}^\star = - \frac{2 N_5 N_2}{\pi} \int_{\mathcal{W}''_\e} \diff t \diff x \diff z \frac{1}{z^3}
	+ \frac{N_5 N_2}{\pi \e^2} \int_{\p \mathcal{W}''_\e} \diff t \diff x.
\end{equation}
The domain of integration is restricted to the cutoff entanglement wedge
\begin{equation}
	\mathcal{W}''_\e = \{t^2 + x^2 + \r^2(z) +z^2 \leq \ell^2\} \cap \{z \geq \e\},
\end{equation}
with \(\r(z)\) given by the solution \eqref{eq:spike_solution}, and \(\p\mathcal{W}''_\e\) denotes the part of the boundary of this surface at \(z=\e\).

The integrals evaluate to
\begin{multline}
	S_\mathcal{W}^\star = 2 N_2 \le( N_5 + \frac{N_2}{2} \ri)
		\ln \le(
		\frac{z_*}{\e}
	\ri)
	\\
	+ \frac{1}{2} N_2 \left[-N_2 \ln \left(1 + \mt z_*^2\right)+2 N_5 \left(\frac{\ell^2}{z_*^2}-1\right)-N_2\right] + \mathcal{O}(\e),
\end{multline}
where \(z_*\) is the maximal value of \(z\) inside the entanglement wedge, given by~\eqref{eq:zmax}.

\begin{figure}[t!]
	\begin{subfigure}{0.5\textwidth}
		\includegraphics[width=\textwidth]{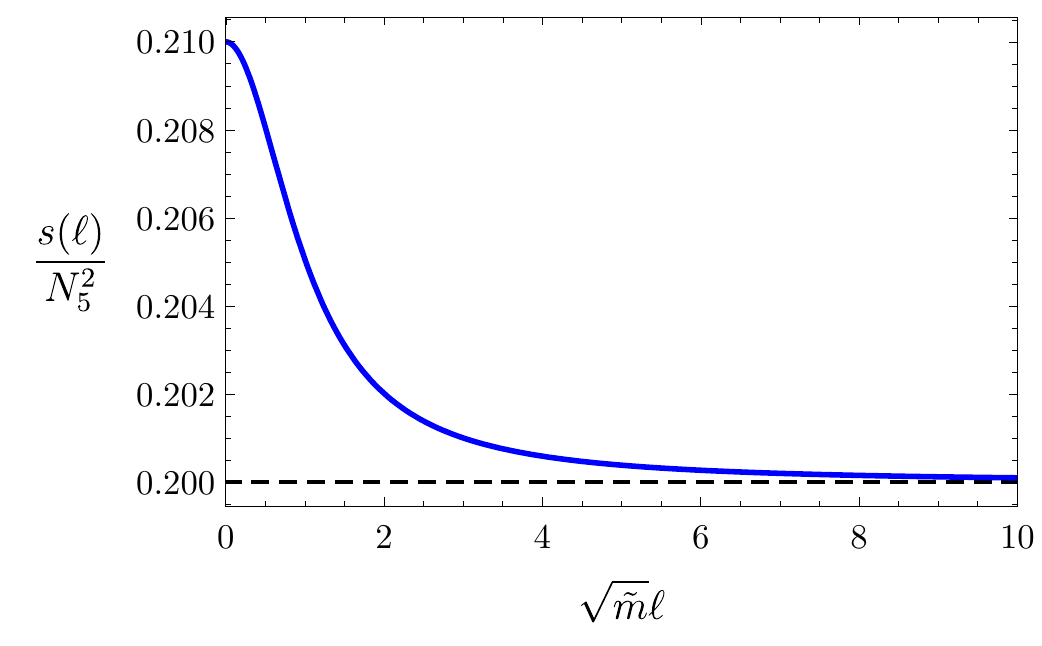}
		\caption{\(\mt > 0,~\dfrac{N_2}{N_5} = \dfrac{1}{10}\)}
	\end{subfigure}
	\begin{subfigure}{0.5\textwidth}
		\includegraphics[width=\textwidth]{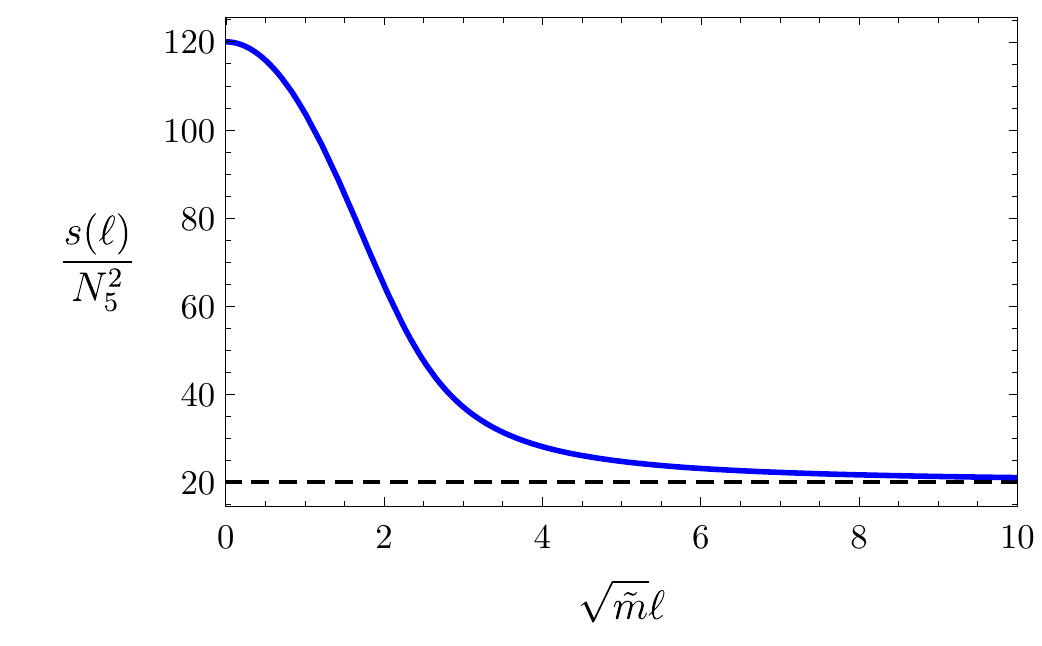}
		\caption{\(\mt > 0,~\dfrac{N_2}{N_5} = 10\)}
	\end{subfigure}
	\\[1em]
	\begin{subfigure}{0.5\textwidth}
		\includegraphics[width=\textwidth]{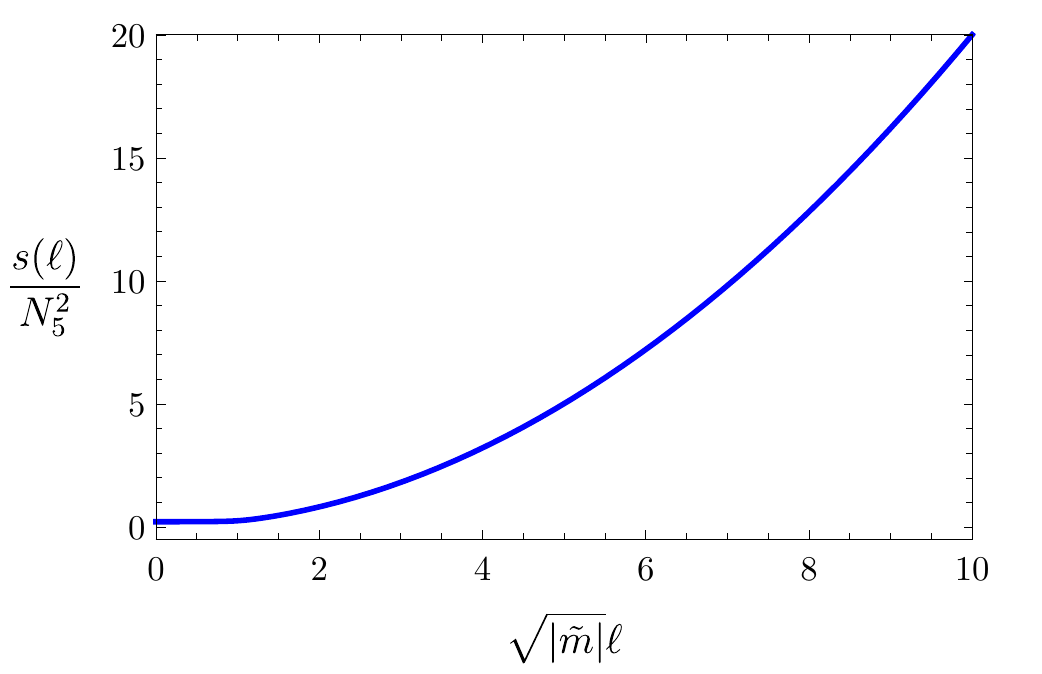}
		\caption{\(\mt < 0,~\dfrac{N_2}{N_5} = \dfrac{1}{10}\)}
	\end{subfigure}
	\begin{subfigure}{0.5\textwidth}
		\includegraphics[width=\textwidth]{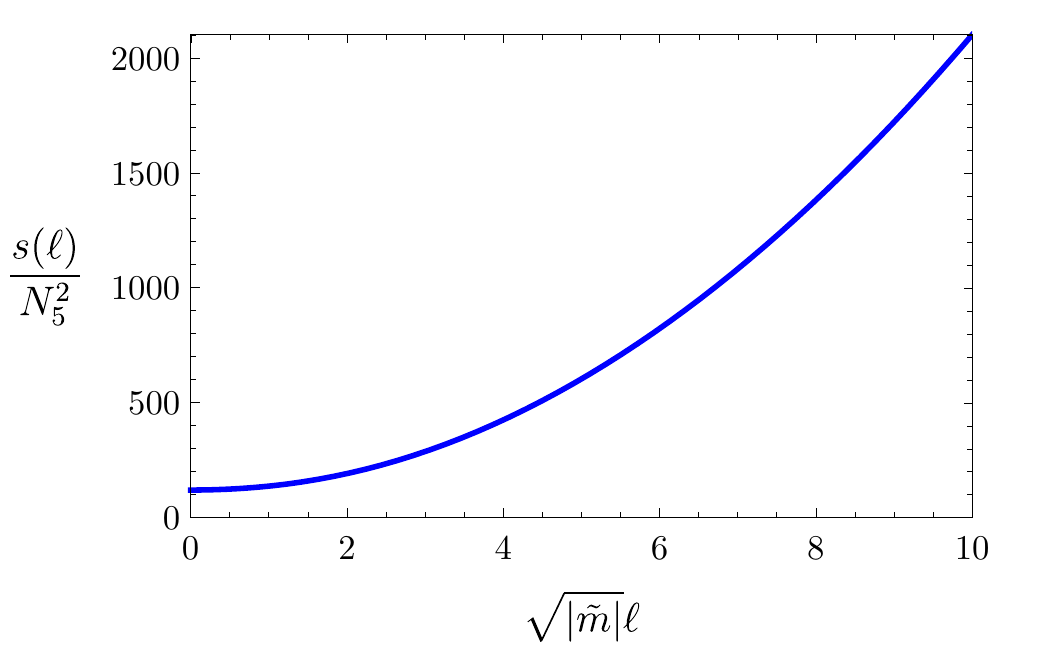}
		\caption{\(\mt < 0,~\dfrac{N_2}{N_5} = 10\)}
	\end{subfigure}
	\caption[The contribution of M5-branes in $\mathrm{AdS}_7$ to the on-shell action in the entanglement wedge.]{
			The derivative with respect to \(\ln \ell\) of the contribution to the entanglement wedge on-shell action of the symmetric flow (top row) and funnel (bottom row) solutions, for sample values of \(N_2/N_5\). For \(\mt < 0\), corresponding to a flow from a symmetric representation Wilson surface in the UV to a bundle of M2-branes in the IR, the derivative interpolates monotonically between the values at the fixed points, given in \eqref{eq:symmetric_flow_s}. The horizontal dashed line shows the value at the IR fixed point. For \(\mt > 0\), corresponding to a funnel solution, the coefficient of the logarithm in the on-shell action increases monotonically without bound.
		}
	\label{fig:symmetric_flow_action}
\end{figure}

Computing the logarithmic derivative with respect to the radius of the entangling region, we obtain the UV finite quantity \(s(\ell)\), which we write as
\begin{equation} \label{eq:symmetric_flow_s}
	s(\ell)
	\equiv \ell \frac{\diff S_\mathcal{W}^\star}{\diff \ell} 
	= N_5 N_2 \le(
		1 + \frac{N_2}{2N_5} - \mt \ell^2
		+ \sqrt{\le(1 + \frac{N_2}{2N_5} - \mt \ell^2 \ri)^2 + 4 \mt \ell^2}
	\ri). 
\end{equation}
We plot the form of this function for sample values of \(N_2/N_5\) in figure~\ref{fig:symmetric_flow_action}. It is bounded from below by \(2 N_5 N_2\),\footnote{
	Rewriting the function as \(s(\ell)
	= N_5 N_2 \le[
		2
		+ \sqrt{\le(1 - \frac{N_2}{2 N_5} + \mt \ell^2 \ri)^2 + \frac{2N_2}{N_5}} - \le(1 - \frac{N_2}{2 N_5} + \mt \ell^2 \ri)
	\ri]\) makes this manifest.}
and it is straightforward to show that it monotonically decreases with \(\ell\) for the symmetric flow solution (\(\mt > 0\)), and monotonically increases for the funnel solution (\(\mt < 0\)). To do so, we note that \(\ell\) appears only in the dimensionless combination \(\mt \ell^2\), and
\begin{equation}
	\frac{\diff s}{\diff (\mt \ell^2)} = \frac{2 N_5 N_2 - s(\ell)}{N_5 N_2 \sqrt{\le(1 - \frac{N_2}{2N_5} + \mt \ell^2 \ri)^2 + \frac{2N_2}{N_5}}} < 0.
\end{equation}

For small \(\ell\), we find that \(s\) is given by the value for a symmetric representation Wilson surface,
\begin{equation}
	s(\ell=0) = 2 N_2 \le(
		N_5 + \frac{N_2}{2}
	\ri).
\end{equation}
The behaviour at large \(\ell\) depends on the sign of \(\mt\),
\begin{equation}
	s(\ell \to \infty) \sim \begin{cases}
		2 N_5 N_2, \quad & \mt > 0,
		\\
		2 N_5 N_2 |\mt| \ell^2, & \mt < 0.
	\end{cases}
\end{equation}
In particular, the large \(\ell\) limit for \(\mt > 0\) is \(N_2\) times the value for a single M2-brane. For both the antisymmetric and symmetric flow solutions, the entanglement wedge on-shell action provides a quantity which decreases monotonically under RG flows.

\section{Discussion}
\label{sec:m5_discussion}

We have computed the contribution to entanglement entropy from a number of defects in \(\mathcal{N}=(2,0)\) SCFT, holographically dual to probe M-theory branes, for a spherical entangling region centred on the defect. Some of these defects were Wilson surfaces, and for these the entanglement entropy reproduces the probe limit of the results of \cite{Estes:2018tnu}.

The contribution of a two-dimensional conformal defect to the entanglement entropy of a spherical subregion takes the same form as the entanglement entropy of a single interval in a two-dimensional CFT, and in particular is logarithmically divergent in the UV. It is therefore tempting to identify the coefficient \(c\) of the logarithm as a central charge measuring degrees of freedom on the defect. Moreover the function \(C(\ell)\) defined in~\eqref{eq:b_function} provides a natural quantity that interpolates between the central charges of the fixed points of a defect RG flow.

However, the M5-brane embeddings we have studied show that \(C(\ell)\) is not necessarily monotonic along RG flows, and in particular \(c\) can be larger in the IR than in the UV. This suggests that the central charge as defined from the entanglement entropy may not provide a measure of the number of massless degrees of freedom on the defect. On the other hand, two of the coefficients of the defect's contribution to the Weyl anomaly, \(b\) and \(d_2\), decreased along all of the RG flows that we studied. A monotonicity theorem for \(b\) has already been proven~\cite{Jensen:2015swa}. It would be be interesting to attempt the proof of a similar theorem for \(d_2\), or alternatively to find examples of flows where \(d_2\) increases, in order to test whether \(d_2\) may also count degrees of freedom.

An alternative quantity, the on-shell action inside the entanglement wedge, decreases monotonically along the flows we study, as well as similar flows involving D-branes dual to one-dimensional defects in \(\cN = 4\) SYM~\cite{Kumar:2017vjv}. This provides another candidate \(C\)-function. It would be interesting to test whether it is monotonic in other holographic examples of RG flows, and to understand what this quantity corresponds to in the dual field theory.

There are several possible directions for future work. For example, we have studied planar defects, and a natural generalisation would be to study the entanglement entropy of defects with more complicated geometries. One example is the spherical Wilson surface, which may be obtained from the planar surface by a conformal transformation \cite{Chen:2007ir}. One could also study different geometries for the entangling region. It is plausible that the entanglement entropy for differently shaped defects or entangling regions may be sensitive to the third Weyl anomaly coefficient \(d_1\)~\cite{Fursaev:2016inw}. It would also be interesting to study defects in holographic examples of six-dimensional SCFTs with \(\mathcal{N} = (1,0)\) supersymmetry \cite{Apruzzi:2013yva, Gaiotto:2014lca}.

The techniques used in this chapter could also be applied to higher dimensional defects. For example, \(\cN = (2,0)\) theory admits four-dimensional defects~\cite{Gaiotto:2009we,Gaiotto:2009hg}, corresponding to intersecting M5-branes. The entanglement entropy of a spherical region is likely to be sensitive to a linear combination of Weyl anomaly coefficients, similar to~\eqref{eq:defect_entropy_weyl_relation}, although the general form of the Weyl anomaly for higher dimensional defects is not known. In addition, computation of entanglement entropy and Weyl anomaly coefficients for holographic examples of defect  RG flows could provide evidence for monotonicity theorems for higher-dimensional defects.

\chapter{Concluding remarks}

We have used gauge/gravity duality to investigate a variety of phenomena in strongly coupled quantum field theories. We close with a summary of our results, and some speculation for the future.

In chapter~\ref{chap:entanglement_density}, we computed entanglement density in a variety of holographic models. Many of these models exhibit area theorem violation when they approach regimes with different scaling symmetry in the IR than the UV. This potentially indicates that there is an enhanced number of low-energy degrees of freedom in such regimes. If this can be made precise, entanglement density may be a useful tool to probe the low-energy effective descriptions of physical systems.

Following this, in chapter~\ref{chap:zero_sound} we studied the spectrum of excitations in a holographic model of compressible quantum matter. As in similar models, the spectrum included a low temperature mode with sound-like dispersion --- holographic zero sound. The attenuation of holographic zero sound in this model is well described by hydrodynamics. Similar behavior has been observed for other modes and in other holographic models of compressible quantum matter~\cite{Davison:2013bxa,Gushterov:2018nht,Davison:2013uha}. Whether this property is special to holographic systems or is more general remains to be seen.

Only relatively recently has the density response of a cuprate been measured~\cite{Mitrano5392}. These initial results show no evidence for zero sound. If this finding is strengthened by further experiments, then this poses a challenge: how do we explain this using holography? Alternatively, if future experiments \textit{do} find a zero sound mode, then its properties will be a valuable input for holographic models. In either case, we hope that understanding the necessary modifications to the holographic models discussed in chapter~\ref{chap:zero_sound} will provide lessons about real non-Fermi liquids.

In chapter~\ref{chap:probe_m5} we used gauge/gravity duality to study properties of two-dimensional supersymmetric defects in the \(\cN = (2,0)\) theory. We found that the entanglement entropy for spherical subregions is not monotonic along RG flows, making it a poor candidate to measure degrees of freedom on the defect. On the other hand, two coefficients appearing in the defect contribution to the Weyl anomaly, \(b\) and \(d_2\), decreased along all of the flows that we studied. One of these coefficients, \(b\), satisfies a monotonicity theorem~\cite{Jensen:2015swa}. It is currently unknown whether the same is true for \(d_2\).

The status of monotonicity theorems and the counting of degrees of freedom on higher dimensional defects is less clear, although some conjectures have been made~\cite{Kobayashi:2018lil, Nozaki:2012qd, Gaiotto:2014gha}.\footnote{In addition, ref.~\cite{Casini:2018nym} found a version of the area theorem for boundary RG flows.} It is likely that gauge/gravity duality will play a role in future attempts to prove new monotonicity theorems. For example, explicit holographic models of defect RG flows may suggest candidate \(C\)-functions to target. It may also be easier to prove new monotonicity theorems first in holography. For example, the \(F\)-theorem was proved holographically before it was proved in general~\cite{Myers:2010xs}.

More than twenty years after the original AdS/CFT proposal, gauge/gravity duality remains an active field of research, with many applications beyond those discussed in this thesis. For instance, holography provides a toolbox for building models of QCD at finite density, with applications to heavy ion collisions and neutron star physics.

A particularly interesting outcome of holography is the connection it has revealed between the structure of spacetime and quantum information theory. For example, locality in the bulk of \aads\ may be understood as arising from quantum error correction in the dual QFT~\cite{Almheiri:2014lwa}.\footnote{See ref.~\cite{Harlow:2018fse} for a pedagogical introduction to this topic.} In the future, quantum information approaches will hopefully shed further light on spacetime in quantum gravity in \aads, such as the nature of the interior of black holes.

More generally, gauge/gravity duality has proved a useful tool for understanding QFT and quantum gravity. We hope that it will continue to provide insight and surprises in the years to come.

\cleardoublepage

\appendix
\part{Appendices}

\chapter{Appendix to chapter~\ref{chap:entanglement_density}}
\label{app:entanglement_density}

\section{Entanglement density for large spheres}
\label{app:entanglement_density_large_subregions}
 
\subsection{Matching expansions}

In this appendix we compute the large-\(\ell\) behaviour of the holographic entanglement density for the sphere geometry in spacetimes of the form~\eqref{eq:aads_metric}, with a horizon at \(z=z_H\). The large \(\ell\) limit is slightly more complicated than small \(\ell\), since in computing the area of the RT surface we must include both the UV divergent contributions from small \(z\), and the near horizon contributions which dominate in the entanglement density. To account for this, we will use the method of matched expansions used in ref.~\cite{Liu:2013una}.

\begin{figure}
	\begin{center}
		\includegraphics{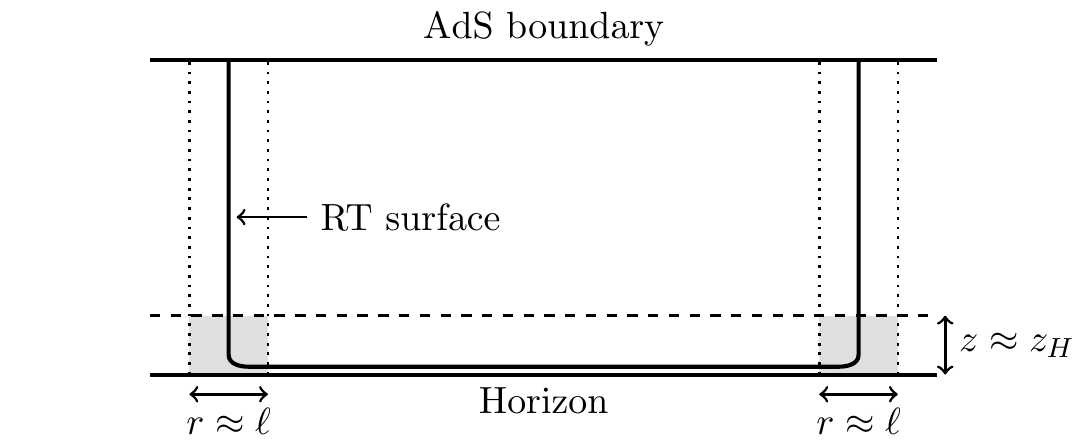}
	\end{center}
	\caption[Regions used in the large-\(\ell\) expansion of the sphere entanglement entropy.]{Diagram showing the two regions used in the matched expansion. The vertical dotted lines indicate the large-radius region \(r \approx \ell\), while the horizontal dashed line indicates the near-horizon region \(z \approx z_H\). For sufficiently large \(\ell\), part of the RT surface lies in the overlap between these two regions, indicated by the grey shading. We match the \(r \approx \ell\) and \(z \approx z_H\) expansions in the overlap.}
	\label{fig:sphere_large_radius}
\end{figure}

We assume that near the horizon, the metric function \(g(z)\) has the Taylor expansion
\begin{equation}
	g(z) = g_1 (z_H-z) + g_2 (z_H - z)^2 + \dots \; ,
\end{equation}
with \(g_1 \neq 0\). It will be useful to define \(\g \equiv \sqrt{(d-1)g_1 / 2 z_H}\). We divide the RT surface into two regimes, a near-horizon region where \(z \approx z_H\), and a large-\(r\) region where \(r \approx \ell\). For large \(\ell\), the RT surface typically drops rapidly from the boundary to the horizon at \(r \approx \ell\), almost lying flat on the horizon for \(r < \ell\). Hence, as sketched in figure~\ref{fig:sphere_large_radius}, for \(\ell \gg z_H\) the near-horizon and large-\(r\) regions overlap. We will find approximate solutions for the embedding of the RT surface in these two regions, and match them in the overlap.

In the near-horizon region, we parameterise the RT surface by the boundary spherical polar coordinates \((r,\q_i)\), so that the surface is specified by \(z(r)\). By symmetry, the maximal extent \(z_*\) of the RT surface into the bulk occurs at \(r=0\), \(z(0) = z_*\). Requiring the surface to be smooth at this point implies that \(z'(0) = 0\).

Let us write the maximal extent of the RT surface into the bulk as \(z_* = z_H(1-\d)\), where \(\d \ll 1\). The precise value of \(\d\) will be determined by matching to the large \(r\) expansion. We make the ansatz \(z(r) \approx z_* - \d z_1(r) + \cO(\d^2)\). The boundary conditions on \(z(r)\) imply \(z_1(0) = z_1'(0) = 0\). The equation of motion for \(z_1\) and the solution which obeys these boundary conditions are
\begin{equation}
	\frac{z_1''}{z_1} - \frac{z_1'^2}{2 z_1^2} + \frac{d-2}{r} \frac{z_1'}{z_1} - 2 \g^2 = 0
	\quad
	\Rightarrow
	\quad
	z_1(r) = \G^2 \le( \frac{d-1}{2} \ri) \le( \frac{\g r}{2} \ri)^{3-d} I^2_{(d-3)/2}(\g r),
\end{equation}
where \(I\) denotes a modified Bessel function of the first kind. The asymptotic form of this solution for large \(r\) is
\begin{equation} \label{eq:z1_large_r}
	z_1(r) = \G^2\left( \frac{d-1}{2} \right) \left( \frac{\g r}{2} \right)^{3-d} \frac{e^{2 \g r}}{2 \pi \g r} \left[1 + \cO(\g^{-2} r^{-2}) \right].
\end{equation}
This will be used in the matching of the two solutions.

In the large-\(r\) region, we parameterise the RT surface by \((z,\q_i)\), so that its shape is specified by \(r(z)\). Requiring that the RT surface tends to the entangling region at the boundary sets the condition \(r(0) = \ell\). We make the ansatz
\begin{equation} \label{eq:sphere_large_l_uv_expansion}
	r(z) = \ell - r_0(z) - \frac{r_1(z)}{\ell} - \dots,
\end{equation}
where the functions \(r_{0,1}(z)\) satisfy the boundary condition \(r_{0,1}(0) = 0\). Substituting this expansion into the equation of motion~\eqref{eq:sphere_equation_of_motion}, we find that \(r_{0,1}\) satisfy
\begin{align}
	r_0'' + \frac{g'}{2 g} r_0' - \frac{d-1}{z} r_0' (1 + g r_0'^2) &= 0,
	\nonumber \\
	r_1'' + \left(\frac{g'}{2 g} - \frac{(d-1)(1 + 3 g r_0'^2)}{z}\right) r_1' + \frac{d-2}{g}\left(1 + g r_0'^2\right) &= 0.
\end{align}
The solutions are 
\begin{align} \label{eq:large_radius_coefficients}
		r_0(z) &=
		\int_0^z \diff u \frac{u^{d-1}}{\sqrt{g(u)(a^{2d-2} - u^{2d-2})}},
		\nonumber \\
		r_1(z) &=
		\int_0^z \diff u \frac{u^{d-1}}{\sqrt{g(u)}[1 - (u/a)^{2d-2}]^{3/2}} \left[
			b + (d-2) \int_u^1 \diff v \frac{\sqrt{1 - (v/a)^{2d-2}}}{\sqrt{g(v)} v^{d-1}}
		\right],
\end{align}
where \(a\) and \(b\) are integration constants. The expansion should break down near the turning point at \(z_*\), since \(r(z_*) = 0\). For large radii, the turning point approaches the horizon, so we should expect that the breakdown will occur for \(z \to z_H\). In order for this to occur, we set
\(
		a = z_H
\).\footnote{To capture subleading terms in the large \(\ell\) expansion one must presumably set \(a = z_*\) instead.} The other integration constant \(b\) is to be fixed by the matching.

The two expansions should match in the overlap region where both \(r \gg z_H\) and \(z_H - z \ll z_H\). Expanding \eqref{eq:z1_large_r} for large \(\ell\) gives
\begin{align}
	z_1 &\approx \Lambda e^{-2 \g (\ell - r)} 
	\biggl(
		1
		+ \underbrace{ \frac{(d-2)(\ell-r)}{\ell} }_{c_{11}(r)} 
		+ \underbrace{ \frac{(d-1)(d-2) (\ell - r)^2}{\ell^2} }_{c_{22}(r)}
		+ \ldots
    \biggr)
    \nonumber \\
	&\equiv \Lambda C_1(r) e^{-2 \g (\ell - r)},
\end{align}
where we have defined
\begin{equation}
    \Lambda = \G^2\left( \frac{d-1}{2} \right) \frac{2^{d-4} e^{2\g \ell}}{\pi  (\g \ell)^{d-2}},
    \quad
    C_1(r) = 1 + c_{11}(r) + c_{22}(r) + \dots \; .
\end{equation}

Expanding~\eqref{eq:large_radius_coefficients} around \(z \approx z_H\) yields
\begin{align} \label{eq:large_radius_coefficients_expansion}
	r_0 &= - \frac{1}{2\g} \ln (1 - z/z_H) + b_{00} + b_{01} (1- z/z_H) + \dots \; ,
	\nonumber \\
	r_1 &= \frac{b}{4(d-1)\g (1 - z/z_H)} - b_\mathrm{\ln} \ln (1 - z/z_H) + b_{10} + b_{11}  (1 - z/z_H) + \dots \; ,
\end{align}
where
\begin{equation}	
	b_{00} = \int_0^{z_H} \diff z \left[\frac{z^{d-1}}{z_H^{d-1}\sqrt{g(z)[1 - (z/z_H)^{2d-2}]}} - \frac{1}{2\g (z_H-z)}\right].
	\label{eq:b0}
\end{equation}
We will not need the values of the other coefficients appearing in~\eqref{eq:large_radius_coefficients_expansion} to perform the matching at the order in which we are interested.

Let us take the matching point to be at \(r = r_m\) and \(z = z_m\). Keeping only the terms necessary to perform the matching at leading order, we require
\begin{subequations}
\begin{align}
    z_m  &= z(r_m) \approx z_H - \d z_1 (r_m)
    \nonumber\\
    &\approx \Lambda e^{-2 \g (\ell - r_m)},
    \\
    r_m &= r(z_m) \approx \ell - r_0(z_m) - \frac{r_1(z_m)}{\ell} 
    \nonumber \\
    &\approx
    \ell + \frac{1}{2\g} \ln \le( \frac{\d \Lambda}{z_H} e^{2 \g (r_m - \ell)} \ri) - b_{00} - \frac{b z_H}{4 (d-1) \g \ell \d \Lambda} e^{2 \g (\ell - r_m)}.
\end{align}
\end{subequations}
These two equations are consistent provided \(b = 0\) and
\begin{equation}
    \d \approx \frac{z_H}{\Lambda} e^{2 \g b_{00}} = \G^{-2} \le( \frac{d-2}{2} \ri) \frac{\pi (\g \ell)^{d-2} z_H}{2^{d-4}} e^{2\g (b_{00} - \ell)}.
\end{equation}

Everything in this appendix so far is a reproduction of the results of ref.~\cite{Liu:2013una}. In the next section we apply these results to compute the large-\(\ell\) behaviour of the entanglement density.

\subsection{Entanglement density}

Let us divide the area of the RT surface into two pieces, in which we use the two different asymptotic expansions,
\begin{gather}
    \mathrm{Area}[\cS] = L^{d-1} \vol(\sph[d-2]) \le( A_\mathrm{UV} + A_\mathrm{IR} \ri)
    \\[0.5em]
    A_\mathrm{UV} \equiv   \int_\e^{z_m} \diff z \frac{r(z)^{d-2}}{z^{d-1}} \sqrt{\frac{1}{g(z)} + r'(z)^2},
    \quad
    A_\mathrm{IR} \equiv  \int_0^{r_m} \diff r \frac{r^{d-2}}{z(r)^{d-1}} \sqrt{1 + \frac{z'(r)^2}{g(z(r))}},
    \nonumber
\end{gather}
where \((r_m, z_m)\) is the matching point and \(\e\) is the UV cutoff.

Substituting the expansion~\eqref{eq:sphere_large_l_uv_expansion} into \(A_\mathrm{UV}\), we find
\begin{equation}
	A_\mathrm{UV} = \ell^{d-3} \left[
		\int_\e^{z_m}\diff z \frac{\ell - (d-2) r_0(z)}{z^{d-1}\sqrt{g(z) [1 - (z/z_H)^{2d-2}]}}
		+ r_1(z_m) + \cO \left(\ell^{-1} \right)
	\right].
\end{equation}
This generalises equation (6.36) of ref.~\cite{Liu:2013una} to arbitrary dimension. We will only compute terms \(\cO(\ell^{d-2})\) or greater, so we simplify this to
\begin{equation}
	A_\mathrm{UV} = \ell^{d-2} 
		\int_\e^{z_m}\diff z \frac{1}{z^{d-1} \sqrt{g(z) [1 - (z/z_H)^{2d-2}]}}.
\end{equation}
Finally, some slight manipulation of the upper limit on the integral, along with the approximation that \(z_m \approx z_H\), leads to the expression
\begin{multline} \label{eq:sphere_large_l_area_uv}
	A_\mathrm{UV} \approx \ell^{d-2} \int_\e^{z_H} \diff z \le[ \frac{1}{z^{d-1} \sqrt{g(z) [1 - (z/z_H)^{2d-2}]}} - \frac{1}{2  \g z_H^{d-1} ( z_H - z )} \ri]
	\\
	 - \frac{\ell^{d-2}}{2 \g z_H^{d-1}} \ln \le(1 - \frac{z_m}{z_H} \ri).
\end{multline}
The IR contribution is simpler,
\begin{equation} \label{eq:sphere_large_l_area_ir}
    A_\mathrm{IR} = \frac{r_m^{d-1}}{(d-1) z_H^{d-1}} + \cO(\d) \approx \frac{\ell^{d-1}}{(d-1)z_H^{d-1}} + \frac{\ell^{d-2}}{z_H^{d-1}} \le[\frac{1}{2\g} \ln \le(1 - \frac{z_m}{z_H} \ri) - b_{00} \ri].
\end{equation}

To obtain the entanglement density, we need to subtract off the area of the RT surface with the same radius \(\ell\) in pure \ads. We write this area as \(\mathrm{Area}[\cS_\mathrm{AdS}] = L^{d-1} \vol(\sph[d-2]) A_\mathrm{AdS}\), where the pure \ads\ area integral is
\begin{align} \label{eq:sphere_large_l_area_ads}
    A_\mathrm{AdS} &= \ell \int_\e^\ell \diff z \frac{1}{z^{d-1}} \le( \ell^2 - z^2 \ri)^{(d-3)/2}
    \nonumber \\
    &\approx \ell^{d-2} \int_\e^{z_H}  \diff z \frac{1}{z^{d-1}}  + \frac{\ell^{d-2}}{(d-2)z_H^{d-2}} + \cO(z_H/\ell).
\end{align}
The first line of~\eqref{eq:sphere_large_l_area_ads} follows from the substitution of the solution \(r(z) = \sqrt{\ell^2 - z^2}\) into the area integral~\eqref{eq:sphere_area}.

\begin{figure}
	\begin{center}
		\includegraphics[width=0.5\textwidth]{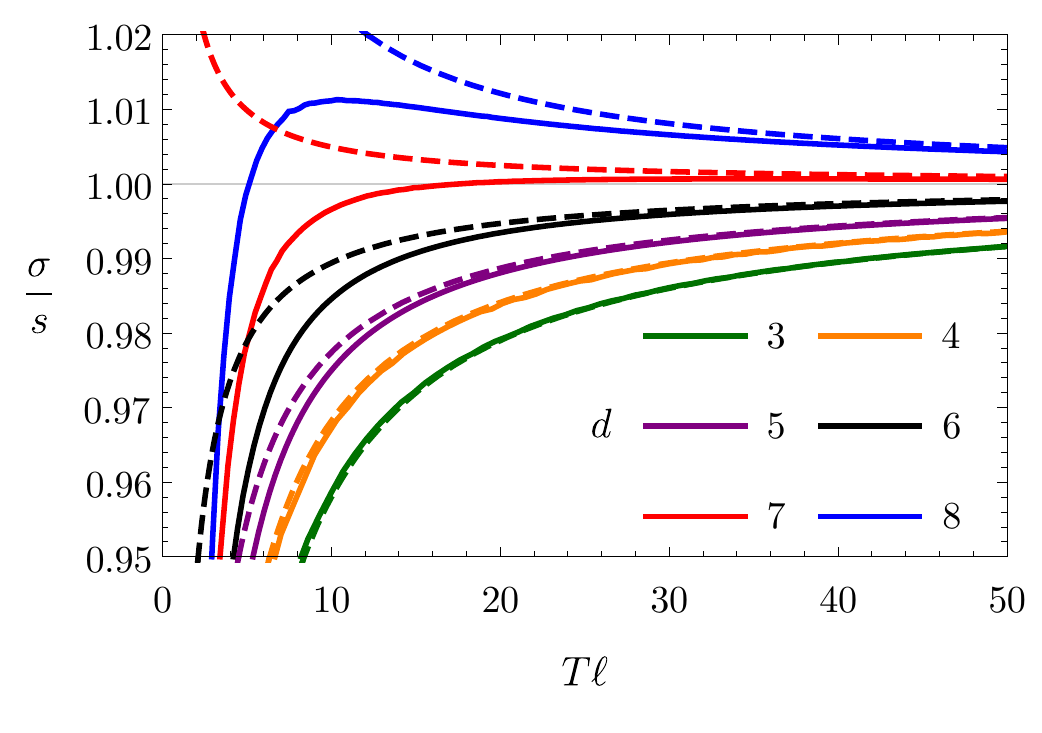}
	\end{center}
	\caption[Comparison of numerical results to large radius approximation for the entanglement density of the sphere geometry in \ads-Schwarzschild.]{Entanglement density for the sphere geometry in \ads[d+1]-Schwarzschild as a function of sphere radius \(\ell\), for different values of \(d\). The solid lines are numerical results, and the dotted lines are the large radius approximation~\eqref{eq:sphere_sigma_large_l_appendix}. The approximation works well for sufficiently large \(T\ell\), where \(T\) is the temperature.}
	\label{fig:ads_schwarzschild_sigma_large_l}
\end{figure}

Combining equations~\eqref{eq:sphere_large_l_area_uv}, \eqref{eq:sphere_large_l_area_ir}, and \eqref{eq:sphere_large_l_area_ads}, we obtain the subtracted area,
\begin{align}
	\D A &\equiv A_\mathrm{UV} + A_\mathrm{IR} - A_\mathrm{AdS}
	\\
	 &= \frac{\ell^{d-1}}{(d-1)z_H^{d-1}}  - \frac{\ell^{d-2}}{z_H^{d-2}} \le(\frac{1}{d-2} + \frac{b_{00}}{z_H} \ri)
	\nonumber \\ & \phantom{=}
	 + \ell^{d-2} \int_\e^{z_H} \diff z \le[ \frac{1}{z^{d-1}} \le( \frac{1}{\sqrt{g(z) [1 - (z/z_H)^{2d-2}]}} - 1 \ri) - \frac{1}{2 \g z_H^{d-1} (z_H - z)}\ri]  + \dots \; .
\end{align}
Noting that the entanglement density is \(\s = (d-1) L^{d-1} \D A / 4 \gn \ell^{d-1}\) and that the thermodynamic entropy density is \(s = L^{d-1}/4 \gn z_H^{d-1}\), and using expression~\eqref{eq:b0} for \(b_{00}\), we find that the entanglement density for large sphere radius is
\begin{equation} \label{eq:sphere_sigma_large_l_appendix}
	\s_\mathrm{sphere} \approx s \le[1 + \frac{(d-1) z_H}{\ell} C(z_H) \ri] + \dots \;,
\end{equation}
where, changing integration variables to \(u = z/z_H\),
\begin{equation}
	C(z_H) \equiv - \frac{1}{d-2} + \int_0^{1} \diff u \frac{1}{u^{d-1}} \le( \sqrt{\frac{1 - u^{2d-2} }{ g(z_H u) } } - 1 \ri).
\end{equation}
We compare the approximation~\eqref{eq:sphere_sigma_large_l_appendix} to numerical results for the entanglement density in \ads[d+1]-Schwarzschild in figure~\ref{fig:ads_schwarzschild_sigma_large_l}, finding good agreement at sufficiently large sphere radii.

\section{Monotonicity of \(C(z_H)\) for \ads-Schwarzschild}
\label{app:monotonicity_of_c_for_ads_schwarzschild}

In this section we wish to show that the coefficient \(C(z_H)\), defined in~\eqref{eq:large_l_coefficient_integral}, increases monotonically with \(d\) for \ads[d+1]-Schwarzschild.

We continue \(d\) to non-integer values, and differentiate~\eqref{eq:large_l_coefficient_integral} with respect to \(d\) to obtain
\begin{equation}
\label{eq:dnschwarz_dd}
	\frac{\partial C(z_H)}{\partial d} = \frac{1}{(d-2)^2} + \int_0^1 \diff u \, \frac{\ln(u)}{u^{d-1}} \left(1 - \frac{1}{2} \frac{2 + u^{3d-2} -  3u^{d}}{(1-u^d)^{3/2}(1-u^{2d-2})^{1/2}}\right),
\end{equation}
where we have used \(g(z) = 1 - (z/z_H)^d\) for \ads-Schwarzschild. We will prove that
\begin{equation}
\label{eq:condition}
\frac{1}{2} \frac{2 + u^{3d-2} -  3u^{d}}{(1-u^d)^{3/2}(1-u^{2d-2})^{1/2}} \geq 1,
\end{equation}
for $u \in [0,1]$. Since $u^{1-d} \ln u\leq 0$ for $u \in [0,1]$, this implies that \(\p C(z_H)/\p d \geq 0\).

Both the numerator and denominator in~\eqref{eq:condition} are positive, so multiplying both sides of~\eqref{eq:condition} by $(1-u^d)^{3/2}(1-u^{2d-2})^{1/2}$, squaring, and re-arranging, we find that the condition~\eqref{eq:condition} is equivalent to
\begin{equation}
\label{eq:monotonicity_inequality}
\left(1 + \frac{1}{2} u^{3d-2} - \frac{3}{2} u^d\right)^2 - (1-u^d)^3 (1-u^{2d-2}) \geq 0.
\end{equation}
Since $u^{2d-2} \geq u^{2d}$ and \(u^{3d-2} \geq u^{3d}\) for $u \in [0,1]$, we have
\begin{align}
\left(1 + \frac{u^{3d-2}}{2}  - \frac{3u^d}{2} \right)^2 -	(1-u^d)^3 (1-u^{2d-2}) &\geq \left(1 + \frac{u^{3d}}{2}  - \frac{3u^d}{2} \right)^2 - (1-u^d)^3 (1-u^{2d}) \nonumber \\ &= \frac{1}{4} u^{2d} (1 - u^d)^4 \geq 0,
\end{align}
which proves the inequality~\eqref{eq:monotonicity_inequality}, and therefore the monotonicity of \(C(z_H)\) with \(d\).

\section{Large-width entanglement entropy for hyperscaling-violating geometries with \(\q = d-2\)}
\label{app:hyperscaling_large_width}

In this appendix we derive the logarithmic violation of the area law~\eqref{eq:arealawviolation}, for the strip entanglement entropy for the models of section~\ref{hyper} with \(\q = d-2\).

Recall that the width of the strip is related to the maximal extension \(z_*\) of the RT surface into the bulk by~\eqref{eq:ldef}. For large \(\ell\) (and hence large \(z_*\)), the integral in~\eqref{eq:ldef} is dominated by contributions coming from large \(z\). From the asymptotic scalings~\eqref{eq:hvscalings}, for \(\q = d-2\) we find that the metric function \(g(z)\) has the large-\(z\) expansion
\begin{equation} \label{eq:hyper_metric_expansion}
	g(z) = \frac{\g}{(\m z)^{2(d-2)}} \le[1 + \cO(\m^{-1} z^{-1}) \ri],
\end{equation}
where \(\m\) is the chemical potential, and \(\g\) is a numerical constant, determined from the solution to the equations of motion. Substituting the metric expansion~\eqref{eq:hyper_metric_expansion} into~\eqref{eq:ldef}, we find
\begin{equation} \label{eq:hyper_z_star}
	\ell = \frac{2 \m^{d-2} z_*^{d-1}}{\sqrt{\g}(d-1)} \le[
		1 + \cO(\m^{-1} z_*^{-1})
	\ri].
\end{equation}

To obtain a large-\(\ell\) approximation to the entanglement density~\eqref{eq:hyper_metric_expansion}, we need to find an approximate expression for \(C(z_*)\), and then replace all factors of \(z_*\) using~\eqref{eq:hyper_z_star}. One subtlety is that the integral appearing in~\eqref{eq:large_l_coefficient_integral} receives significant contributions from near the boundary, not just from large \(z\), due to the factor of \(1/u^{d-1}\). To make progress, we rewrite \(C(z_*)\) as
\begin{align} \label{eq:hyper_c_z_rewrite}
	C(z_*) &= -\frac{1}{d-2} + z_*^{d-2} \int_0^{z_*} \diff z \le[ 
		\frac{1}{z^{d-1}} \le( \sqrt{\frac{1 - (z/z_*)^{2d-2}}{g(z)}} - 1 \ri)
		- \frac{1}{\sqrt{\g}}\frac{\m^{2d-3} z^{d-2}}{1 + \m^{d-1} z^{d-1}}
	\ri]
	\nonumber \\
	& + z_*^{d-2} \int_0^{z_*} \diff z \frac{1}{\sqrt{\g}}\frac{\m^{2d-3} z^{d-2}}{1 + \m^{d-1} z^{d-1}}.
\end{align}
The integral on the first line has a finite limit as \(z_* \to \infty\),
\begin{equation}
	\mathcal{I} \equiv \frac{1}{\m^{d-2}} \lim_{z_* \to \infty} \int_0^{z_*} \diff z \le[ 
		\frac{1}{z^{d-1}} \le( \sqrt{\frac{1 - (z/z_*)^{2d-2}}{g(z)}} - 1 \ri)
		- \frac{1}{\sqrt{\g}}\frac{\m^{2d-3} z^{d-2}}{1 + \m^{d-1} z^{d-1}}
	\ri],
\end{equation}
which we will evaluate numerically. The integral on the second line of~\eqref{eq:hyper_c_z_rewrite} is
\begin{equation}
	z_*^{d-2} \int_0^{z_*} \diff z \frac{1}{\sqrt{\g}}\frac{\m^{2d-3} z^{d-2}}{1 + \m^{d-1} z^{d-1}} = \frac{(\m z_*)^{d-2}}{\sqrt{\g} (d-1)} \ln \le[1 + (\m z_*)^{d-1} \ri].
\end{equation}
We thus arrive at an approximate expression for \(C(z_*)\) at large \(z_*\),
\begin{equation}
	C(z_*) \approx \mathcal{I} (\m z_*)^{d-2} + \frac{(\m z_*)^{d-2}}{\sqrt{\g} (d-1)} \le[ \ln(\m z_*)^{d-1} +  \frac{1}{(\m z_*)^{d-1}} \ri].
\end{equation}

Substituting this result into~\eqref{eq:hyper_entanglement_density}, we find
\begin{equation}
	\s \approx \frac{\m^{d-1} L^{d-1}}{2 \gn} \le\{
		\frac{1}{\sqrt{\g}(d-1) \m \ell} \ln \le[ \frac{\sqrt{\g} (d-1) \m \ell}{2} \ri] + \le[\frac{1}{\sqrt{\g}(d-1)} + \mathcal{I} \ri] \frac{1}{\m \ell}
	\ri\},
\end{equation}
which is the anticipated logarithmic violation of the area law. For the solution with \(d=3\), \(\z=3\) and \(\q=1\) discussed in section~\ref{hyper}, we find \(\g = 2.068\) and \(\mathcal{I} = -0.2614\), leading to~\eqref{eq:arealawviolation}.

\chapter{Appendix to chapter~\ref{chap:zero_sound}}
\label{app:zero_sound}

\section{Equations of motion for fluctuations}
\label{app:zero_sound_eom}

In this appendix, we list the coefficients appearing in the equations of motion~\eqref{eq:fluctuation_eoms} for \(Z_1\) and \(Z_2\). So simplify the equations slightly, we define the combination
\begin{equation} \label{eq:hzs_curly_F}
	\mathcal{F}(z) = \sqrt{1 - \talpha^2 z^4 F_{tz}^2}.
\end{equation}
The coefficients are
\begin{align}
	A_1 &= \frac{1}{f \mathcal{F}^2 z \left(f k^2 \mathcal{F}^2-\omega ^2\right) \left(k^2 \left(z f'-4 f\right)+4 \omega ^2\right)}
	\nonumber \\ & \phantom{=}
		\times \biggl\{
			k^4 f \mathcal{F}
		\left[  \tau \left(\mathcal{F}^2-1\right) \left(2 f \left(\mathcal{F}^2+1\right)-z f'\right)-3 z^6 f  \mathcal{F}^2  \mathcal{F}' \left(f/z^4\right)'\right]
		+ 4 f^2 \mathcal{F} \omega ^4 z \left(\mathcal{F}/f\right)'
	\nonumber \\ & \phantom{= \times \biggl[}
		+ k^2 \omega ^2 \mathcal{F}  \left[-4 z f^2  \mathcal{F}'\left(3 \mathcal{F}^2+1\right) +f \left(z^2 f' \mathcal{F}'+4 \mathcal{F} z f'-4 \mathcal{F}^2 \tau +4 \tau \right)-\mathcal{F} z^2
		   f'^2\right]
	\biggr\},
	\\
	A_2 &= \frac{k \mathcal{F}^2}{z^4\talpha^2 F_{tz}\mathcal{F}^3 \left(f k^2 \mathcal{F}^2-\omega ^2\right) \left(k^2 \left(z f'-4 f\right)+4 \omega ^2\right)^2}
	\nonumber \\ & \phantom{=}
	\times \biggl\{
		k^4 \Bigl[
			 \left(1-\mathcal{F}^2\right) \Bigl(-4 f^2 \mathcal{F} \left(\mathcal{F}^4+3
			   \mathcal{F}^2-6\right)+2 f \left(-3 z \mathcal{F} f'+\tau -\tau  \mathcal{F}^4\right)
	\nonumber \\ & \hspace{1.5cm}
				+z f' \left(z \mathcal{F}^3
			   f'+\tau  \left(\mathcal{F}^2-1\right)\right)\Bigr)-f \mathcal{F} \left(\mathcal{F}^2-3\right) \left(z f'-4
			   f\right) \left(z \mathcal{F} \mathcal{F}'-2 \mathcal{F}^2+2\right)
		\Bigr]
	\nonumber \\ & \phantom{= \times \biggl[}
		+ k^2 \omega ^2 \Bigl[
			   -2 \left(\mathcal{F}^2-1\right) \left(\mathcal{F} \left(\left(2 \mathcal{F}^2-1\right) z f'+f \left(6 \mathcal{F}^2+2\right)+2 \mathcal{F} \tau \right)-2 \tau
			      \right)
			  \nonumber \\ & \hspace{6cm}
			  - z \mathcal{F}' \left(\left(\mathcal{F}^2+1\right) z f'+4 f \left(\mathcal{F}^4-4 \mathcal{F}^2-1\right)\right)
		   \Bigr]
	\nonumber \\ & \phantom{= \times \biggl[}
		-4 \omega ^4 \left(\left(\mathcal{F}^2+1\right) z \mathcal{F}'-2 \mathcal{F} \left(\mathcal{F}^2-1\right)\right)
	\biggr\},
\end{align}
\begin{align}
	A_3 &= \frac{1}{f^2 \mathcal{F}^3 z^2 \left(f k^2 \mathcal{F}^2-\omega ^2\right) \left(k^2 \left(z f'-4 f\right)+4 \omega ^2\right)^2}
	\nonumber \\ & \phantom{=}
	\times \biggl\{
		2  k^6  \tau f^2 \mathcal{F} \Bigl[4 f^2 \mathcal{F}^3 \left(\mathcal{F}^4-1\right) +z f' \left(\mathcal{F}^2-1\right) \left(2  z f'\mathcal{F}^3 + \tau\left(\mathcal{F}^2-1\right) \right)
	\nonumber \\ & \phantom{= \biggl\{}
		+z f  \mathcal{F}^2  \mathcal{F}' \left(\mathcal{F}^2-3\right) \left(z f'-4 f\right)-2 f \left(\mathcal{F}^2-1\right) \left( z
				   f' \mathcal{F}^3\left(\mathcal{F}^2+4\right)+ \tau \left(\mathcal{F}^4-1\right)\right)\Bigr]
	\nonumber \\ & \phantom{= \times \biggl[}
		-k^8 z^{12}  f^2 \mathcal{F}^7 \left(f/z^4\right)'^2
		+ 2  k^6 \omega ^2 z^7 f \mathcal{F}^5  \left(f/z^4\right)' \left(z f'-4 f \left(\mathcal{F}^2+1\right)\right)
	\nonumber \\ & \phantom{= \times \biggl[}
		+ 2 k^4 \omega ^2  \tau  f  \mathcal{F} \Bigl[z f \mathcal{F}' \Bigl( z f'\left(\mathcal{F}^2+1\right)+4 f \left(\mathcal{F}^4-4
		   \mathcal{F}^2-1\right)\Bigr)
		  \nonumber \\ & \hspace{3.1cm}
		   -\left(\mathcal{F}^2-1\right) \Bigl(-2 z f' f \mathcal{F} \left(7 \mathcal{F}^2+2\right) +\mathcal{F} z^2 f'^2
		   \nonumber \\ & \hspace{5.5cm}
		   +4 f \left\{f
		   \left(2 \mathcal{F}^5+5 \mathcal{F}^3+\mathcal{F}\right)+\tau \left( \mathcal{F}^2 + 1\right) \right\}\Bigr)\Bigr]
	\nonumber \\ & \phantom{= \times \biggl[}
		+ 8 k^2 \omega ^4 \tau  f \mathcal{F}   \left[2 \mathcal{F} \left(\mathcal{F}^2-1\right) \left(f \left(3 \mathcal{F}^2+2\right)-z f'\right)+ z f \mathcal{F}' \left(\mathcal{F}^2+1\right)
		  \right]
	\nonumber \\ & \phantom{= \times \biggl[}
		-k^4  \omega ^4 z^2 \mathcal{F}^3 \left[16 f^2 \left(\mathcal{F}^4+4 \mathcal{F}^2+1\right)-8 z f' f \left(2 \mathcal{F}^2+1\right)+z^2
		   f'^2\right]
	\nonumber \\ & \phantom{= \times \biggl[}
		-32 \omega^6 \tau  f \mathcal{F}^2 \left(\mathcal{F}^2-1\right) 
		+ 8 \omega ^6 k^2 z^2  \mathcal{F}^3 \left[4 f \left(\mathcal{F}^2+1\right)-z f'\right]
		-16 \omega ^8  z^2 \mathcal{F}^3
	\biggr\},
\end{align}
\begin{align}
	A_4 &= \frac{\mathcal{F}^2}{2 \talpha^2 F_{tz} z^5  f k \mathcal{F}^3 \left(\omega ^2-f k^2 \mathcal{F}^2\right) \left(k^2 \left(z f'-4 f\right)+4 \omega ^2\right)^2}
	\nonumber \\ & \phantom{=}
		\times \biggl\{
			k^6 f\Bigl[
				4 z f'\left(\mathcal{F}^2-1\right)  \left\{f\mathcal{F}^3(3\mathcal{F}^2-1) - \tau(\mathcal{F}^2-1) \right\}-2 z^2 f'^2 \mathcal{F}^3 \left(\mathcal{F}^4-1\right)
				 \nonumber \\ & \phantom{= \times \biggl\{ }
				  + z^6 \mathcal{F}^2 \mathcal{F}' \left(f/z^4\right)' \left(\mathcal{F}^2 z f'+f \left(12-8 \mathcal{F}^2\right)\right)
				 +8 f
				   \left(\mathcal{F}^2-1\right) \left(4 f \mathcal{F}^3+\left(\mathcal{F}^4-1\right) \tau \right)
			\Bigr]
		\nonumber \\ & \phantom{= \times \biggl\{}
			+ 2 k^8 z^7 f \mathcal{F}^3 \left(\mathcal{F}^2-1\right)^2 \left(f/z^4\right)'
			- k^4  \omega ^4 z^2 \mathcal{F} \left(\mathcal{F}^2-1\right)^2
						+ 16\omega ^6 \left(-2 \mathcal{F}^3+z \mathcal{F}'+2 \mathcal{F}\right)
		\nonumber \\ & \phantom{= \times \biggl\{}
			-k^4 \omega ^2 \Bigl[
				16 f^2 \Bigl(2 \mathcal{F} \left(-2 \mathcal{F}^4+\mathcal{F}^2+1\right)+ z \mathcal{F}' \left(-4 \mathcal{F}^4+6 \mathcal{F}^2+1\right) \Bigr)
				\nonumber \\ & \phantom{= \times \biggl\{ }
				-4 f \mathcal{F} \left(\mathcal{F}^2-1\right) 
				   \Bigl( z f' \left(2 \mathcal{F}^4+7 \mathcal{F}^2-5\right)+2 \tau \mathcal{F} \left(\mathcal{F}^2+2\right)  \Bigr)- z^2 f' \mathcal{F}' \mathcal{F} \left(3 \mathcal{F}^2+1\right)
				   \nonumber \\ & \hspace{2.5cm}
				    -6 \tau \mathcal{F}+z f' \left(2 z f' \left(3 \mathcal{F}^5-4 \mathcal{F}^3+\mathcal{F}\right) -\mathcal{F}^2 z^2 f' \mathcal{F}'+4 \tau \left(\mathcal{F}^2-1\right)^2 \right)\Bigr]
		\nonumber \\ & \phantom{= \times \biggl\{}
			+2 k^6 \omega^2 z^2 \mathcal{F} \left(\mathcal{F}^2-1\right)^2  \left[4 f \left(\mathcal{F}^2+1\right)-z f'\right]
		\nonumber \\ & \phantom{= \times \biggl\{}
			+ 4 k^2 \omega ^4 \Bigl[
				2 \left(\mathcal{F}^2-1\right) \left\{\mathcal{F} \left(\left(3 \mathcal{F}^2-2\right) z f'+4 f \left(\mathcal{F}^2+2\right)+2 \mathcal{F} \tau \right)-2 \tau
				   \right\}
				\nonumber \\ & \hspace{3.1cm}
				+z \mathcal{F}' \left\{4 f \left(\mathcal{F}^2-2\right) \left(2 \mathcal{F}^2+1\right)- z f'\left(\mathcal{F}^2-1\right)\right\}
			\Bigr]
		\biggr\},
\end{align}
\begin{align}
	B_1 &= \frac{ \tau k  \tilde{\alpha}^2 z^2 F_{t z} \left(k^2 \left(z f'-2 f \left(\mathcal{F}^2+1\right)\right)+4 \omega ^2\right)}{\mathcal{F}(f
	   k^2 \mathcal{F}^2- \omega ^2)},
	\\[1em]
	B_2 &= \frac{1}{f \mathcal{F} z \left(f k^2 \mathcal{F}^2-\omega ^2\right) \left(k^2 \left(z f'-4 f\right)+4 \omega ^2\right)}
	\nonumber \\ & \phantom{=}
	\times \biggl[
		k^4 f \left(8 f^2 \mathcal{F}^3-2 f \left(\mathcal{F}^3 z^2 f''(z)+\left(\mathcal{F}^4-1\right) \tau \right)+z f' \left(\mathcal{F}^3 z f'+\left(\mathcal{F}^2-1\right)
		   \tau \right)\right)
	\nonumber \\ & \phantom{= \times \biggl[}
		+k^2 \omega ^2 \left(-8 f^2 \left(\mathcal{F}^3+\mathcal{F}\right)-\mathcal{F} z^2 f'^2+2 f \left(\mathcal{F} \left(z^2 f''(z)+2 \mathcal{F} \left(\mathcal{F} z
		   f'+\tau \right)\right)-2 \tau \right)\right)
	\nonumber \\ & \phantom{ = \times \biggl[}
	+ 4  \omega ^4 \mathcal{F}  \left(2 f-z f'\right)
	\biggr],
	\\[1em]
	B_3 &= \frac{\tau k \talpha^2 z F_{tz}}{f \mathcal{F}^2 \left(\omega ^2-f k^2 \mathcal{F}^2\right) \left(k^2 \left(z f'-4 f\right)+4 \omega ^2\right)}
	\nonumber \\ & \phantom{=}
	\times \biggl\{
		2 k^4 f  \left[z f' \Bigl(2 \mathcal{F}^3 z f'+ \tau \left(\mathcal{F}^2-1\right) \Bigr)-2 f \Bigl(\mathcal{F}^3 z
		   \left(z f''+f'\right)+ \tau (\mathcal{F}^4 - 1)\Bigr)\right]
	\nonumber \\ &\phantom{= \times \biggl[}
		+ 4 k^2\omega ^2   \left[f \Bigl(4 \mathcal{F}^3 z f'+\mathcal{F} z \left(z f''+f'\right)+2  \tau (\mathcal{F}^2 - 1) \Bigr)-z^2 f'^2 \mathcal{F} \right]
		-16 \omega ^4  z f' \mathcal{F}
	\biggr\},
	\\[1em]
	B_4 &= \frac{1}{f^2 \mathcal{F} z^2 \left(f k^2 \mathcal{F}^2-\omega ^2\right) \left(k^2 \left(z f'-4 f\right)+4 \omega ^2\right)}
	\nonumber \\ & \phantom{=}
	\times \biggl\{
		k^4 f^2  \left[4 z f \left(z^3  \mathcal{F}^3 \left(f'/z^2\right)'+ \tau (\mathcal{F}^4-1)\right)-z f' \Bigl(z f' \mathcal{F}^3 +2 \tau 
		   \left(\mathcal{F}^2-1\right) \Bigr)\right]
	\nonumber \\ & \phantom{= \biggl[ \times}
		+ k^6 z^7  f^2  \mathcal{F}^3\left(f/z^4\right)'
		+ k^4 \omega ^2 z^2 f \mathcal{F} \left[z f'\left(\mathcal{F}^2+1\right) -4 f \left(2 \mathcal{F}^2+1\right)\right]
	\nonumber \\ & \phantom{= \biggl[ \times}
		+k^2 \omega ^2 f \Bigl[z f' \Bigl(z f' \mathcal{F}+2 \tau \left(\mathcal{F}^2-1\right)  \Bigr)
	\nonumber \\ & \hspace{3cm} -4 f \left(z^2 f''\left(\mathcal{F}^3+\mathcal{F}\right) -2
		   z f' \Bigl(\mathcal{F}^3+\mathcal{F}\right) + \tau \left(\mathcal{F}^2+2\right) \mathcal{F}^2  -3 \tau \Bigr)\Bigr]
	\nonumber \\ & \phantom{= \biggl[ \times}
		+ 4 f \omega ^4 \left[z^4 \mathcal{F} \left(f'/z^2\right)'+2 \tau \mathcal{F}^2  -2 \tau \right]
	\nonumber \\ & \phantom{= \biggl[ \times}
		+ k^2 \mathcal{F} \omega ^4 z^2 \left[4 f \left(\mathcal{F}^2+2\right)-z f'\right]
		-4 \omega ^6 \mathcal{F} z^2
	\biggr\}.
\end{align}

\section{Numerical methods}
\label{app:zero_sound_numerics}

In this appendix we discuss technical details of our holographic calculations of the retarded Green's functions, their poles, and the spectral functions. We use the shooting method developed in ref.~\cite{Kaminski:2009dh}.

The poles in the Green's functions are holographically dual to the quasinormal modes of the black brane solution~\eqref{bgsol}, which we determine numerically as follows. For a given \(\w\) and \(k\), we form two independent solutions to the equations of motion~\eqref{eq:fluctuation_eoms}, which we label \(Z_{a,I}\), where \(a \in \{1,2\}\) labels the fluctuations defined in~\eqref{zdef}, while \(I \in \{1,2\}\) labels the independent solutions. The solutions are constructed by integrating the equations of motion~\eqref{eq:fluctuation_eoms} from the horizon, with boundary conditions \(c_a^\mathrm{out} = 0\) for all \(a\) and \(I\), and
\begin{alignat}{2}
		&c_1^\mathrm{in} = c_2^\mathrm{in} = 1, & I = 1,
		\nonumber \\
		& c_1^\mathrm{in}= - c_2^\mathrm{out}= 1,\qquad  &  I = 2.
		\label{eq:hzs_horizon_boundary_conditions}
\end{alignat}
Since the equations of motion~\eqref{eq:fluctuation_eoms} are linear, any solution obeying ingoing boundary conditions may be written as a linear superposition of the \(Z_{a,I}\).

We now construct a matrix from the solutions,
\begin{equation} \label{eq:hzs_solution_matrix}
		 \mathbf{Z}_{aI} (z) = \begin{pmatrix}
			Z_{1,1}(z) & Z_{1,2} (z)
			\\
			Z_{2,1}(z) & Z_{2,2} (z)
		\end{pmatrix}.
\end{equation}
In the limit \(z\to0\), the elements of this matrix reduce to the boundary values, \(Z_{a,I}^{(0)}\) as defined in~\eqref{eq:hzs_Z_near_boundary},
\begin{equation}
		\lim_{z\to 0} \mathbf{Z}_{aI} (z) = \begin{pmatrix}
			Z_{1,1}^{(0)} & Z_{1,2}^{(0)}
			\\
			Z_{2,1}^{(0)} & Z_{2,2}^{(0)}
		\end{pmatrix}.
		\label{eq:hzs_solution_matrix_boundary}
\end{equation}
The determinant of~\eqref{eq:hzs_solution_matrix_boundary} vanishes when there exists a linear combination of the solutions  which is normalisable at the boundary. The quasinormal modes are defined as the frequencies for which such a solution exists, so we determine the quasinormal modes for a given \(k\) by numerically searching for the values of \(\w\) for which the determinant of~\eqref{eq:hzs_solution_matrix_boundary} vanishes.

The Green's functions are determined from the on-shell action for the fluctuations, obtained by expanding~\eqref{eq:hzs_model} up to quadratic order in the fluctuations. Writing the on-shell action as
\begin{equation} \label{eq:on_shell_action}
	S_\mathrm{grav}^\star = \int_{\e}^{z_H} \diff z \int \frac{\diff \omega \diff^2 k}{(2\pi)^3}   C_{ab}(z)  \partial_z Z_a(z,-\omega,-k)  \partial_z Z_b(z,\omega,k) + \dots \; ,
\end{equation}
where the ellipsis denotes terms containing at most one derivative with respect to \(z\), as well as counterterms from holographic renormalisation. These terms are cumbersome, and cannot be written purely in terms of the gauge invariant variables~\(Z_a\). The coefficients \(C_{ab}(z)\) appearing in~\eqref{eq:on_shell_action} are
\begin{align}
	C_{11}(z) &= \frac{1}{16 \pi \gn} \frac{\tau  \tilde{\alpha}^2 \, f}{\mathcal{F} \left(\omega ^2-f \mathcal{F}^2 k^2 \right)},
	\nonumber \\
	C_{12}(z) &= - C_{21} = - \frac{1}{16 \pi \gn}\frac{i \tau \talpha^2 L^2 \, z f^2  k \, F_{tz}  }{\mathcal{F} \left(\omega ^2-f \mathcal{F}^2 k^2 \right) \left[k^2 \left(z f'(z)-4
	   f\right)+4 \omega ^2\right]},
	\\
	C_{22}(z) &= \frac{1}{16 \pi \gn} \frac{f^3 L^2 \left[2 \mathcal{F} (f \mathcal{F}^2 k^2  - \omega^2 ) - k^2 z^4 \tau \talpha^2 F_{tz}^2 \right]}{z^2 \left(f k^2 \mathcal{F}^3-\mathcal{F} \omega^2\right)
	   \left[k^2 \left(z f'(z)-4 f\right)+4 \omega ^2\right]^2},
	\nonumber
\end{align}
where \(\mathcal{F}\) was defined in~\eqref{eq:hzs_curly_F}. 

If we define the matrix\footnote{We note that \(F\) is independent of the choice of boundary conditions~\eqref{eq:hzs_horizon_boundary_conditions}. To see this, note that if we choose a new pair of independent boundary conditions at the horizon, the resulting solutions for \(Z_{1,2}\) may be expressed as a linear combination of the solutions with boundary conditions~\eqref{eq:hzs_horizon_boundary_conditions}, by linearity of the equations of motion. In terms of the matrix~\eqref{eq:hzs_solution_matrix}, the change in boundary conditions therefore amounts to replacing \(\mathbf{Z}\) with \(\mathbf{Z} M\), for some constant matrix \(M\). The combination \(\mathbf{Z}(z) \mathbf{Z}^{-1}(\e)\) is manifestly invariant under this replacement.}
\begin{equation}
F(z) \equiv \mathbf{Z}(z) \mathbf{Z}^{-1}(\e),
\end{equation}
then applying the procedure outlined in section~\ref{sec:quasinormal_modes}, we can write the retarded Green's functions of the functions holographically dual to \(Z_a\) as~\cite{Kaminski:2009dh}
\begin{equation}
\label{eq:retgreen}
G_{ab}(\w,k) = - \lim_{\e\to0} \left[\left(C_{ac}(\e)+C^*_{ca} (\e) \right) F_{cb}'(\e) + \ldots\right],
\end{equation}
where $F' \equiv \partial_z F$. The ellipsis denotes terms descending from the ellipsis in~\eqref{eq:on_shell_action}. We will not need the explicit form of these terms.

The retarded Green's functions for the sound-channel components of \(J^\m\) and \(T^{\m\n}\) may be obtained from~\eqref{eq:retgreen}, as well as the relationship~\eqref{zdef} between the gauge invariant variables and \(\d A_\m\) and \(\d T_{\m\n}\). For example, we find
\begin{align}
	G_{J^t J^t}(\w,k) &= \frac{k^2 \t \talpha^2}{8 \pi \gn (k^2 - \w^2)} \frac{
		Z_{2,2}^{(0)} Z_{1,1}^{(1)} - Z_{2,1}^{(0)} Z_{1,2}^{(1)}
	}{
		Z_{1,1}^{(0)} Z_{2,2}^{(0)} - Z_{1,2}^{(0)} Z_{2,1}^{(0)}
	} + \dots \; ,
	\nonumber \\
	G_{T^{tt} T^{tt}}(\w,k) &= - \frac{3 L^2 k^4}{64 \pi \gn (k^2 - \w^2)^2} \frac{
		Z_{1,1}^{(0)} Z_{2,2}^{(3)} - Z_{1,2}^{(0)} Z_{2,1}^{(3)}
	}{
		Z_{1,1}^{(0)} Z_{2,2}^{(0)} - Z_{1,2}^{(0)} Z_{2,1}^{(0)}
	} + \dots \; ,
	\label{eq:greens_function_examples}
\end{align}
where \(Z_{a,I}^{(n)}\) is the \(\cO(z^n)\) term in the near-boundary expansion of \(Z_a\), with horizon boundary conditions \(I\). The ellipses in~\eqref{eq:greens_function_examples} denote contact terms, which arise due to the ellipsis in~\eqref{eq:on_shell_action}. The contact terms are analytic in \(\w\), so do not affect the poles of the Green's functions, and real \cite{Edalati:2010pn,Davison:2011uk}, so do not contribute to the spectral functions~\eqref{eq:hzs_spectral_functions}.

The combination~\(Z_{1,1}^{(0)} Z_{2,2}^{(0)} - Z_{1,2}^{(0)} Z_{2,1}^{(0)}\), appearing in the denominators of~\eqref{eq:greens_function_examples}, is precisely the determinant of~\eqref{eq:hzs_solution_matrix_boundary}. This is an explicit example of the fact that the quasinormal modes are the poles of the Green's functions.\footnote{The apparent pole at \(\w = k\) in~\eqref{eq:greens_function_examples} is spurious~\cite{Edalati:2010pn}, and is compensated by a zero in the numerators.} Near a given pole, which we label as \(\w_*(k)\), the Green's functions take the form
\begin{equation}
		G_{ab}(\w,k) \approx \frac{\mathcal{R}_{ab}(k)}{\w - \w_*(k)}.
\end{equation}
Ref.~\cite{Kaminski:2009dh} provides a formula for numerically computing the residue,
\begin{equation}
\label{eq:residue}
\mathcal{R}_{ab}(k) = - \lim_{\e\to0} \left . \frac{\textrm{det}\left[\mathbf{Z}(\epsilon)\right]}{\partial_{\omega} \textrm{det}\left[\mathbf{Z}(\epsilon)\right]} \left[C_{ac}(\e) + C_{ca}^*(\e)\right]  F'(\e)_{cb} \right |_{\omega_*(k)}.
\end{equation}

The reason why this works is that by construction \(\textrm{det}\left[\mathbf{Z}(0)\right]\) vanishes at \(\w = \w_*\), so we can Taylor expand to find \(\frac{\textrm{det}\left[\mathbf{Z}(\epsilon)\right]}{\partial_{\omega} \textrm{det}\left[\mathbf{Z}(\epsilon)\right]} \approx (\w - \w_*)\) (where we assume that the zero of \(\textrm{det}\left[\mathbf{Z}(0)\right]\) is first order). Hence, the right-hand side of~\eqref{eq:residue} is equivalent to
\(
	\lim_{\w \to \w_*} (\w-\w_*) G_{ab}(\w,k)	
\), which is manifestly the residue \(\mathcal{R}_{ab}\). The form~\eqref{eq:residue} is more convenient for numerical computation, since it doesn't require a priori knowledge of the quasinormal mode frequency \(\w_*\).

\chapter{Appendix to chapter~\ref{chap:probe_m5}}

\section{Entanglement entropy of the antisymmetric flow}
\label{app:antisymmetric_flow}

Differentiating the off-shell action \eqref{eq:antisymmetric_flow_hyperbolic_action} with respect to the inverse temperature \(\b\), and making use of \eqref{eq:probe_generalised_gravitational_entropy_beta} we find that the entanglement entropy may be written as
\begin{subequations}
\begin{equation}
	\see^{(1)} = \see^\mathrm{horizon} + \see^\mathrm{bulk},
\end{equation}
with horizon and bulk contributions given respectively by 
\begin{align}
	\see^\mathrm{horizon} &= \frac{1}{5} \int_0^{2\pi} \diff \t \int_0^{u_c} \diff u \le. \mathcal{L} \ri|_{v=v_H=1},
	\\
	\see^\mathrm{bulk} &= - \frac{1}{5} \int_0^{2\pi} \diff \t \int_1^\infty \diff v \int_0^{u_c} \diff u \lim_{v_H \to 1} \p_{v_H} \mathcal{L},
\end{align}
\end{subequations}
where \(\mathcal{L}\) is given by \eqref{eq:antisymmetric_flow_hyperbolic_lagrangian}, and \(\q\) is to be taken on-shell after the differentiation with respect to \(v_H\) is performed.

By performing the coordinate transformation back to flat slicing using the inverse of the map \eqref{eq:map_to_hyperbolic}, the combination of derivatives appearing in \(\mathcal{L}\) may be written as (for \(v_H = 1\))
\begin{equation}
	4 + \frac{1}{f(v)} (\p_\t \q)^2 + f(v) (\p_v \q)^2 + \frac{1}{v^2} (\p_u \q)^2 = 4 + z^2 \q'(z)^2 = 4 \frac{D(\q)^2 + \sin^6 \q}{\le(
		D(\q) \cos \q + \sin^4 \q
	\ri)^2},	
\end{equation}
where we have made use of the BPS condition \eqref{eq:flow_bps}. This simplifies the integrands slightly, so that
\begin{subequations}
\begin{align}
	\see^\mathrm{horizon} &= \frac{N_5^2}{5} \int_0^{u_c} \diff u 
	\frac{
		D(\q)^2 + \sin^6 \q
	}{
		D(\q) \cos \q + \sin^4 \q
	},
	\\
	\see^\mathrm{brane} &= \frac{N_5^2}{20\pi} \int_0^{2\pi} \diff \t \int_1^\infty \diff v \int_0^{u_c} \diff u \frac{1}{v^3}\le[
		(\p_v \q)^2 - \frac{(\p_\t \q)^2}{(v^2 - 1)^2}
	\ri] \le[
		D(\q) \cos \q + \sin^4 \q
	\ri].
\end{align}
\end{subequations}
We now change integration variables back from \((\t,v,u)\) to the flat slicing coordinates \((x^0,x,z)\). Once more making use of the BPS condition \eqref{eq:flow_bps}, we find that the integrals may be written as
\begin{subequations}
	\label{eq:antisymmetric_flow_ee_intermediate}
	\begin{align}
	\see^\mathrm{horizon} &= \frac{2N_5^2}{5} \int_\e^\ell \diff z \frac{\ell}{z \sqrt{\ell^2 - z^2}}
	\frac{
		D(\q)^2 + \sin^6 \q
	}{
		D(\q) \cos \q + \sin^4 \q
	},
	\\
	\see^\mathrm{brane} &= \frac{4 N_5^2}{5\pi} \int \diff x^0 \diff x \diff z \frac{1}{z^5 v^2 (v^2 - 1)}\le[
		(v^2 - 1)^2 (\p_v z)^2 - (\p_\t z)^2
	\ri]
	\nonumber \\ &\hspace{6cm}
	\times \frac{
		\le( D(\q) \sin\q - \cos\q \sin^3\q \ri)^2
	}{
		D(\q) \cos\q + \sin^4 \q
	}.
	\label{eq:antisymmetric_flow_ee_intermediate_bulk}
	\end{align}
\end{subequations}
In Euclidean signature, the image of the inverse of the map \eqref{eq:map_to_hyperbolic} is all of local \ads, rather than the region \(x_0^2 + x^2 + \r^2 + z^2 \leq \ell^2\) as is the case in Lorentzian signature. The integration region in the bulk contribution \eqref{eq:antisymmetric_flow_ee_intermediate_bulk} is therefore the entirety of local \ads, with the cutoff region at~\(z<\e\) excised.

The derivatives of \(z\) with respect to hyperbolic slicing coordinates may be written as 
\begin{align}
	\p_\t z = \frac{x^0 z}{\ell},
	\quad
	(v^2 - 1)\p_v z = \frac{1}{2 \ell z} \frac{
		z^4 - \le[x_0^2 + (\ell - x)^2\ri]\le[x_0^2 + (\ell + x)^2\ri]
	}{
		\sqrt{(\ell^2 - x_0^2 - x^2 - z^2) + 4\ell^2 (x_0^2 + z^2)}
	}.
\end{align}
Substituting these into the bulk integral in \eqref{eq:antisymmetric_flow_ee_intermediate} yields the final form \eqref{eq:antisymmetric_flow_ee}.

\section{Entanglement entropy for M5-branes wrapping \sph[3] \(\subset\) \ads[7]}
\label{app:spike_entanglement}

In the hyperbolic slicing of \ads~\eqref{eq:hyperbolic_slicing}, parameterising the M5-brane by  \((\t,v,u,\f_1,\f_2,\f_3)\) the solution~\eqref{eq:spike_solution} becomes
\begin{equation} \label{eq:spike_solution_hyperbolic}
    \sin\f_0 \equiv \Phi= \frac{\kappa}{ v \sinh u \sqrt{1 + \mt \ell^2 \Omega^{-2} }},
    \quad
    \Omega = v \cosh u + \sqrt{v^2 - 1} \cos \t,
\end{equation}
where \(\kappa^2 \equiv N_2/2N_5\).

We again split the entanglement entropy into horizon and bulk contributions, \(\see^{(1)} = \see^\mathrm{horizon} + \see^\mathrm{bulk}\), with
\begin{subequations} \label{eq:spike_ee_split}
\begin{align}
	\see^\mathrm{horizon} &= \frac{1}{5} \int_0^{2\pi} \diff \t \int_{u_\mathrm{min}}^{u_c} \diff u \le. \mathcal{L} \ri|_{v=v_H=1},
	\\
	\see^\mathrm{bulk} &= - \frac{1}{5} \int_0^{2\pi} \diff \t \int_1^\infty \diff v \int_{u_\mathrm{min}}^{u_c} \diff u \lim_{v_H \to 1} \p_{v_H} \mathcal{L},
\end{align}
\end{subequations}
where \(\mathcal{L}\) is given by \eqref{eq:symmetric_off_shell_lagrangian}, \(u_\mathrm{min}\) is determined by the requirement that \(\F \leq 1\), and \(\F\) is to be taken on-shell only after the differentiation with respect to \(v_H\) is performed.

The solution~\eqref{eq:spike_solution_hyperbolic} satisfies,
\begin{equation}
    1 - \F^2  + v^2 \sinh^2 u \, \hat{g}^{ab} \p_a \F \p_b \F
    =
    1 + \frac{\F^6 v^6 \sinh^6 u}{\kappa^4},
\end{equation}
which simplifies the integrands appearing in~\eqref{eq:spike_entanglement_integral}
\begin{subequations}
\begin{align}
    \see^\mathrm{horizon} &= \frac{8 N_5 N_2}{5} \int_{u_\mathrm{min}}^\infty \diff u
    \le. \frac{1}{\sqrt{1 - \F^2} }
    \le(
        1 + \frac{\F^6 v^6 \sinh^6 u}{\kappa^4}
    \ri) \ri|_{v=1},
    \\
    \see^\mathrm{bulk} &= \frac{8 N_5^2}{5 \pi} \int \diff \t \diff v \diff u \frac{1}{\sqrt{1 - \F^2}} \le\{
        \frac{\kappa^2 \sinh^2 u}{v} \le[
            (\p_v \F)^2 - \frac{(\p_\t \F)^2}{(v^2 - 1)^2}
        \ri]
        + 6 \sinh^4 u \F^3 \p_v \F
    \ri\}.
\end{align}
\end{subequations}

As for the antisymmetric flow solution, it will be convenient to perform a coordinate transformation back to flat slicing. We could use the chain rule to transform the derivatives appearing in the bulk integral, but it is simpler to note that the solution~\eqref{eq:spike_solution_hyperbolic} satisfies 
\begin{subequations}
\begin{align}
    \p_\t \F &= - \frac{\kappa \sqrt{v^2 - 1} \sin \t}{\sqrt{\mt} \ell v \sinh u} \le[
        1 - \le(
            \frac{\F v \sinh u}{\kappa}
        \ri)^2
    \ri]^{3/2},
    \\
    \p_v \F &= \frac{\kappa \le(
        \sqrt{v^2 - 1} \cosh u + v \cos \t
    \ri)}{
        \sqrt{\mt} \ell v \sqrt{v^2 - 1} \sinh u
    } \le[
        1 - \le(
            \frac{\F v \sinh u}{\kappa}
        \ri)^2
    \ri]^{3/2} - \frac{\F}{v}.
\end{align}
\end{subequations}
In flat slicing, the solution becomes
\begin{equation} \label{eq:phi_flat_slicing}
    \F = \frac{\kappa z}{
        \sqrt{(1 + \mt z^2)x^2 + \kappa^2 z^2}.
    }
\end{equation}
Performing the inverse of the transformation~\eqref{eq:map_to_hyperbolic} and substituting the solution~\eqref{eq:phi_flat_slicing}, we find that the derivatives become
\begin{subequations}
    \begin{align}
        \p_\t \F &= - \frac{ \kappa \mt x^0 z^3 }{
            \ell (1 + \mt z^2) \sqrt{(1 + \mt z^2) x^2 + \kappa^2 z^2}
        },
        \\
        \p_v \F &= -\frac{\kappa z}{v} \frac{
            \le( \ell^2 - x_0^2 - x^2 - \r^2 \ri)^2
            + 4 \ell^2 x_0^2 - z^4
            - \frac{2 \kappa^2 z^4}{\r^2} \le(
                \ell^2 - x_0^2 - x^2 - \r^2 - z^2
            \ri)
        }{
            \sqrt{(1 + \mt z^2)x^2 + \kappa^2 z^2}
            ( 1 + \mt z^2 )
            \le[
                4 \ell^2 x_0^2 + (\ell^2 - x_0^2 - x^2 - \r^2 - z^2)^2
            \ri]
        }.
    \end{align}
\end{subequations}
Plugging these into the integrals in~\eqref{eq:spike_ee_split} and performing the coordinate transformation \((\t,v,u) \to (x^0,x,z)\), we obtain~\eqref{eq:spike_entanglement_integral}.

\printbibliography

\end{document}